\documentclass [a4paper, twoside, onecolumn]{book}

\usepackage{a4wide, amsmath, amsfonts, amssymb, geometry, fancyhdr}
\usepackage{fancyhdr, dsfont}
\usepackage[latin1]{inputenc}
\usepackage[pdftex]{graphicx}
\usepackage{float}
\usepackage{pstricks}
\usepackage{color}

\def\beq{\begin{equation}}
\def\eeq{\end{equation}}

\newcommand{\vecb}{\left(\begin{array}{c}}
\newcommand{\vece}{\end{array}\right)}
\newcommand{\ccb}{\left(\begin{array}{cc}}
\newcommand{\cce}{\end{array}\right)}
\newcommand{\cccb}{\left(\begin{array}{ccc}}
\newcommand{\ccce}{\end{array}\right)}
\newcommand{\ccccb}{\left(\begin{array}{cccc}}
\newcommand{\cccce}{\end{array}\right)}
\newcommand{\cccccb}{\left(\begin{array}{ccccc}}
\newcommand{\ccccce}{\end{array}\right)}

\newcommand{\vph}{\varphi}
\newcommand{\vth}{\vartheta}
\newcommand{\ve}{\vec}
\newcommand{\pa}{\partial}

\newcommand{\al}{\alpha}
\newcommand{\be}{\beta}
\newcommand{\ga}{\gamma}
\newcommand{\de}{\delta}
\newcommand{\ep}{\epsilon}
\newcommand{\vep}{\varepsilon}
\newcommand{\si}{\sigma}
\newcommand{\la}{\lambda}
\newcommand{\ka}{\kappa}
\newcommand{\om}{\omega}
\newcommand{\Ga}{\Gamma}

\newcommand{\Si}{\Sigma}
\newcommand{\La}{\Lambda}
\newcommand{\Om}{\Omega}

\newcommand{\Psib}{\overline{\Psi}}
\newcommand{\dd}{\mathrm{d}}
\newcommand{\mto}{\rightarrow}
\newcommand{\te}{\textrm}
\newcommand{\inn}{ \ {\cal 3} \ } 
\newcommand{\eq}{ \ \ = \ \ }
\newcommand{\so}{ \ \ \ \Longrightarrow \ \ \ }
\newcommand{\co}{\ , \ \ \ \ \ \ }

\newcommand{\thb}{\bar{\theta}}
\newcommand{\vac}{|\te{vac} \rangle}

\newcommand{\da}{\dot{\alpha}}
\newcommand{\db}{\dot{\beta}}
\newcommand{\dg}{\dot{\gamma}}
\newcommand{\dl}{\dot{\lambda}}
\newcommand{\g}{\gamma}
\newcommand{\s}{\sigma}
\newcommand{\bs}{\bar{\sigma}}
\newcommand{\mQ}{\mathcal{Q}}
\newcommand{\bmQ}{\bar{\mathcal{Q}}}
\newcommand{\bt}{\bar{\theta}}
\newcommand{\bc}{\bar{\chi}}
\newcommand{\bet}{\bar{\eta}}
\newcommand{\mL}{\mathcal{L}}
\newcommand{\vp}{\varphi}
\newcommand{\bp}{\bar{\psi}}
\newcommand{\bl}{\bar{\lambda}}

\newcommand{\bmD}{\bar{\mathcal{D}}}
\newcommand{\bpa}{\bar{\partial}}
\newcommand{\bep}{\bar{\epsilon}}

\begin{document}

\title{\begin{flushright} \vspace{-7cm}
{\large DAMTP-2010-90}\vspace{-5mm} \\
{\large IC/2010/xx}\vspace{-5mm} \\
{\large  MPP-2010-143}
\end{flushright} \vspace{5cm} \Huge \textbf{Cambridge Lectures on Supersymmetry and \\ Extra Dimensions}}
\author{Lectures by: Fernando Quevedo$^{1,2}$,    Notes by: Sven Krippendorf$^{1}$, Oliver Schlotterer$^{3}$ \vspace{1cm}\\
$^{1}$ { \it DAMTP, University of Cambridge, Wilberforce Road, Cambridge, CB3 0WA, UK.}\\
$^2$ {\it ICTP, Strada Costiera 11, Trieste  34151, Italy.}\\
$^{3}$ {\it Max-Planck-Institut f\"ur Physik, F\"ohringer Ring 6, 80805 M\"unchen, Germany.}\\
\vspace{15pt}\\}

\date{\today}

\maketitle \setcounter{chapter}{0}
\tableofcontents
\newpage

\begin{center}
{\bf Guide through the notes}
\end{center}

\noindent
These lectures on supersymmetry and extra dimensions are aimed at finishing undergraduate and beginning postgraduate students with a background in quantum field theory and group theory. Basic knowledge in general relativity might be advantageous for the discussion of extra dimensions.

\noindent
This course was taught as a 24+1 lecture course in Part III of the Mathematical Tripos in recent years. The first six chapters give an introduction to supersymmetry in four spacetime dimensions, they fill about two thirds of the lecture notes and are in principle self-contained. The remaining two chapters are devoted to extra spacetime dimensions which are in the end combined with the concept of supersymmetry. Understanding the interplay between supersymmetry and extra dimensions is essential for modern research areas in theoretical and mathematical physics such as superstring theory.

\noindent
Videos from the course lectured in 2006 can be found online at:
\begin{center}
\texttt{http://www.sms.cam.ac.uk/collection/659537}
\end{center}

\noindent There are a lot of other books, lecture notes and reviews on supersymmetry, supergravity and extra dimensions, some of which are listed in the bibliography \cite{all}.

\vspace{3cm}
\begin{center}
{\bf Acknowledgments}
\end{center}
\noindent
We are very grateful to Ben Allanach for enumerous suggestions to these notes from teaching this course during the last year, Joe Conlon for elaborating most solutions during early years, Shehu AbdusSalam for filming and Bj\"orn Ha\"sler for editing and publishing the videos on the web. We wish to thank lots of students from Part III 2005 to 2010 and from various other places for pointing out various typos and for making valuable suggestions for improvement.\vspace{2cm} \\

\chapter{Physical Motivation for supersymmetry and extra dimensions}
\label{sec:PhysicalMotivationForSupersymmetryAndExtraDimensions}

Let us start with a simple question in high energy physics:
What do we know so far about the universe we live in? 
\section{Basic Theory: QFT}

Microscopically we have {\em quantum mechanics} and {\em special relativity} as our two basic theories.

\noindent
The framework to make these two theories consistent with each other is {\em quantum field theory (QFT)}. In this theory the fundamental entities are quantum fields. Their excitations correspond to the physically observable elementary particles which are the basic constituents of matter as well as the mediators of all the known interactions. Therefore, fields have particle-like character. Particles can be classified in two general classes: bosons (spin $s = n \ \in \ \mathbb Z$) and fermions ($s = n + \frac{1}{2} \in \mathbb Z + \frac{1}{2}$). Bosons and fermions have very different physical behaviour. The main difference is that fermions can be shown to satisfy the \textsc{Pauli} ''exclusion principle'' , which states that two identical fermions cannot occupy the same quantum state, and therefore explaining the vast diversity of atoms.

\noindent
All elementary matter particles: the leptons (including electrons and neutrinos) and quarks (that make protons, neutrons and all other hadrons) are fermions. Bosons on the other hand include the photon (particle of light and mediator of electromagnetic interaction), and the mediators of all the other interactions. They are not constrained by the Pauli principle and therefore have very different physical properties as can be appreciated in a laser for instance.
 As we will see, {\em supersymmetry} is a symmetry that unifies bosons and fermions despite all their differences.

\section{Basic Principle: Symmetry}
\label{sec:SymmetryAsOurBasicPrinciple}

If QFT is the basic framework to study elementary processes, the basic tool to learn about these processes is the concept of {\em symmetry}.

\noindent
A symmetry is a transformation that can be made to a physical system leaving the physical observables unchanged. Throughout the history of science symmetry has played a very important role to better understand nature.
Let us try to classify the different classes of symmetries and their physical implications.
\subsection{Classes of symmetries}

There are several ways to classify symmetries.
Symmetries can be discrete or continuous. They can also be global or local.
For elementary particles, we can define two general classes of symmetries:
\begin{itemize}
\item {\it Spacetime symmetries:} These symmetries correspond to
 transformations on a field theory acting explicitly on the spacetime
 coordinates,
	\[x^\mu \ \ \mapsto \ \ x'^{\mu}\left(x^\nu\right) \co \mu,\nu=0,1,2,3 \ .
\]
Examples are rotations, translations and, more generally, {\em Lorentz- and Poincaré transformations} defining the global symmetries of special relativity as well as {\em general coordinate transformations} that are the local symmetries that define {\em general relativity}.
\item {\it Internal symmetries:} These are symmetries that correspond to transformations of the different fields in a field theory,
	\[\Phi^a(x)  \ \ \mapsto \ \  M^a \, _b \, \Phi^b(x) \ .
\]
Roman indices $a,b$ label the corresponding fields. If $M^a \, _b$ is constant
then the symmetry is a {\it global symmetry}; in case of spacetime dependent $M^a \, _b(x)$ the symmetry is called a {\it local symmetry}.
 \end{itemize}

\subsection{Importance of symmetries}

Symmetries are  important for various reasons:
\begin{itemize}
\item {\it Labelling and classifying particles:} Symmetries label and classify particles according to the different conserved quantum numbers identified by the spacetime and internal symmetries (mass, spin, charge, colour, etc.). 
This is a consequence of Noether's theorem that states that  each continuous symmetry implies a conserved quantity. In this regard symmetries actually ``define'' an elementary particle according to the behaviour of the corresponding field with respect to the corresponding symmetry. This property was used to classify particles not only as fermions and bosons but also to group them in multiplets with respect to approximate internal symmetries as in the eightfold way that was at the origin of the quark model of strong interactions.
\item Symmetries determine the {\em interactions} among particles by means of the {\em gauge principle}. By promoting a global symmetry to a local symmetry  gauge fields (bosons) and interactions have to be introduced accordingly defining the interactions among particles with gauge fields as mediators of interactions. As an illustration, consider the Lagrangian
    \[{\cal L} \eq \pa_{\mu} \phi \, \pa^{\mu} \phi^{*} \ - \ V(\phi, \phi^{*})
\]
which is invariant under rotation in the complex plane
    \[\phi \ \ \mapsto \ \ \exp(i\al) \, \phi \ ,
\]
as long as $\al$ is a constant (global symmetry). If $\al = \al(x)$, the kinetic term is no longer invariant:
    \[\pa_{\mu} \phi \ \ \mapsto \ \ \exp(i\al) \, \bigl(\pa_{\mu} \phi \, + \, i(\pa_{\mu} \al) \phi \bigr) \,.
\]
However, the covariant derivative $D_{\mu}$, defined as
    \[D_{\mu} \phi \ \ := \ \ \pa_{\mu} \phi \ + \ iA_{\mu} \, \phi \ ,
\]
transforms like $\phi$ itself, if the gauge - potential $A_{\mu}$ transforms to $A_{\mu} - \pa_{\mu} \al$:
    \[D_{\mu} \ \ \mapsto \ \ \exp(i\al) \, \bigl(\pa_{\mu} \phi \, + \, i(\pa_{\mu} \al) \phi  \ + \ i(A_{\mu} - \pa_{\mu} \al) \, \phi \bigr) \eq \exp(i\al) \, D_{\mu} \phi \ ,
\]
so rewrite the Lagrangian to ensure gauge - invariance:
    \[{\cal L} \eq D_{\mu} \phi \, D^{\mu} \phi^{*} \ - \ V(\phi, \phi^{*})\, .
\]
The scalar field $\phi$ couples to the gauge - field $A_{\mu}$ via $A_{\mu} \phi A^{\mu} \phi$, similarly, the Dirac Lagrangian
    \[{\cal L} \eq \Psib  \, \ga^{\mu} \, D_{\mu} \Psi
\]
has an interaction term $\Psib A_{\mu} \Psi$. This interaction provides the three point vertex that describes interactions of electrons and photons and illustrate how photons mediate the electromagnetic interactions.
\item Symmetries can hide or be {\em spontaneously
  broken:} Consider the potential $V(\phi , \phi^{*})$ in the
  scalar field Lagrangian above.

\begin{figure}[ht]
    \centering
        \includegraphics[width=0.70\textwidth]{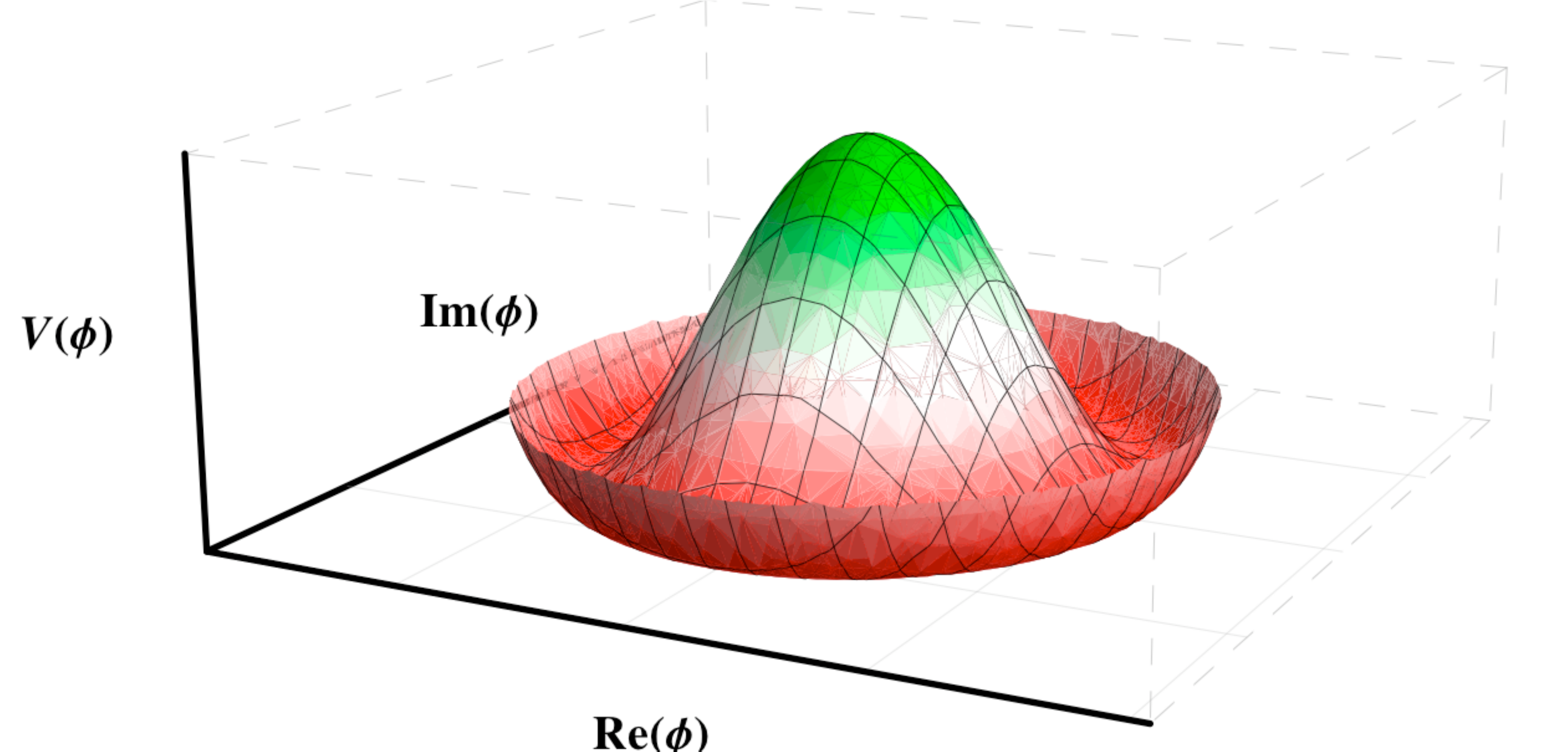}
        \caption{\footnotesize{The Mexican hat potential for $V=\Bigl( a \ - \ b \, | \phi |^2   \Bigr)^2$ with
 $a,b\geq 0.$}}
\end{figure}
If $V(\phi,\phi^*)= V(\vert\phi\vert^2)$, then it is
 symmetric for $\phi \mapsto \exp(i\al) \phi$.
If the potential is of the type 
	\[ V \eq a \, | \phi |^2 \ + \
 b \, | \phi | ^4 \co a,b\geq 0 \ ,
\]
 then the minimum is at $\langle \phi \rangle =0$ (here
 $\langle\phi\rangle\equiv\langle 0\vert \phi\vert 0\rangle$ denotes
 the {\em vacuum expectation value (vev)} of the field $\phi$). The vacuum
 state is then also symmetric under the symmetry since the origin is
 invariant.
However if the potential is of the form 
	\[V \eq \Bigl( a \ - \ b \, | \phi |^2   \Bigr)^2
\co a,b\geq 0 \ ,
\]
the symmetry of $V$ is lost in the ground state $\langle \phi \rangle \neq 0$. 
The existence of hidden symmetries is important for at least two
reasons:
\begin{itemize}
\item [(i)] This is a natural way to introduce an energy scale in
the system, determined by the nonvanishing vev. In particular, we will see that for the standard model $M_{\te{ew}} \approx 10^3$ GeV, defines the basic scale of mass for the particles of the standard model, the electroweak gauge
bosons and the matter fields, through their Yukawa couplings, obtain their mass from this effect. 
\item [(ii)] The existence of hidden symmetries implies that the
fundamental symmetries of nature may be huge despite the fact that we observe a
limited amount of symmetry. This is because the only manifest symmetries we can 
observe are the symmetries of the vacuum we live in and not those of the
full underlying theory. This opens-up an essentially unlimited resource to consider physical theories with  an indefinite number of symmetries even though they are
not explicitly realised in nature. The standard model is the typical
example and supersymmetry and theories of extra dimensions are further examples.
\end{itemize}
\end{itemize}

\section{Basic example: The Standard Model}

The concrete example is the particular QFT known as {\em The Standard Model} which describes all known particles and interactions in 4 dimensional spacetime.
\begin{itemize}
\item{\it Matter particles}: Quarks and leptons. They come in three
  identical families differing only by their mass. Only the first
  family participate in making the atoms and all composite matter we
  observe. Quarks and leptons are fermions of spin $\frac{\hbar}{2}$ and
  therefore satisfy Pauli's exclusion principle. Leptons include the
  electron $e^-$, muon $\mu$ and $\tau$ as well as the three
  neutrinos. Quarks come in three colours and are the building blocks
  of strongly interacting particles such as the proton and neutron in
  the atoms.
\item{\it Interaction particles}: The three non-gravitational
  interactions (strong, weak and electromagnetic) are described by
  a gauge theory based on an internal symmetry:
    \[G_{\te{SM}} \eq \underbrace{SU(3)_{C}}_{\te{strong}} \ \otimes \ \underbrace{SU(2)_{L} \otimes U(1)}_{\te{electroweak}}
\]
Here $SU(3)_C$ refers to quantum chromodynamics part of the standard
model describing the strong interactions, the subindex $C$ refers to
colour. Also $SU(2)_L\otimes U(1)$ refers to the electroweak part of
the standard model, describing the electromagnetic and weak
interactions. The subindex $L$ in $SU(2)_L$ refers to the fact that
the Standard Model does not preserve parity and differentiates between
left-handed and right-handed particles. In the Standard Model only
left-handed particles transform non-trivially under $SU(2)_L$. The
gauge particles have all spin $s=1 \hbar$ and mediate each of the
three forces: photons ($\gamma$) for $U(1)$ electromagnetism, gluons
for $SU(3)_C$ of strong interactions, and the massive $W^{\pm}$ and
$Z$ for the weak interactions.
\item {\it The Higgs particle}: This is the spin $s=0$ particle that
  has a potential of the ``Mexican hat'' shape (see figure ) and is responsible for the
  breaking of the Standard Model gauge symmetry:
    \[SU(2)_{L} \ \otimes \ U(1) \ \ \ \stackrel{\langle \phi \rangle \approx 10^{3} \te{GeV}}{\longrightarrow} \ \ \ U_{EM}(1)
\]
For the gauge particles this is the {\em Higgs effect}, that explains how the $W^{\pm}$ and $Z$
 particles get a mass and therefore the weak interactions are short
 range. This is also the source of masses for all quarks and leptons.
\item
{\it Gravity}:  Gravity can also be understood as a gauge theory in the sense that the global
spacetime symmetries of special relativity, defined by the Poincar\'e group, when made local give 
rise to the general coordinate transformations of general relativity. However the corresponding gauge particle, the graviton,
corresponds to a massless particle of spin $s=2\hbar$ and there is not a QFT that describes these particles to arbitrarily small distances. Therefore, contrary to gauge theories which are
consistent quantum mechanical theories, the Standard Model only describes gravity at
the classical level. 
\end{itemize}

\section{Problems of the Standard Model}
\label{sec:ProblemsOfTheStandardModel}

The Standard Model is one of the cornerstones of all science and one of the great triumphs of the XX century. It has been carefully experimentally verified in many ways, especially during the past 20 years, but there are many questions it cannot answer:
\begin{itemize}
\item {\it Quantum Gravity}: The Standard Model describes three of the four fundamental interactions at the quantum level and therefore microscopically. However, gravity is only treated classically and any quantum discussion of gravity has to be considered as an effective field theory valid at scales smaller than the Planck scale ($M_{\te{pl}} = \sqrt{\frac{Gh}{c^{3}}} \approx 10^{19} \te{GeV}$). At this scale quantum effects of gravity have to be included and then Einstein theory has the problem of being non-renormalizable and therefore it cannot provide proper answers to observables beyond this scale.
\item Why $G_{\te{SM}} = SU(3) \otimes SU(2) \otimes U(1)$? Why there are four interactions and three families of fermions? Why 3 + 1 spacetime - dimensions? Why there are some 20 parameters (masses and couplings between particles) in the Standard Model for which their values are only determined to fit experiment without any theoretical understanding of these values?
\item {\it Confinement}: Why quarks can only exist confined in hadrons such as protons and neutrons? The fact that the strong interactions are asymptotically free (meaning that the value of the coupling increases with decreasing energy) indicates that this is due to the fact that at the relatively low energies we can explore the strong interactions are so strong that do not allow quarks to separate. This is an issue about our ignorance to treat strong coupling field theories which are not well understood because standard (\textsc{Feynman} diagrams) perturbation theory cannot be used.
\item {\it The hierarchy problem}: Why there are totally different energy scales
    \[M_{\te{ew}} \ \ \approx \ \ 10^{2} \te{GeV} \co M_{\te{pl}} \eq \sqrt{\frac{Gh}{c^{3}}} \ \ \approx \ \ 10^{19} \te{GeV} \so \frac{M_{\te{ew}}}{M_{\te{pl}}} \ \ \approx \ \ 10^{-15}
\]
This problem has two parts. First why these fundamental scales are so different which may not look that serious. The second part refers to a naturalness issue. A fine tuning of many orders of magnitude has to be performed order by order in perturbation theory in order to avoid the electroweak scale $M_{\te{ew}}$ to take the value of the ''cutoff'' scale which can be taken to be $M_{\te{pl}}$.
\item {\it The strong CP problem}: There is a coupling in the Standard Model of the form $\theta F^{\mu\nu}\tilde F_{\mu\nu}$ where $\theta$ is a parameter, $F^{\mu\nu}$ refers to the field strength of quantum chromodynamics (QCD) and $\tilde{F_{\mu\nu}}=\epsilon_{\mu\nu\rho\sigma} F^{\rho\sigma}$. This term breaks the symmetry $CP$ (charge conjugation followed by parity). The problem refers to the fact that the parameter $\theta$ is unnaturally  small $\theta<10^{-8}$. A parameter can be made naturally small by the \textsc{t'Hooft} ''naturalness criterion'' in which a parameter is naturally small if setting it to zero implies there is a symmetry protecting its value. For this problem, there is a concrete proposal due to \textsc{Peccei} and \textsc{Quinn} in which, adding a new particle,
the {\em axion} $a$, with coupling  $a F^{\mu\nu}\tilde F_{\mu\nu}$, then the corresponding Lagrangian will be symmetric under $a\rightarrow a+c$ which is the {\em PQ symmetry}. This solves the strong CP problem because non-perturbative QCD effects introduce a potential for $a$ with minimum at $a=0$ which would correspond to $\theta=0$.
\item {\it The cosmological constant problem}: Observations about the accelerated expansion of the universe indicate that the cosmological constant interpreted as the energy of the vacuum is near zero, $\La \approx 10^{-120} M_{\te{pl}}^{4}$
    \[\frac{M_{\La}}{M_{\te{ew}}} \ \ \approx \ \ 10^{-15}
\]
This is probably the biggest puzzle in theoretical physics. The problem, similar to the hierarchy problem, is the issue of naturalness. There are many contributions within the Standard Model to the value of the vacuum energy and they all have to cancel to 60-120 orders of magnitude (since the relevant quantity is $M_{La}^4$, in order to keep the cosmological constant small after quantum corrections for vacuum fluctuations are taken into account.
\end{itemize}
All of this indicates that the Standard Model is not the fundamental theory of the universe but only an effective theory describing the fundamental one at low energies. We need to find an extension that could solve some or all of the problems mentioned above in order to generalize the Standard Model. 

\noindent
In order to go beyond the Standard Model we can follow several avenues.
\begin{itemize}
\item Experiments: This is the traditional way of making progress in science. We need experiments to explore energies above the currently attainable scales and discover new particles and underlying principles that generalize the Standard Model. This avenue is presently important due to the current explorations of the Large Hadron Collider (LHC) at CERN, Geneva. This experiment is exploring physics above the weak scale with a center of mass energy of up to $14$ TeV and may discover the last remaining
particle of the Standard Model,  the {\em Higgs particle}, as well as new physics beyond the Standard Model. Until the present day in September 2010, the Standard Model has been tested even more accurately with collisions at  about $7$ TeV center of mass energy, and new physics can be potentially found in the next months already due to increasing luminosity. But still, exploring energies closer to the Planck scale $M_{\te{pl}} \approx 10^{18}$ GeV is out of the reach for many years to come.
\item Add new particles and/or interactions. This {\it ad hoc} technique is not well guided but it is possible to follow if by doing this we are addressing some of the questions mentioned before.
\item More general symmetries. As we understand by now the power of symmetries in the foundation of the Standard Model, it is then natural to use this as a guide and try to generalize it by adding more symmetries. These can be of the two types mentioned before:
\begin{itemize}
\item [(i)] More general internal symmetries leads to consider {\em grand unified theories (GUTs)} in which the symmetries of the Standard Model are themselves the result of the breaking of yet a larger symmetry group:
    \[G_{\te{GUT}} \ \ \ \stackrel{M \approx 10^{17} \te{GeV}}{\longrightarrow} \ \ \ G_{\te{SM}} \ \ \ \stackrel{M \approx 10^{2} \te{GeV}}{\longrightarrow} \ \ \ SU(3) \ \otimes \ U(1) \ 
\]
This proposal is very elegant because it unifies, in one single symmetry, the three gauge interactions of the Standard Model. It leaves unanswered most of the open questions above, except for the fact that it reduces the number of independent parameters due to the fact that there is only one gauge coupling at large energies. This is expected to ``run''  at low energies and give rise to the three different couplings of the Standard Model (one corresponding to each group factor). Unfortunately, with our present precision understanding of the gauge couplings and spectrum of the Standard Model, the running of the three gauge couplings does \textbf{not} unify at a single coupling at higher energies but they cross each other at different energies.
\item [(ii)] More general  spacetime symmetries  1: {\em  Extra spacetime dimensions}. If we  add more dimensions to spacetime, therefore the Poincaré symmetries of the Standard Model and more generally the general coordinate transformations of general relativity, get substantially enhanced. This is the well known {\em Kaluza Klein theory} in which our observation of a 4 dimensional universe is only due to the fact that we have limitations about ``seeing'' other dimensions of spacetime that may be hidden to our experiments.

\noindent
In recent years this has been extended to the {\em brane world scenario} in which our 4 dimensional universe is only a brane or surface inside a larger dimensional universe. These ideas approach very few of the problems of the Standard Model. They may lead to a different perspective of the hierarchy problem and also about the possibility to unify internal and spacetime symmetries.

\item [(iii)] More general spacetime symmetries 2:  {\em supersymmetry}. If we do keep to the standard four spacetime dimensions, there is another way to enhance the spacetime symmetries.  Supersymmetry is a 
symmetry under the exchange of  bosons and fermions. As we will see, it is a spacetime symmetry, despite the fact that it is seen only as a transformation that exchanges bosons and fermions. Supersymmetry solves the naturalness issue (the most important part)
of the hierarchy problem due to cancellations between the contributions of bosons and fermions to the electroweak scale, defined by the Higgs mass. Combined with the GUT idea, it solves the unification of the three gauge couplings at one single point at larger energies. Supersymmetry also provides the best example for dark matter candidates. Moreover, it provides well defined QFTs in which issues of strong coupling can be better studied than in the non-supersymmetric models.
\end{itemize}
\item Beyond QFT: Supersymmetry and extra dimensions do not address the most fundamental problem mentioned above, that is the problem of quantizing gravity. For this purpose we may have to look for a generalisation 
of QFT to a more general framework. Presently the best hope is string theory which goes beyond our basic framework of QFT. It so happens that for its consistency, string theory requires both supersymmetry and extra dimensions also. This gives a further motivation to study these two areas which are the subject of this course.
\end{itemize}

\chapter{Supersymmetry algebra and representations}
\label{sec:SupersymmetryAlgebraAndRepresentation}

\section{Poincaré symmetry and spinors}
\label{sec:PoincareSymmetryAndSpinors}

The Poincaré group corresponds to the basic symmetries of special relativity, it acts on spacetime coordinates $x^{\mu}$ as follows:
    \[x^{\mu} \ \ \mapsto \ \ x'^{\mu} \eq \underbrace{\La^{\mu}\,_{\nu}}_{\te{Lorentz}} x^{\nu} \ + \ \underbrace{a^{\mu}}_{\te{translation}}
\]
Lorentz transformations leave the metric tensor $\eta_{\mu \nu} = \te{diag}(1, \ -1 , \ -1 , \ -1)$ invariant:
    \[\La^{T} \, \eta \, \La \eq \eta
\]
They can be separated between those that are connected to the identity and those that are not (like parity for which $\Lambda = \te{diag}(1 , \ -1, \ -1, \ -1)$). We will mostly discuss those $\La$ connected to identity, i.e. the {\em proper orthochronous group} $SO(3,1)^{\uparrow}$. Generators for the Poincaré group are the $M^{\mu \nu}$, $P^{\si}$ with algebra
\begin{align*}
\Bigl[P^{\mu} \ , \ P^{\nu} \Bigr] \ \ &= \ \ 0 \\
\Bigl[M^{\mu \nu} \ , \ P^{\si} \Bigr] \ \ &= \ \ i \, \bigl(P^{\mu} \, \eta^{\nu \si} \ - \ P^{\nu} \, \eta^{\mu \si} \bigr)\\
\Bigl[M^{\mu \nu} \ , \ M^{\rho \si} \Bigr] \ \ &= \ \ i\, \bigl(M^{\mu \si} \, \eta^{\nu \rho} \ + \ M^{\nu \rho} \, \eta^{\mu \si} \ - \ M^{\mu \rho} \, \eta^{\nu \si} \ - \ M^{\nu \si} \, \eta^{\mu \rho} \bigr)
\end{align*}
A 4-dimensional matrix representation for the $M^{\mu \nu}$ is
    \[(M^{\rho \si})^{\mu}\,_{\nu} \eq i\, \bigl(\eta^{\mu \nu} \, \de^{\rho}\,_{\nu} \ - \ \eta^{\rho \mu} \, \de^{\si}\,_{\nu} \bigr) \ .
\]

\subsection{Properties of the Lorentz group}
\label{sec:PropertiesOfLorentzGroup}

\begin{itemize}
\item{} Locally, we have a correspondence
    \[SO(3,1) \ \ \cong \ \ SU(2) \oplus SU(2) \ ,
\]
the generators $J_{i}$ of rotations and $K_{i}$ of Lorentz boosts can be expressed as
    \[J_{i} \eq \frac{1}{2} \; \ep_{ijk} \, M_{jk} \co K_{i} \eq M_{0i} \ ,
\]
and their linear combinations (which are neither hermitian nor antihermitian)
    \[A_{i} \eq \frac{1}{2} \; \bigl(J_{i} \, + \, iK_{i} \bigr) \co B_{i} \eq \frac{1}{2} \; \bigl(J_{i} \, - \, iK_{i} \bigr)
\]
satisfy $SU(2)$ commutation relations (following from $[J_{i},J_{j}] = i\ep_{ijk}J_{k}$ as well as $[J_{i},K_{j}] = i\ep_{ijk}K_{k}$ and $[K_{i},K_{j}] = -i\ep_{ijk}K_{k}$):
    \[\Bigl[ A_{i} \ , \ A_{j} \Bigr] \eq i \ep_{ijk} \, A_{k} \co \Bigl[B_{i} \ , \ B_{j} \Bigr] \eq i \ep_{ijk} \, B_{k} \co \Bigl[A_{i} \ , \ B_{j} \Bigr] \eq 0 
\]
Under parity P ($x^{0} \mapsto x^{0}$ and $\ve{x} \mapsto -\ve{x}$) we have
    \[J_{i} \ \ \mapsto \ \ J_{i} \co K_{i} \ \ \mapsto \ \ -K_{i} \so A_{i} \ \ \ \leftrightarrow \ \ \ B_{i} \ .
\]
We can interpret $\ve{J} = \ve{A} + \ve{B}$ as the physical spin.

\item{}On the other hand, there is a homeomorphism (not an isomorphism)
    \[SO(3,1) \ \ \cong \ \ SL(2,\mathbb C) \ .
\]
To see this, take a 4 vector $X$ and a corresponding $2 \times 2$ - matrix $\tilde{x}$,
    \[X \eq x_{\mu} \, e^{\mu} \eq (x_{0} \ , \ x_{1} \ , \ x_{2} \ , \ x_{3}) \co \tilde{x} \eq x_{\mu} \, \si^{\mu} \eq \ccb x_{0} + x_{3} &x_{1} - ix_{2} \\ x_{1} + ix_{2} &x_{0} - x_{3} \cce \ ,
\]
where $\si^{\mu}$ is the 4 vector of {\em Pauli matrices}
    \[\si^{\mu} \eq \left\{ \ccb 1 &0 \\ 0 &1 \cce \ , \ \ccb 0 &1 \\ 1 &0 \cce \ , \ \ccb 0 &-i \\ i &0 \cce \ , \ \ccb 1 &0 \\ 0 &-1 \cce \right\} \ .
\]
Transformations $X \mapsto \La X$ under $SO(3,1)$ leaves the square
    \[|X|^{2} \eq x_{0}^{2} \ - \ x_{1}^{2} \ - \ x_{2}^{2} \ - \ x_{3}^{2}
\]
invariant, whereas the action of $SL(2,\mathbb C)$ mapping $\tilde{x} \mapsto N\tilde{x} N^\dag$ with $N \in SL(2,\mathbb C)$ preserves the determinant
    \[\det \tilde{x} \eq x_{0}^{2} \ - \ x_{1}^{2} \ - \ x_{2}^{2} \ - \ x_{3}^{2} \ .
\]
The map between $SL(2,\mathbb C)$ is 2-1, since $N = \pm \mathds{1}$ both correspond to $\La = \mathds{1}$, but $SL(2,\mathbb C)$ has the advantage to be simply connected, so $SL(2,\mathbb C)$ is the universal covering group.
\end{itemize}

\subsection{Representations and invariant tensors of $SL(2,\mathbb C)$}
\label{sec:RepresentationsAndInvariantTensorsOfSL2C}

The basic representations of $SL(2,\mathbb C)$ are:
\begin{itemize}
\item The fundamental representation
    \[\psi'_{\al} \eq N_{\al}\,^{\be} \, \psi_{\be} \co \al,\be = 1,2
\]
The elements of this representation $\psi_\al$ are called {\em left-handed Weyl spinors}.
\item The conjugate representation
    \[\bar{\chi}'_{\dot{\al}} \eq N^{*}_{\dot{\al}}\,^{\dot{\be}} \, \bar{\chi}_{\dot{\be}} \co \dot{\al},\dot{\be} = 1,2
\]
Here $\bar{\chi}_{\dot{\be}}$ are called {\em right-handed Weyl spinors}.
\item The contravariant representations
    \[\psi'^{\al} \eq \psi^{\be} \, (N^{-1})_{\be}\,^{\al} \co \bar{\chi}'^{\dot{\al}} \eq \bar{\chi}^{\dot{\be}} \, (N^{*-1})_{\dot{\be}}\,^{\dot{\al}}
\]
\end{itemize}
The fundamental and conjugate representations are the basic representations of
$SL(2, \mathbb C)$ and the Lorentz group, giving then the importance to spinors as the basic objects of special relativity, a fact that could be missed by not realizing the connection of the Lorentz group and $SL(2, \mathbb C)$. We will see next that the contravariant representations are however not independent.

\noindent
To see this we will consider now the different ways to raise and lower indices.
\begin{itemize}
\item The metric tensor $\eta^{\mu \nu} = (\eta_{\mu \nu})^{-1}$ is invariant under $SO(3,1)$.
\item The analogy within $SL(2,\mathbb C)$ is
    \[\ep^{\al \be} \eq \ep^{\dot{\al} \dot{\be}} \eq \ccb 0 &1 \\ -1 &0 \cce \eq -\ep_{\al \be} \eq - \ep_{\dot{\al} \dot{\be}} \ ,
\]
since
    \[\ep'^{\al \be} \eq \ep^{\rho \si} \, N_{\rho}\,^{\al} \, N_{\si}\,^{\be} \eq \ep^{\al \be} \cdot \det N \eq \ep^{\al \be} \ .
\]
That is why $\ep$ is used to raise and lower indices
    \[\psi^{\al} \eq \ep^{\al \be} \, \psi_{\be} \co \bar{\chi}^{\dot{\al}} \eq \ep^{\dot{\al} \dot{\be}} \, \bar{\chi}_{\dot{\be}} \ ,
\]
so contravariant representations are not independent.
\item To handle mixed $SO(3,1)$- and $SL(2,\mathbb C)$ indices, recall that the transformed components $x_{\mu}$ should look the same, whether we transform the vector $X$ via $SO(3,1)$ or the matrix $\tilde{x} = x_{\mu} \si^{\mu}$
    \[(x_{\mu} \, \si^{\mu})_{\al \dot{\al}} \ \ \mapsto \ \ N_{\al}\,^{\be} \, (x_{\nu} \, \si^{\nu})_{\be \dot{\ga}} \, N^{*}_{\dot{\al}}\,^{\dot{\ga}} \eq \La_{\mu}\,^{\nu} \, x_{\nu} \, \si^{\mu} \ ,
\]
so the right transformation rule is
    \[(\si^{\mu})_{\al \dot{\al}} \eq N_{\al}\,^{\be} \, (\si^{\nu})_{\be \dot{\ga}} \, (\La^{-1})^{\mu}\,_{\nu} \, N^{*}_{\dot{\al}}\,^{\dot{\ga}} \ .
\]
Similar relations hold for
    \[(\bar{\si}^{\mu})^{\dot{\al}\al} \ \ := \ \ \ep^{\al \be} \, \ep^{\dot{\al} \dot{\be}} \, (\si^{\mu})_{\be \dot{\be}} \eq (\mathds{1} , \ -\ve{\si}) \ .
\]
\end{itemize}

\paragraph{Exercise 2.1:} {\it Show that the $SO(3,1)$ rotation matrix $\La$ corresponding to the $SL(2, \mathbb C)$ transformation $N$ is given by}
\[ \La^\mu \, _\nu \eq \frac{1}{2} \, \te{Tr} \Bigl\{ \bar \si^\mu \, N \, \si_\nu \, N^\dag \Bigr\} \ .
\]
{\bf See appendix \ref{sec:UsefulSpinorIdentities} for some $\si^\mu$ matrix identities.}

\subsection{Generators of $SL(2,\mathbb C)$}
\label{sec:GeneratorsOfSL2C}

Let us define tensors $\si^{\mu \nu}$, $\bar{\si}^{\mu \nu}$ as antisymmetrized products of $\si$ matrices:
\begin{align*}
(\si^{\mu \nu})_{\al}\,^{\be} \ \ &:= \ \ \frac{i}{4} \; \bigl(\si^{\mu} \, \bar{\si}^{\nu} \ - \ \si^{\nu} \, \bar{\si}^{\mu} \bigr)_{\al}\,^{\be} \\
(\bar{\si}^{\mu \nu})_{\dot{\al}}\,^{\dot{\be}} \ \ &:= \ \ \frac{i}{4} \;  \bigl(\bar{\si}^{\mu} \, \si^{\nu} \ - \ \bar{\si}^{\nu} \, \si^{\mu}  \bigr)^{\dot{\al}}\,_{\dot{\be}}
\end{align*}
which satisfy the Lorentz algebra
\[\Bigl[ \si^{\mu \nu} \ , \ \si^{\la \rho} \Bigr] \eq i \, \Bigl( \eta^{\mu \rho} \, \si^{\nu \la} \ + \ \eta^{\nu \la} \, \si^{\mu \rho} \ - \ \eta^{\mu \la} \, \si^{\nu \rho} \ - \ \eta^{\nu \rho} \, \si^{\mu \la} \Bigr)
\ . \]
\paragraph{Exercise 2.2:} {\it Verify this by means of the Dirac algebra $\si^\mu \bar \si^\nu + \si^\nu \bar \si^\mu = 2 \eta^{\mu \nu}$.}

\bigskip
\noindent
Under a finite Lorentz transformation with parameters $\om_{\mu \nu}$, spinors transform as follows:
\begin{align*}
\psi_{\al} \ \ &\mapsto \ \ \exp\left(-\frac{i}{2} \om_{\mu \nu} \si^{\mu \nu}\right)_{\al}\,^{\be} \, \psi_{\be} &\te{(left-handed)} \\
\bar{\chi}^{\dot{\al}} \ \ &\mapsto \ \ \exp\left(-\frac{i}{2} \om_{\mu \nu} \bar{\si}^{\mu \nu}\right)^{\dot{\al}}\,_{\dot{\be}} \, \bar{\chi}^{\dot{\be}} &\te{(right-handed)}
\end{align*}
Now consider the spins with respect to the $SU(2)$s spanned by the $A_{i}$ and $B_{i}$:
\begin{align*}
&\psi_{\al}: &(A, \ B) = \left(\frac{1}{2} , \ 0\right) \ \ \ &\Longrightarrow \ \ \ J_{i} \eq \frac{1}{2} \; \si_{i} \co K_{i} \eq -\frac{i}{2} \; \si_{i} \\
&\bar{\chi}^{\dot{\al}}: &(A , \ B) = \left(0, \ \frac{1}{2}\right) \ \ \ &\Longrightarrow \ \ \ J_{i} \eq \frac{1}{2} \; \si_{i} \co K_{i} \eq +\frac{i}{2} \; \si_{i}
\end{align*}
Some useful identities concerning the $\si^{\mu}$ and $\si^{\mu \nu}$ can be found in appendix \ref{sec:UsefulSpinorIdentities}. For now, let us just mention the identities
\begin{align*}
\si^{\mu \nu} \ \ &= \ \ \frac{1}{2i} \; \ep^{\mu \nu \rho \si} \, \si_{\rho \si} \\
\bar{\si}^{\mu \nu} \ \ &= \ \ -\frac{1}{2i} \; \ep^{\mu \nu \rho \si} \, \bar{\si}_{\rho \si} \ ,
\end{align*}
known as {\em self duality} and {\em anti self duality}. They are important because naively $\si^{\mu\nu}$ being antisymmetric seems to have
$\frac{4 \times 3}{2}$ components, but the self duality conditions reduces this by half. A reference book illustrating many of the calculations for two - component spinors is \cite{MKW}.

\subsection{Products of Weyl spinors}
\label{sec:WeylSpinors}

Define the product of two Weyl spinors as
\begin{align*}
\chi \psi \ \ &:= \ \ \chi^{\al} \, \psi_{\al} \ \ = \ \ -\chi_{\al} \, \psi^{\al} \\
\bar{\chi} \bar{\psi} \ \ &:= \ \ \bar{\chi}_{\dot{\al}} \, \bar{\psi}^{\dot{\al}} \ \ = \ \ -\bar{\chi}^{\dot{\al}} \, \bar{\psi}_{\dot{\al}} \ ,
\end{align*}
particularly,
    \[\psi \psi \eq \psi^{\al} \, \psi_{\al} \eq \ep^{\al \be} \, \psi_{\be} \, \psi_{\al} \eq \psi_{2} \, \psi_{1}\ - \ \psi_{1} \, \psi_{2} \ .
\]
Choose the $\psi_{\al}$ to be {\em anticommuting Grassmann numbers}, $\psi_{1} \psi_{2} = - \psi_{2} \psi_{1}$, so $\psi \psi = 2\psi_{2} \psi_{1}$.

\noindent
From the definitions
    \[\psi_{\al}^\dag \ \ := \ \ \bar{\psi}_{\dot{\al}} \co \bar{\psi}^{\dot{\al}} \ \ := \ \ \psi^{*}_{\be} \, (\si^{0})^{\be \dot{\al}}
\]
it follows that
	\[
(\chi \psi)^\dag \ \ = \ \ \bar{\chi} \bar{\psi} \co
(\psi \, \si^{\mu} \, \bar{\chi})^\dag \ \ = \ \ \chi \, \si^{\mu} \, \bar{\psi}
 \]
which justifies the $\nearrow$ contraction of dotted indices in contrast to the $\searrow$ contraction of undotted ones.
 
\noindent
In general we can generate all higher dimensional representations of the Lorentz group by products of the fundamental representation $(\frac{1}{2}, \, 0)$ and its conjugate $(0, \, \frac{1}{2})$. The computation of tensor products $(\frac{r}{2}, \, \frac{s}{2}) = (\frac{1}{2}, \, 0)^{\otimes r} \otimes (0, \, \frac{1}{2})^{\otimes s}$ can be reduced to successive application of the elementary $SU(2)$ rule $(\frac{j}{2}) \otimes (\frac{1}{2}) = (\frac{j-1}{2}) \oplus (\frac{j+1}{2})$ (for $j\neq 0$).

\noindent
Let us give two examples for tensoring Lorentz representations:
\begin{itemize}
\item $(\frac{1}{2}, \, 0)\otimes (0, \, \frac{1}{2})\ = \ (\frac{1}{2}, \, \frac{1}{2}) $

\noindent
Bispinors with different chiralities can be expanded in terms of the $\si^\mu_{\al \dot \al}$. Actually, the $\si$ matrices form a complete orthonormal set of $2 \times 2$ matrices with respect to the trace Tr$\{ \si^\mu \bar \si^\nu \} =2 \eta^{\mu \nu}$:
\[ \psi_\al \, \bar{\chi}_{\dot{\al}}\ \ = \ \ \frac{1}{2} \; \left(\psi \, \si_\mu \, \bar{\chi}\right)\, \si^\mu_{\al\dot\al}  \]
Hence, two spinor degrees of freedom with opposite chirality give rise to a Lorentz vector $\psi  \si_\mu \bar{\chi}$.
\item $(\frac{1}{2}, \, 0)\otimes (\frac{1}{2}, \, 0)\ = \ (0,0) \oplus (1,0) $

\noindent
Alike bispinors require a different set of matrices to expand, $\ep_{\al \be}$ and $(\si^{\mu \nu})_{\al} \,^\ga \ep_{\ga \be} =: (\si^{\mu \nu} \ep^T)_{\al \be}$. The former represents the unique antisymmetric $2 \times 2$ matrix, the latter provides the symmetric ones. Note that the (anti-)self duality reduces the number of linearly independent $\si^{\mu \nu}$'s (over $\mathbb C$) from 6 to 3:
\[\psi_\al \, \chi_\be \ \ = \ \ \frac{1}{2}\; \ep_{\al\be} \, \left(\psi\chi\right) \
+ \ \frac{1}{2} \; \left(\si^{\mu\nu} \, \ep^T\right)_{\al\be}\, \left(\psi \, \si_{\mu\nu} \, \chi\right) \]
The product of spinors with alike chiralities decomposes into two Lorentz irreducibles, a scalar $\psi\chi$ and a self-dual antisymmetric rank two tensor $\psi \, \si_{\mu\nu} \, \chi$. The counting of independent components of $\si^{\mu\nu}$ from its self-duality property precisely provides the right number of three components for the $(1,0)$ representation. Similarly, there is an anti-self dual tensor $\bar{\chi} \bar{\si}^{\mu \nu} \bar{\psi}$ in $(0,1)$.
\end{itemize}
These expansions are also referred to as \textsc{Fierz} identities. Their most general form and some corollories can be found in appendix \ref{sec:UsefulSpinorIdentities}.

\section{Supersymmetry algebra}
\label{sec:SupersymmetryAlgebra}

\subsection{History of supersymmetry}
\label{sec:HistoryOfSupersymmetry}

\begin{itemize}
\item In the 1960's, from the study of strong interactions, many hadrons have been discovered and were successfully organized in multiplets of $SU(3)_{f}$, the $f$ referring to flavour. This procedure was known as the {\em eightfold way} of \textsc{Gell-Mann} and \textsc{Neeman}. Questions arouse about bigger multiplets including particles of different spins.
\item {\em No-go theorem} (\textsc{Coleman, Mandula} 1967): most general symmetry of the $S$ - matrix is Poincaré $\otimes$ internal, that cannot mix different spins
\item \textsc{Golfand}, \textsc{Likhtman} (1971): extended the Poincar\'e algebra to include spinor generators $Q_{\al}$, where $\al = 1,2$.
\item \textsc{Ramond}, \textsc{Neveu-Schwarz}, \textsc{Gervais}, \textsc{Sakita} (1971): supersymmetry in 2 dimensions (from string theory).
\item \textsc{Volkov}, \textsc{Akulov} (1973): neutrinos as Goldstone particles ($m = 0$)
\item \textsc{Wess}, \textsc{Zumino} (1974): supersymmetric field theories in 4 dimensions. They opened the way to many other contributions to the field. This is generally seen as the actual starting point in the systematic study of supersymmetry.
\item \textsc{Haag}, \textsc{Lopuszanski}, \textsc{Sohnius} (1975): Generalized Coleman
  Mandula theorem including spinor generators $Q_{\al}^{A}$ ($\al
  = 1,2$ and $A = 1,...,{\cal N}$) corresponding to spins $(A , \ B) =
  \left( \frac{1}{2} , \ 0 \right)$ and $\bar{Q}_{\dot{\al}}^{A}$
  with $(A , \ B) = \left( 0 , \ \frac{1}{2} \right)$ in addition
  to $P^{\mu}$ and $M^{\mu \nu}$; but no further generators
  transforming in higher dimensional representations of the Lorentz
  group such as $\left( 1 , \ \frac{1}{2} \right)$, etc.
\end{itemize}

\subsection{Graded algebra}
\label{sec:GradedAlgebra}

In order to have a supersymmetric extension of the Poincaré algebra, we need to introduce the concept of {\em graded algebras}. Let $O_{a}$ be operators of a Lie algebra, then
    \[O_{a} \, O_{b} \ - \ (-1)^{\eta_{a} \eta_{b}} \, O_{b} \, O_{a} \eq iC^{e}\,_{ab} \, O_{e} \ ,
\]
where {\em gradings} $\eta_{a}$ take values
    \[\eta_{a} \eq \left\{ \begin{array}{ll} 0 &: O_{a} \ \te{bosonic generator} \\ 1 &: O_{a} \ \te{fermionic generator} \end{array} \right. \ .
\]
For supersymmetry, generators are the Poincaré generators $P^{\mu}$, $M^{\mu \nu}$ and the spinor generators $Q_{\al}^{A}$, $\bar{Q}_{\dot{\al}}^{A}$, where $A = 1,...,{\cal N}$. In case ${\cal N} = 1$ we speak of a simple SUSY, in case ${\cal N} > 1$ of an extended SUSY. In this chapter, we will only discuss ${\cal N} = 1$.

\noindent
We know the commutation relations $[P^{\mu} , P^{\nu}]$, $[P^{\mu} , M^{\rho \si}]$ and $[M^{\mu \nu} , M^{\rho \si}]$ from Poincaré - algebra, so we need to find
    \[\begin{array}{ll} (\te{a}) \ \Bigl[Q_{\al} \ , \ M^{\mu \nu} \Bigr] \ , &(\te{b}) \ \Bigl[Q_{\al} \ , \ P^{\mu} \Bigr] \ , \\ (\te{c}) \ \Bigl\{Q_{\al} \ , \ Q_{\be} \Bigr\} \ , &(\te{d}) \ \Bigl\{Q_{\al} \ , \ \bar{Q}_{\dot{\be}} \Bigr\} \ , \end{array}
\]
also (for internal symmetry generators $T_{i}$)
    \[(\te{e}) \ \Bigl[Q_{\al} \ , \ T_{i} \Bigr] \ .
\]

\begin{itemize}
\item (a) $\ \Bigl[Q_{\al} \ , \ M^{\mu \nu} \Bigr]$

Since $Q_{\al}$ is a spinor, it transforms under the exponential of the $SL(2 , \mathbb C)$ generators $\si^{\mu \nu}$:
    \[Q'_{\al} \eq \exp\left(-\frac{i}{2} \om_{\mu \nu} \si^{\mu \nu}\right)_{\al}\,^{\be} \, Q_{\be} \ \ \approx \ \ \left(\mathds{1} \ - \ \frac{i}{2} \; \om_{\mu \nu} \, \si^{\mu \nu}\right)_{\al}\,^{\be} \, Q_{\be} \ ,
\]
but $Q_{\al}$ is also an operator transforming under Lorentz transformations $U = \exp\left(-\frac{i}{2} \om_{\mu \nu} M^{\mu \nu}\right)$ to
    \[Q'_{\al} \eq U^\dag \, Q_{\al} \, U \ \ \approx \ \ \left(\mathds{1} \ + \ \frac{i}{2} \; \om_{\mu \nu} \, M^{\mu \nu}\right) \, Q_{\al} \, \left(\mathds{1} \ - \ \frac{i}{2} \; \om_{\mu \nu} \, M^{\mu \nu}\right) \ .
\]
Compare these two expressions for $Q'_{\al}$ up to first order in $\om_{\mu \nu}$,
    \[Q_{\al} \ - \ \frac{i}{2} \; \om_{\mu \nu} \, \left(\si^{\mu \nu}\right)_{\al}\,^{\be} \, Q_{\be} \eq Q_{\al} \ - \ \frac{i}{2} \; \om_{\mu \nu} \, \bigl(Q_{\al} \, M^{\mu \nu}\ - \ M^{\mu \nu} \, Q_{\al} \bigr)\ + \ {\cal O}(\om^{2})
\]
    \[\Longrightarrow \ \ \ \fbox{$\displaystyle \Bigl[Q_{\al} \ , \ M^{\mu \nu} \Bigr] \eq (\si^{\mu \nu})_{\al}\,^{\be} \, Q_{\be} $ }
\]
\item (b) $\ \Bigl[Q_{\al} \ , \ P^{\mu} \Bigr]$

$c \cdot(\si^{\mu})_{\al \dot{\al}} \bar{Q}^{\dot{\al}}$ is the only
 way of writing a sensible term with free indices $\mu$, $\al$ which
 is linear in $Q$. To fix the constant $c$, consider
 $[\bar{Q}^{\dot{\al}},P^{\mu}] = c^{*} \cdot
 (\bar{\si})^{\dot{\al}\be} Q_{\be}$ (take adjoints using
 $(Q_{\al})^\dag = \bar{Q}_{\dot{\al}}$ and $(\si^{\mu} \bar{Q})^\dag_{\al}
 = ({Q}\si^{\mu})_{\dot{\al}}$). The Jacobi identity for
 $P^{\mu}$,
 $P^{\nu}$ and $Q_{\al}$
\begin{align*}
0 \ \ &= \ \ \Biggl[ P^{\mu} \ , \ \Bigl[P^{\nu} \ , \ Q_{\al} \Bigr]
 \Biggr] \ + \ \Biggl[ P^{\nu} \ , \ \Bigl[Q_{\al} \ , \ P^{\mu} \Bigr]
 \Biggr] \ + \ \Biggl[ Q_{\al} \ , \ \underbrace{\Bigl[P^{\mu} \ , \ P^{\nu} \Bigr]}_{0} \Biggr] \\
&= \ \ -c \, (\si^{\nu})_{\al \dot{\al}} \, \Bigl[ P^{\mu} \ , \
 \bar{Q}^{\dot{\al}} \Bigr] \ + \  c \, (\si^{\mu})_{\al \dot{\al}} \, \Bigl[
P^{\nu} \ , \ \bar{Q}^{\dot{\al}} \Bigr] \\
&= \ \ |c|^{2} \, (\si^{\nu})^{\al \dot{\al}} \, (\bar{\si}^{\mu})^{\dot{\al}
 \be} \, Q_{\be} \ - \ |c|^{2} \, (\si^{\mu})_{\al \dot{\al}}
 \, (\bar{\si}^{\nu})^{\dot{\al} \be} \, Q_{\be} \\
&= \ \ |c|^{2} \, \underbrace{(\si^{\nu} \, \bar{\si}^{\mu} \ - \ \si^{\mu} \, \bar{\si}^{\nu})_{\al}\,^{\be}}_{\neq 0} \, Q_{\be}
\end{align*}
can only hold for general $Q_{\be}$, if $c = 0$, so
    \[ \fbox{$\displaystyle \Bigl[Q_{\al} \ , \ P^{\mu} \Bigr] \eq \Bigl[\bar{Q}^{\dot{\al}} \ , \ P^{\mu} \Bigr] \eq 0 $}
\]
\item (c) $\ \Bigl\{Q_{\al} \ , \ Q_{\be} \Bigr\}$

Due to index structure, that commutator should look like
    \[\Bigl\{Q_{\al} \ , \ Q^{\be} \Bigr\} \eq k \, (\si^{\mu \nu})_{\al}\,^{\be} \, M_{\mu \nu} \ .
\]
Since the left hand side commutes with $P^{\mu}$ and the right hand side doesn't, the only consistent choice is $k = 0$, i.e.
    \[ \fbox{$\displaystyle \Bigl\{Q_{\al} \ , \ Q_{\be} \Bigr\} \eq 0 $}
\]
\item (d) $\ \Bigl\{Q_{\al} \ , \ \bar{Q}_{\dot{\be}} \Bigr\}$

This time, index structure implies an ansatz
    \[\Bigl\{Q_{\al} \ , \ \bar{Q}_{\dot{\be}} \Bigr\} \eq t \, (\si^{\mu})_{\al \dot{\be}} \, P_{\mu} \ .
\]
There is no way of fixing $t$, so, by convention, set $t = 2$:
    \[\fbox{$\displaystyle \Bigl\{Q_{\al} \ , \ \bar{Q}_{\dot{\be}} \Bigr\} \eq 2 \,(\si^{\mu})_{\al \dot{\be}} \, P_{\mu} $}
\]
\end{itemize}
Notice that two symmetry transformations $Q_{\al} \bar{Q}_{\dot{\be}}$ have the effect of a translation. Let $|B \rangle$ be a bosonic state and $|F \rangle$ a fermionic one, then
    \[Q_{\al} \, |F \rangle \eq |B \rangle \co \bar{Q}_{\dot{\be}}  \, |B \rangle \eq |F \rangle \so Q\bar{Q}: \ |B \rangle \ \ \mapsto \ \ |B \ \te{(translated)} \rangle \ .
\]
\begin{itemize}
\item (e) $\ \Bigl[Q_{\al} \ , \ T_{i} \Bigr]$

Usually, this commutator vanishes, exceptions are $U(1)$
automorphisms of the supersymmetry algebra known as {\em $R$ symmetry}.
    \[Q_{\al} \ \ \mapsto \ \ \exp(i\la) \, Q_{\al} \co \bar{Q}_{\dot{\al}} \ \ \mapsto \ \ \exp(-i\la) \, \bar{Q}_{\dot{\al}} \ .
\]
Let $R$ be a $U(1)$ generator, then
    \[\Bigl[ Q_{\al} \ , \ R \Bigr] \eq Q_{\al} \co \Bigl[\bar{Q}_{\dot{\al}} \ , \ R \Bigr] \eq -\bar{Q}_{\dot{\al}} \ .
\]
\end{itemize}

\subsection{Representations of the Poincaré group}
\label{sec:RepresentationsOfPoincareGroup}

Recall the rotation group $\{J_{i}: \ i=1,2,3 \}$ satisfying
    \[\Bigl[J_{i} \ , \ J_{j} \Bigr] \eq i\ep_{ijk}\, J_{k} \ .
\]
The Casimir operator
    \[J^{2} \eq \sum^{3}_{i = 1} J_{i}^{2}
\]
commutes with all the $J_{i}$ and labels irreducible representations by eigenvalues $j(j+1)$ of $J^{2}$. Within these representations, diagonalize $J_{3}$ to eigenvalues $j_{3} = -j,-j+1,...,j-1,j$. States are labelled like $|j,j_{3} \rangle$.

\noindent
Also recall the two Casimirs in Poincaré group, one of which involves the {\em Pauli Ljubanski vector} $W_{\mu}$,
    \[W_{\mu} \eq \frac{1}{2} \; \ep_{\mu \nu \rho \si} \, P^{\nu} \, M^{\rho \si}
\]
(where $\ep_{0123} = -\ep^{0123} = +1$).

\paragraph{Exercise 2.3:} {\it Prove that the Pauli Ljubanski vector satisfies the following commutation relations:}
\begin{align*}
\Bigl[ W_\mu \ , \ P_\nu \Bigr] \ \ &= \ \ 0 \\
\Bigl[ W_\mu \ , \ M_{\rho \si} \Bigr] \ \ &= \ \ i \eta_{\mu \rho} \, W_\si \ - \ i \eta_{\mu \si} \, W_\rho \\
\Bigl[ W_\mu \ , \ W_\nu \Bigr] \ \ &= \ \ - \, i\ep_{\mu \nu \rho \si} \, W^\rho \, P^\si \\
\Bigl[ W_\mu \ , \ Q_\al \Bigr] \ \ &= \ \ - \, i \, P_\nu \, (\si^{\mu \nu})_\al \, ^\be \, Q_\be \\
\end{align*}
{\it In intermediate steps one might need}
\[
\ep^{\mu \nu \rho \si} \, \ep_{\si \al \be \ga} \eq 6 \, \de^{[\mu}_\al \, \de^\nu_\be \, \de^{\rho]}_\ga \co \ep^{\mu \nu \rho \si} \, \ep_{\rho \si \al \be} \eq - \, 4 \, \de^{[\mu}_\al \, \de^{\nu]}_\be \ ,
\]
{\it and it is useful to note that $\ep^{\mu \nu \rho \si} W_\si = 3 M^{[\mu \nu} P^{\rho]}$.

\bigskip
\noindent
The Poincaré Casimirs are then given by}
    \[C_{1} \eq P^{\mu} \, P_{\mu} \co C_{2} \eq W^{\mu} \, W_{\mu} \ .
\]
{\it the $C_{i}$ commute with all generators.}

\paragraph{Exercise 2.4:} {\it Show that $C_2$ indeed commutes with the Poincaré generators but not with the extension $Q_\be$ to super Poincaré.}

\bigskip
\noindent
Poincaré multiplets are labelled $|m,\om \rangle$, eigenvalues $m^{2}$ of $C_{1}$ and eigenvalues of $C_{2}$. States within those irreducible representations carry the
eigenvalue $p^{\mu}$ of the generator $P^{\mu}$ as a label. Notice
that at this level the Pauli Ljubanski vector only provides a short
way to express the second Casimir. Even though $W_\mu$ has standard
commutation relations with the generators of the Poincar\'e group $
M_{\mu\nu}, P_\mu$ stating that it transform as a vector under
Lorentz transformations and commutes with $P_\mu$ (invariant under
translations), the commutator $[W_\mu,W_\nu]\sim
\epsilon_{\mu\nu\rho\si}W^\rho P^\sigma$ implies that the $W_\mu$'s by
themselves are not generators of any algebra.

\noindent
To find more labels, take $P^{\mu}$ as given and look for all elements of the Lorentz group that commute with $P^{\mu}$. This defines little groups:

\begin{itemize}
\item Massive particles, $p^{\mu} = (m , \ \underbrace{0  , \ 0  , \ 0}_{\te{invariant under rot.}})$, have rotations as their little group. Due to the antisymmetric $\ep_{\mu \nu \rho \si}$ in the $W_{\mu}$, it follows
    \[W_{0} \eq 0 \co W_{i} \eq -m \, J_{i} \ .
\]
Every particle with nonzero mass is an irreducible representation of Poincaré group with labels $|m,j;p^{\mu},j_{3} \rangle$.
\item Massless particles' momentum has the form $p^{\mu} = (E  , \ 0  , \ 0  , \ E)$ which implies
    \[(W_{0}, \ W_{1} , \ W_{2}, \ W_{3}) \eq E \, \bigl( J_{3}, \ -J_{1} \, + \, K_{2}, \ -J_{2} \, - \, K_{1}, \ -J_{3} \bigr)
\]
    \[\Longrightarrow \ \ \ \Bigl[W_{1} \ , \ W_{2} \Bigr] \eq 0 \co \Bigl[W_{3} \ , \ W_{1} \Bigr] \eq -iE \, W_{2} \co \Bigl[W_{3} \ , \ W_{2} \Bigr] \eq iE \, W_{1} \ .
\]
Commutation relations are those for Euclidean group in two dimensions. For finite dimensional representations, $SO(2)$ is a subgroup and $W_{1}$, $W_{2}$ have to be zero. In that case, $W^{\mu} = \la P^{\mu}$ and states are labelled $|0,0;p^{\mu},\la \rangle =: |p^{\mu} , \la \rangle$, where $\la$ is called {\em helicity}. Under CPT, those states transform to $|p^{\mu} , -\la \rangle$. The relation
    \[\exp(2\pi i \la) \, |p^{\mu} , \la \rangle \eq \pm |p^{\mu} , \la \rangle
\]
requires $\la$ to be integer or half integer $\la = 0,\frac{1}{2},1,...$, e.g. $\la = 0$ (Higgs), $\la = \frac{1}{2}$ (quarks, leptons), $\la = 1$ ($\ga$, $W^{\pm}$, $Z^{0}$, $g$) and $\la = 2$ (graviton).
\end{itemize}

\subsection{${\cal N} = 1$ supersymmetry representations}
\label{sec:N1SupersymmetryRepresentation}

For ${\cal N} = 1$ supersymmetry, $C_{1} = P^{\mu} P_{\mu}$ is still a good Casimir, $C_{2} = W^{\mu} W_{\mu}$, however, is not. So one can have particles of different spin within one multiplet. To get a new Casimir $\tilde{C}_{2}$ (corresponding to superspin), define
    \[B_{\mu} \ \ := \ \ W_{\mu} \ - \ \frac{1}{4} \; \bar{Q}_{\dot{\al}} \, (\bar{\si}_{\mu})^{\dot{\al} \be}  \, Q_{\be} \co C_{\mu \nu} \ \ := \ \ B_{\mu} \, P_{\nu} \ - \ B_{\nu} \, P_{\mu}
\]
    \[\tilde{C}_{2} \ \ := \ \ C_{\mu \nu} \, C^{\mu \nu} \ .
\]

\subsubsection{Proposition}
\label{sec:Proposition1}

In any supersymmetric multiplet, the number $n_{B}$ of bosons equals the number $n_{F}$ of fermions,
    \[n_{B} \eq n_{F} \ .
\]

\subsubsection{Proof}
\label{sec:Proof1}

Consider the {\em fermion number operator} $(-1)^{F} = (-)^{F}$, defined via
    \[(-)^{F} \, |B \rangle \eq |B \rangle \co (-)^{F} \, |F \rangle \eq -|F \rangle \ .
\]
This new operator $(-)^{F}$ anticommutes with $Q_{\al}$ since
    \[(-)^{F} \, Q_{\al} \, |F \rangle \eq (-)^{F} \, |B \rangle \eq |B \rangle \eq Q_{\al} \, |F \rangle \eq -Q_{\al} \, (-)^{F} \, |F \rangle \so \Bigl\{ (-)^{F} \ , \ Q_{\al} \Bigr\} \eq 0 \ .
\]
Next, consider the trace
\begin{align*}
\te{Tr} \Biggl\{ (-)^{F} \, \Bigl\{ Q_{\al} \ , \ \bar{Q}_{\dot{\be}} \Bigr\} \Biggr\} \ \ &= \ \ \te{Tr} \Bigl\{ \underbrace{(-)^{F} \, Q_{\al}}_{\te{anticommute}} \, \bar{Q}_{\dot{\be}} \ + \ \underbrace{(-)^{F} \, \bar{Q}_{\dot{\be}} \, Q_{\al}}_{\te{cyclic perm.}} \Bigr\} \\
&= \ \ \te{Tr} \Bigl\{ -Q_{\al}\, (-)^{F} \, \bar{Q}_{\dot{\be}} \ + \ Q_{\al} \, (-)^{F} \, \bar{Q}_{\dot{\be}} \Bigr\} \eq 0 \ .
\end{align*}
On the other hand, it can be evaluated using $\{Q_{\al} , \bar{Q}_{\dot{\be}} \} = 2(\si^{\mu})_{\al \dot{\be}} P_{\mu}$,
    \[\te{Tr} \Biggl\{ (-)^{F} \, \Bigl\{ Q_{\al} \ , \ \bar{Q}_{\dot{\be}} \Bigr\} \Biggr\} \eq \te{Tr} \Biggl\{ (-)^{F} \, 2 \,(\si^{\mu})_{\al \dot{\be}} \, P_{\mu} \Biggr\} \eq 2 \, (\si^{\mu})_{\al \dot{\be}} \, p_{\mu} \, \te{Tr} \Bigl\{ (-)^{F} \Bigr\} \ ,
\]
where $P^{\mu}$ is replaced by its eigenvalues $p^{\mu}$ for the specific state. The conclusion is
    \begin{align*}
    0 \ \ &= \ \ \te{Tr} \Bigl\{ (-)^{F} \Bigr\} \eq \sum_{\te{bosons}} \langle B | \, (-)^{F} \, | B \rangle \ + \ \sum_{\te{fermions}} \langle F | \, (-)^{F} \, | F \rangle \\ &= \ \ \sum_{\te{bosons}} \langle B | B \rangle \ - \ \sum_{\te{fermions}} \langle F | F \rangle \eq n_{B} \ - \ n_{F} \ .
\end{align*}

\subsection{Massless supermultiplet}
\label{sec:MasslessSupermultiplet}

States of massless particles have $P_{\mu}$ - eigenvalues $p_{\mu} = (E , \ 0 , \  0 , \ E)$. The Casimirs $C_{1} = P^{\mu} P_{\mu}$ and $\tilde{C}_{2} = C_{\mu \nu} C^{\mu \nu}$ are zero. Consider the algebra
    \[\Bigl\{ Q_{\al} \ , \ \bar{Q}_{\dot{\be}} \Bigr\} \eq 2 \, (\si^{\mu})_{\al \dot{\be}} \, P_{\mu} \eq 2 \, E \, \bigl(\si^{0} \, + \, \si^{3} \bigr)_{\al \dot{\be}} \eq 4 \, E \, \ccb 1 &0 \\ 0 &0 \cce_{\al \dot{\be}} \ ,
\]
which implies that $Q_{2}$ is zero in the representation:
    \[\Bigl\{ Q_{2} \ , \ \bar{Q}_{\dot{2}} \Bigr\} \eq 0 \so \langle p^{\mu},\la | \, \bar{Q}_{\dot{2}} \, Q_{2} \, | \tilde{p}^{\mu},\tilde{\la} \rangle \eq 0 \so Q_{2} \eq 0
\]
The $Q_{1}$ satisfy $\{Q_{1} , \bar{Q}_{\dot{1}} \} = 4E$, so defining creation- and annihilation operators $a$ and $a^\dag$ via
    \[a \ \ := \ \ \frac{Q_{1}}{2\sqrt{E}} \co a^\dag \ \ := \ \ \frac{\bar{Q}_{\dot{1}}}{2\sqrt{E}} \ ,
\]
get the anticommutation relations
    \[\Bigl\{ a \ , \ a^\dag \Bigr\} \eq 1 \co \Bigl\{ a \ , \ a \Bigr\} \eq \Bigl\{ a^\dag \ , \ a^\dag \Bigr\} \eq 0 \ .
\]
Also, since $[a , J^{3}] = \frac{1}{2} (\si^{3})_{11}a = \frac{1}{2}a$,
    \[J^{3}  \, \bigl( a \, |p^{\mu} , \la \rangle \bigr) \eq \Biggl( a \, J^{3} \ - \ \Bigl[ a \ , \ J^{3} \Bigr] \Biggr) \, |p^{\mu} , \la \rangle \eq \left(a \, J^{3} \ - \ \frac{a}{2}\right) \, |p^{\mu} , \la \rangle \eq \left(\la \, - \, \frac{1}{2} \right) \, a \, |p^{\mu} , \la \rangle \ .
\]
$a |p^{\mu} , \la \rangle$ has helicity $\la - \frac{1}{2}$, and by similar reasoning, find that the helicity of $a^\dag |p^{\mu} , \la \rangle$ is $\la + \frac{1}{2}$. To build the representation, start with a vacuum state of minimum helicity $\la$, let's call it $|\Om \rangle$. Obviously $a|\Om \rangle = 0$ (otherwise $|\Om \rangle$ would not have lowest helicity) and $a^\dag a^\dag |\Om \rangle = 0 |\Om \rangle = 0$, so the whole multiplet consists of
    \[|\Om \rangle \eq |p^{\mu} , \la \rangle \co a^\dag \, |\Om \rangle \eq |p^{\mu} , \la + \tfrac{1}{2} \rangle \ .
\]
Add the CPT conjugate to get
    \[|p^{\mu} , \pm \la \rangle \co |p^{\mu} , \pm \left(\la + \tfrac{1}{2} \right) \rangle \ .
\]
There are, for example, chiral multiplets with $\la = 0,\frac{1}{2}$, vector- or gauge multiplets ($\la = \frac{1}{2},1$ gauge and gaugino)
    \[\begin{array}{r|l} \la = 0 \ \te{scalar} & \la = \frac{1}{2} \ \te{fermion} \\\hline
    \te{squark} & \te{quark} \\ \te{slepton} & \te{lepton} \\ \te{Higgs} & \te{Higgsino} \end{array} \ \ \ \ \ \ \ \begin{array}{r|l} \la = \frac{1}{2} \ \te{fermion} & \la = 1 \ \te{boson} \\\hline
    \te{photino} & \te{photon} \\ \te{gluino} & \te{gluon} \\ W\te{ino} , \ Z\te{ino} & W , \ Z \end{array} \ ,
\]
as well as the graviton with its partner:
    \[\begin{array}{r|l} \la = \frac{3}{2} \ \te{fermion} & \la = 2 \ \te{boson} \\\hline
    \te{gravitino} & \te{graviton} \end{array}
\]

\subsection{Massive supermultiplet}
\label{sec:MassiveSupermultiplet}

In case of $m \neq 0$, there are $P^{\mu}$ - eigenvalues $p^{\mu} = (m , \ 0 , \ 0 , \ 0)$ and Casimirs
    \[C_{1} \eq P^{\mu} \, P_{\mu} \eq m^{2} \co \tilde{C}_{2} \eq C_{\mu \nu} \, C^{\mu \nu} \eq 2 \, m^{4} \,Y^{i} \, Y_{i} \ ,
\]
where $Y_{i}$ denotes superspin
    \[Y_{i} \eq J_{i} \ - \ \frac{1}{4m} \, \bar{Q} \, \bar{\si}_{i} \, Q \eq \frac{B_{i}}{m} \co \Bigl[Y_{i} \ , \ Y_{j} \Bigr] \eq i\ep_{ijk} \, Y_{k} \ .
\]
Eigenvalues to $Y^{2} = Y^{i} Y_{i}$ are $y(y+1)$, so label irreducible representations by $|m , y \rangle$. Again, the anticommutation - relation for $Q$ and $\bar{Q}$ is the key to get the states:
    \[\Bigl\{ Q_{\al} \ , \ \bar{Q}_{\dot{\be}} \Bigr\} \eq 2 \, (\si^{\mu})_{\al \dot{\be}} \, P_{\mu} \eq 2 \, m \, (\si^{0})_{\al \dot{\be}} \eq 2 \, m \, \ccb 1 &0 \\ 0 &1 \cce_{\al \dot{\be}}
\]
Since both $Q$'s have nonzero anticommutators with their $\bar{Q}$ - partner, define two sets of ladder operators
    \[a_{1,2} \ \ := \ \ \frac{Q_{1,2}}{\sqrt{2m}} \co a^\dag_{1,2} \ \ := \ \ \frac{\bar{Q}_{\dot{1},\dot{2}}}{\sqrt{2m}} \ ,
\]
with anticommutation relations
    \[\Bigl\{ a_{p} \ , \ a^\dag_{q} \Bigr\} \eq \de_{pq} \co \Bigl\{ a_{p} \ , \ a_{q} \Bigr\} \eq \Bigl\{ a^\dag_{p} \ , \ a^\dag_{q} \Bigr\} \eq 0 \ .
\]
Let $|\Om \rangle$ be the vacuum state, annihilated by $a_{1,2}$. Consequently,
    \[Y_{i} \, |\Om \rangle \eq J_{i} \, |\Om \rangle \ - \ \frac{1}{4m} \; \bar{Q} \, \bar{\si}_{i} \, \sqrt{2m} \, \underbrace{a |\Om \rangle}_{0} \eq J_{i} \, |\Om \rangle \ ,
\]
i.e. for  $|\Om \rangle$ the spin number $j$ and superspin - number
$y$ are the same. So for given $m,y$:
    \[|\Om \rangle \eq |m , j=y ; p^{\mu} , j_{3} \rangle
\]
Obtain the rest of the multiplet using
\begin{align*}
&a_{1} \, |j_{3} \rangle \eq |j_{3} - \tfrac{1}{2} \rangle \co a_{1}^\dag \, |j_{3} \rangle \eq |j_{3} + \tfrac{1}{2} \rangle \\
&a_{2} \, |j_{3} \rangle \eq |j_{3} + \tfrac{1}{2} \rangle \co a_{2}^\dag \, |j_{3} \rangle \eq |j_{3} - \tfrac{1}{2} \rangle \ ,
\end{align*}
where $a^\dag_{p}$ acting on $|\Om \rangle$ behave like coupling of
two spins $j$ and $\frac{1}{2}$. This will yield a linear combination
of two possible total spins $j + \frac{1}{2}$ and $j - \frac{1}{2}$
with Clebsch Gordan coefficients $k_{i}$ (recall $j\otimes
\frac{1}{2}= (j-\frac{1}{2}) \oplus (j+\frac{1}{2})$):
\begin{align*}
&a_{1}^\dag \, |\Om \rangle \eq k_{1} \, |m , j = y + \tfrac{1}{2} ; p^{\mu} , j_{3} + \tfrac{1}{2} \rangle \ + \ k_{2} \, |m , j = y - \tfrac{1}{2} ; p^{\mu} , j_{3} + \tfrac{1}{2} \rangle \\
&a_{2}^\dag \, |\Om \rangle \eq k_{3} \, |m , j = y + \tfrac{1}{2} ; p^{\mu} , j_{3} - \tfrac{1}{2} \rangle \ + \ k_{4} \, |m , j = y - \tfrac{1}{2} ; p^{\mu} , j_{3} - \tfrac{1}{2} \rangle \ .
\end{align*}
The remaining states
    \[a_{2}^\dag \, a_{1}^\dag \, |\Om \rangle \eq -a_{1}^\dag \, a_{2}^\dag \, | \Om \rangle \ \ \propto \ \ |\Om \rangle
\]
represent spin $j$ - objects. In total, we have
    \[\underbrace{2 \cdot |m , j=y ; p^{\mu} , j_{3} \rangle}_{(4y + 2) \ \te{states}} \co \underbrace{1 \cdot |m , j = y + \tfrac{1}{2} ; p^{\mu} , j_{3} \rangle}_{(2y + 2) \ \te{states}} \co \underbrace{1 \cdot |m , j = y - \tfrac{1}{2} ; p^{\mu} , j_{3} \rangle}_{(2y) \ \te{states}} \ ,
\]
in a $|m , y \rangle$ multiplet, which is of course an equal number
of bosonic and fermionic states.
Notice that in labelling the states we have the value of $m$ and $y$
\textbf{fixed} throughout the multiplet and the values of $j$
change state by state, as it should since in a supersymmetric multiplet
there are states of different spin. 

\noindent
The case $y = 0$ needs to be treated separately:
\begin{align*}
|\Om \rangle \ \ &= \ \ |m,j = 0; p^{\mu}, j_{3} = 0 \rangle \\
a^\dag_{1,2} \, |\Om \rangle \ \ &= \ \ |m,j = \tfrac{1}{2}; p^{\mu}, j_{3} = \pm \tfrac{1}{2} \rangle \\
a^\dag_{1} \, a^\dag_{2} \, |\Om \rangle \ \ &= \ \ |m,j = 0 ; p^{\mu}, j_{3} = 0 \rangle \ \ =: \ \ |\Om' \rangle
\end{align*}
Parity interchanges $(A , \ B) \leftrightarrow (B , \ A)$, i.e. $(\frac{1}{2} , \ 0) \leftrightarrow (0 , \ \frac{1}{2})$. Since $\{Q_{\al} , \bar{Q}_{\dot{\be}} \} = 2 (\si^{\mu})_{\al \dot{\be}} P_{\mu}$, need the following transformation - rules for $Q_{\al}$ and $\bar{Q}_{\dot{\al}}$ under parity $P$ (with phase factor $\eta_{P}$ such that $|\eta_{P}| = 1$):
\begin{align*}
P \, Q_{\al} \, P^{-1} \ \ &= \ \ \eta_{P} \, (\si^{0})_{\al \dot{\be}} \, \bar{Q}^{\dot{\be}}
\\
P \, \bar{Q}^{\dot{\al}} \, P^{-1} \ \ &= \ \ \eta_{P}^{*} \, (\bar{\si}^{0})^{\dot{\al} \be} \, Q_{\be} 
\end{align*}
That ensures $P^{\mu} \mapsto (P^{0} \ , \ -\ve{P})$ and has the interesting effect $P^{2} Q P^{-2} = -Q$. Moreover, consider the two $j = 0$ states $|\Om \rangle$ and $|\Om' \rangle$: The first is annihilated by $a_{i}$, the second one by $a^\dag_{i}$. Due to $Q \leftrightarrow \bar{Q}$, parity interchanges $a_{i}$ and $a^\dag_{i}$ and therefore $|\Om \rangle \leftrightarrow |\Om' \rangle$. To get vacuum states with a defined parity, we need linear combinations
    \[|\pm \rangle \ \ := \ \ |\Om \rangle \ \pm \ |\Om' \rangle \co P \, |\pm \rangle \eq \pm \, |\pm \rangle \ .
\]
Those states are called scalar ($|+ \rangle$) and pseudoscalar ($|- \rangle$).

\section{Extended supersymmetry}
\label{sec:ExtendedSupersymmetry}

Having discussed the algebra and representations of simple
(${\cal N}=1$) supersymmetry, we will turn now to the more general case of
{\em extended supersymmetry} ${\cal N}>1$.

\subsection{Algebra of extended supersymmetry}
\label{sec:AlgebraOfExtendedSupersymmetry}

Now, the spinor generators get an additional label $A,B = 1,2,...,{\cal N}$. The algebra is the same as for ${\cal N} = 1$ except for

\medskip

\framebox{\begin{minipage}{5.9in}
\begin{align*}
\Bigl\{Q_{\al}^{A} \ , \ \bar{Q}_{\dot{\be}B} \Bigr\} \ \ &= \ \ 2 \, (\si^{\mu})_{\al \dot{\be}} \, P_{\mu} \, \de^{A}\,_{B} \\
\Bigl\{Q_{\al}^{A} \ , \ Q_{\be}^{B} \Bigr\} \ \ &= \ \ \ep_{\al \be} \, Z^{AB}
\end{align*}
\end{minipage}}

\medskip

\noindent
with antisymmetric {\em central charges} $Z^{AB} = -Z^{BA}$ commuting with all the generators
    \[\Bigl[Z^{AB} \ , \ P^{\mu} \Bigr] \eq \Bigl[Z^{AB} \ , \ M^{\mu \nu} \Bigr] \eq \Bigl[Z^{AB} \ , \ Q_{\al}^{A} \Bigr] \eq \Bigl[Z^{AB} \ , \ Z^{CD} \Bigr] \eq \Bigl[Z^{AB} \ , \ T_{a} \Bigr] \eq 0 \ .
\]
They form an abelian invariant subalgebra of internal symmetries. Recall that $[T_{a},T_{b}] = i C_{abc} T_{c}$. Let $G$ be an internal symmetry group, then define the {\em R symmetry} $H \subset G$ to be the set of $G$ elements that do not commute with the supersymmetry generators, e.g. $T_{a} \in G$ satisfying
    \[\Bigl[Q_{\al}^{A} \ , \ T_{a} \Bigr] \eq S_{a}\,^{A}\,_{B} \, Q_{\al}^{B} \ \ \neq \ \ 0
\]
is an element of $H$. If $Z^{AB} = 0$, then the R symmetry is $H =
U(N)$, but with $Z^{AB} \neq 0$, $H$ will be a subgroup.
The existence of central charges is the main new ingredient of
extended supersymmetries. The derivation of the previous algebra is a
straightforward generalization of the one for ${\cal N}=1$ supersymmetry.

\subsection{Massless representations of ${\cal N} > 1$ supersymmetry}
\label{sec:MasslessRepresentationsOfN1Supersymmetry}

As we did for ${\cal N}=1$, we will proceed now to discuss massless and
massive representations. We will start with the massless case which is
simpler and has very important implications.

\noindent
Let $p_{\mu} = (E , \ 0 , \ 0 , \ E)$, then (similar to ${\cal N} = 1$).
    \[\Bigl\{Q_{\al}^{A} \ , \ \bar{Q}_{\dot{\be}B} \Bigr\} \eq 4 \, E \, \ccb 1 &0 \\ 0 &0 \cce_{\al \dot{\be}} \de^{A}\,_{B} \so Q_{2}^{A} \eq 0
\]
We can immediately see from this that the central charges $Z^{AB}$
vanish since $Q_{2}^A=0$ implies $Z^{AB}=0$ from the anticommutators
$\Bigl\{Q_{\al}^{A} \ , \ Q_{\be}^{B} \Bigr\} = \ep_{\al \be}
Z^{AB}$.

\noindent
In order to obtain the full representation, define $N$ creation- and annihilation - operators
    \[a^{A} \ \ := \ \ \frac{Q_{1}^{A}}{2\sqrt{E}} \co a^{A \dag} \ \
    :=
 \ \ \frac{\bar{Q}_{\dot{1}}^{A}}{2\sqrt{E}} \so \Bigl\{a^{A} \ , 
\ a_{B}^\dag \Bigr\} \eq \de^{A}\,_{B} \ ,
\]
to get the following states (starting from vacuum $|\Om \rangle$, which is annihilated by all the $a^{A}$):
    \[\begin{array}{l|c|r} \te{states} &\te{helicity} &\te{number of states} \\\hline |\Om \rangle &\la_{0} &1 = \left( \begin{smallmatrix} {\cal N} \\ 0 \end{smallmatrix} \right) \\ a^{A\dag} |\Om \rangle &\la_{0} + \frac{1}{2} &{\cal N} = \left( \begin{smallmatrix} {\cal N} \\ 1 \end{smallmatrix} \right) \\ a^{A\dag} a^{B \dag} |\Om \rangle &\la_{0} + 1 &\frac{1}{2!}{\cal N} ({\cal N} - 1) = \left( \begin{smallmatrix} {\cal N} \\ 2 \end{smallmatrix} \right) \\ a^{A\dag} a^{B \dag} a^{C \dag} |\Om \rangle &\la_{0} + \frac{3}{2} &\frac{1}{3!}{\cal N} ({\cal N} - 1) ({\cal N} - 2) = \left( \begin{smallmatrix} {\cal N} \\ 3 \end{smallmatrix} \right) \\ \ \ \ \ \vdots &\vdots &\vdots \ \ \ \ \\ a^{{\cal N} \dag} a^{({\cal N} - 1) \dag} ... a^{1 \dag} |\Om \rangle &\la_{0} + \frac{{\cal N}}{2} &1 = \left( \begin{smallmatrix} {\cal N} \\ {\cal N} \end{smallmatrix} \right) \end{array}
\]
Note that the total number of states is given by
    \[\sum^{{\cal N}}_{k = 0} \vecb {\cal N} \\ k \vece \eq \sum^{{\cal N}}_{k = 0} \vecb {\cal N} \\ k \vece \, 1^{k} \, 1^{{\cal N} - k} \eq 2^{{\cal N}} \ .
\]
Consider the following examples:
\begin{itemize}
\item ${\cal N} = 2$ vector - multiplet ($\la_{0} = 0$)
    \[\begin{array}{ccc} &\la = 0 & \\ \la = \frac{1}{2} & &\la = \frac{1}{2} \\ &\la = 1 & \end{array}
\]
We can see that this ${\cal N}=2$ multiplet can be decomposed in terms of
${\cal N}=1$ multiplets: one ${\cal N}=1$ vector and one ${\cal N}=1$ chiral multiplet.
\item ${\cal N} = 2$ hyper - multiplet ($\la_{0} = -\frac{1}{2}$)
    \[\begin{array}{ccc} &\la = -\frac{1}{2} & \\ \la = 0 & &\la = 0 \\ &\la = \frac{1}{2} & \end{array}
\]
Again this can be decomposed in terms of two ${\cal N}=1$ chiral multiplets.
\item ${\cal N} = 4$ vector - multiplet ($\la_{0} = -1$)
    \[\begin{array}{cc} 1 \times &\la = -1 \\ 4 \times &\la = -\frac{1}{2} \\ 6 \times &\la = \pm 0 \\ 4 \times &\la = +\frac{1}{2} \\ 1 \times &\la = +1 \end{array}
\]
This is the single ${\cal N}=4$ multiplet with states of helicity $\lambda
<2$. It consists of one ${\cal N}=2$ vector multiplet and two ${\cal N}=2$ hypermultiplets
plus their CPT conjugates (with opposite helicities). Or one ${\cal N}=1$ vector and three ${\cal N}=1$ chiral
multiplets plus their CPT conjugates.
\item ${\cal N} = 8$ maximum - multiplet ($\la_{0} = -2$)
    \[\begin{array}{cc} 1 \times &\la = \pm 2 \\ 8 \times &\la = \pm \frac{3}{2} \\ 28 \times &\la = \pm 1 \\ 56 \times &\la = \pm \frac{1}{2} \\ 70 \times &\la = \pm 0 \end{array}
\]
\end{itemize}
From these results we can extract very important general conclusions:
\begin{itemize}
\item In every multiplet: $\la_{\max} - \la_{\min} = \frac{{\cal N}}{2}$
\item Renormalizable theories have $|\la| \leq 1$ implying ${\cal N} \leq
  4$. Therefore ${\cal N}=4$ supersymmetry is the largest supersymmetry for
  renormalizable field theories. Gravity is not renormalizable!
\item {\it The maximum number of supersymmetries is ${\cal N}=8$}.
There is a strong belief that no massless particles of helicity $|\la|
> 2$ exist (so only have ${\cal N} \leq 8$). One argument is the fact that
massless particle of $|\la| > \frac{1}{2}$ and low momentum couple to
some conserved currents ($\pa_{\mu} j^{\mu} = 0$ in $\la = \pm 1$ -
electromagnetism, $\pa_{\mu} T^{\mu \nu}$ in $\la = \pm 2$ -
gravity). But there are no further conserved currents for $|\la| > 2$
(something that can also be seen from the Coleman Mandula
theorem). Also, ${\cal N} > 8$ would imply that there is more than one
graviton. See chapter 13 in \cite{WB1} on soft photons for a detailed
discussion of this
and the extension of his argument to supersymmetry in an article \cite{pend} by \textsc{Grisaru} and \textsc{Pendleton} (1977). Notice this is not a full no-go theorem, in
particular the constraint of low momentum has to be used.
\item {\it ${\cal N} > 1$ supersymmetries are non-chiral}. 
 We know that the Standard Model
  particles live on complex fundamental
representations. They are chiral since right handed quarks and leptons
 do not feel the weak interactions whereas left-handed ones do feel it
 (they are doublets under $SU(2)_L$). 
All ${\cal N} > 1$ multiplets, except for the ${\cal N} = 2$ hypermultiplet,
have $\la = \pm 1$ particles transforming in the adjoint
  representation which is real (recall that in $SU(N)$ theories the
 adjoint representation is obtained from the product of fundamental and
 complex conjugate representations and so is real)
and therefore non-chiral. Then
the $\la = \pm \frac{1}{2}$ particle within the multiplet would
  transform in the same representation and
 therefore be non-chiral. The only exception is 
 the ${\cal N} = 2$ hypermultiplets - for this
  the previous argument doesn't work because they do not include
  $\la=\pm 1$ states, but since $\la = \frac{1}{2}$- and $\la =
  -\frac{1}{2}$ states are in the same
 multiplet, there can't be chirality either in this multiplet.
Therefore only ${\cal N} = 1,0$ can be chiral, 
for instance ${\cal N} = 1$ with $\left( \begin{smallmatrix} \frac{1}{2} \\ 0 \end{smallmatrix} \right)$ predicting at least one extra
 particle for each Standard Model particle. But they have not been observed.
 Therefore the only hope for a realistic supersymmetric theory is:
  broken ${\cal N} = 1$ supersymmetry at low
energies $E \approx 10^{2} \ \te{GeV}$.
\end{itemize}

\subsection{Massive representations of ${\cal N}>1$ supersymmetry and BPS states}
\label{sec:MassiveRepresentationsOfExtendedSupersymmetry}

Now consider $p_{\mu} = (m  , \ 0  , \ 0  , \ 0)$, so
    \[\Bigl\{ Q_{\al}^{A} \ , \ \bar{Q}_{\dot{\be}B} \Bigr\} \eq 2 \, m \, \ccb 1 &0 \\ 0 &1 \cce \, \de^{A}\,_{B} \ .
\]
Contrary to the massless case, here the central charges can be
non-vanishing. Therefore we have to distinguish two cases:
\begin{itemize}
\item $Z^{AB}$ = 0

There are $2{\cal N}$ creation- and annihilation operators
    \[a_{\al}^{A} \ \ := \ \ \frac{Q_{\al}^{A}}{\sqrt{2m}} \co a^{A\dag}_{\dot{\al}} \ \ := \ \ \frac{\bar{Q}_{\dot{\al}}^{A}}{\sqrt{2m}}
\]
leading to $2^{2N}$ states, each of them with dimension $(2y + 1)$. In the ${\cal N} = 2$ case, we find:
    \[\begin{array}{rc} |\Om \rangle &1 \times \te{spin} \ 0 \\ a_{\dot{\al}}^{A\dag} \, |\Om \rangle &4 \times \te{spin} \ \frac{1}{2} \\ a_{\dot{\al}}^{A\dag} \, a_{\dot{\be}}^{B\dag} \, |\Om \rangle &3\times \te{spin} \ 0 \ ,\  3\times \te{spin} \ 1 \\ a_{\dot{\al}}^{A\dag} \, a_{\dot{\be}}^{B\dag} \, a_{\dot{\ga}}^{C\dag} \, |\Om \rangle  & 4 \times \te{spin} \ \frac{1}{2} \\ a_{\dot{\al}}^{A\dag} \, a_{\dot{\be}}^{B\dag} \, a_{\dot{\ga}}^{C\dag} \, a_{\dot{\de}}^{D\dag} \, |\Om \rangle & 1 \times \te{spin} \ 0 \end{array} \ ,
\]
i.e. as predicted $16 = 2^{4}$ states in total. Notice that these
multiplets are much larger than the massless ones with only $2^{\cal N}$
states, due to the fact that in that case, half of the supersymmetry
generators vanish ($Q_2^A=0$).

\item $Z^{AB} \neq 0$

Define the scalar quantity ${\cal H}$ to be
    \[{\cal H} \ \ := \ \ (\bar{\si}^{0})^{\dot{\be} \al} \, \Bigl\{ Q_{\al}^{A} \ - \ \Ga_{\al}^{A} \ , \ \bar{Q}_{\dot{\be}A} \ - \ \bar{\Ga}_{\dot{\be} A} \Bigr\} \ \ \geq \ \ 0 \ .
\]
As a sum of products $A A^\dag$, ${\cal H}$ is semi-positive, and the $\Ga_{\al}^{A}$ are defined as
    \[\Ga_{\al}^{A} \ \ := \ \ \ep_{\al \be} \, U^{AB} \, \bar{Q}_{\dot{\ga}} \, (\bar{\si}^{0})^{\dot{\ga}\be}
\]
for some unitary matrix $U$ (satisfying $UU^\dag = \mathds{1}$). Anticommutators $\{ Q_{\al}^{A}, \bar{Q}_{\dot{\be}}^{B} \}$ imply
    \[{\cal H} \eq 8 \, m \, {\cal N} \ - \ 2 \, \te{Tr} \Bigl\{Z \, U^\dag \ + \  U \, Z^\dag \Bigr\} \ \ \geq \ \ 0 \ .
\]
Due to the polar decomposition theorem, each matrix $Z$ can be written as a product $Z = HV$ of a positive hermitian $H = H^\dag$ and a unitary phase matrix $V = (V^\dag)^{-1}$. Choose $U = V$, then
    \[{\cal H} \eq 8 \, m \, {\cal N} \ - \ 4 \, \te{Tr} \Bigl\{ H \Bigr\} \eq 8\, m \, {\cal N} \ - \ 4 \, \te{Tr} \Bigl\{ \sqrt{Z^\dag Z} \Bigr\} \ \ \geq \ \ 0 \ .
\]
This is the BPS - bound for the mass $m$:
    \[\fbox{$\displaystyle m \ \ \geq \ \ \frac{1}{2{\cal N}} \; \te{Tr} \Bigl\{ \sqrt{Z^\dag Z} \Bigr\} $}
\]
States of minimal $m = \frac{1}{2{\cal N}} \te{Tr} \Bigl\{ \sqrt{Z^\dag Z}
\Bigr\}$ are called {\em BPS states} (due to \textsc{Bogomolnyi}, \textsc{Prasad} and \textsc{Sommerfeld}). They are characterized by a vanishing combination $Q_{\al}^{A} - \Ga_{\al}^{A}$, so the multiplet is shorter (similar to the massless case in which $Q_2^a=0$) having only $2^{{\cal N}}$ instead of $2^{2{\cal N}}$ states.

\noindent
In ${\cal N} = 2$, define the components of the antisymmetric $Z^{AB}$ to be
    \[Z^{AB} \eq \ccb 0 &q_{1} \\ -q_{1} &0 \cce  \so m \ \ \geq \ \ \frac{q_{1}}{2} \ .
\]
More generally, if ${\cal N} > 2$ (but ${\cal N}$ even)
    \[Z^{AB} \eq \left( \begin{array}{rrrrrrrr} 0 &q_{1} &0 &0 &0 &\cdots & & \\ -q_{1} &0 &0 &0 &0 &\cdots & & \\ 0 &0 &0 &q_{2} &0 &\cdots & & \\ 0 &0 &-q_{2} &0 &0 &\cdots & & \\ 0 &0 &0 &0 &\ddots & & & \\ \vdots &\vdots &\vdots &\vdots & &\ddots & & \\ & & & & & &0 &q_{\frac{{\cal N}}{2}} \\ & & & & & &-q_{\frac{{\cal N}}{2}} &0 \end{array} \right) \ ,
\]
the BPS conditions holds block by block: $2m \geq q_{i}$. To see that, define an $\cal H$ for each block. If $k$ of the $q_{i}$ are equal to $2m$, there are $2{\cal N} - 2k$ creation operators and $2^{2({\cal N} - k)}$ states.
\begin{align*}
k \eq 0 \ \ \ &\Longrightarrow \ \ \ 2^{2{\cal N}} \ \te{states, long multiplet} \\
0 \ \ < \ \ k \ \ < \ \ \frac{{\cal N}}{2} \ \ \ &\Longrightarrow \ \ \ 2^{2({\cal N} - k)} \ \te{states, short multiplets} \\
k \eq \frac{{\cal N}}{2} \ \ \ &\Longrightarrow \ \ \ 2^{{\cal N}} \ \te{states, ultra - short multiplet}
\end{align*}
Let us conclude this section about non-vanishing central charges with some remarks:

\begin{itemize}
\item [(i)] BPS states and bounds started in {\em soliton} (monopole-)
  solutions of \textsc{Yang Mills} systems, which are localized
  finite energy solutions of the classical equations of motion. The bound refers to an energy bound.

\item [(ii)] The BPS states are stable since they are the lightest charged particles.
\item [(iii)] The equivalence of mass and charge reminds that us charged black holes.
Actually, extremal black holes (which are the end points of the
\textsc{Hawking} evaporation and therefore stable) happen to be BPS states for
extended supergravity theories.
\item [(iv)] BPS states are important in understanding strong-weak
  coupling dualities in field- and string theory.
In particular the fact that they correspond to short multiplets allows
  to extend them from
weak to strong coupling since the size of a multiplet is not expected
  to change by continuous changes in the coupling from weak to strong.
\item [(v)] In string theory some of the extended objects known as {\em D branes} are BPS.
\end{itemize}
\end{itemize}

\chapter{Superfields and superspace}
\label{sec:SuperfieldsAndSuperspace}

So far, we have just considered 1 particle states in supermultiplets. The goal is a supersymmetric field theory describing interactions. Recall that particles are described by fields $\vph(x^{\mu})$ with properties:
\begin{itemize}
\item function of coordinates $x^{\mu}$ in Minkowski spacetime
\item transformation of $\vph$ under Lorentz group
\end{itemize}
In the supersymmetric case, we want to deal with objects $\Phi(X)$,
\begin{itemize}
\item function of coordinates $X$ in superspace
\item transformation of $\Phi$ under super Poincaré
\end{itemize}
But what is that superspace? In any case, it should not be confused with 'stuperspace' \cite{stup}.

\section{Basics about superspace}

\subsection{Groups and cosets}
\label{sec:GroupsAndCosets}

We know that every continuous group $G$ defines a manifold ${\cal M}_{G}$ via
    \[\La: \ G \ \ \longrightarrow \ \ {\cal M}_{G} \co \Bigl\{ g = \exp(i\al_{a}T^{a}) \Bigr\} \ \ \longrightarrow \ \ \Bigl\{ \al_{a} \Bigr\} \ ,
\]
where $\dim G = \dim {\cal M}_{G}$. Consider for example:
\begin{itemize}
\item $G = U(1)$ with elements $g = \exp(i\al Q)$, then $\al \in [0, 2\pi]$, so the corresponding manifold is the 1 - sphere (a circle) ${\cal M}_{U(1)} = S^{1}$.
\item $G = SU(2)$ with elements $g = \left( \begin{smallmatrix} \al &\be \\ -\be^{*} &\al^{*} \end{smallmatrix} \right)$, where complex parameters $\al$ and $\be$ satisfy $|\al |^{2} + |\be |^{2} = 1$. Write $\al = x_{1} + ix_{2}$ and $\be = x_{3} + ix_{4}$ for $x_{k} \in \mathbb R$, then the constraint for $p$, $q$ implies $\sum^{4}_{k = 1} x_{k}^{2} = 1$, so ${\cal M}_{SU(2)} = S^{3}$
\item $G = SL(2, \mathbb C)$ with elements $g = H \cdot V$, $V \in SU(2)$ and
 $H = H^\dag$ positive, $\det H = 1$. Writing the generic element $h
 \in H$
 as $h = x_{\mu} \si^{\mu} = \left( \begin{smallmatrix} x_{0} + x_{3} & x_{1} + ix_{2} \\ 
x_{1} - ix_{2} & x_{0} - x_{3} \end{smallmatrix} \right)$, the determinant - constraint is $(x_{0})^{2} - \sum^{3}_{k = 1} (x_{k})^{2} = 1$, so ${\cal M}_{SL(2, \mathbb C)} = \mathbb R^{3} \times S^{3}$.
\end{itemize}
To be more general, let's define a coset $G/H$ where $g \in G$ is identified with $g \cdot h \ \forall \ h \in H$, e.g.
\begin{itemize}
\item $G = U_{1}(1) \times U_{2}(1) \, \inn \, g = \exp \bigl(i(\al_{1}Q_{1} + \al_{2}Q_{2}) \bigr)$, $H = U_{1}(1) \, \inn \, h = \exp(i\be Q_{1})$. In $G / H = \bigl(U_{1}(1) \times U_{2}(1) \bigr) / U_{1}(1)$, the identification is
    \[g\, h \eq \exp \Bigl\{i \, \bigl((\al_{1} \, + \, \be) \, Q_{1} \ + \ \al_{2} \, Q_{2} \bigr) \Bigr\} \eq \exp \bigl(i \, (\al_{1} \,Q_{1} \ + \ \al_{2} \, Q_{2}) \bigr) \eq g \ ,
\]
so only $\al_{2}$ contains an effective information, $G/H = U_{2}(1)$.
\item $G/H = SU(2)/U(1) \cong SO(3)/SO(2)$: Each $g\in SU(2)$ can be written as $g= \left( \begin{smallmatrix} \alpha &\beta \\ -\beta^* &
  \alpha^* \end{smallmatrix} \right)$, identifying this by a $U(1)$ element
  $\te{diag}(e^{i\gamma}, e^{-i\gamma})$ makes $\alpha$ effectively 
real. Hence, the parameter space is the 2 sphere
  ($\beta_1^2+\beta_2^2 + \alpha^2=1$), i.e. ${\cal M}_{SU(2)/U(1)}= S^2$.
\item More generally, ${\cal M}_{SO(n+1)/SO(n)}= S^n$.

\begin{figure}[ht]
    \centering
        \includegraphics[width=0.70\textwidth]{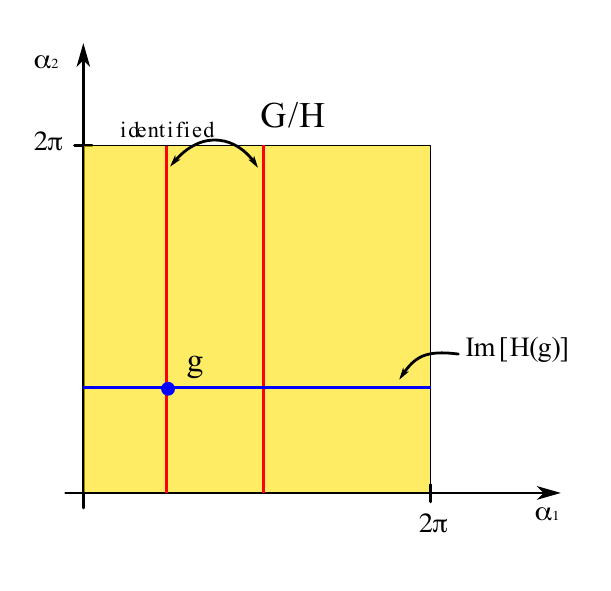}
        \caption{\footnotesize{Illustration of the coset identity $G/H = \bigl(U_{1}(1) \times U_{2}(1) \bigr) / U_{1}(1) = U_2(1)$: The blue horizontal line shows the orbit of some $G= U_{1}(1) \times U_{2}(1)$ element $g$ under the $H=U_1(1)$ group which is divided out. All its points are identified in the coset. Any red vertical line contains all the distinct coset elements and is identified with its neighbours in $\al_1$ direction.}}
\end{figure}
\item $\te{Minkowski}=\te{Poincaré} \ / \ \te{Lorentz} = \{ \om^{\mu \nu} , a^{\mu} \} / \{ \om^{\mu \nu} \}$ simplifies to the translations $\{a^{\mu} = x^{\mu} \}$ which can be identified with Minkowski space.
\end{itemize}
We define ${\cal N} = 1$ {\em superspace} to be the coset
    \[\te{Super Poincaré} \ / \ \te{Lorentz} \eq \Bigl\{ \om^{\mu \nu} , a^{\mu} , \theta^{\al} , \bar{\theta}_{\dot{\al}} \Bigr\} \ / \ \Bigl\{ \om^{\mu \nu} \Bigr\} \ .
\]
Recall that the general element $g$ of super Poincaré group is given by
    \[g \eq \exp \bigl( i \, (\om^{\mu \nu} \, M_{\mu \nu} \ + \ a^{\mu} \, P_{\mu} \ + \ \theta^{\al} \, Q_{\al} \ + \ \thb_{\dot{\al}} \, \bar{Q}^{\dot{\al}}) \bigr) \ ,
\]
where Grassmann parameters $\theta^{\al}$, $\thb_{\dot{\be}}$ reduce anticommutation relations for $Q_{\al}$, $\bar{Q}^{\dot{\be}}$ to commutators:
    \[\Bigl\{ Q_{\al} \ , \ \bar{Q}_{\dot{\al}} \Bigr\} \eq 2 \, (\si^{\mu})_{\al \dot{\al}} \, P_{\mu} \so \Bigl[ \theta^{\al} \, Q_{\al} \ , \ \thb^{\dot{\be}} \, \bar{Q}_{\dot{\be}} \Bigr] \eq 2  \, \theta^{\al} \, (\si^{\mu})_{\al \dot{\be}} \, \thb^{\dot{\be}} \, P_{\mu}
\]

\subsection{Properties of Grassmann variables}
\label{sec:PropertiesOfGrassmannVariables}

Superspace was first introduced in 1974 by \textsc{Salam} and \textsc{Strathdee} \cite{Salam:1974yz,Salam:1974jj}. Recommendable books about this subject are \cite{BZ} and \cite{BDW}.

\noindent
Let us first consider one single variable $\theta$. When trying to expand a generic (analytic) function in $\theta$ as a power series, the fact that $\theta$ squares to zero, $\theta^{2} = 0$, cancels all the terms except for two,
    \[f(\theta) \eq \sum^{\infty}_{k = 0} f_{k} \, \theta^{k} \eq f_{0} \ + \ f_{1} \, \theta \ + \ f_{2} \, \underbrace{\theta^{2}}_{0} + \underbrace{...}_{0} \eq f_{0} \ + \ f_{1} \, \theta \ .
\]
So the most general function $f(\theta)$ is linear. Of course, its derivative is given by $\frac{\dd f}{\dd \theta} = f_{1}$. For integrals, define
    \[\int \dd \theta \ \frac{\dd f}{\dd \theta} \ \ := \ \ 0 \so \int \dd \theta \eq 0 \ ,
\]
as if there were no boundary terms. Integrals over $\theta$ are left to talk about: To get a non-trivial result, define
    \[\int \dd \theta \ \theta \ \ := \ \ 1 \so \de(\theta) \eq \theta \ .
\]
The integral over a function $f(\theta)$ is equal to its derivative,
    \[\int \dd \theta \ f(\theta) \eq \int \dd \theta \ (f_{0} \ + \ f_{1}  \, \theta) \eq f_{1} \eq \frac{ \dd f}{\dd \theta} \ .
\]
Next, let $\theta^{\al}$, $\thb_{\dot{\al}}$ be spinors of Grassmann numbers. Their squares are defined by
    \[\theta \theta \ \ := \ \ \theta^{\al} \, \theta_{\al} \co \thb \thb \ \ := \ \ \thb_{\dot{\al}} \, \thb^{\dot{\al}}
\]
    \[\Longrightarrow \ \ \ \theta^{\al} \, \theta^{\be} \eq -\frac{1}{2} \; \ep^{\al \be} \, \theta \theta \co \thb^{\dot{\al}} \, \thb^{\dot{\be}} \eq \frac{1}{2} \; \ep^{\dot{\al} \dot{\be}} \, \thb \thb \ .
\]
Derivatives work in analogy to Minkowski coordinates:
    \[\frac{\pa \theta^{\be}}{\pa \theta^{\al}} \eq \de_{\al}\,^{\be} \so \frac{\pa \thb^{\dot{\be}}}{\pa \thb^{\dot{\al}}} \eq \de_{\dot{\al}}\,^{\dot{\be}}
\]
As to multi integrals,
    \[\int \dd \theta^{1} \int \dd \theta^{2} \ \theta^{2} \, \theta^{1} \eq \frac{1}{2} \int \dd \theta^{1} \int \dd \theta^{2} \ \theta \theta \eq 1 \ ,
\]
which justifies the definition
    \[\frac{1}{2} \int \dd \theta^{1} \int \dd \theta^{2} \ \ =: \ \ \int \dd^{2}\theta \co \int \dd^{2} \theta \ \theta \theta \eq 1 \co \int \dd^{2} \theta \int \dd^{2} \thb \ (\theta \theta) \, (\thb \thb) \eq 1 \ .
\]
Written in terms of $\ep$:
    \[\dd^{2} \theta \eq -\frac{1}{4} \; \dd \theta^{\al} \, \dd \theta^{\be} \, \ep_{\al \be} \co \dd^{2} \thb \eq \frac{1}{4} \; \dd\thb^{\dot{\al}} \, \dd\thb^{\dot{\be}} \, \ep_{\dot{\al} \dot{\be}} \ .
\]
One can again identify integration and differentiation:
    \[\int \dd^{2} \theta \eq \frac{1}{4} \; \ep^{\al \be} \; \frac{\pa}{\pa \theta^{\al}} \; \frac{\pa}{\pa \theta^{\be}} \co \int \dd^{2} \thb \eq -\frac{1}{4} \; \ep^{\dot{\al} \dot{\be}} \; \frac{\pa}{\pa \thb^{\dot{\al}}} \; \frac{\pa}{\pa \thb^{\dot{\be}}}\, .
\]

\subsection{Definition and transformation of the general scalar superfield}
\label{sec:DefinitionAndTranformationOfTheScalarSuperfield}

To define a {\em superfield}, recall properties of scalar fields $\vph(x^{\mu})$:
\begin{itemize}
\item function of spacetime coordinates $x^{\mu}$
\item transformation under Poincaré, e.g. under translations:

\noindent
Treating $\vph$ as an operator, a translation with parameter $a_{\mu}$ will change it to
    \[\vph \ \ \mapsto \ \ \exp(-ia_{\mu} \, P^{\mu}) \, \vph \, \exp(ia_{\mu} \, P^{\mu}) \ .
\]
But $\vph(x^{\mu})$ is also a Hilbert vector in some function space ${\cal F}$, so
    \[\vph(x^{\mu})  \ \ \mapsto \ \  \exp(-ia_{\mu} \, {\cal P}^{\mu}) \, \vph(x^{\mu}) \ \ =: \ \ \vph(x^{\mu} \, - \, a^{\mu}) \so {\cal P}_{\mu} \eq -i\pa_{\mu} \ .
\]
${\cal P}$ is a representation of the abstract operator $P^{\mu}$ acting on ${\cal F}$. Comparing the two transformation rules to first order in $a_{\mu}$, get the following relationship:
    \[\bigl(1 \ - \ ia_{\mu} \, P^{\mu} \bigr) \, \vph \, \bigl(1 \ + \ ia_{\mu} \, P^{\mu} \bigr) \eq
 \bigl(1 \ - \ ia_{\mu} \, {\cal P}^{\mu} \bigr) \, \vph \so i \, \Bigl[\vph \ ,
 \ a_{\mu} \, P^{\mu} \Bigr] \eq -ia^{\mu} \, {\cal P}_{\mu} \, \vph \eq - a^{\mu} \,
 \pa_{\mu} \, \vph
\]
\end{itemize}
For a general scalar superfield $S(x^{\mu}, \theta_{\al}, \thb_{\dot{\al}})$, one can do an expansion in powers of $\theta_{\al}$, $\thb_{\dot{\al}}$ with a finite number of nonzero terms:
\begin{align*}
S(x^{\mu}, \theta_{\al}, \thb_{\dot{\al}}) \ \ &= \ \ \vph(x) \ + \ \theta \psi(x) \ + \ \thb \bar{\chi}(x) \ + \ \theta \theta \, M(x) \ + \ \thb \thb \, N(x) \ + \ (\theta \, \si^{\mu} \, \thb) \, V_{\mu}(x) \\
&\ \ \ \ \ \ + \ (\theta \theta) \, \thb \bar{\la}(x) \ + \ (\thb \thb) \, \theta \rho(x) \ + \ (\theta \theta) \, (\thb \thb) \, D(x)
\end{align*}
Transformation of $S(x^{\mu}, \theta_{\al}, \thb_{\dot{\al}})$ under super Poincaré, firstly as a field operator
    \[S(x^{\mu}, \theta_{\al}, \thb_{\dot{\al}}) \ \ \mapsto \ \ \exp \bigl(-i  \, (\ep Q \, + \, \bar{\ep} \bar{Q}) \bigr) \, S \, \exp \bigl(i \, (\ep Q \, + \, \bar{\ep} \bar{Q}) \bigr) \ ,
\]
secondly as a Hilbert vector
    \[S(x^{\mu}, \theta_{\al}, \thb_{\dot{\al}}) \ \ \mapsto \ \ \exp \bigl(i \, (\ep {\cal Q} \, + \, \bar{\ep} \bar{{\cal Q}}) \bigr) \, S(x^{\mu}, \theta_{\al}, \thb_{\dot{\al}}) \eq S \bigl(x^{\mu} \, - \, ic(\ep \si^{\mu}\thb) \, + \, ic^{*}(\theta \si^{\mu} \bar{\ep}), \, \theta \, + \, \ep, \, \thb \, + \, \bar{\ep} \bigr) \ .
\]
Here, $\ep$ denotes a parameter, ${\cal Q}$ a representation of the spinorial generators $Q_{\al}$ acting on functions of $\theta$, $\thb$, and $c$ is a constant to be fixed later, which is involved in the translation
    \[x^{\mu} \ \ \mapsto \ \ x^{\mu} \ - \ ic \, (\ep \, \si^{\mu} \, \thb) \ + \ ic^{*} \, (\theta \, \si^{\mu} \, \bar{\ep}) \ .
\]
The translation of arguments $x^{\mu}$, $\theta_{\al}$, $\thb_{\dot{\al}}$ implies,

\medskip    
\framebox{\begin{minipage}{5.9in}
\begin{align*}
    {\cal Q}_{\al} \ \ &= \ \ -i \; \frac{\pa}{\pa \theta^{\al}} \ - \
    c \, (\si^{\mu})_{\al \dot{\be}} \, \thb^{\dot{\be}} \; \frac{\pa}{\pa
      x^{\mu}} \\
\bar{{\cal Q}}_{\dot{\al}} \ \ &= \ \ +i \; \frac{\pa}{\pa \thb^{\dot{\al}}} \ + \ c^{*} \,
      \theta^{\be} \, (\si^{\mu})_{\be \dot{\al}} \; \frac{\pa}{\pa x^{\mu}} \\
 {\cal P}_{\mu} \ \ &= \ \ -i\pa_{\mu} \ ,
\end{align*}
\end{minipage}}

\medskip
\noindent
where $c$ can be determined from the commutation relation which, of course, holds in any representation:
    \[\Bigl\{ {\cal Q}_{\al} \ , \ \bar{{\cal Q}}_{\dot{\al}} \Bigr\} \eq 2 \, (\si^{\mu})_{\al \dot{\al}} \, {\cal P}_{\mu} \so \te{Re} \{c \} \eq 1
\]
It is convenient to set $c = 1$. Again, a comparison of the two expressions (to first order in $\ep$) for the transformed superfield $S$ is the key to get its commutation relations with $Q_{\al}$:
    \[ \fbox{$\displaystyle  i \, \Bigl[ S \ , \ \ep Q \, + \, \bar{\ep} \bar{Q} \Bigr] \eq i \, \bigl(\ep {\cal Q} \ + \ \bar{\ep} \bar{{\cal Q}} \bigr) \, S \eq \de S $}
\]
Knowing the ${\cal Q}$, $\bar{{\cal Q}}$ and $S$, we get explicit terms for the change in the different parts of $S$:
\begin{align*}
\de \vph \ \ &= \ \ \ep \psi \ + \ \bar{\ep} \bar{\chi} \\
\de \psi \ \ &= \ \ 2 \, \ep \, M \ + \ \si^{\mu} \, \bar{\ep} \, (i \pa_{\mu} \vph \ + \ V_{\mu}) \\
\de \bar{\chi} \ \ &= \ \ 2 \, \bar{\ep} \, N \ - \ \ep \, \si^{\mu} \, (i\pa_{\mu} \vph \ - \ V_{\mu}) \\
\de M \ \ &= \ \ \bar{\ep} \bar{\la} \ - \ \frac{i}{2} \; \pa_{\mu} \psi \, \si^{\mu} \, \bar{\ep} \\
\de N \ \ &= \ \ \ep \rho \ + \ \frac{i}{2} \; \ep \, \si^{\mu} \, \pa_{\mu} \bar{\chi} \\
\de V_{\mu} \ \ &= \ \ \ep \, \si_{\mu} \, \bar{\la} \ + \ \rho \, \si_{\mu} \, \bar{\ep} \ + \ \frac{i}{2} \; (\pa^{\nu} \psi \, \si_{\mu} \ \bar{\si}_{\nu} \, \ep \ - \ \bar{\ep} \, \bar{\si}_{\nu} \, \si_{\mu} \, \pa^{\nu} \bar{\chi}) \\
\de \bar{\la} \ \ &= \ \ 2 \, \bar{\ep} \, D \ + \ \frac{i}{2} \; (\bar{\si}^{\nu} \, \si^{\mu} \, \bar{\ep}) \, \pa_{\mu} V_{\nu} \ + \ i\bar{\si}^{\mu} \, \ep \, \pa_{\mu} M \\
\de \rho \ \ &= \ \ 2 \, \ep \, D \ - \ \frac{i}{2} \; (\si^{\nu} \, \bar{\si}^{\mu} \, \ep) \, \pa_{\mu} V_{\nu} \ + \ i\si^{\mu} \, \bar{\ep} \, \pa_{\mu} N \\
\de D \ \ &= \ \ \frac{i}{2} \; \pa_{\mu} \, (\ep \, \si^{\mu} \, \bar{\la} \ - \ \rho \, \si^{\mu} \, \bar{\ep})
\end{align*}
Note that $\de D$ is a total derivative.

\noindent
\paragraph{Exercise 3.1:} {\it Derive these transformation rules. It might be useful to note that $\frac{\pa \theta^\al}{\pa \theta^\be} = +\de^\al_\be$ implies $\frac{\pa \theta_\al}{\pa \theta_\be} = -\de_\al^\be$ and similarly $\frac{\pa \bar \theta^{\dot \al}}{\pa \bar \theta^{\dot \be}} = +\de^{\dot \al}_{\dot \be} \ \Rightarrow \ \frac{\pa \bar \theta_{\dot \al}}{\pa \bar \theta_{ \dot \be}} = -\de_{\dot \al}^{\dot \be}$.}

\subsection{Remarks on superfields}
\label{sec:RemarksOnSuperfields}

\begin{itemize}
\item If $S_{1}$ and $S_{2}$ are superfields then so is the product $S_{1}S_{2}$:
\begin{align*}
\de (S_{1} \, S_{2}) \ \ &= \ \ i \, \Bigl[S_{1} \, S_{2} \ , \ \ep Q \, + \, \bar{\ep} \bar{Q} \Bigr] \eq i S_{1} \, \Bigl[S_{2} \ , \ \ep Q \, + \, \bar{\ep} \bar{Q} \Bigr] \ + \ i \, \Bigl[S_{1} \ , \ \ep Q \, + \, \bar{\ep} \bar{Q} \Bigr] \, S_{2} \\
&= \ \ S_{1} \, \bigl( i \, ( \ep {\cal Q} \, + \, \bar{\ep} \bar{{\cal Q}}) \, S_{2} \bigr) \ + \ \bigl( i \, ( \ep {\cal Q} \, + \, \bar{\ep} \bar{{\cal Q}}) \, S_{1} \bigr) \, S_{2} \\
&= \ \ i \, ( \ep {\cal Q} \, + \, \bar{\ep} \bar{{\cal Q}}) \, (S_{1} \, S_{2})
\end{align*}
In the last step, we used the Leibnitz property of the ${\cal Q}$ and $\bar{{\cal Q}}$ as differential operators.
\item Linear combinations of superfields are superfields again
  (straightforward proof).
\item $\pa_{\mu} S$ is a superfield but $\pa_{\al} S$ is not:
    \begin{align*}
    \de(\pa_{\al}S) \ \ &= \ \ i \, \Bigl[ \pa_{\al} S \ , \ \ep Q \, + \, \bar{\ep} \bar{Q} \Bigr] \eq i \pa_{\al} \, \Bigl[ S \ , \ \ep Q \, + \, \bar{\ep} \bar{Q} \Bigr] \\
    &= \ \ i \pa_{\al} \, ( \ep {\cal Q} \ + \ \bar{\ep} \bar{{\cal Q}}) \, S \ \ \neq \ \ i \, ( \ep {\cal Q} \ + \ \bar{\ep} \bar{{\cal Q}}) \, (\pa_{\al} S)
\end{align*}
The problem is $[\pa_{\al}, \ep {\cal Q} + \bar{\ep} \bar{{\cal Q}}] \neq 0$. We need to define a covariant derivative,
    \[{\cal D}_{\al} \ \ := \ \ \pa_{\al} \ + \ i(\si^{\mu})_{\al \dot{\be}} \, \thb^{\dot{\be}} \, \pa_{\mu} \co \bar{{\cal D}}_{\dot{\al}} \ \ := \ \ -\bar{\pa}_{\dot{\al}} \ - \ i \theta^{\be} \, (\si^{\mu})_{\be \dot{\al}} \, \pa_{\mu}
\]
which satisfies
    \[\Bigl\{ {\cal D}_{\al} \ , \ {\cal Q}_{\be} \Bigr\} \eq \Bigl\{ {\cal D}_{\al} \ , \ \bar{{\cal Q}}_{\dot{\be}} \Bigr\} \eq \Bigl\{ \bar{{\cal D}}_{\dot{\al}} \ , \ {\cal Q}_{\be} \Bigr\} \eq \Bigl\{ \bar{{\cal D}}_{\dot{\al}} \ , \ \bar{{\cal Q}}_{\dot{\be}} \Bigr\} \eq 0
\]
and therefore
    \[\Bigl[ {\cal D}_{\al} \ , \ \ep {\cal Q} \, + \, \bar{\ep} \bar{{\cal Q}} \Bigr] \eq 0 \so {\cal D}_{\al} S \ \ \ \te{is superfield} \ .
\]
Also note that supercovariant derivatives satisfy anticommutation relations
\[ \Bigl\{ {\cal D}_{\al} \ , \ \bar{{\cal D}}_{\dot{\be}} \Bigr\} \eq - 2i \, (\si^{\mu})_{\al \dot{\be}} \, \pa_{\mu} \co  \Bigl\{ {\cal D}_{\al} \ , \ {\cal D}_{\be} \Bigr\} \eq \Bigl\{ \bar{{\cal D}}_{\dot{\al}} \ , \ \bar{{\cal D}}_{\dot{\be}} \Bigr\} \eq 0 \ . \]
\item $S = f(x)$ is a superfield only if $f = const$, otherwise, there would be some $\de \psi \propto \ep \pa^{\mu} f$. For constant spinor $c$, $S = c \theta$ is not a superfield due to $\de \phi = \ep c$.
\end{itemize}
$S$ is \textbf{not} an irreducible representation of supersymmetry, so we
can eliminate some of its components keeping it still as a
superfield. In general we can impose consistent constraints on $S$,
leading to smaller superfields that can be irreducible representations
of the supersymmetry algebra. To give a list of some relevant superfields:
\begin{itemize}
\item chiral superfield $\Phi$ such that $\bar{{\cal D}}_{\dot{\al}} \Phi = 0$
\item antichiral superfield $\bar{\Phi}$ such that ${\cal D}_{\al} \bar{\Phi} = 0$
\item vector (or real) superfield $V = V^\dag$
\item linear superfield $L$ such that ${\cal D} {\cal D} L = 0$ and $L = L^\dag$.
\end{itemize}

\section{Chiral superfields}
\label{sec:ChiralSuperfields}

We want to find the components of a superfields $\Phi$ satisfying $\bar{{\cal D}}_{\dot{\al}} \Phi = 0$. Define
    \[y^{\mu} \ \ := \ \ x^{\mu} \ \ + \ \ i \theta \, \si^{\mu} \, \thb \ .
\]
If $\Phi = \Phi(y , \theta , \thb)$, then
\begin{align*}
\bar{{\cal D}}_{\dot{\al}} \Phi \ \ &= \ \ -\bar{\pa}_{\dot{\al}} \Phi \ - \ \frac{\pa \Phi}{\pa y^{\mu}}  \; \frac{\pa y^{\mu}}{\pa \thb^{\dot{\al}}} \ - \ i \theta^{\be} \, (\si^{\mu})_{\be \dot{\al}} \, \pa_{\mu} \Phi \\
&= \ \ -\bar{\pa}_{\dot{\al}} \Phi \ - \ \pa_{\mu} \Phi \, (-i\theta  \, \si^{\mu})_{\dot{\al}} \ - \ i \theta^{\be} \, (\si^{\mu})_{\be \dot{\al}}  \ \pa_{\mu} \Phi \\
&= \ \ -\bar{\pa}_{\dot{\al}} \Phi \eq 0 \ ,
\end{align*}
so there is no $\thb^{\dot{\al}}$ - dependence and $\Phi$ depends only
on $y$ and $\theta$. In components, one finds
    \[\Phi(y^{\mu} , \theta^{\al}) \eq \vph(y^{\mu}) \ + \ \sqrt{2} \, \theta \psi(y^{\mu}) \ + \ \theta \theta \, F(y^{\mu}) \ ,
\]
where the left handed supercovariant derivative acts as ${\cal D}_\al = \pa_\al + 2i(\si^{\mu} \bar{\theta})_{\al} \frac{\pa}{\pa y^\mu}$ on $\Phi(y^{\mu} , \theta^{\al})$.

\noindent
The physical components of a chiral superfield are: $\vph$ represents
a scalar part (squarks, sleptons, Higgs), $\psi$ some $s =
\frac{1}{2}$ particles (quarks, leptons, Higgsino) and $F$ is an
{\em auxiliary field} in a way to be defined later. Off shell, there are 4 bosonic
(complex $\vph$, $F$) and 4 fermionic (complex $\psi_{\al}$)
components.
Reexpress $\Phi$ in terms of $x^{\mu}$:

\medskip
\framebox{\begin{minipage}{5.9in}
\begin{align*}
    \Phi(x^{\mu} , \theta^{\al} , \thb^{\dot{\al}}) \ \ &= \ \ \vph(x) \ + \
    \sqrt{2} \, \theta \psi(x) \ + \ \theta \theta \, F(x) \ +
\ i \theta \, \si^{\mu} \, \thb  \, \pa_{\mu} \vph(x) \\
& \ \ \ \ - \ \frac{i}{\sqrt{2}} \; (\theta \theta) \, \pa_{\mu} \psi(x) \, \si^{\mu} \, \thb \ -
 \ \frac{1}{4} \; (\theta \theta) \, (\thb \thb) \, \pa_{\mu} \pa^{\mu} \vph(x)
\end{align*}
\end{minipage}}

\paragraph{Exercise 3.2:}
{\it Verify by explicit computation that this component expression for $\Phi$ satisfies $\bar{ {\cal D}}_{\dot \alpha} \Phi = 0$.

\bigskip
\noindent
Under supersymmetry transformation
    \[\de \Phi \eq i \, \bigl(\ep  {\cal Q} \ + \ \bar{\ep} \bar{{\cal Q}} \bigr) \, \Phi \ ,
\]
find for the change in components}

\medskip
\framebox{\begin{minipage}{5.9in}
\begin{align*}
\de \vph \ \ &= \ \ \sqrt{2} \, \ep \psi \\
\de \psi \ \ &= \ \ i\sqrt{2} \, \si^{\mu} \, \bar{\ep} \, \pa_{\mu} \vph \ + \ \sqrt{2} \, \ep \, F \\
\de F \ \ &= \ \ i\sqrt{2} \, \bar{\ep} \, \bar{\si}^{\mu} \, \pa_{\mu} \psi \ .
\end{align*}
\end{minipage}}

\bigskip
\noindent
So $\de F$ is another total derivative term, just like $\de D$ in a general superfield. Note that:
\begin{itemize}
\item The product of chiral superfields is a chiral superfield. In
  general, any {\it holomorphic} function $f(\Phi)$ of chiral $\Phi$ is chiral.
\item If $\Phi$ is chiral, then $\bar{\Phi} = \Phi^\dag$ is antichiral.
\item $\Phi^\dag \Phi$ and $\Phi^\dag + \Phi$ are real superfields but neither chiral nor antichiral.
\end{itemize}

\section{Vector superfields}
\label{sec:VectorSuperfields}

\subsection{Definition and transformation of the vector superfield}
\label{sec:DefinitionAndTransformationOfTheVectorSuperfield}

The most general vector superfield $V(x,\theta, \thb) = V^\dag(x,\theta,\thb)$ has the form
\begin{align*}
V(x,\theta,\thb) \ \ &= \ \ C(x) \ + \ i\theta \chi(x) \ - \ i \thb \bar{\chi}(x) \ + \ \frac{i}{2} \; \theta \theta \, \bigl(M(x) \, + \, iN(x) \bigr) \ - \ \frac{i}{2} \; \thb \thb \, \bigl(M(x) \, - \, iN(x) \bigr) \\
&\ \ \ \ \ \ + \ \theta \, \si^{\mu} \, \thb \, V_{\mu}(x) \ + \ i \theta \theta \, \thb \left( -i \bar{\la}(x) \, + \, \frac{i}{2} \bar{\si}^{\mu} \pa_{\mu} \chi(x) \right) \\
&\ \ \ \ \ \ - \ i\thb \thb \, \theta \left( i\la(x) \, - \, \frac{i}{2} \si^{\mu} \pa_{\mu} \bar{\chi}(x) \right) \ + \ \frac{1}{2} \; (\theta \theta) \, (\thb \thb) \, \left(  D \, - \, \frac{1}{2} \pa_{\mu} \pa^{\mu} C \right) \ .
\end{align*}
These are 8 bosonic components $C$, $M$, $N$, $D$, $V_{\mu}$ and 4 + 4 fermionic ones $(\chi_{\al} , \ \la_{\al})$.

\noindent
If $\La$ is a chiral superfield, then $i(\La - \La^\dag)$ is a vector superfield. It has components:
\begin{align*}
C \ \ &= \ \ i \, \bigl(\vph \ - \ \vph^\dag \bigr) \\
\chi \ \ &= \ \ \sqrt{2} \, \psi \\
\frac{1}{2} \; (M \ + \ iN) \ \ &= \ \ F \\
V_{\mu} \ \ &= \ \ -\pa_{\mu} \bigl(\vph \ + \ \vph^\dag \bigr) \\
\la \ \ = \ \ D \ \ &= \ \ 0
\end{align*}
We can define a generalized gauge transformations to vector fields via
    \[V \ \ \mapsto \ \ V \  - \ \frac{i}{2} \; \bigl(\La \ - \ \La^\dag \bigr) \ ,
\]
which induces a standard gauge transformation for the vector component of $V$
    \[V_{\mu} \ \ \mapsto \ \ V_{\mu} \ + \ \pa_{\mu} \, \bigl[ \, \te{Re}(\vph) \, \bigr] \ \ =: \ \ V_{\mu} \ - \ \pa_{\mu} \al \ .
\]
Then we can choose $\vph$, $\psi$, $F$ within $\La$ to gauge away some of the components of $V$.

\subsection{Wess Zumino gauge}
\label{sec:WessZuminoGauge}

We can choose the components of $\Lambda$ above: $\varphi, \psi, F$
in such a way to set $C=\chi=M=N=0$. This defines the {\em Wess Zumino (WZ) gauge}.
A vector superfield in Wess Zumino gauge reduces to the form
    \[\framebox{$ \displaystyle \Bigl. \Bigr. V_{\te{WZ}}(x,\theta, \thb) \eq (\theta \, \si^{\mu} \, \thb) \, V_{\mu}(x) \ + \ (\theta \theta) \, \bigl(\thb \bar{\la}(x) \bigr) \ + \ (\thb \thb) \, \bigl(\theta \la(x) \bigr) \ + \ \frac{1}{2} \; (\theta \theta) \, (\thb \thb) \, D(x) \ . $}
\]
The physical components of a vector superfield are:
 $V_{\mu}$ corresponding  to gauge particles ($\ga$, $W^{\pm}$, $Z$,
 gluon), the $\la$ and $\bar{\la}$ to gauginos and $D$ is an
 auxiliary field in a way to be defined later. Powers of $V_{\te{WZ}}$ are given by
    \[V_{\te{WZ}}^{2} \eq \frac{1}{2} \; (\theta \theta) \, (\thb \thb) \, V^{\mu} \, V_{\mu} \co V_{\te{WZ}}^{2 + n} \eq 0 \ \forall \ n \in \mathbb N \ .
\]
Note that the Wess Zumino gauge is not supersymmetric, since $V_{\te{WZ}} \mapsto V'_{\not{\te{WZ}}}$ under supersymmetry. However, under a combination of supersymmetry and generalized gauge transformation $V'_{\not{\te{WZ}}} \mapsto V''_{\te{WZ}}$ we can end up with a vector field in Wess Zumino gauge.

\subsection{Abelian field strength superfield}
\label{sec:FieldStrengthSuperfield}

Recall that a non-supersymmetric complex scalar field $\vph$ coupled to a gauge field $V_{\mu}$ via covariant derivative $D_{\mu} = \pa_{\mu} - iq V_{\mu}$ transforms like
    \[\vph(x) \ \ \mapsto \ \ \exp \bigl(iq\al(x) \bigr) \, \vph(x) \co V_{\mu}(x) \ \ \mapsto \ \ V_{\mu}(x) \ + \ \pa_{\mu} \al(x)
\]
under local $U(1)$ with charge $q$ and local parameter $\al(x)$.

\noindent
Under supersymmetry, these concepts generalize to chiral superfields $\Phi$ and vector superfields $V$. To construct a gauge invariant quatitiy out of $\Phi$ and $V$, we impose the following transformation properties:
    \[\left. \begin{array}{rll}  \Phi &\mapsto &\exp(iq\La) \, \Phi \\ V &\mapsto &V \ - \ \frac{i}{2} \; \bigl(\La \, - \, \La^\dag \bigr) \end{array} \right\} \ \ \ \Rightarrow \ \ \ \Phi^\dag \, \exp(2qV) \, \Phi \ \ \ \te{gauge invariant}
\]
Here, $\La$ is the chiral superfield defining the generalized gauge
transformations. Note that $\exp(iq\La ) \Phi$ is also chiral if $\Phi$ is.

\noindent
Before supersymmetry, we defined
    \[F_{\mu \nu} \eq \pa_{\mu} V_{\nu} \ - \ \pa_{\nu} V_{\mu}
\]
as an abelian field - strength. The supersymmetric analogy is
    \[W_{\al} \ \ := \ \ -\frac{1}{4} \; (\bar{{\cal D}}\bar{{\cal D}}) \, {\cal D}_{\al}V
\]
which is both chiral and invariant under generalized gauge transformations.

\paragraph{Exercise 3.3:} {\it Demonstrate these properties.}\\

\noindent
To obtain $W_{\al}$ in components, it is most convenient to rewrite $V$ in the shifted $y^\mu = x^\mu + i \theta \si^\mu \bar{\theta}$ variable (where $\theta \si^{\mu} \bar{\theta} V_\mu(x) = \theta \si^{\mu} \bar{\theta} V_\mu(y) - \frac{i}{2} \theta^2 \bar{\theta}^2 \pa_\mu V^\mu(y)$), then the supercovariant derivatives simplify to ${\cal D}_\al = \pa_\al + 2i (\si^\mu \bar{\theta})_\al \pa_\mu$ and $\bar{{\cal D}}_{\dot{\al}} =- \pa_{\dot{\al}}$:
    \[\framebox{ $ \displaystyle \Bigl. \Bigr. W_{\al}(y, \theta) \eq \la_{\al}(y) \ + \  \theta_{\al} \, D(y) \ + \ (\si^{\mu \nu} \, \theta)_{\al} \, F_{\mu \nu}(y) \ - \ i(\theta \theta) \, (\si^{\mu})_{\al \dot{\be}} \, \pa_{\mu} \bar{\la}^{\dot{\be}}(y)  $}
\]
\paragraph{Exercise 3.4:} {\it Verify this component expansion.}

\subsection{Non - abelian field strength}

In this section supersymmetric $U(1)$ gauge theories are generalized to nonabelian gauge groups. The gauge degrees of freedom then take values in the associated Lie algebra spanned by hermitian generators $T^a$:
\[ \La \eq \La_a \, T^a \co V \eq V_a \, T^a \co \Bigl[ T^a \ , \ T^b \Bigr] \eq i f^{abc} \, T_c \]
Just like in the abelian case, we want to keep $\Phi^\dag e^{2qV} \Phi$ invariant under the gauge transformation $\Phi \mapsto e^{iq\La} \Phi$, but the non-commutative nature of $\La$ and $V$ enforces a nonlinear transformation law $V \mapsto V'$:
\begin{align*}
 \exp(2qV') \ \ &= \ \ \exp(iq \La^\dag) \, \exp(2qV) \, \exp(-iq\La) \\ \Rightarrow \ \ \ V' \ \ &= \ \ V \ - \ \frac{i}{2} \; (\La \, - \, \La^\dag) \ - \ \frac{iq}{2} \; \Bigl[ V \ , \ \La \, + \, \La^\dag \Bigr] \ + \ ... \end{align*}
The commutator terms are due to the Baker Campbell Hausdorff formula for matrix exponentials
\[ \exp (X) \, \exp(Y) \eq \exp \left( X \ + \ Y \ + \ \frac{1}{2} \; \Bigl[X \ , \ Y \Bigr] \ + \ ... \right) \ . \]
The field strength superfield $W_\al$ also needs some modification in nonabelian theories. Recall that the field strength tensor $F_{\mu \nu}$ of non-supersymmetric Yang Mills theories transforms to $U F_{\mu \nu} U^{-1}$ under unitary transformations. Similarly, we define
\[ W_{\al} \ \ := \ \ -\frac{1}{8 \,q} \; (\bar{{\cal D}}\bar{{\cal D}}) \, \bigl( \exp(-2qV) \,  {\cal D}_{\al} \, \exp(2qV) \bigr) \]
and obtain a gauge covariant quantity.

\paragraph{Exercise 3.5:} {\it Check $(\bar {\cal D} \bar {\cal D}) {\cal D}_\al e^{iq\La} = 0$ and use this to prove the transformation law
\[
W_\al \ \ \mapsto \ \ e^{iq\La} \, W_\al \, e^{-iq\La}
\]
under gauge transformations $e^{2qV} \mapsto e^{iq\La^\dag} e^{2qV} e^{-iq\La}$.}

\bigskip
\noindent
In Wess Zumino gauge, the supersymmetric field strength can be evaluated as

\medskip
\framebox{\begin{minipage}{5.9in}
\begin{align*} W^a_{\al}(y, \theta) \ \ &= \ \ -\frac{1}{4} \; (\bar{{\cal D}}\bar{{\cal D}}) \, {\cal D}_{\al} \, \bigl( V^a(y,\theta,\bar{\theta}) \ + \ i \, V^b(y,\theta,\bar{\theta}) \, V^c(y,\theta,\bar{\theta}) \, f^a\,_{bc} \bigr) \\
&= \ \  \la^a_{\al}(y) \ + \  \theta_{\al} \, D^a(y) \ + \ (\si^{\mu \nu} \, \theta)_{\al} \, F^a_{\mu \nu}(y) \ - \ i(\theta \theta) \, (\si^{\mu})_{\al \dot{\be}} \, D_{\mu} \bar{\la}^{a\dot{\be}}(y) 
\end{align*}
\end{minipage}}

\medskip

\noindent
where
\begin{align*}
F_{\mu \nu}^a \ \ &:= \ \ \pa_\mu V_\nu^a \ - \ \pa_\nu V_\mu^a \ + \ q \, f^{a}\,_{bc} \, V^b_\mu \, V^c_\nu \\
D_\mu \bar{\la}^a \ \ &:= \ \ \pa_\mu \bar{\la}^a \ + \ q \, V_\mu^b \, \bar{\la}^c \, f_{bc} \, ^a
\end{align*}

\chapter{Four dimensional supersymmetric Lagrangians}
\label{sec:4DSupersymmetricLagrangians}

\section{${\cal N} = 1$ global supersymmetry}
\label{sec:N1GlobalSupersymmetry}

We want to determine couplings among superfields $\Phi$'s, $V$'s and
$W_{\al}$ which include the particles of the Standard Model. For this
we need a prescription to build Lagrangians which are invariant (up to
a total derivative) under a supersymmetry transformation. We will
start with the simplest case of only chiral superfields.

\subsection{Chiral superfield Lagrangian}
\label{sec:ChiralSuperfieldLagrangian}

In order to find an object ${\cal L}(\Phi)$ such that $\de {\cal L}$ is a
total derivative under supersymmetry transformation, we can exploit:
that
\begin{itemize}
\item For a general scalar superfield $S = ... + (\theta \theta )( \thb \thb )
  D(x)$, the $D$ term transforms as:
    \[\de D \eq \frac{i}{2} \; \pa_{\mu} \, \bigl(\ep \, \si^{\mu} \, \bar{\la} \ - \ \rho \, \si^{\mu} \, \bar{\ep} \bigr)
\]
\item For a chiral superfield $\Phi = ... + (\theta \theta) F(x)$, the
  $F$ term transforms as:
    \[\de F \eq i\sqrt{2} \, \bar{\ep} \, \bar{\si}^{\mu} \, \pa_{\mu} \psi \ ,
\]
\end{itemize}
Therefore, the most general Lagrangian for a chiral superfield
$\Phi$'s can be written as:
    \[{\cal L} \eq \underbrace{K(\Phi,\Phi^\dag)}_{\te{K\"ahler - potential}} \Bigl. \Bigr|_{D} \ + \ \Biggl( \underbrace{W(\Phi)}_{\te{super - potential}} \Bigl. \Bigr|_{F} + h.c. \Biggr) \ .
\]
Where $|_{D}$ refers to the $D$ term of the corresponding superfield
and similar for $F$ terms. The function $K$ is known as the {\it K\"ahler
potential}, it is a real function of $\Phi$ and
$\Phi^\dagger$. $W(\Phi)$ is known as the {\it superpotential}, it is
a holomorphic function of the chiral superfield $\Phi$ (and therefore
is a chiral superfield itself).

\noindent
In order to obtain a renormalizable theory, we need to construct a
Lagrangian in terms of operators of dimensionality such that the
Lagrangian has dimensionality 4. We know $[\vph] = 1$ (where the
square brackets stand for dimensionality of the field) and want $[{\cal L}] = 4$ .
Terms of dimension 4, such as $\pa^{\mu} \vph \pa_{\mu} \vph^{*}$,
$m^{2} \vph \vph^{*}$ and $g|\vph|^{4}$, are renormalizable,
but $\frac{1}{M^{2}} |\vph|^{6}$ is not. The dimensionality of the
superfield $\Phi$ is the same as that of its scalar component and that
of $\psi$ is as any standard fermion, that is
    \[[\Phi] \eq [\vph] \eq 1 \co [\psi] \eq \frac{3}{2}
\]
From the expansion $\Phi = \vph + \sqrt{2} \theta \psi + \theta \theta F + ...$ it follows that
    \[[\theta] \eq -\frac{1}{2} \co [F] \eq 2 \ .
\]
This already hints that $F$ is not a standard scalar field.
In order to have $[{\cal L}] = 4$ we  need:
\begin{align*}
[K_{D}] \ \ &\leq \ \ 4 \ \ \te{in} \ \ K \eq ... + \ (\theta \theta) \, (\thb \thb) \, K_{D} \\
[W_{F}] \ \ &\leq \ \ 4 \ \ \te{in} \ \ W \eq ... +  \ (\theta \theta) \, W_{F} \\
\so &[K] \ \ \leq \ \ 2 \co [W] \ \ \leq \ \ 3 \ .
\end{align*}
A possible term for $K$ is $\Phi^\dag \Phi$, but no $\Phi + \Phi^\dag$
 nor $\Phi \Phi$ since those are linear combinations of chiral
 superfields.
 
\noindent 
Therefore we are lead to the following general expressions for $K$ and $W$:
    \[K \eq \Phi^\dag \, \Phi \co W \eq \al \ + \ \la \, \Phi \ + \ \frac{m}{2} \; \Phi^{2} \ + \  \frac{g}{3} \; \Phi^{3} \ ,
\]
whose Lagrangian is known as {\em Wess Zumino model}:
\begin{align*}
{\cal L} \ \ &= \ \ \Phi^\dag \, \Phi \Bigl. \Bigr|_{D} \ + \ \Biggl( \left(\al \ + \ \la \, \Phi \ + \ \frac{m}{2} \; \Phi^{2} \ + \ \frac{g}{3} \; \Phi^{3} \right) \Bigl. \Bigr|_{F} \ +  \ h.c. \Biggr) \\
&= \ \ \pa^{\mu} \vph^{*} \, \pa_{\mu} \vph \ - \ i\bar{\psi} \, \bar{\si}^{\mu} \, \pa_{\mu} \psi \ + \ F \, F^{*} \ + \ \left(\frac{\pa W}{\pa \vph} \; F \ + \ h.c. \right) \\
& \ \ \ \ \ \ - \ \frac{1}{2} \; \left( \frac{\pa^{2} W}{\pa \vph^{2}} \; \psi \psi \ + \ h.c. \right)
\end{align*}

\paragraph{Exercise 4.1:} {\it Verify that $\pa^{\mu} \vph^{*} \, \pa_{\mu} \vph \, - \, i\bar{\psi} \, \bar{\si}^{\mu} \, \pa_{\mu} \psi \, + \, F \, F^{*}$ are due to the $D$ term of $\Phi^\dag \, \Phi$ after integration by parts.}

\paragraph{Exercise 4.2:} {\it Determine the $F$ term of the superpotential $W = \frac{m}{2} \Phi^2 + \frac{g}{3} \Phi^3$.}

\bigskip
\noindent
Note that
\begin{itemize}
\item The expression for $\Phi^\dag \Phi \Bigl. \Bigr|_{D}$ is justified by
    \[\Phi \eq \vph \ + \ \sqrt{2} \, \theta \psi \ + \ \theta \theta \, F \ + \ i \theta \, \si^{\mu} \, \thb \, \pa_{\mu} \vph \ - \ \frac{i}{\sqrt{2}} \; (\theta \theta) \, \pa_{\mu} \psi \, \si^{\mu} \, \thb \ - \ \frac{1}{4} \; (\theta \theta) \, (\thb \thb) \, \pa_{\mu} \pa^{\mu} \vph
\]
\item In general, the procedure to obtain the expansion of the
  Lagrangian in terms of the components of the superfield is to
  perform a Taylor expansion around $\Phi = \vph$, for instance (where $\frac{\pa W}{\pa \vph} = \frac{\pa W}{\pa \Phi} \Bigl. \Bigr|_{\Phi = \vph}$):
    \[W(\Phi) \eq W(\vph) \ + \ \underbrace{(\Phi \, - \, \vph)}_{...\, +\, \theta \theta F\, + \, ...} \; \frac{\pa W}{\pa \vph} \ + \  \underbrace{\frac{1}{2} \; (\Phi \, - \, \vph)^{2}}_{... \, + \, (\theta \psi)\, (\theta \psi) \, + \, ...} \; \frac{\pa^{2}W}{\pa \vph^{2}}
\]
\end{itemize}
The part of the Lagrangian depending on the auxiliary field $F$ takes
the simple form:
    \[{\cal L}_{(F)} \eq F \, F^{*} \ + \ \frac{\pa W}{\pa \vph} \; F \ + \ \frac{\pa W^{*}}{\pa \vph^{*}} \, F^{*}
\]
Notice that this is quadratic and without any derivatives. This means
that the field $F$ does not propagate. Also, we can easily eliminate $F$ using the field equations
\begin{align*}
\frac{\de {\cal S}_{(F)}}{\de F} \eq 0 \ \ \ &\Longrightarrow \ \ \ F^{*} \ + \ \frac{\pa W}{\pa \vph} \eq 0 \\
\frac{\de {\cal S}_{(F)}}{\de F^{*}} \eq 0 \ \ \ &\Longrightarrow \ \ \ F \ + \ \frac{\pa W^{*}}{\pa \vph^{*}} \eq 0
\end{align*}
and substitute the result back into the Lagrangian,
    \[{\cal L}_{(F)} \ \ \mapsto \ \ - \left| \frac{\pa W}{\pa \vph} \right|^{2} \ \ =: \ \ -V_{(F)}(\vph) \ ,
\]
This defines the scalar potential. From its expression we can easily
see that it is a positive definite scalar potential $V_{(F)}(\vph)$.

\noindent
We finish the section about chiral superfield Lagrangian with two remarks,
\begin{itemize}
\item The ${\cal N} = 1$ Lagrangian is a particular case of standard ${\cal N} =
  0$ Lagrangians: the scalar potential is semipositive ($V \geq
  0$). Also the mass for scalar field $\vph$ (as it can be read from
  the quadratic term in the scalar potential) equals the one for the
  spinor $\psi$ (as can be read from the term $\frac{1}{2} \frac{\pa^{2} W}{\pa
  \vph^{2}} \psi \psi$).
 Moreover, the coefficient $g$ of Yukawa coupling $g(\vph \psi
  \psi)$ also determines the scalar
self coupling, $g^{2} |\vph|^{4}$. This is the source of ''miraculous'' cancellations in SUSY perturbation theory.
Divergences are removed from diagrams:
\begin{figure}[ht]
    \centering
     \includegraphics[width=0.85\textwidth]{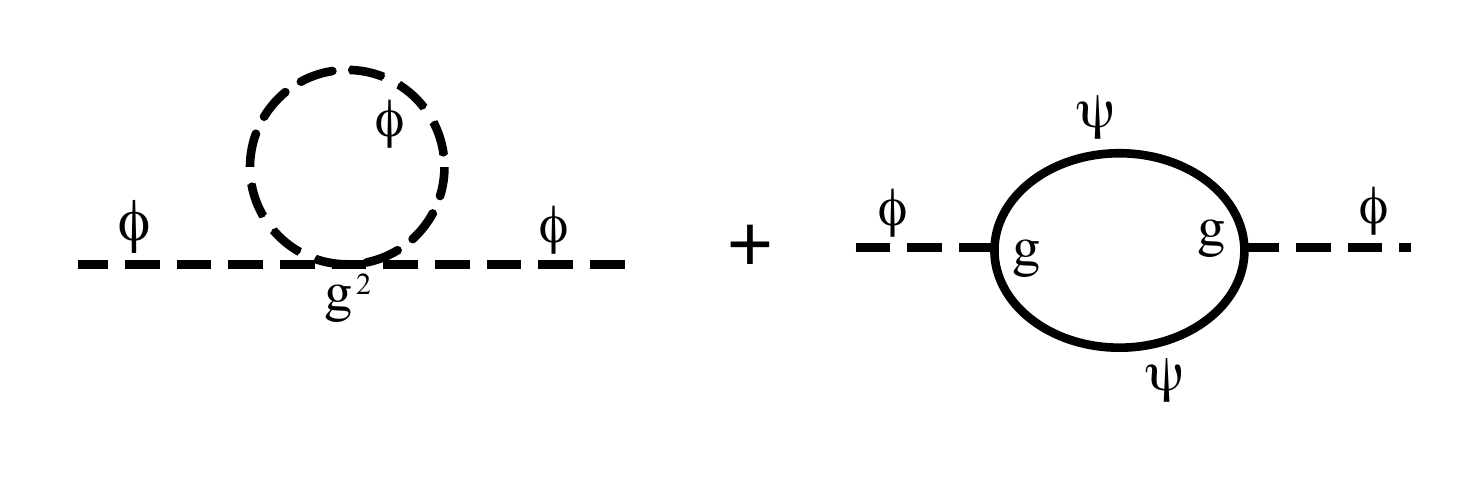}
       \caption{\footnotesize{One loop diagrams which yield a corrections to the scalar mass. SUSY relates the $\phi^4$ coupling to the Yukawa couplings $\phi (\psi \bar \psi)$ and therefore ensures cancellation of the leading divergence.}}
\end{figure}
\item In general, expand $K(\Phi^{i} , \Phi^{j^\dag})$ and $W(\Phi^{i})$ around $\Phi^{i} = \vph^{i}$, in components
    \[\left(\frac{\pa^{2}K}{\pa \vph^{i} \pa \vph^{\bar{j}*}}\right)
   \; \pa_{\mu} \vph^{i} \, \pa^{\mu}
\vph^{\bar{j}*} \eq K_{i\bar{j}} \, \pa_{\mu} \vph^{i} \, \pa^{\mu} \vph^{\bar{j}*} \ .
\]
$K_{i\bar{j}}$ is a metric in a space with coordinates $\vph^{i}$ which is a complex K\"ahler - manifold:
    \[g_{i\bar{j}} \eq K_{i{\bar{j}}} \eq \frac{\pa^{2}K}{\pa \vph^{i} \pa \vph^{\bar{j}*}}
\]
\end{itemize}

\subsection{Miraculous cancellations in detail}
\label{sec:MiraculousCancellationsInDetail}

In this subsection, we want to show in detail how virtual bosons and fermions contribute to cancel their contributions to observables such as the Higgs mass. Using suitable redefinitions, the most general cubic superpotential can be reduced to
	\[W \eq \frac{m}{2} \; \Phi^{2} \ + \  \frac{g}{3} \; \Phi^{3} \ .
\]
Together with the standard K\"ahler potential $K = \Phi^\dag \Phi$, it yields a Lagrangian
\begin{align*}
{\cal L} \ \ &= \ \ \pa^{\mu} \vph^{*} \, \pa_{\mu} \vph \ + \ i\bar{\psi} \, \bar{\si}^{\mu} \, \pa_{\mu} \psi \ - \ \bigl| m\, \vph \ + \ g \, \vph^{2} \bigr|^{2} \ - \ \left( \frac{m}{2} \ + \  g \, \vph \right) \, \psi \psi \ - \ \left( \frac{m}{2} \ + \  g \, \vph^{*} \right) \, \bar{\psi} \bar{\psi} \\
&= \ \ \frac{1}{2} \; \pa^{\mu} A \, \pa_{\mu} A \ - \ \frac{1}{2} \; m^{2} \, A^{2} \ + \ \frac{1}{2} \; \pa^{\mu} B \, \pa_{\mu} B \ - \ \frac{1}{2} \; m^{2} \, B^{2} \ + \ \frac{1}{2} \; \Psib \, \bigl(i \! \not{\! \pa} \ - \ m \bigr) \Psi  \\
& \ \ \ \ \ \ - \frac{m\, g}{\sqrt{2}} \; A \, (A^{2} \ + \ B^{2}) \ - \ \frac{g^{2}}{4} \; \bigl( A^{4} \ + \ B^{4} \ + \ 2 \, A^{2} \, B^{2} \bigr) \ - \ \frac{g}{\sqrt{2}} \; \Psib \, \bigl( A \ - \ i B \, \ga^{5} \bigr) \, \Psi
\end{align*}
with cubic and quartic interactions for the complex scalar $\vph = \frac{A + i B}{\sqrt{2}}$ and the 4 spinor $\Psi = (\psi , \, \bar{\psi})$.

\noindent
Let us compute the 1 loop corrections to the mass of the scalar $A$, given by the following diagrams:
\begin{figure}[ht]
    \centering
     \includegraphics[width=0.3\textwidth]{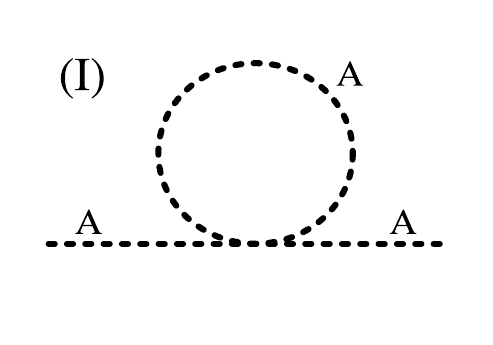}\includegraphics[width=0.3\textwidth]{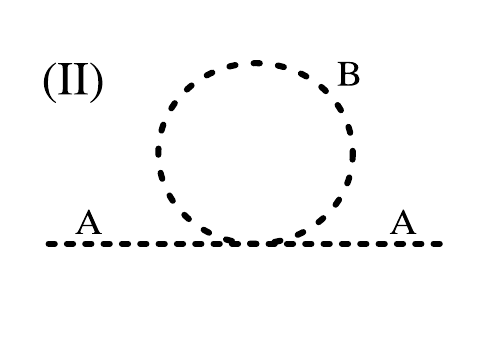}\\
     \includegraphics[width=0.3\textwidth]{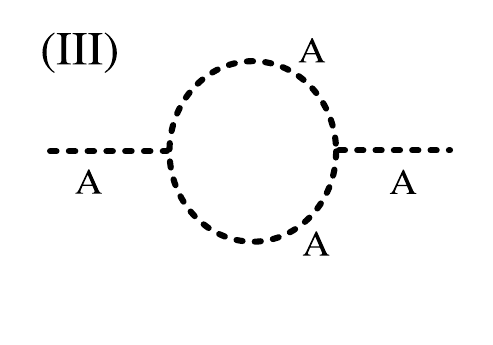}\includegraphics[width=0.3\textwidth]{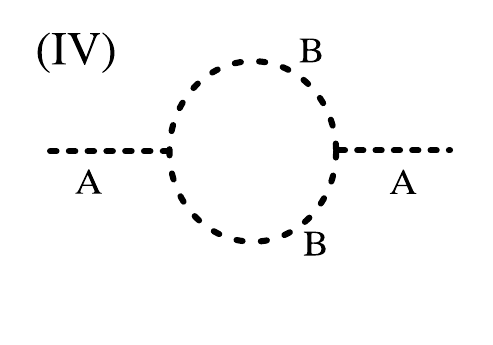}\\
     \includegraphics[width=0.3\textwidth]{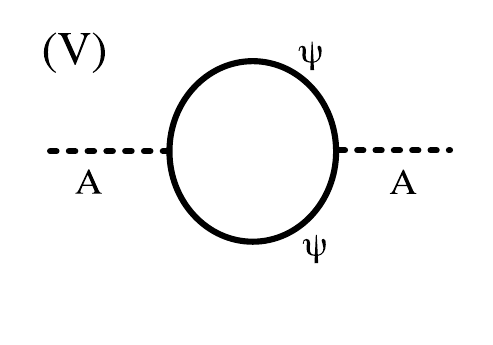}
       \caption{\footnotesize{One loop diagrams that give corrections to the mass of the scalar $A.$}}
\end{figure}

\noindent
The usual Feynman rules from non-supersymmetric field theory allow to evaluate them as follows:
\begin{align*}
(I) \ \ &= \ \ -\frac{ig^{2}}{4} \; 4 \cdot 3 \int \frac{\dd^{4}k}{(2\pi)^{4}} \; \frac{i}{k^{2} \, - \, m^{2}} \eq 3 \, g^{2} \int \frac{\dd^{4}k}{(2\pi)^{4}} \; \frac{1}{k^{2} \, - \, m^{2}} \\
(II) \ \ &= \ \ -\frac{ig^{2}}{2} \; 2 \int \frac{\dd^{4}k}{(2\pi)^{4}} \; \frac{i}{k^{2} \, - \, m^{2}} \eq g^{2} \int \frac{\dd^{4}k}{(2\pi)^{4}} \; \frac{1}{k^{2} \, - \, m^{2}} \\
(III) \ \ &= \ \ \left(-\frac{im\,g}{\sqrt{2}} \right)^{2} \; 3 \cdot 2 \int \frac{\dd^{4}k}{(2\pi)^{4}} \; \frac{i}{k^{2} \, - \, m^{2}} \; \frac{i}{(k \, - \, p)^{2} \, - \, m^{2}} \\
&= \ \ 3 \, g^{2} \, m^{2} \int \frac{\dd^{4}k}{(2\pi)^{4}} \; \frac{1}{(k^{2} \, - \, m^{2}) \, \bigl((k \, - \, p)^{2} \, - \, m^{2} \bigr)} \\
(IV) \ \ &= \ \ \left(-\frac{im\,g}{\sqrt{2}} \right)^{2} \;  2 \int \frac{\dd^{4}k}{(2\pi)^{4}} \; \frac{i}{k^{2} \, - \, m^{2}} \; \frac{i}{(k \, - \, p)^{2} \, - \, m^{2}} \\
&= \ \  g^{2} \, m^{2} \int \frac{\dd^{4}k}{(2\pi)^{4}} \; \frac{1}{(k^{2} \, - \, m^{2}) \, \bigl((k \, - \, p)^{2} \, - \, m^{2} \bigr)} \\
(V) \ \ &= \ \ - \left( -\frac{ig}{\sqrt{2}} \right)^{2} \; 2 \int \frac{\dd^{4}k}{(2\pi)^{4}} \; \te{Tr} \left\{ \frac{i (\not{ \! k} \, + \, m)}{k^{2} \, - \, m^{2}} \; \frac{i (\not{\! k} \, - \, \not{ \! p} \, + \, m)}{(k \, - \, p)^{2} \, - \, m^{2}} \right\} \\
&= \ \ -2 \, g^{2} \left( \int \frac{\dd^{4}k}{(2\pi)^{4}} \; \frac{1}{k^{2} \, - \, m^{2}} \ + \ \int \frac{\dd^{4}k}{(2\pi)^{4}} \; \frac{1}{(k \, - \, p)^{2} \, - \, m^{2}} \right. \\
& \ \ \ \ \ \ \left. + \ \int \frac{\dd^{4}k}{(2\pi)^{4}} \; \frac{4 \, m^{2} \ - \ p^{2}}{(k^{2} \, - \, m^{2} ) \, \bigl((k \, - \, p)^{2} \, - \, m^{2} \bigr)} \right)
\end{align*}
In total, we arrive at a mass correction of
	\[2\, g^{2} \, \left\{ \int \frac{\dd^{4}k}{(2\pi)^{4}} \; \frac{1}{k^{2} \, - \, m^{2}} \ - \ \int \frac{\dd^{4}k}{(2\pi)^{4}} \; \frac{1}{(k \, - \, p)^{2} \, - \, m^{2}} \ + \ \int \frac{\dd^{4}k}{(2\pi)^{4}} \; \frac{p^{2} \ - \ 2 \, m^{2}}{(k^{2} \, - \, m^{2}) \, \bigl((k \, - \, p)^{2} \, - \, m^{2} \bigr)} \right\} \ .
\]
The important lesson is the relative sign between the bosonic diagrams $(I)$ to $(IV)$ and the fermionic one $(V)$. UV divergent pieces of the first two integrals cancel, and the cutoff $\La$ only enters logarithmically
	\[\int_{\La} \frac{\dd^{4}k}{(2\pi)^{4}} \; \frac{1}{(k^{2} \, - \, m^{2}) \, \bigl((k \, - \, p)^{2} \, - \, m^{2} \bigr)} \ \ \approx \ \ \int^{\La} \limits_{0} \frac{2 \, \pi^{2} \, k^{3} \, \dd k }{(2\pi)^{4}} \; \frac{1}{k^{4}} \ \ \sim \ \ \int^{\La} \limits_{0} \frac{\dd k}{k} \ \ \sim \ \ \ln \La
\]
whereas non-supersymmetric theories usually produce quadratic divergences such as
	\[\int_{\La} \frac{\dd^{4}k}{(2\pi)^{4}} \; \frac{1}{k^{2} \, - \, m^{2}} \ \ \approx \ \ \int^{\La} \limits_{0} \frac{2 \, \pi^{2} \, k^{3} \, \dd k }{(2\pi)^{4}} \; \frac{1}{k^{2}} \ \ \sim \ \ \int^{\La} \limits_{0} k \, \dd k \ \ \sim \ \ \La^{2} \ .
\]

\subsection{Abelian vector superfield Lagrangian}
\label{sec:VectorSuperfieldLagrangian}

Before attacking vector superfield Lagrangians, let us first discuss how we ensured gauge invariance of $\pa^{\mu} \vph \pa_{\mu} \vph^{*}$ under local transformations $\vph \mapsto \exp \bigl(iq\al(x) \bigr)$ in the non-supersymmetric case.
\begin{itemize}
\item Introduce covariant derivative $D_{\mu}$ depending on gauge potential $A_{\mu}$
    \[D_{\mu} \vph \ \ := \ \ \pa_{\mu} \vph \ - \ iq \, A_{\mu} \, \vph \co A_{\mu} \ \ \mapsto \ \ A_{\mu} \ + \ \pa_{\mu} \al
\]
and rewrite kinetic term as
    \[{\cal L} \eq D^{\mu} \vph \, (D_{\mu} \vph)^{*} + ...
\]
\item Add kinetic term for $A_{\mu}$ to ${\cal L}$
    \[{\cal L} \eq ... \ + \ \frac{1}{4g^{2}} \; F_{\mu \nu} \, F^{\mu \nu} \co F_{\mu \nu} \eq \pa_{\mu} A_{\nu} \ - \ \pa_{\nu} A_{\mu} \ .
\]
\end{itemize}
With SUSY, the K\"ahler potential $K = \Phi^\dag \Phi$ is not invariant under
    \[\Phi \ \ \mapsto \ \ \exp(iq\La) \, \Phi \co \Phi^\dag \, \Phi \ \ \mapsto \ \ \Phi^\dag \, \exp \bigl(iq(\La \, - \, \La^\dag) \bigr) \, \Phi
\]
for chiral $\La$. Our procedure to construct a suitable Lagrangian is
analogous to the non-supersymmetric case (although the expressions look slightly different):
\begin{itemize}
\item Introduce $V$ such that
    \[K \eq \Phi^\dag \, \exp(2qV) \, \Phi \co V \ \ \mapsto \ \ V \ - \ \frac{i}{2} \, \bigl(\La \ - \ \La^\dag \bigr) \ ,
\]
i.e. $K$ is invariant under general gauge transformation.
\item Add kinetic term for $V$ with coupling $\tau$
    \[{\cal L}_{kin} \eq f(\Phi) \, (W^{\al} \, W_{\al}) \Bigl. \Bigr|_{F} \ + \ h.c.
\]
which is renormalizable if $f(\Phi)$ is a constant $f=\tau$. For general $f(\Phi)$, however, it is non-renormalizable. We will call $f$ the {\it gauge kinetic function}.
\item A new ingredient of supersymmetric theories is that an extra
  term can be added to ${\cal L}$. It is also invariant (for $U(1)$ gauge
  theories) and known as the {\it Fayet Iliopoulos term}:
    \[{\cal L}_{FI} \eq \xi \, V \Bigl. \Bigr|_{D} \eq \frac{1}{2} \; \xi \, D
\]
The parameter $\xi$ is a constant. Notice that the FI term is gauge invariant for
a $U(1)$ theory because the corresponding gauge field is not charged
under $U(1)$ (the photon is chargeless), whereas for a non-abelian
gauge theory the gauge fields (and their corresponding $D$ terms) would transform under the gauge group and therefore have to be forbidden. This is the reason the FI term only exists for abelian gauge theories.
\end{itemize}
The renormalizable Lagrangian of super QED involves $f = \tau = \frac{1}{4}$:
    \[{\cal L} \eq \bigl( \Phi^\dag \, \exp(2qV) \, \Phi \bigr)
    \Bigl. \Bigr|_{D} \ + \
 \Biggl( W(\Phi) \Bigl. \Bigr|_{F} \ + \ h.c. \Biggr) \ + \ 
\Biggl( \frac{1}{4} \;  W^{\al} \, W_{\al} \Bigl. \Bigr|_{F} \ + \ h.c. \Biggr) \ + \ \xi \, V \Bigl. \Bigr|_{D} \ .
\]
If there were only one superfield $\Phi$ charged under $U(1)$ then
$W=0$. For several superfields the superpotential $W$ is constructed
out of holomorphic combinations of the superfields which are gauge invariant.
In components (using Wess Zumino gauge):
\begin{align*}
\bigl(\Phi^\dag \, \exp(2qV) \, \Phi \bigr) \Bigl. \Bigr|_{D} \ \ &= \ \ F^{*} \, F \ + \ \pa_{\mu} \vph \, \pa^{\mu} \vph^{*} \ + \ i\bar{\psi} \, \bar{\si}^{\mu} \, \pa_{\mu} \psi \ + \ q \, V^{\mu} \, \bigl( \bar{\psi} \, \bar{\si}_{\mu} \, \psi \ + \ i \vph^{*} \, \pa_{\mu} \vph \  -  \ i \vph \, \pa_{\mu} \vph^{*} \bigr) \\
&\ \ \ \ \ \ + \ \sqrt{2} \, q \, \bigl(\vph \, \bar{\la} \bar{\psi} \ + \ \vph^{*} \, \la \psi \bigr) \ + \ q \, \left(D \ + \ q \, V_{\mu} \, V^{\mu} \right) \, | \vph |^{2}
\end{align*}
Note that
\begin{itemize}
\item $V^{n \geq 3} = 0$ due to Wess Zumino gauge
\item can complete $\pa_{\mu}$ to $D_{\mu}=\pa_{\mu} - iqV_{\mu}$ using the terms $\sim qV_{\mu}$
\end{itemize}
In gauge theories, need $W(\Phi) = 0$ if there is only one $\Phi$. In case of several $\Phi_{k}$, only chargeless combinations of products of $\Phi_{k}$ contribute, since $W(\Phi)$ has to be invariant under $\Phi_k \mapsto \exp(iq_k \La) \Phi_k$.

\noindent
Let us move on to the $W^{\al} W_{\al}$- term:
    \[W^{\al} \; W_{\al} \Bigl. \Bigr|_{F} \eq  D^{2} \ - \ \frac{1}{2} \; F_{\mu \nu} \, F^{\mu \nu} \ - \ 2i \, \la \, \si^{\mu} \, \pa_{\mu} \bar{\la} \ + \ \frac{i}{2} \; F_{\mu \nu} \, \tilde{F}^{\mu \nu} \ .
\]

\paragraph{Exercise 4.3:} {\it Verify the $F$ term of $W_\al W^\al$ using $\te{Tr}\{ \si^{\mu \nu} \si^{\ka \tau} \} = \frac{1}{2} \bigl( \eta^{\mu \ka} \eta^{\nu \tau} - \eta^{\mu \tau} \eta^{\nu \ka} + i\ep^{\mu \nu \ka \tau} \bigr)$.}

\bigskip
\noindent
In the QED choice $f= \frac{1}{4}$, the kinetic terms for the vector superfields are given by
	\[{\cal L}_{kin} \eq \frac{1}{4} \;  W^{\al} \, W_{\al} \Bigl. \Bigr|_{F} \ + \ h.c. \eq \frac{1}{2} \; D^{2} \ - \ \frac{1}{4} \; F_{\mu \nu} \, F^{\mu \nu} \ - \ i \la \, \si^{\mu} \, \pa_{\mu} \bar{\la} \ .
\]
The last term in $W^{\al}  W_{\al} \bigl. \bigr|_{F}$ involving $\tilde{F}_{\mu \nu} = \ep_{\mu \nu \rho \si} F^{\rho \si}$ drops out whenever $f(\Phi)$ is chosen to be real. Otherwise, it couples as $\frac{1}{2} \te{Im} \{ f(\Phi) \} F_{\mu \nu} \tilde{F}^{\mu \nu}$ where $F_{\mu \nu} \tilde{F}^{\mu \nu}$ itself is a total derivative without any local physics.

\noindent
With the FI contribution $\xi \, V \bigl. \bigr|_{D} = \frac{1}{2} \xi D$, the collection of the $D$ dependent terms in ${\cal L}$
    \[{\cal L}_{(D)} \eq q\,  D \, |\vph|^{2} \ + \ \frac{1}{2} \, D^{2} \ + \ \frac{1}{2} \; \xi \, D
\]
yields field equations
    \[\frac{\de {\cal S}_{(D)}}{\de D} \eq 0 \so D \eq -\frac{\xi}{2} \ - \ q \, |\vph|^{2} \ .
\]
Substituting those back into ${\cal L}_{(D)}$,
    \[{\cal L}_{(D)} \eq -\frac{1}{8} \; \Bigl( \xi \ + \ 2 \, q\, |\vph|^{2} \Bigr)^{2} \ \ =: \ \ -V_{(D)}(\vph) \ ,
\]
get a positive semidefinite scalar potential $V_{(D)}(\vph)$. Together with $V_{(F)}(\vph)$ from the previous section, the total potential is given by
    \[V(\vph) \eq V_{(F)}(\vph) \ + \ V_{(D)}(\vph) \eq \left|\frac{\pa W}{\pa \vph}\right|^{2}\ + \ \frac{1}{8} \; \Bigl( \xi \ + \ 2 \, q\, |\vph|^{2} \Bigr)^{2} \ .
\]

\subsection{Action as a superspace integral}
\label{sec:ActionAsASuperspaceIntegral}

Without SUSY, the relationship between the action ${\cal S}$ and ${\cal L}$ is
    \[{\cal S} \eq \int \dd^{4}x \ {\cal L} \ .
\]
To write down a similar expression for SUSY - actions, recall
    \[\int \dd^{2} \theta \ (\theta \theta) \eq 1 \co \int \dd^{4} \theta \ (\theta \theta) \, (\thb \thb) \eq 1 \ .
\]
This provides elegant ways of expressing $K \Bigl. \Bigr|_{D}$ and so on:
    \begin{align*}
    {\cal L} \ \ &= \ \ K\Bigl. \Bigr|_{D} \ + \ \left( W\Bigl. \Bigr|_{F} \ + \ h.c. \right) \ + \ \left( W^{\al} \, W_{\al} \Bigl. \Bigr|_{F} \ + \ h.c. \right) \\
    &= \ \ \int \dd^{4} \theta \ K \ + \ \left( \int \dd^{2} \theta \ W \ + \ h.c. \right) \ + \ \left( \int \dd^{2}\theta \ W^{\al} \, W_{\al} \ + \ h.c. \right)
\end{align*}
We end up with the most general action

\medskip
\framebox{\begin{minipage}{5.9in}
\begin{align*}
    {\cal S}\Bigl[ K \bigl(\Phi_{i}^\dag, \exp(2qV), \Phi_{i} \bigr),W\bigl(\Phi_{i} \bigr), f\bigl(\Phi_{i} \bigr), \xi \Bigr] \ \ &= \ \ \int \dd^{4}x \int \dd^{4} \theta \ \bigl(K \ + \ \xi \, V \bigr) \\
    & \ \ \ \ \ \ + \ \int \dd^{4}x \int \dd^{2} \theta \ \bigl(W \ + \ f \, W^{\al} \, W_{\al} \ + \ h.c. \bigr) \ .
\end{align*}
\end{minipage}}

\medskip

\noindent
Recall that the FI term $\xi V$ can only appear in abelian $U(1)$ gauge theories and that the non-abelian generalization of the $W^{\al} W_{\al}$ term requires an extra trace to keep it gauge invariant:
\[\te{Tr} \Bigl\{ W^{\al} \, W_{\al} \Bigr\} \ \ \mapsto \ \ \te{Tr} \Bigl\{e^{iq\La} \, W^{\al} \, W_{\al} \, e^{-iq\la} \Bigr\} \eq \te{Tr} \Bigl\{ W^{\al} \, W_{\al} \, \underbrace{e^{-iq\la} \, e^{iq\La} }_{= \ 1} \Bigr\}
\]

\section{Non-renormalization theorems}
\label{sec:NonRenormalizationTheorems}

We have seen that in general the functions $K,W,f$ and the FI constant
$\xi$ determine the full structure of ${\cal N}=1$ supersymmetric theories
(up to two derivatives of the fields as usual). If we know their
expressions we know all the interactions among the fields.

\noindent
In order to understand the important properties of supersymmetric
theories under quantization, we must address the following question:
How do $K$, $W$, $f$ and $\xi$ behave under quantum corrections? We
will show now that:
\begin{itemize}
\item $K$ gets corrections order by order in perturbation theory
\item only one loop - corrections for $f(\Phi)$
\item $W(\Phi)$ and $\xi$ {\it not} renormalized in perturbation theory.
\end{itemize}
The non-renormalization of the superpotential is one of the most
important results of supersymmetric field theories. The simple
behaviour of $f$ and the non-renormalization of $\xi$ have also
interesting consequences. We will proceed now to address these issues.

\subsection{History}

\begin{itemize}
\item In 1977 \textsc{Grisaru}, \textsc{Siegel}, \textsc{Rocek} showed using ''supergraphs'' that, except for 1 loop corrections in $f$, quantum corrections only come in the form
    \[\int \dd^{4}x \int \dd^{4} \theta \ \Bigl\{...\Bigr\} \ .
\]
\item In 1993, \textsc{Seiberg} (based on string theory arguments by \textsc{Witten}
  1985) used symmetry and holomorphicity arguments to establish these
  results in a simple an elegant way \cite{sei}. We will follow here this
  approach following closely the discussion \cite{WB3} (section 27.6)
\end{itemize}

\subsection{Proof of the non-renormalization theorem}
\label{sec:ProofOfTheNonRenormalizationTheorem}

Let us follow \textsc{Seiberg}'s path of proving the non-renormalization theorem. For that purpose, introduce ''spurious'' superfields $X$, $Y$
    \[X \eq (x, \psi_{x}, F_{x}) \co Y \eq (y, \psi_{y}, F_{y})
\]
involved in the action
    \[{\cal S} \eq \int \dd^{4}x \int \dd^{4}\theta \ \bigl( K \ + \ \xi \, V_{U(1)} \bigr) \ + \ \int \dd^{4}x \int \dd^{2}\theta \ \Bigl( Y \, W(\Phi_{i}) \ + \ X \, W^{\al} \, W_{\al} \ + \ h.c. \Bigr) \ .
\]
We will use:
\begin{itemize}
\item symmetries
\item holomorphicity
\item limits $X \mto \infty$ and $Y \mto 0$
\end{itemize}

\subsubsection{Symmetries}

\begin{itemize}
\item SUSY and gauge - symmetries
\item R - symmetry $U(1)_{R}$: Fields have different $U(1)_{R}$ charges determining how they transform under that group
    \[\begin{array}{l|ccccccc} \te{fields} &\Phi_{i} &V &X &Y &\theta &\thb &W^{\al} \\\hline U(1)_{R} \ \te{- charge} &0 &0 &0 &2 &-1 &1 &1 \end{array}
\]
\medskip
    \[\te{e.g.} \ \ \ \ \ \ Y \ \ \mapsto \ \ \exp(2i\al) \, Y \co \theta \ \ \mapsto \ \ \exp(-i\al) \, \theta \ , \ \te{etc.}
\]
\item Peccei Quinn symmetry
    \[X \ \ \mapsto \ \ X \ + \ ir \co r \in \mathbb R
\]
Since $XW^{\al} W_{\al}$ involves terms like
    \[\te{Re} \{X\} \, F_{\mu \nu} \, F^{\mu \nu} \ + \ \te{Im} \{X\} \, F_{\mu \nu} \, \tilde{F}^{\mu \nu} \ ,
\]
a change in the imaginary part of $X$ would only add total derivatives to ${\cal L}$,
    \[{\cal L} \ \ \mapsto \ \ {\cal L} \ + \ r \, F_{\mu \nu} \, \tilde{F}^{\mu \nu}
\]
without any local physics. Call $X$ an {\em axion field}.
\end{itemize}

\subsubsection{Holomorphicity}
\label{sec:Holomorphicity}

Consider the quantum corrected {\em Wilsonian action}
\[\exp (i {\cal S}_{\la}) \eq \int_{|p|>\la} {\cal D} \varphi \, \exp (i {\cal S}) \]
where the path integral is understood to go for all the fields in the
system and the integration is only over all momenta greater than
$\lambda$ in the standard Wilsonian formalism (different from the 1PI
action in which the integral is over all momenta). If supersymmetry is
preserved by the quantization process, we can write the
effective action as:
    \begin{align*}
    {\cal S}_{\la} \ \ &= \ \ \int \dd^{4}x \int \dd^{4} \theta \ \Bigl[ J\bigl(\Phi, \Phi^\dag, e^{V}, X,Y, {\cal D}... \bigr) \ + \ \xi(X,X^\dag,Y,Y^\dag) \, V_{U(1)} \Bigr] \\
    & \ \ \ \ \ \ + \ \int \dd^{4}x \int \dd^{2}\theta \ \Bigl[ \underbrace{H(\Phi,X,Y,W^{\al})}_{\te{holomorphic}} \ + \ h.c. \Bigr] \ .
\end{align*}
Due to $U(1)_{R}$ transformation invariance, $H$ must have the form
    \[H \eq Y \, h(X, \Phi) \ + \ g(X, \Phi) \, W^{\al} \, W_{\al} \ .
\]
Invariance under shifts in $X$ imply that $h=h(\Phi)$ (independent of
$X$). But a  linear $X$ dependence is allowed in front of $W^{\al}
W_{\al}$ (due to $F_{\mu \nu} \tilde{F}^{\mu \nu}$ as a total derivative). So the $X$ dependence in $h$ and $g$ is restricted to
    \[H \eq Y \, h(\Phi) \ + \ \bigl(\al \, X \ + \ g(\Phi) \bigr) \, W^{\al} \, W_{\al} \ .
\]

\subsubsection{Limits}
\label{sec:Limits}

In the limit $Y \mto 0$, there is an equality $h(\Phi) = W(\Phi)$ at tree level, so $W(\Phi)$ is not renormalized! The gauge kinetic function $f(\Phi)$, however, gets a 1 loop correction
    \[f(\Phi) \eq \underbrace{\al \, X}_{\te{tree level}} \ + \ \underbrace{g(\Phi)}_{1 \ \te{loop}} \ .
\]
Note that gauge field propagators are proportional to
$\frac{1}{x}$ (since gauge couplings behave as $\sim xF^{\mu \nu} F_{\mu \nu}$ $\sim X \pa^{[\mu} A^{\nu]} \pa_{[\mu} A_{\nu]}$, gauge self couplings to $X^{3}$ corresponding to a vertex of 3 $X$ lines).
\begin{figure}[ht]
    \centering
    \includegraphics[width=0.45\textwidth]{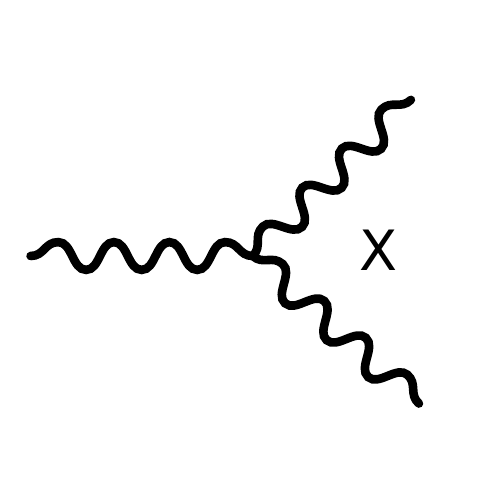}
        \caption{\footnotesize{Three vertex of $X$ fields multiplying the gauge field self couplings}}
\end{figure}
\subsubsection{}
Count the number $N_{x}$ of $x$ - powers in any diagram; it is given by
    \[N_{x} \eq V_{W} \ - \ I_{W}
\]
and is therefore related to the numbers of loops $L$:
    \[L \eq I_{W} \ - \ V_{W} \ + \ 1 \eq -N_{x} \ + \ 1 \so N_{x} \eq 1 \ - \ L
\]
\begin{align*}
&L \eq 0 \ \te{(tree level)}: \ N_{x} \eq 1 \co \al \eq 1 \\
&L \eq 1 \ \te{(one loop)}: \ N_{x} \eq 0
\end{align*}
Therefore the gauge kinetic term $X + g(\Phi)$ is corrected only at
1 loop! (All other (infinite) loop corrections just cancel.)

\noindent
On the other hand, the K\"ahler potential, being non-holomorphic,
 is corrected to all orders $J(Y,Y^\dag,X + X^\dag,...)$. For
 the FI term $\xi(X,X\dag,Y,Y^\dag) V_{U(1)}
 \Bigl. \Bigr|_{D}$, gauge invariance
under $V \mapsto V + i(\La - \La^\dag)$ implies that $\xi$ is a constant. The only contributions are proportional to
\begin{figure}[ht]
    \centering
        \includegraphics[width=0.50\textwidth]{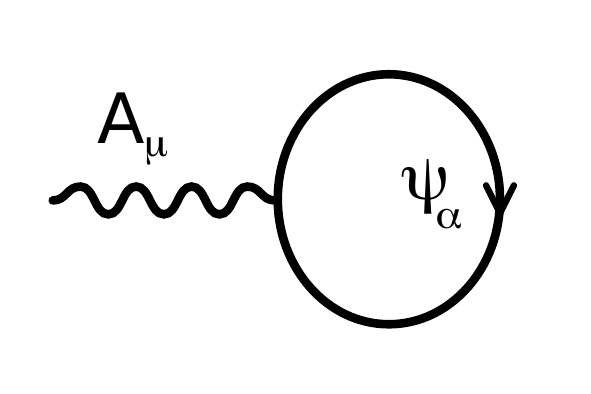}
        \caption{\footnotesize{Loop correction to the K\"ahler potential due to a FI term.}}
\end{figure}
    \[\sum_{i} q_{i} \eq \te{Tr}\Bigl\{Q_{U(1)} \Bigr\} \ .
\]
But if $\te{Tr}\{Q\} \neq 0$, the theory is ''inconsistent'' due to gravitational anomalies:
\begin{figure}[ht]
    \centering
    \includegraphics[width=0.70\textwidth]{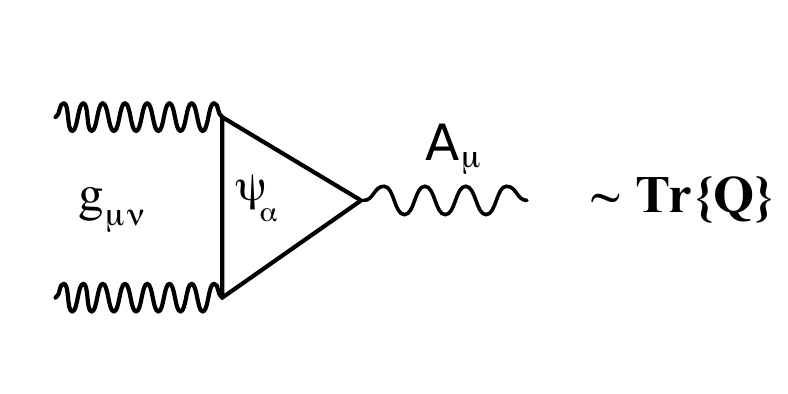}
        \caption{\footnotesize{Gravitational anomalies due to $U(1)$ charged fields running in the loop.}}
\end{figure}

\noindent
Therefore, if there are no gravitational anomalies, there are no
corrections to the FI term.

\section{${\cal N} = 2,4$ global supersymmetry}
\label{sec:N24GlobalSupersymmetry}

For ${\cal N}= 1$ SUSY, we had an action ${\cal S}$ depending on $K$, $W$, $f$ and $\xi$. What will the ${\cal N} \geq 2$ actions depend on?

\noindent
We know that in global supersymmetry, the ${\cal N}=1$ actions are particular
cases of non-supersymmetric actions (in which some of the couplings
are related, the potential is positive, etc.). In the same way, actions
for extended supersymmetries are particular cases of ${\cal N}=1$
supersymmetric actions and will therefore be determined by $K$, $W$,
$f$ and $\xi$. The extra supersymmetry will put constraints to these
functions and therefore the corresponding actions will be more
rigid. The larger the number of supersymmetries the more constraints on
actions arise.

\subsection{${\cal N} = 2$}
\label{sec:N2}

Consider the ${\cal N} = 2$ vector multiplet
    \[\begin{array}{ccc} &A_{\mu} & \\ \la & &\psi \\ &\vph & \end{array}
\]
where the $A_{\mu}$ and $\la$ are described by a vector superfield $V$ and the $\vph$, $\psi$ by a chiral superfield $\Phi$.

\noindent
We need $W = 0$ in the ${\cal N} = 2$ action. $K$, $f$ can be written in terms of a single holomorphic function ${\cal F}(\Phi)$ called {\em prepotential}:
    \[f(\Phi) \eq \frac{\pa^{2}{\cal F}}{\pa \Phi^{2}} \co K(\Phi , \Phi^\dag) \eq \frac{1}{2i} \; \left( \Phi^\dag \, \exp(2V) \; \frac{\pa {\cal F}}{\pa \Phi} \ - \ h.c. \right)
\]
The full perturbative action does not contain any corrections for more than 1 loop,
\[
{\cal F} \ \ = \ \ \left\{ \begin{array}{ll} \Phi^{2} &: \ (\te{tree level}) \\
\Phi^{2} \; \ln \left(\frac{\Phi^{2}}{\La^{2}} \right) &: \ (\te{1  loop}) \end{array} \right.
\]
where $\La$ denotes some cutoff. These statements apply to the Wilsonian effective action. Note that:
\begin{itemize}
\item Perturbative processes usually involve series $\sum_{n} a_{n} g^{n}$ with small coupling $g \ll 1$.
\item $\exp \left(-\frac{c}{g^{2}} \right)$ is a non-perturbative example (no expansion in powers of $g$ possible).
\end{itemize}
There are obviously more things in QFT than Feynman diagrams can tell, e.g. instantons and monopoles.

\noindent
Decompose the ${\cal N} = 2$ prepotential ${\cal F}$ as
    \[{\cal F}(\Phi) \eq {\cal F}_{1 \te{loop}} \ + \ {\cal F}_{\te{non-pert}}
\]
where ${\cal F}_{\te{non-pert}}$ for instance could be the
{\em instanton expansion} $\sum_{k} a_{k} \exp\left(-\frac{c}{g^{2}}
k \right)$. In 1994, \textsc{Seiberg} and \textsc{Witten} achieved such an expansion in ${\cal N}=2$ SUSY \cite{Seiberg:1994rs}.

\noindent
Of course, there are still vector- and hypermultiplets in ${\cal N} = 2$, but
those are much more complicated. We will now consider a particularly
simple combination of these multiplets.

\subsection{${\cal N} = 4$}
\label{sec:N4}

As an $N = 4$ example, consider the vector multiplet,
    \[\underbrace{\cccb &A_{\mu} & \\ \la & &\psi_{1} \\ &\vph_{1} & \ccce}_{{\cal N} = 2 \ \te{vector}} \ + \ \underbrace{\cccb &\vph_{2} & \\ \psi_{3} & &\psi_{2} \\ &\vph_{3} & \ccce}_{{\cal N} = 2 \ \te{hyper}} \ .
\]
We are more constrained than in above theories, there are no free functions at all, only 1 free parameter:
    \[f \eq \tau \eq \underbrace{\frac{\Theta}{2\pi}}_{F_{\mu \nu} \, \tilde{F}^{\mu \nu}} \ + \ \underbrace{\frac{4\pi i}{g^{2}}}_{F_{\mu \nu} \, F^{\mu \nu}}
\]
$N = 4$ is a finite theory, moreover its $\be$ function vanishes. Couplings remain constant at any scale, we have {\em conformal invariance}. There are nice transformation properties under modular {\em $S$ duality},
    \[\tau \ \ \mapsto \ \ \frac{a\tau \, + \, b}{c\tau \, + \, d} \ ,
\]
where $a$, $b$, $c$, $d$ form a $SL(2, \mathbb Z)$ matrix.

\noindent
Finally, as an aside, major developments in string and field theories
have led to the realization that certain theories of gravity in Anti
de Sitter space are ''dual'' to field theories (without gravity) in one
less dimension, that happen to be invariant under conformal
transformations. This is the {\em AdS/CFT correspondence} allowing
to describe gravity (and string) theories to domains where they are not
well understood (and the same benefit applies to field theories as well). The prime example of this correspondence is AdS in 5 dimensions dual to a conformal field theory in 4 dimensions that happens to be ${\cal N}=4$ supersymmetry. 

\subsection{Aside on couplings}
\label{sec:AsideOnCouplings}

For all kinds of renormalization, couplings $g$ depend on a scale $\mu$. The coupling changes under RG transformations scale by scale. Define the $\be$ function to be
    \[\mu \; \frac{\dd g}{\dd \mu} \eq \be(g) \eq \underbrace{-b \, g^{3}}_{1 - \te{loop}} \ + \ ... \ .
\]
The theory's cutoff depends on the particle content.

\noindent
Solve for $g(\mu)$ up to 1 loop order:
    \[\int^{M}\limits_{m} \frac{\dd g}{g^{3}} \eq -b \int^{+\infty}\limits_{-\infty} \frac{\dd \mu}{\mu} \so -\frac{1}{2} \, \left(\frac{1}{g_{M}^{2}} \ - \ \frac{1}{g_{m}^{2}} \right) \eq  - b \, \ln \left(\frac{M}{m} \right)
\]
    \[\so g_{m}^{2} \eq \frac{1}{\frac{1}{g_{M}^{2}}\  + \ b \,\ln \left(\frac{m^{2}}{M^{2}} \right)}
\]
The solution has a pole at
    \[m_{0} \ \ =: \ \ \La \eq M \, \exp \left(-\frac{b}{2g^{2}} \right)
\]
\begin{figure}[ht]
    \centering
     \includegraphics[width=0.70\textwidth]{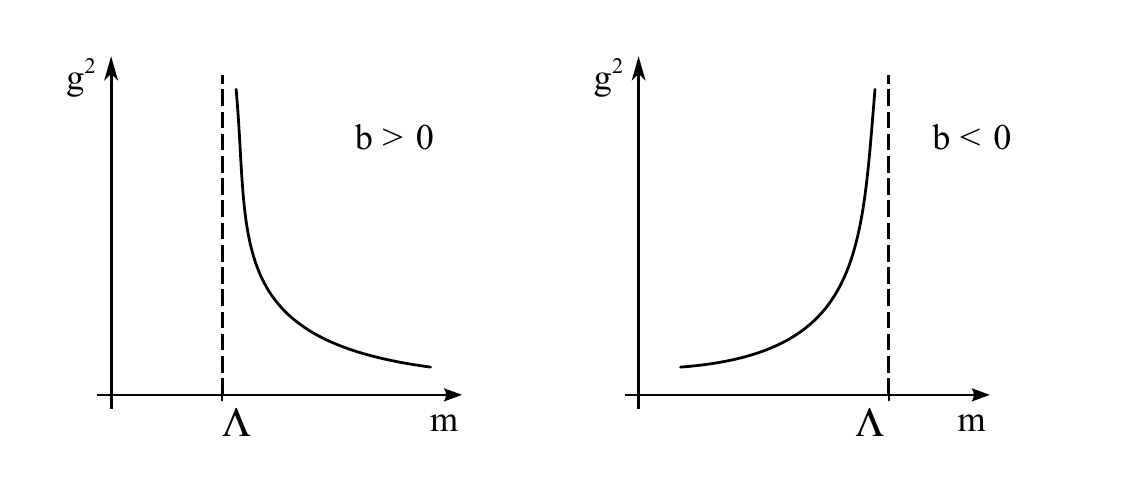}
      \caption{ \footnotesize{Landau pole in the running of the gauge coupling depending on the sign of the $-b g^3$ contribution to the $\be$ function: If $b>0$, the strength of the interaction monotonically decreases towards higher mass scales $m$. Negative values $b<0$, on the other hand, give rise to a pole in the renormalized coupling $g_m$ when $m$ grows towards some threshold scale $\La = M \cdot \exp \left( - \frac{1}{2 b g_M^2} \right).$
      }}
\end{figure}
which is the natural scale of the theory. For $m \mto \infty$, get asymptotic freedom as long as $b > 0$, i.e. $\lim_{m  \mto \infty} g_{m} = 0$. This is the case in QCD. If $b < 0$, however, a {\em Landau pole} emerges at some scale which is an upper bound for the energies where we can trust the theory. QED breaks down in that way.

\section{Supergravity: an Overview}
\label{sec:Supergravity}

This chapter provides only a brief overview of the main ideas and results on ${\cal{N}}=1$ supergravity. A detailed description is beyond the scope of the lectures.

\subsection{Supergravity as a gauge theory}
\label{sec:SupergravityAsAGaugeTheory}

We have seen that a superfield $\Phi$ transforms under supersymmetry like
    \[\de \Phi \eq i \, (\ep  {\cal Q} \ + \ \bar{\ep} \bar{{\cal Q}} ) \, \Phi \ .
\]
The questions arises if we can make $\ep$ a function of spacetime coordinates $\ep(x)$, i.e. extend SUSY to a local symmetry. The answer is yes, the corresponding theory is {\em supergravity}.

\noindent
How did we deal with local $\al(x)$ in internal symmetries? We introduced a gauge field $A_{\mu}$ coupling to a current $J^{\mu}$ via interaction term $A_{\mu} J^{\mu}$. That current $J^{\mu}$ is conserved and the corresponding charge $q$ constant
    \[q \eq \int \dd^{3}x \ J^{0} \eq const \ .
\]
For spacetime symmetries, local Poincaré parameters imply the equivalence principle which is connected with gravity. The metric $g_{\mu \nu}$ as a gauge field couples to the ''current'' $T^{\mu \nu}$ via $g_{\mu \nu} T^{\mu \nu}$. Conservation $\pa_{\mu} T^{\mu \nu} = 0$ implies constant total momentum
    \[P^{\mu} \eq \int \dd^{3}x \ T^{\mu 0} \eq const \ .
\]
Now consider local SUSY. The gauge field of that supergravity is the {\em gravitino} $\Psi^{\mu}_{\al}$ with associated {\em supercurrent} ${\cal J}^{\mu}_{\al}$ and SUSY charge
    \[Q_{\al} \eq \int \dd^{3}x \ {\cal J}^{0}_{\al} \ .
\]
Let us further explain the role of the gravitino gauge field and its embedding into a supermultiplet in the following subsections.

\subsection{The linear supergravity multiplet with global supersymmetry}

Recall that the vector field $A_\mu$ associated with a local internal symmetry has a gauge freedom under $\de^g A_\mu = \pa_\mu \al$ for some local parameter $\al(x)$ which is a scalar under the Lorentz group. The analogue for a spinorial gauge parameter $\eta_\al(x)$ is the gravitino field $\Psi^\mu_\al$ which carries both a vector and a spinor index and can be gauge transformed as
\[ \de^g_\eta \Psi^\mu_\al \eq \pa^\mu \eta_\al \ .
\]
The gravitino's dynamics is described by the gauge invariant \textsc{Rarita Schwinger} action
\[ {\cal S}_{\te{RS}}[\Psi] \ \ := \ \ \frac{1}{2} \int \dd^4 x \ \ep^{\mu \nu \rho \si} \, \Psib_\mu \, \ga_5 \, \ga_\nu \, \pa_{\rho} \Psi_\si 
\]
which we give in Dirac spinor notation (see appendix \ref{sec:Dirac}). The gravitino can be easily combined with a linearized graviton excitation
\[ g_{\mu \nu} \eq \eta_{\mu \nu} \ + \ \ka \, h_{\mu \nu} \co \ka^2 \eq \frac{8\pi}{M_{\te{pl}}^2} \]
into the {\em linearized supergravity multiplet} $(h_{\mu \nu} , \Psi_\mu)$. The latter is governed by the linearized Einstein Hilbert action
\begin{align*}
S_{\te{EH}}[h] \ \ &:= \ \ -\, \frac{1}{2} \int \dd^4 x \ h^{\mu \nu} \, \left( R_{\mu \nu}^{\te{L}} \ - \ \frac{1}{2} \; \eta_{\mu \nu} \, R^{\te{L}} \right) \\
R_{\mu \nu}^{\te{L}} \ \ &:= \ \ \frac{1}{2} \; \Bigl( \pa_\mu \pa_\la h^\la \, _\nu \ + \ \pa_\nu \pa_\la h^\la\, _\mu \ - \ \pa_\mu \pa_\nu h^\la \,_\la \ - \ \pa^2 h_\mu \nu \Bigr) \\
R^{\te{L}} \ \ &:= \ \ \eta^{\mu \nu} \, R_{\mu \nu}^{\te{L}}
\end{align*}
in terms of the linearized Ricci tensor $R_{\mu \nu}^{\te{L}}$ and Ricci scalar $R^{\te{L}}$. It enjoys the spin two gauge invariance under
\[ \de^g_\xi h_{\mu \nu} \eq \pa_\mu \xi_\nu \ + \ \pa_\nu \xi_\mu \ . \]
By adding ${\cal S}_{\te{RS}} + {\cal S}_{\te{EH}}$, we arrive at a field theory with global supersymmetry under variations
\begin{align*}
\de_{\vep} \Psi_\mu \ \ &= \ \ \frac{1}{2} \; \Bigl[ \ga^\rho \ , \ \ga^\si \Bigr] \, \vep \, \pa_\rho h_{\mu \si} \\
\de_{\vep} h_{\mu \nu} \ \ &= \ \ -\, \frac{i}{2} \; \bar \vep \, \bigl( \ga_\mu \Psi_\nu \ + \ \ga_\nu \, \Psi_\mu \bigr) \ .
\end{align*}
However, their algebra closes up to gauge transformations only,
\begin{align*}
\Bigl[ \de_{\vep_1} \ , \ \de_{\vep_2} \Bigr] \, \Psi_\mu \ \ &= \ \ - \, 2i \, (\bar \vep_1 \, \ga^\nu \, \vep_2) \, \pa_\nu \Psi_\mu \ + \ i \pa_\mu \, \left( (\bar \vep_1 \, \ga^\nu \, \vep_2) \, \Psi_\nu \ + \ \frac{1}{4} \, (\bar \vep_1 \, \ga^\nu \, \vep_2) \, \ga_\nu \, \ga_\rho \, \Psi^\rho \right)  \\
&=: \ \ 2 \, (\bar \vep_1 \, \ga^\nu \, \vep_2) \, {\cal P}_\nu \, \Psi_\mu \ + \ \de^g_\eta \Psi_\mu \\
\Bigl[ \de_{\vep_1} \ , \ \de_{\vep_2} \Bigr] \, h_{\mu \nu} \ \ &= \ \ - \, 2i \, (\bar \vep_1 \, \ga^\rho \, \vep_2) \, \pa_\rho h_{\mu \nu} \ + \ i \, (\bar \vep_1 \, \ga^\rho \, \vep_2) \, \bigl( \pa_\mu h_{\rho \nu} \ + \ \pa_\nu h_{\rho \mu} \bigr)  \\
&=: \ \ 2 \, (\bar \vep_1 \, \ga^\rho \, \vep_2) \, {\cal P}_\rho \, h_{\mu \nu} \ + \ \de^g_\xi h_{\mu \nu}
\end{align*}
The commutator of two supersymmetry transformations with parameters $\vep_1, \vep_2$ not only yields the translation ${\cal P}_\nu$ familiar from the Wess Zumino model but also a gravitino gauge transformation with spinor parameter $\eta = (\bar \vep_1 \ga^\nu  \vep_2 ) \Psi_\nu \, + \, \frac{1}{4}  (\bar \vep_1  \ga^\nu \vep_2) \ga_\nu \ga_\rho \Psi^\rho$ and a graviton gauge transformation with vectorial parameter $\xi_\mu = i (\bar \vep_1 \ga^\rho \vep_2) h_{\rho \mu}$.

\subsection{The supergravity multiplet with local supersymmetry}

If the supersymmetry transformation parameters of the linear supergravity multiplet $\de_{\vep}(\Psi_\mu, h_{\mu \nu})$ are promoted to spacetime functions $\vep = \vep(x)$, then its free action is modified as
\[
\de_\vep \bigl( {\cal S}_{\te{RS}}[\Psi] \ + \ {\cal S}_{\te{EH}}[h]  \bigr) \ \ = \ \ \int \dd^4 x \ {\cal J}^\mu  \, \pa_\mu \vep 
\]
from which we can read off the supercurrent
\[
{\cal J}^\mu \ \ = \ \ \frac{1}{4} \; \ep^{\mu \nu \rho \si} \, \Psib_\rho \ga_5 \, \ga_\nu \, \Bigl[ \ga^\la \ , \ \ga^\tau \Bigr] \, \pa_\la h_{\tau \si} \ .
\]
One can now proceed in close analogy to electromagnetism and apply the \textsc{Noether} procedure to maintain invariance of the overall action under local transformations. Suppose we want to achieve local $U(1)$ symmetry $\de \psi = i\al (x) \psi$ into the electron's \textsc{Dirac} action
\[ {\cal S}_{\te{D}}[\psi] \eq i \int \dd^4 x \ \bar \psi \, \ga^\mu \, \pa_\mu \psi \co \de {\cal S}_{\te{D}}[\psi] \eq - \int \dd^4 x \ J^\mu \, \pa_\mu \al \ ,
\]
then the extra contribution
\[
{\cal S}_{\te{int}}[\psi,A] \eq - \int \dd^4 x \ J^\mu \, A_\mu 
\]
restores invariance in the total action if the gauge field $A_\mu$ obeys the transformation law $\de A_\mu =  -\pa_\mu \al$. As a result, the photon is coupled to the electric current $J^\mu = \bar \psi \ga^\mu \psi$. This interaction can be expressed more elegantly as
\[
{\cal S}_{\te{D}}[\psi] \ + \  {\cal S}_{\te{int}}[\psi,A] \eq i \int \dd^4 x \ \bar \psi \, \ga^\mu \, D_\mu \psi \co D_\mu \ \ := \ \ \pa_\mu + iA_\mu
\]
in terms of a covariant derivative $D_\mu$.

\noindent
The variation in the action ${\cal S}_{\te{RS}}[\Psi] + {\cal S}_{\te{EH}}[h]$ of the linear supergravity multiplet can be compensated by an interaction term
\[
{\cal S}_{\te{int}}[\Psi,h] \eq - \, \frac{\ka}{2} \int \dd^4 x \ {\cal J}^\mu \, \Psi_\mu  
\]
provided that the supersymmetry transformation of $\Psi_\mu$ is enriched by
\[
\de_\vep \Psi_\mu \eq \frac{2}{\ka} \; \pa_\mu \vep \ + \ \frac{1}{2} \; \Bigl[ \ga^\rho \ , \ \ga^\si \Bigr] \, \vep \, \pa_\rho h_{\mu \si} \ + \ ...
\]
Just like in electrodynamics, the graviton-gravitino interaction can be absorbed into ${\cal S}_{\te{RS}}$ by replacing the ordinary derivative by an appropriately defined covariant one $\pa_\rho \mapsto D_\rho$. However, to achieve local invariance to all orders in $\ka$, some bilinear terms in $\Psi$ are required in the full transformation $\de_\vep \Psi_\mu$. In the non-linear theory of ${\cal N} = 1$ supergravity, the covariantized \textsc{Rarita Schwinger} action $\sim \int \dd^4 x \, \ep^{\mu \nu \rho \si} \Psib_\mu \ga_5 \ga_\nu D_{\rho} \Psi_\si$ is in fact quite involved with all these extra terms in $D_\rho$ and therefore beyond the scope of these lectures.

\noindent
Historically, the first local supergravity actions were constructed by \textsc{Ferrara}, \textsc{Freedman} and \textsc{van Niewenhuizen}, followed closely by \textsc{Deser} and \textsc{Zumino} in 1976.
\subsection{${\cal N} = 1$ Supergravity in Superspace}
\label{sec:SupergravitySuperspace}

There is a very convenient formulation of supergravity in terms of superfields, generalising the superfield formulation of global supersymmetry. For this the superspace coordinates
$z^M=\{x^\mu, \theta_{\alpha}, \bar{\theta}_{\dot{\alpha}} \}$ are subject to supersymmetric generalisations of general coordinate transformations $z'^M=z^M+\zeta^M$. The supergravity multiplet is included into a superfield
with components $\{e^\mu_a, \psi^\mu_\alpha, M, b_a\}$ where $e^\mu_a$ is the vierbein describing the metric $g_{\mu\nu}=e^a_\mu e_{a\nu}$, $\psi^\mu_\alpha$ the gravitino, $M$ a complex scalar auxiliary field and $b_a$ a real vector auxiliary field. The vierbein has a superspace generalisation $E^M_A$. A superspace density (generalising $\sqrt{-g}=e=\det{e^\mu_a}$)  is given by $\det{E^M_A}\equiv {\bf E}$. The supergravity action 
(in Planck units $M_{\te{pl}}^2=1$) can be written in a compact way as:
\[
{\cal{S}}_{\te{SG}} \ = \ -3 \, \int \dd^8z \ {\bf E} \eq -\frac{1}{2} \int \dd^4 x \ e \, \biggl\{ \,  R \ - \ \frac{1}{3} \; \bar{M} \, M \  + \ \frac{1}{3} \;  b^a \, b_a \ + \ \frac{1}{2} \;   \epsilon^{\mu\nu\rho\sigma} \, \left(\bar{\psi}_\mu \,  \bar{\sigma}_\nu \, {\cal{D}}_\rho  \psi_\sigma \ - \ \psi_\mu \, \sigma_\nu \, {\cal{D}}_\rho\bar{\psi}_\sigma\right) \, \biggr\}
\]
Here $\dd^8z= \dd^4x \, \dd^4\theta$ and $\cal{D}$ is a covariant derivative. The non-propagating auxiliary fields complete the supergravity multiplet providing an off-shell invariant action. Integrating them out by their field equations give rise to the Einstein plus Rarita-Schwinger actions.

\subsection{${\cal N} = 1$ supergravity coupled to matter}
\label{sec:N1SupergravityCoupledToMatter}

Here we will provide, without a full derivation from first principles, some relevant properties of ${\cal N}=1$ supergravity actions coupled to matter.

\noindent
The total Lagrangian is a sum of supergravity contribution ${\cal L}_{\te{SG}}$ and the SUSY Lagrangian discussed before,
    \[{\cal L} \eq {\cal L}_{\te{SG}} \ + \ {\cal L}(K , W,f,\xi) \ .
\]
where the second term is understood to be covariantized under general coordinate transformations. We are interested in the scalar potential of supergravity, for this we focus on the chiral scalar part of the action which can be written as
\[
{\cal S} \eq - \,\frac{3}{\kappa^2}\int \dd^4x \ \dd^4 \theta \ {\bf E} \,e^{-\frac{\kappa^2}{3}K} \ + \ \left(\int \dd^4x \ \dd^4\theta \ {\cal{E}} \, W \ + \ {\rm h.c.}\right)
\]
where we have restored $M_{\te{pl}}$ and $\kappa^2:=8\pi G_N=1/M_{\te{pl}}^2$. As above,  ${\bf E}$ is the determinant of the supervielbein and ${\cal{E}}$ is defined by $2{\cal{R}} {\cal{E}}=\bf{E}$ where ${\cal{R}}$ is the curvature superfield (having components $R, \psi^\mu, M, b_a$). Notice that the first term of this action, when expanded in powers of $\kappa^2$ includes the pure supergravity action plus the standard kinetic term for matter fields. This can be seen by writing:
\[
e^{-\frac{\kappa^2}{3}K} \ \ = \ \ 1 \ - \ \frac{\kappa^2}{3}\,  K \ + \  {\cal{O}}(\kappa^4)
\]
The flat space limit corresponds to $\kappa\to 0,$ $\int d^2\bar{\theta}\,{\cal{E}}\to 1,$ and  ${\bf E}\to 1$ and the flat space global supersymmetric action from in terms of $K$ and $W$ is reproduced. Actually the condition to reproduce the flat limit together with general supercoordinate invariance singles out the apparent unusual dependence of the action on $K$ above. 
  
\noindent
For any finite value of $\kappa$ the fact that $K$ appears explicitly in the pure supergravity part of the action implies 
  that the coefficient of the \textsc{Einstein} term, which is the effective Planck mass, depends on the chiral matter fields as in \textsc{Brans-Dicke} theories. In order to go to the \textsc{Einstein} frame (constant Planck mass) a rescaling of the metric needs to be done, this in turn requires a rescaling of the fermionic fields in the theory, by supersymmetry, complicating substantially the derivation of the action in components. In order to avoid these complications an extra superfield $\varphi$ is usually introduced, known as the {\it Weyl compensator} field. This field is not a physical field, since it does not propagate. It is introduced in such a way that it makes the action invariant under scale and conformal transformations. After the component action is computed, $\varphi$ is fixed to a value such that the \textsc{Einstein} term is canonical, breaking the scale invariance and reproducing the wanted action in components. The action above is then modified as:
  \[
{\cal S} \eq -3\int \dd^4x \ \dd^4 \theta \ {\bf E} \,  \varphi \, \bar{\varphi}\,e^{-K/3} \ + \ \left(\int \dd^4x \ \dd^4\theta\,  {\cal{E}} \, \varphi^3 \, W \ + \ {\rm h.c.}\right)
\]
This action is invariant under 'rescalings' of the metric $\bf{E}\to e^{2(\tau+\bar{\tau})}$ and ${\cal{E}}\to e^{6\tau}\, {\cal{E}} +\cdots$ with $\tau$ a chiral superfield (and all matter fields invariant) if $\varphi\to e^{-2\tau} \varphi$. 
Notice that in order to obtain the standard Einstein action the lowest component of $\varphi$ has to be fixed to $\varphi\bar{\varphi} e^{-K/3}= M_{\te{pl}}^2$ thus breaking explicitly the (artificial)
conformal  invariance and leaving the physical fields properly normalised with standard kinetic terms.

\noindent
Deriving  the full  component action from the superfield action above is then straightforward but tedious. Here we are interested in  obtaining the scalar potential which plays a very important role in supersymmetric theories.
For this  it is sufficient to consider flat spacetime, which leads to ${\bf E}=1, \int d^2\bar{\theta}\, {\cal{E}}=1$ and the covariant derivatives reduce to the global covariant derivatives.
\[
{\cal S} \eq -3 \int \dd^4x \ \dd^4 \theta\ \varphi \, \bar{\varphi}\,e^{-K/3} \ + \ \left(\int \dd^4x\ \dd^2\theta\ \varphi^3 \, W \ + \ {\rm h.c.}\right)
\]
In similar fashion to the global supersymmetric case, one can obtain the scalar potential in supergravity
  \[\framebox{  $ \displaystyle V_{F} \eq \exp \left( \frac{K}{M_{\te{pl}}^{2}} \right)  \, \left\{ (K^{-1})^{i\bar{j}}  \, D_{i}W \, D_{\bar{j}}W^{*} \ - \ 3 \; \frac{|W|^{2}}{M_{\te{pl}}^{2}} \right\} $} 
\]
	\[D_{i} W \ \ := \ \ \pa_{i}W\ + \ \frac{1}{M_{\te{pl}}^{2}} \; (\pa_{i} K) \, W \ .
\]
In the $M_{\te{pl}} \mto \infty$ limit, gravity is decoupled and the global supersymmetric scalar potential $V_{F} = (K^{-1})^{i\bar{j}} \pa_{i}W
\pa_{\bar{j}}W^{*}$ restored. Notice that for finite values of the Planck mass, the potential $V_{F}$ above is no longer positive. The extra (negative) factor proportional to $-3|W|^2$ comes from the auxiliary fields of the gravity 
(or compensator) multiplet.

\paragraph{Exercise 4.4:}{\it Derive the equations of motion for the auxiliary F-terms in the above action. To simplify the expression use the following covariant derivative $D_i W= \partial_i W + W \partial_i K.$ Using the expression for the F-term, derive in analogy to the global supersymmetric case the F-term scalar potential for the above action.}

\noindent
Some important things should be stated here:
\begin{itemize}
\item{}
This action has a so-called {\em K\"ahler invariance} under
\begin{align*}
K \ \ &\mapsto \ \ K \ + \ h(\Phi) \ + \ h^{*}(\Phi^{*}) \\
W \ \ &\mapsto \ \ \exp \bigl( -h(\Phi) \bigr) \, W
\end{align*}
This can be seen directly since this transformation can be compensated by transforming the Weyl compensator by $\varphi\to e^{h(\Phi)}\, \varphi$. The scalar F-term potential then becomes
\[
V\eq e^G \, (G^{i\bar{j}} \, G_i \, G_{\bar{j}} \ - \ 3) \co G\ \ := \ \ K \ + \ \ln|W|^2 \ .
\]
This implies that, contrary to global supersymmetry, $K$ and $W$ are not totally independent since the action depends only on the invariant combination $G$. In particular, $W$ can in principle be absorved into the 
K\"ahler potential. This is true as long as $W$ is not zero nor singular and therefore in practice it is more convenient to still work with $K$ and $W$ rather than $G$. 
\item{}
So far we have not included the gauge fields couplings to supergravity. These are just as in global supersymmetry except for three important points:
First, the Weyl symmetry introduced above is valid only classically and develops an anomaly at one-loop. In order to cancel the anomaly a shift in the gauge kinetic function is needed:
$f\to f+c\log\varphi$, see \cite{Kaplunovsky:1994fg} for further reference. 
Secondly, the D-term $D^a$ in supergravity is given by
\[
D^a \eq {\rm Tr}(\partial_i K \, T^a \, \phi_i)\, .
\]
Using the K\"ahler invariance from above, this exhibits an interesting relation between the D-term and F-term in supergravity:
\[
D^a \eq  G_i \,  T^a \, \phi_i \eq e^{-G/2} \, F_i \, T^a \, \phi_i
\]
Since the F-term is proportional to $G_i$ (see exercise 4.4). As long as $W$ is non zero, F- and D terms are proportional. One immediate consequence of this relation, that we will see later on,  is that in supergravity, there is no single D-term or F-term supersymmetry breaking.

\item{}
Finally, again contrary to global supersymmetry, there is a correlation between the existence of a (constant) Fayet-Iliopoulos term $\xi$ and the existence of a global symmetry. If there are no global symmetries there cannot be Fayet-Iliopoulos terms \cite{seiberg}.
\end{itemize}

\chapter{Supersymmetry breaking}

\section{Basics}

We know that fields $\vph_{i}$ of gauge theories transform as
    \[\vph_{i} \ \ \mapsto \ \ \bigl( \exp(i\al^{a} T^{a}) \bigr)_{i}\,^{j} \, \vph_{j} \co \de \vph_{i} \eq i\al^{a} \, (T^{a})_{i}\,^{j} \, \vph_{j}
\]
under finite and infinitesimal group elements. Gauge symmetry is broken if the vacuum state $(\vph_{\te{vac}})_{i}$ transforms in a non-trivial way, i.e.
    \[(\al^{a} T^{a})_{i}\,^{j} \, (\vph_{\te{vac}})_{j} \ \ \neq \ \ 0 \ .
\]
In $U(1)$, let $\vph = \rho \exp(i\vth)$ in complex polar coordinates, then infinitesimally
    \[\de \vph \eq i\al \, \vph \so \de \rho \eq 0 \co \de \vth \eq \al \ ,
\]
the last of which corresponds to a Goldstone boson.

\noindent
Similarly, we speak of broken SUSY if the vacuum state $|\te{vac} \rangle$ satisfies
    \[Q_{\al} \, \vac \ \ \neq \ \ 0 \ .
\]
Let us consider the anticommutation relation $\{Q_{\al} , \bar{Q}_{\dot{\be}} \} = 2(\si^{\mu})_{\al \dot{\be}} P_{\mu}$ contracted with $(\bar{\si}^{\nu})^{\dot{\be}\al}$,
    \[(\bar{\si}^{\nu})^{\dot{\be}\al} \, \Bigl\{Q_{\al} \ , \ \bar{Q}_{\dot{\be}} \Bigr\} \eq 2 \, (\bar{\si}^{\nu})^{\dot{\be}\al} \, (\si^{\mu})_{\al \dot{\be}} \, P_{\mu} \eq 4 \, \eta^{\mu \nu}  \, P_{\mu} \eq 4 \, P^{\nu} \ ,
\]
in particular the $(\nu = 0)$ component using $\bar{\si}^{0} = \mathds{1}$:
    \[(\bar{\si}^{0})^{\dot{\be}\al} \, \Bigl\{Q_{\al} \ , \ \bar{Q}_{\dot{\be}} \Bigr\} \eq \sum^{2}_{\al = 1} \bigl( Q_{\al} \, Q_{\al}^\dag \ + \ Q_{\al}^\dag \, Q_{\al} \bigr) \eq 4 \, P^{0} \eq 4 \, E
\]
This has two very important implications:
\begin{itemize}
\item $E \geq 0$ for any state, since $Q_{\al} Q_{\al}^\dag + Q_{\al}^\dag Q_{\al}$ is positive definite
\item $\langle \te{vac} | Q_{\al} Q_{\al}^\dag + Q_{\al}^\dag Q_{\al} \vac > 0$, so in broken SUSY, the energy is strictly positive, $E > 0$
\end{itemize}

\section{$F$- and $D$ breaking}
\label{sec:FAndDBreaking}

\subsection{$F$ term breaking}
\label{sec:FTerm}

Consider the transformation - laws under SUSY for components of a chiral superfield $\Phi$,
\begin{align*}
\de \vph \ \ &= \ \ \sqrt{2} \, \ep \psi \\
\de \psi \ \ &= \ \ \sqrt{2} \, \ep \, F \ + \ i \sqrt{2} \, \si^{\mu} \, \bar{\ep} \, \pa_{\mu}\vph \\
\de F \ \ &= \ \ i \sqrt{2} \, \bar{\ep} \, \bar{\si}^{\mu} \, \pa_{\mu} \psi \ .
\end{align*}
If one of $\de \vph$, $\de \psi$, $\de F \neq 0$, then SUSY is broken. But to preserve Lorentz invariance, need
    \[\langle \psi \rangle \eq \langle \pa_{\mu} \vph \rangle \eq 0
\]
as they would both transform under some representation of Lorentz group. So our SUSY breaking condition simplifies to
    \[\not{\!\!\!\!\!\!\! \te{SUSY}} \ \ \ \Longleftrightarrow \ \ \ \langle F \rangle \ \ \neq \ \ 0 \ .
\]
Only the fermionic part of $\Phi$ will change,
    \[\de \vph \eq \de F \eq 0 \co \de \psi \eq \sqrt{2} \, \ep \, \langle F \rangle \ \ \neq \ \ 0 \ ,
\]
so call $\psi$ a {\em Goldstone fermion} or the {\em goldstino} (although it is not the SUSY partner of some Goldstone boson). Remember that the $F$ term of the scalar potential is given by
    \[V_{(F)} \eq K_{i\bar{j}}^{-1} \; \frac{\pa W}{\pa \vph_{i}} \;  \frac{\pa W^{*}}{\pa \vph^{*}_{\bar{j}}}  \ ,
\]
so SUSY breaking is equivalent to a positive vacuum expectation value
    \[\not{\!\!\!\!\!\!\! \te{SUSY}} \ \ \ \Longleftrightarrow \ \ \ \langle V_{(F)} \rangle \ \ > \ \ 0 \ .
\]
\begin{figure}[ht]
    \centering
    \includegraphics[width=0.950\textwidth]{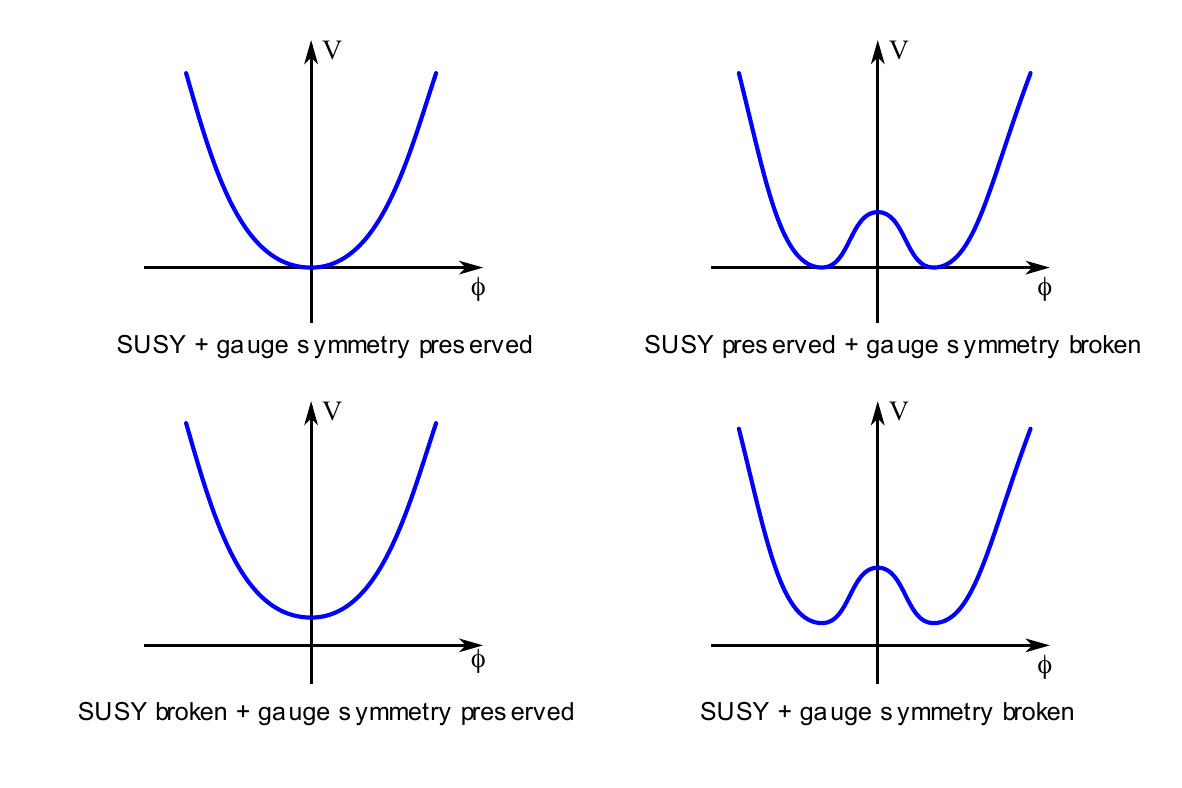}
        \caption{ \footnotesize{Various symmetry breaking scenarions: SUSY is broken, whenever the minimum potential energy $V(\vph_{\te{min}})$ is nonzero. Gauge symmetry is broken whenever the potential's minimum is attained at a nonzero field configuration $\vph_{\te{min}} \neq 0$ of a gauge non-singlet.}}
\end{figure}

\subsection{O'Raifertaigh model}
\label{sec:ORaifertaighModel}

The {\em O'Raifertaigh model} involves a triplet of chiral superfields $\Phi_{1}$, $\Phi_{2}$, $\Phi_{3}$ for which K\"ahler- and superpotential are given by
    \[K \eq \Phi_{i}^\dag \, \Phi_{i} \co W \eq g \, \Phi_{1} \, (\Phi_{3}^{2} \ - \ m^{2}) \ + \ M \,\Phi_{2} \, \Phi_{3} \co M \ \ \gg \ \ m \ .
\]
From the $F$ equations of motion, if follows that
\begin{align*}
-F_{1}^{*} \ \ &= \ \ \frac{\pa W}{\pa \vph_{1}} \eq g \, (\vph_{3}^{2} \ - \ m^{2}) \\
-F_{2}^{*} \ \ &= \ \ \frac{\pa W}{\pa \vph_{2}} \eq M \, \vph_{3} \\
-F_{3}^{*} \ \ &= \ \ \frac{\pa W}{\pa \vph_{3}} \eq 2 \, g \, \vph_{1} \, \vph_{3} \ + \ M \, \vph_{2} \ .
\end{align*}
We cannot have $F_{i}^{*} = 0$ for $i = 1,2,3$ simultaneously, so the form of $W$ indeed breaks SUSY. Now, determine the spectrum:
    \[V \eq \left( \frac{\pa W}{\pa \vph_{i}} \right) \, \left( \frac{\pa W}{\pa \vph_{j}} \right)^{*} \eq g^{2} \, \bigl| \vph_{3}^{2} \ - \ m^{2} \bigr|^{2} \ + \ M^{2} \, |\vph_{3}|^{2} \ + \ \bigl|2 \,g \, \vph_{1} \, \vph_{3} \ + \ M \, \vph_{2} \bigr|^{2}
\]
If $m^{2} < \frac{M^{2}}{2g^{2}}$, then the minimum is at
    \[\langle \vph_{2} \rangle \eq \langle \vph_{3} \rangle \eq 0 \co \langle \vph_{1} \rangle \ \te{arbitrary} \ .
\]
\begin{figure}[ht]
    \centering
        \includegraphics[width=0.45\textwidth]{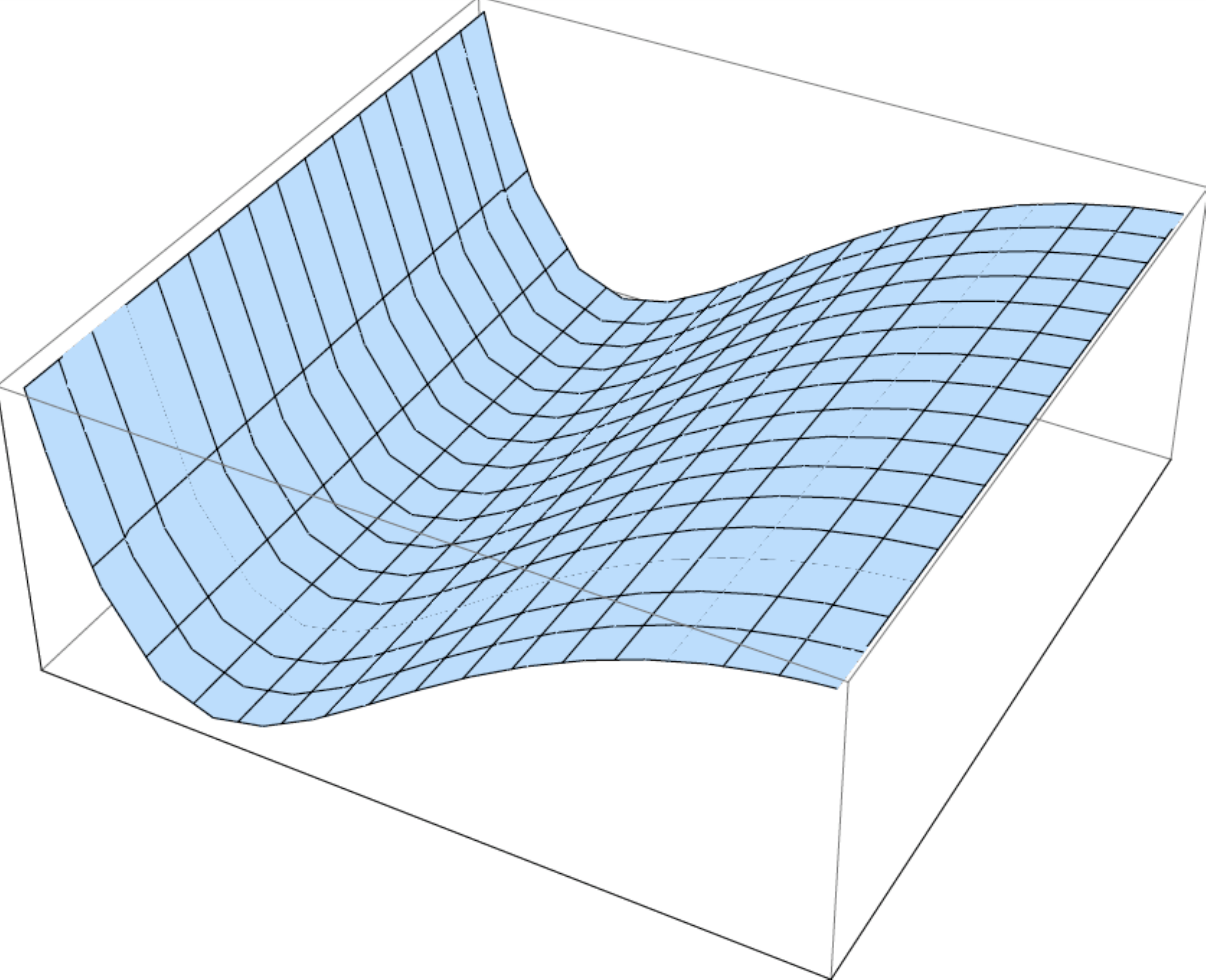}
       \caption{ \footnotesize{Example of a flat direction: If the potential takes its minimum for a continuous range of field configurations (here: for any $\vph_1 \in \mathbb C$), then it is said to have a flat direction. As a result, the scalar field $\vph_1$ will be massless.}}
\end{figure}
    \[\Longrightarrow \ \ \ \langle V \rangle \eq g^{2} \, m^{4} \ \ > \ \ 0 \ .
\]
This arbitrariness of $\vph_{1}$ implies zero mass, $m_{\vph_{1}} = 0$. For simplicity, set $\langle \vph_{1} \rangle = 0$ and compute the spectrum of fermions and scalars. Consider the fermion mass term
    \[\left\langle \frac{\pa^{2}W}{\pa \vph_{i} \pa \vph_{j}}\right\rangle \; \psi_{i} \psi_{j} \eq \cccb 0 &0 &0 \\ 0 &0 &M \\ 0 &M &0 \ccce \, \psi_{i} \psi_{j}
\]
in the Lagrangian, which gives $\psi_{i}$ masses
    \[m_{\psi_{1}} \eq 0 \co m_{\psi_{2}} \eq m_{\psi_{3}} \eq M \ .
\]
$\psi_{1}$ turns out to be the goldstino (due to $\de \psi_{1} \propto \langle F_{1} \rangle \neq 0$ and zero mass). To determine scalar masses, look at the quadratic terms in $V$:
    \[V_{\te{quad}} \eq -m^{2} \, g^{2} \, (\vph_{3}^{2} \, + \, \vph_{3}^{*2}) \ + \ M^{2} \, |\vph_{3}|^{2} \ + \ M^{2} \, |\vph_{2}|^{2} \so m_{\vph_{1}} \eq 0 \co m_{\vph_{2}} \eq M
\]
Regard $\vph_{3}$ as a complex field $\vph_{3} = a + ib$ where real- and imaginary part have different masses,
    \[m_{a}^{2} \eq M^{2} \ - \ 2 \,g^{2} \, m^{2} \co m_{b}^{2} \eq M^{2} \ + \ 2 \, g^{2} \, m^{2} \ .
\]
This gives the following spectrum:
\begin{figure}[ht]
    \centering
        \includegraphics[width=0.70\textwidth]{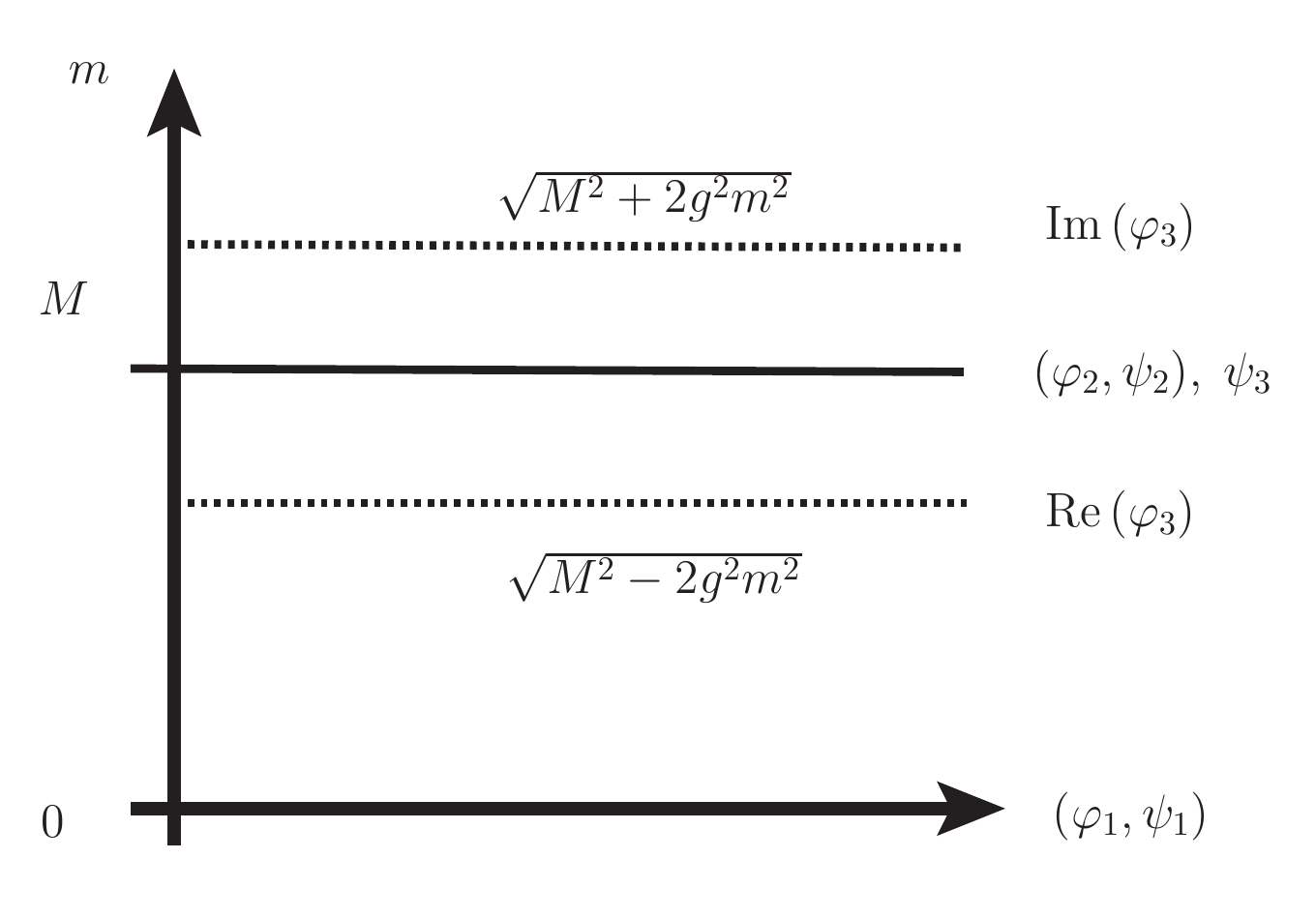}
        \caption{ \footnotesize{Mass splitting of the real- and imaginary part of the third scalar $\vph_3$ in the O'Raifertaigh model.}}
\end{figure}
\subsubsection{}
We generally get heavier and lighter superpartners, the
{\em supertrace} of $M$ (treating bosonic and fermionic parts
differently) vanishes. This is generic for tree level of broken SUSY. Since $W$ is not
renormalized to all orders in perturbation theory, we have an
important result: If SUSY is unbroken at tree level, then it also
unbroken to all orders in perturbation theory. This means that in order to
break supersymmetry we need to consider non-perturbative effects:
    \[\Longrightarrow \ \ \ \ \ \ \ \not{\!\!\!\!\!\!\! \te{SUSY}} \ \te{non-perturbatively}
\]

\paragraph{Exercise 5.1:}{\it By analysing the mass matrix for scalars and fermions, verify for the O'Raifertaigh model
$$ {\rm STr} \bigl\{ M^2 \bigr\} \ \ := \ \ \sum_j (-1)^{2j+1} \, (2j+1) \, m_j^2 \eq 0\ ,$$
where $j$ represents the 'spin' of the particles.
}
\subsection{$D$ term breaking}
\label{sec:DTerm}

Consider a vector superfield $V = (\la , \ A_{\mu}, \ D)$,
    \[\de \la \ \ \propto \ \ \ep \, D \so \langle D \rangle \ \ \neq \ \ 0 \so \ \ \ \ \not{\!\!\!\!\!\!\! \te{SUSY}} \ .
\]
$\la$ is a goldstino (which, again, is not the fermionic partner of any Goldstone boson). More on that in the examples.

\paragraph{Exercise 5.2:}{\it Consider a chiral superfield $\Phi$ of charge $q$ coupled to an Abelian vector superfield $V.$ Write down the D-term part of the scalar potential. Show that a non-vanishing vacuum expectation value of $D,$ the auxiliary field of $V,$ can break supersymmetry. Find the condition that the Fayet-Iliopoulos term and the charge $q$ have to satisfy for supersymmetry to be broken. Find the spectrum of this model after supersymmetry is broken and discuss the mass splitting of the multiplet.}
\paragraph{Exercise 5.3}{\it Consider a renormalisable ${\cal N}=1$ supersymmetric theory with chiral superfields $\Phi_i=(\vp_i,\psi_i,F_i)$ and vector superfields $V_a=(\lambda_a,A_a^\mu,D_a)$ with both D and F term supersymmetry breaking ($F_i\neq0$ and $D_a\neq0$). Show that in the vacuum
$$\frac{\pa V}{\pa \vp_i}=F^j\frac{\pa^2 W}{\pa \vp_i\vp_j}+g^a D^a \vp_j^\dagger (T^a)_{i}^j=0\, .$$
Here $g^a$ and $T^a$ refer to the gauge coupling and generators of the gauge group respectively. Also, since the superpotential $W$ is gauge invariant, the gauge variation of $W$ is
$$\de^{(a)}_{\rm gauge} W= \frac{\pa W}{\pa \vp^i}\de^{(a)}_{\rm gauge}\vp^i=-F_i^\dagger (T^a)^i_j\vp^j\, .$$
Write these two conditions in the form of a matrix $M$ acting on a 'two-vector' with components $\langle F^j\rangle$ and $\langle D^a\rangle.$ Identify this matrix and show that it is the same as the fermion mass matrix. Argue that it has one zero eigenvalue which can be identified with the Goldstone fermion.
}
\section{Supersymmetry breaking in ${\cal N} = 1$ supergravity}
\label{sec:SupersymmetryBreakingInN1Supergravity}

\begin{itemize}
\item Supergravity multiplet adds new auxiliary - fields $F_{g}$ with nonzero $\langle F_{g} \rangle$ for broken SUSY.
\item The $F$ - term is proportional to
    \[F \ \ \propto \ \ D W \eq \frac{\pa W}{\pa \vph} \ + \ \frac{1}{M_{\te{pl}}^{2}} \; \frac{\pa K}{\pa \vph} \; W  \ .
\]
\item The scalar potential $V_{(F)}$ has a negative gravitational term,
    \[V_{(F)} \eq \exp \left( \frac{K}{M_{\te{pl}}^{2}} \right) \, \left\{ (K^{-1})^{i\bar{j}} \, D_{i}W \, D_{\bar{j}}W^{*} \ - \ 3 \; \frac{|W|^{2}}{M_{\te{pl}}^{2}} \right\} \ .
\]
That is why both $\langle V \rangle = 0$ and $\langle V \rangle \neq
0$ are possible after SUSY breaking in supergravity, whereas
broken SUSY in the global case required $\langle V \rangle > 0$. This
is very important for the cosmological constant problem (which is the
lack of understanding of why the vacuum energy today is almost zero).
The vacuum energy essentially corresponds to the value of the scalar potential at
the minimum. In global supersymmetry, we know that the breaking of
supersymmetry implies this vacuum energy to be large. In supergravity
it is possible to break supersymmetry at a physically allowed scale
and still to keep the vacuum energy zero. This does not solve the
cosmological constant problem, but it makes supersymmetric theories
still viable.

\item
{\it The super Higgs effect:} Spontaneously broken gauge theories
realize the Higgs mechanism in which the corresponding Goldstone boson
is ''eaten'' by the corresponding gauge field to get a mass. A similar
phenomenon happens in supersymmetry. The goldstino field joins the
originally massless gravitino field (which is the gauge field of ${\cal N}=1$
supergravity) and gives it a mass, in this sense the gravitino receives its mass by ''eating'' the goldstino. A massive gravitino (keeping a massless
graviton) illustrates the breaking of supersymmetry. The super Higgs
effect should not be confused with the supersymmetric
extension of the standard Higgs effect in which a massless vector
superfield eats a chiral superfield to receive a mass turning it into
a massive vector multiplet.
\end{itemize}

\paragraph{Exercise 5.4:}{\it Consider ${\cal N}=1$ supergravity with three chiral superfields $S,\, T,$ and $C.$ In Planck units, the K\"ahler potential and superpotential are given by
\begin{eqnarray*}
K&=&-\log{(S+S^*)} \ - \ 3 \, \log{(T+T^*-CC^*)}\\
W&=& C^3 \ + \ a \,  e^{-\alpha S}  \ + \ b\ ,
\end{eqnarray*}
where $a,\, b$ are arbitrary complex numbers and $\alpha>0.$ Compute the scalar potential. Find the auxiliary field for $S,T,C$ and verify that supersymmetry is broken. Assuming that $C$ denotes a matter field with vanishing vev, find a minimum of the potential. Are there flat directions? 
}

\chapter{The MSSM - basic ingredients}
\label{sec:TheMSSM}

In this chapter, we will shed light on some aspects of the minimally supersymmetric extension of the Standard Model which is obviously called {\em MSSM} in shorthand.

\section{Particles}
\label{sec:Particles}

First of all, we have vector fields transforming under $SU(3)_{C} \times SU(2)_{L} \times U(1)_{Y}$, secondly there are chiral superfields representing
\begin{itemize}
\item quarks
    \[\underbrace{Q_{i} \eq \left(3 , \ 2 , \ -\tfrac{1}{6} \right)}_{\te{left-handed}} \co \underbrace{\bar{u}_{i}^{c} \eq \left(\bar{3} , \ 1 , \ \tfrac{2}{3} \right) \co \bar{d}_{i}^{c} \eq \left(\bar{3} , \ 1 , \ -\tfrac{1}{3} \right)}_{\te{right-handed}}
\]
\item leptons
    \[\underbrace{L_{i} \eq \left(1 , \ 2 , \ \tfrac{1}{2} \right)}_{\te{left-handed}} \co \underbrace{\bar{e}_{i}^{c} \eq \left(1 , \ 1 , \ -1 \right) \co \bar{\nu}_{i}^{c} \eq \left(1 , \ 1 , \ 0 \right)}_{\te{right-handed}}
\]
\item higgses
    \[H_{1} \eq \left(1 , \ 2 , \ \tfrac{1}{2} \right) \co H_{2} \eq \left(1 , \ 2 , \ -\tfrac{1}{2} \right)
\]
the second of which is a new particle, not present in the Standard
Model. It is needed in order to avoid anomalies, like the one shown below.
\end{itemize}
The sum of $Y^{3}$ over all the MSSM particles must vanish
 (i.e. multiply the third quantum number
 with the product of the first two to cover all the distinct particles).
\begin{figure}[ht]
    \centering
       \includegraphics[width=0.45\textwidth]{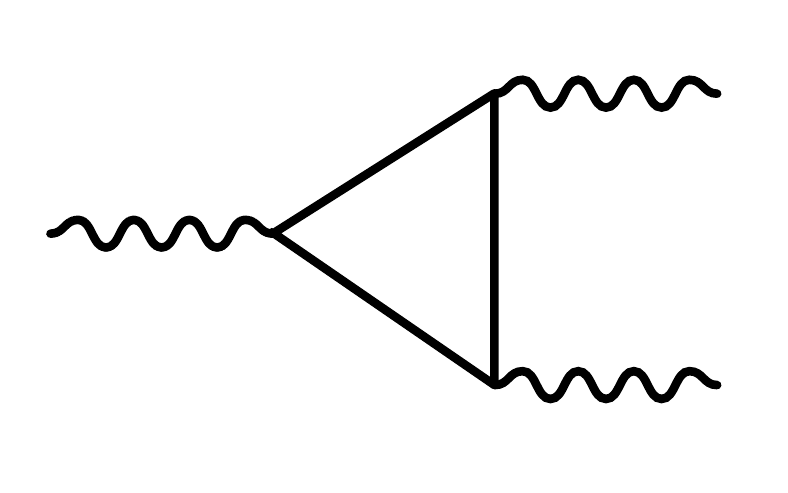}
        \caption{ \footnotesize{Anomalous graph proportional to Tr$\{ Y^3 \}$ which must vanish.}}
\end{figure}

\section{Interactions}
\label{sec:Interactions}

\begin{itemize}
\item $K = \Phi^\dag \exp(2qV) \Phi$ is renormalizable.
\item $f_{a} = \tau_{a}$ where $\te{Re} \{ \tau_{a} \} =
  \frac{4\pi}{g_{a}^{2}}$ determines the gauge coupling
  constants. These coupling constants change with energy as mentioned
  before. The precise way they run is determined by the low energy
  spectrum of the matter fields in the theory. We know from precision
  tests of the Standard Model that with its spectrum, the running of
  the three gauge couplings is such that they do not meet at a single
  point at higher energies (which would signal a gauge coupling unification).
  
\noindent  
However, with the matter field spectrum of the MSSM, the three
  different couplings evolve in such a way that they meet at a large
  energy $E$. This is considered to be the main phenomenological
  success of supersymmetric theories and it hints to a supersymmetric
  grand unified theory at large energies.
\begin{figure}[ht]
    \centering
     \includegraphics[width=0.70\textwidth]{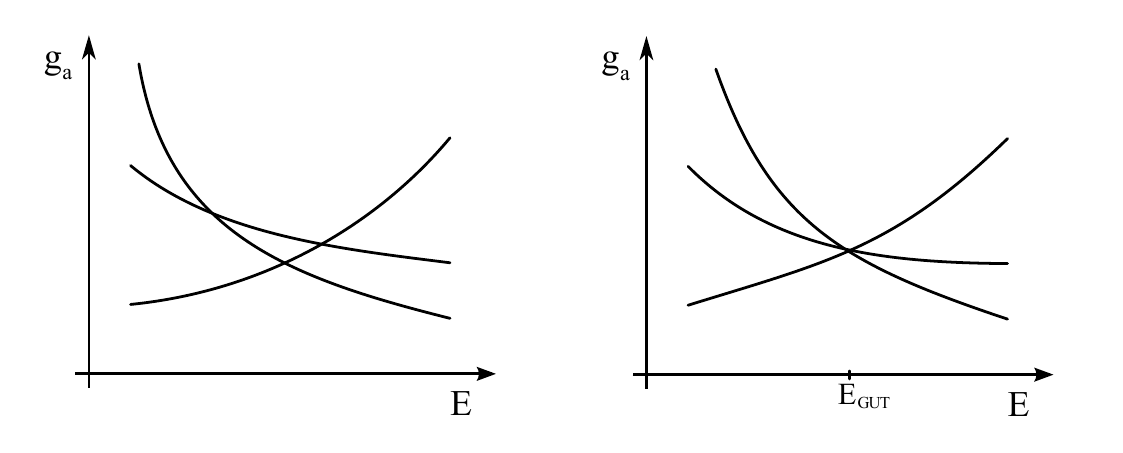}
       \caption{ \footnotesize{Running of the gauge couplings associated with the $SU(3)_C, \, SU(2)_L, \, U(1)_Y$ factors: The left plot refers to the non-supersymmetric Standard Model where all three couplings never coincide for any energy. On the right, the running of the MSSM gauge couplings is plotted, they are unified at some large energy $E=E_{\te{GUT}} \sim 10^{16}$ GeV.
        }}
\end{figure}
\item FI term: need $\xi = 0$, otherwise break charge and colour.
\item The superpotential $W$ is given by
    \begin{align*}
    W \ \ &= \ \ y_{1} \,Q \, H_{2} \, \bar{u}^{c} \ + \ y_{2} \, Q \, H_{1} \, \bar{d}^{c} \ + \ y_{3} \, L \, H_{1} \, \bar{e}^{c} \ + \ \mu \, H_{1} \, H_{2} \ + \ W_{\not{\!\! \te{BL}}} \ ,
\\
W_{\not{\!\! \te{BL}}} \ \ &= \ \ \la_{1} \, L \, L \, \bar{e}^{c} \ + \ \la_{2} \, L \, Q \, \bar{d}^{c} \ + \ \la_{3} \, \bar{u}^{c} \, \bar{d}^{c} \, \tilde{d}^{c} \ + \ \mu' \, L \,H_{2}
\end{align*}
The first three terms in $W$ correspond to standard Yukawa couplings
giving masses to up quarks, down quarks and leptons. The fourth term is
a mass term for the two Higgs fields. But each of the $\not{\!\! \te{BL}}$ terms breaks
baryon- or lepton number. These couplings are not present in the
Standard Model that automatically preserves baryon and lepton number
(as accidental symmetries). The shown interaction would allow proton decay $p \mto e^{+} + \pi^{0}$ within seconds.
\begin{figure}[ht]
    \centering
    \includegraphics[width=0.70\textwidth]{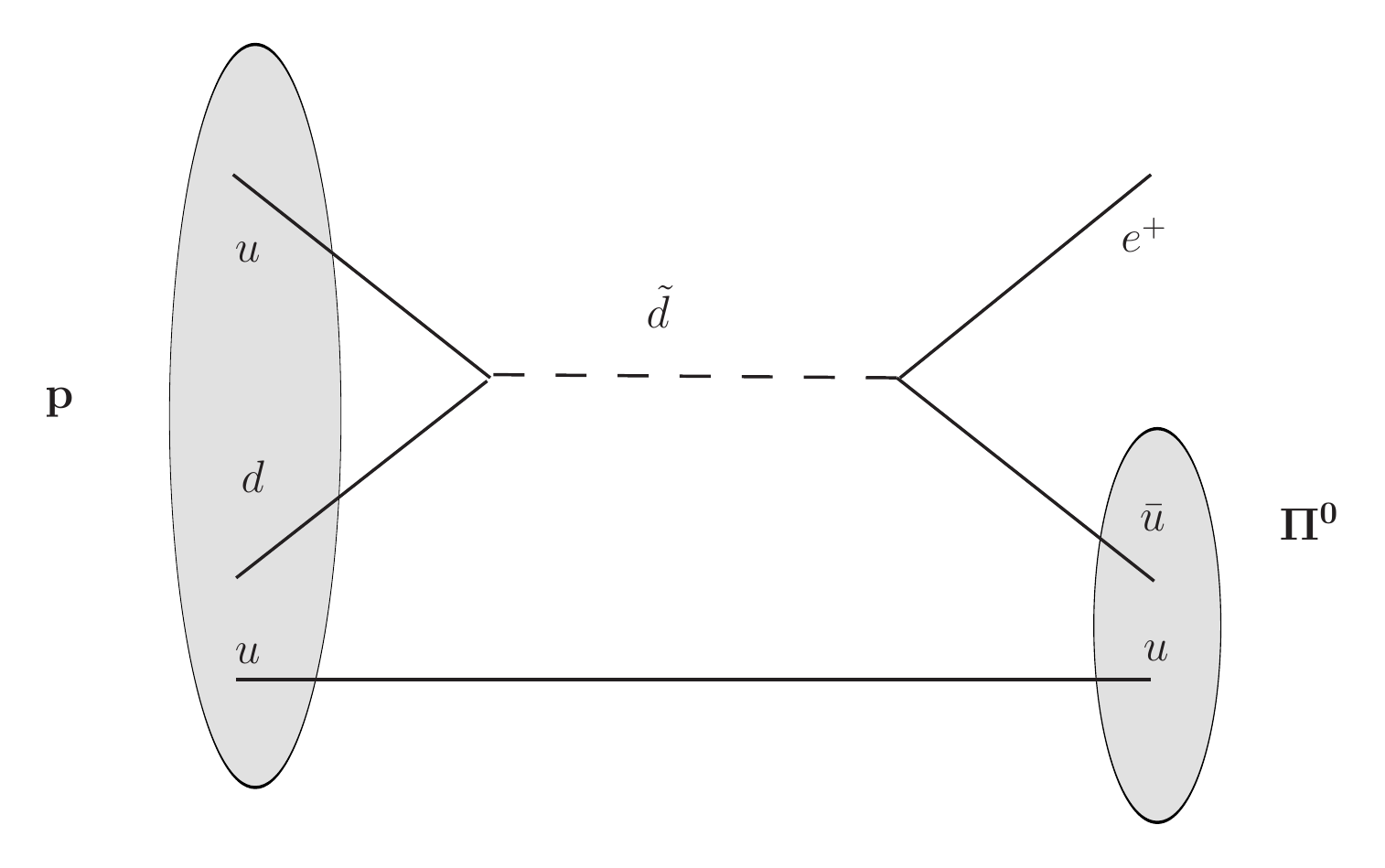}
    \caption{ \footnotesize{Proton decay due to baryon- and lepton number violating interactions.}}
\end{figure}
In order to forbid those couplings, an extra symmetry should be
imposed. The simplest one that works is {\it R parity} defined as
    \[R \ \ := \ \ (-1)^{3(B - L) + 2S} \eq \left\{ \begin{array}{l} +1 \ : \ \te{all observed particles} \\ -1 \ : \ \te{superpartners} \end{array} \right. \ .
\]
It forbids all the terms in $W_{\not{\!\! BL}}$.
\end{itemize}
The possible existence of R parity would have important physical implications:
\begin{itemize}
\item The lightest superpartner (LSP) is stable.
\item Usually, LSP is neutral (higgsino, photino), the neutralino is best candidate for dark matter (WIMP).
\item In colliders, superparticles are produced in pairs which then decay to LSP and give a signal of ''missing energy''.
\end{itemize}

\section{Supersymmetry breaking in the MSSM}
\label{sec:SupersymmetryBreaking}

Recall the two sectors of the Standard Model:
    \[\left( \begin{array}{c} \te{observable} \\ \te{sector (quarks)} \end{array} \right) \ \ \ \stackrel{\te{Yukawa}}{\longleftrightarrow} \ \ \ \left( \begin{array}{c} \te{symmetry -} \\ \te{breaking (Higgs)} \end{array} \right)
\]
Supersymmetry has an additional {\em messenger sector}
    \[\left( \begin{array}{c} \te{observable} \\ \te{sector} \end{array} \right) \ \ \ \longleftrightarrow \ \ \ \left( \begin{array}{c} \te{messenger -} \\ \te{sector} \end{array} \right) \ \ \ \longleftrightarrow  \ \ \ \left( \begin{array}{c} \te{SUSY -} \\ \te{breaking} \end{array} \right)
\]
involving three types of mediation
\begin{itemize}
\item gravity mediation

If the mediating field couples with gravitational strength to the standard model, the couplings will be suppressed by
the inverse Planck mass $M_{\te{pl}}$  which is the natural scale of gravity. We must include some mass square to get the right dimension for the mass splitting in the observable sector. That will be the square of SUSY breaking mass $M_{\not{ \te{SUSY}}}$:
    \[\Delta m \eq \frac{M_{\not{\te{ SUSY}}}^{2}}{M_{\te{pl}}} \ .
\]
We want $\Delta m \approx 1 \ \te{TeV}$ and know $M_{\te{pl}} \approx 10^{18} \ \te{GeV}$, so
    \[M_{\not{\te{SUSY}}} \eq \sqrt{\Delta m \cdot M_{\te{pl}}} \ \ \approx \ \ 10^{11} \ \te{GeV} \ .
\]
The gravitino gets a mass $m_{\frac{3}{2}}$ of $\Delta m$ order TeV from the super Higgs mechanism.
\item gauge mediation
    \[G \eq \bigl(SU(3) \ \times \ SU(2) \ \times \ U(1) \bigr) \ \times \ G_{\not{\te{SUSY}}} \ \ =: \ \ G_{0} \ \times \ G_{\not{\te{SUSY}}}
\]
Matter fields are charged under both $G_{0}$ and $G_{\not{ \te{SUSY}}}$ which gives a $M_{\not{ \te{SUSY}}}$ of order $\Delta m$, i.e. TeV. In that case, the gravitino mass $m_{\frac{3}{2}}$ is given by $\frac{M^{2}_{\not{ \te{ SUSY }}}}{M_{\te{pl}}} \approx 10^{-3} \ \te{eV}.$
\item anomaly mediation

In this case, auxiliary fields of supergravity (or Weyl compensator) get a vacuum expectation value. The effects are always present but suppressed by loop effects.
\end{itemize}
Each if these scenarios has phenomenological advantages and disadvantages and solving their problems is an acting fields of research at the moment. In all scenarios, the Lagrangian for the observable sector has contributions
   \[
     {\cal L} \ \ = \ \ {\cal L}_{\te{SUSY}} \ + \ {\cal L}_{\not{\te{SUSY}}} 
     \]
     Where:
     \[
\framebox{    $ \displaystyle   {\cal L}_{\not{\te{SUSY}}} \ \ = \ \ \underbrace{m_{0}^{2}  \, \vph^{*} \, \vph}_{\te{scalar masses}} \ + \  \left(\underbrace{M_{\la}  \, \la  \la}_{\te{gaugino masses}} \ + \ h.c. \right) \ + \ (A \, \vph ^{3} \ + \ h.c.) $}
\]
$M_\la, m_0^2, A$ are called {\em soft breaking terms}. They
determine the amount by which supersymmetry is expected to be broken
in the observable sector and are the main parameters to follow in the
attempts to identify supersymmetric theories with potential experimental observations.

\section{The hierarchy problem}
\label{sec:HierarchyProblem}

In high energy physics there are at least two fundamental scales - the Planck mass $M_{\te{pl}} \approx 10^{19}$ GeV defining the scale 
of quantum gravity and the electroweak scale $M_{\te{ew} } \approx 10^{2}$ GeV,
defining the symmetry breaking scale of the Standard
Model. Understanding why these two scales are so different is the
{\em hierarchy problem}. Actually the problem can be formulated in two
parts:
\begin{itemize}
\item [(i)] Why $M_{\te{ew}} \ll M_{\te{pl}}$? Answering this question is the proper hierarchy problem.
\item [(ii)] Is this hierarchy stable under quantum corrections? This is the
''naturalness'' part of the hierarchy problem which is the one that
  presents a bigger challenge.
\end{itemize}
Let us try to understand the naturalness part of the hierarchy problem.

\noindent
In the Standard Model we know that:
\begin{itemize}
\item Gauge particles are massless due to gauge invariance, that
  means, a direct mass term for the gauge particles $MA_\mu A^\mu$
is not allowed by
  gauge invariance ($A_\mu \rightarrow A_\mu +\partial_\mu \alpha$ for
  a $U(1)$ field).
\item Fermions masses $m\psi \psi$ are also forbidden for all quarks
  and leptons by gauge invariance. Recall that these particles receive a mass only through the
  Yukawa couplings to the Higgs (e.g. $H\psi\psi$ giving a mass to $\psi$
  after $H$ gets a non-zero value).
\item The Higgs is the only scalar particle in the Standard Model. They are the only
  ones that can have a mass term in the Lagrangian $m^2 H\bar{H}$. So
  there is not a symmetry that protects the scalars from becoming very
  heavy. Actually, if the Standard Model is valid up to a fixed
  cutoff scale
  $\La$ (for instance $\La\sim M_{\te{pl}}$ as an extreme case), it is
  known that loop corrections to the scalar mass $m^2$ induce 
values of order $\Lambda^2$ to the scalar mass. These corrections come
  from both bosons and fermions running in loops. These would make
  the Higgs to be as heavy as $\La$. This is unnatural since $\La$ can
  be much larger than the electroweak scale $\approx 10^2$ GeV. Therefore
  even if we start with a Higgs mass of order the electroweak scale,
  loop corrections would bring it up to the highest scale in the
  theory,
$\La$. This would ruin the hierarchy between large and small scales.
It is possible to adjust or ''fine tune'' the loop corrections such as
  to keep the Higgs light, but this would require adjustments to many 
decimal figures on each order of perturbation theory.
 This fine tuning is considered unnatural and an
  explanation of why the Higgs mass (and the whole electroweak scale)
  can be naturally maintained to be hierarchically smaller than the
  Planck scale or any other large cutoff scale $\La$ is required.
\end{itemize}
In SUSY, bosons have the same masses as fermions, so no problem about
hierarchy for all squarks and sleptons since the fermions have their
mass protected by gauge invariance.

\noindent
Secondly, we have seen in subsection \ref{sec:MiraculousCancellationsInDetail} that explicit computation of loop diagrams cancel boson against fermion loops due to the fact that the
couplings defining the vertices on each case are determined by the
same quantity ($g$ in the Yukawa coupling of fermions to scalar and
$g^2$ in the quartic couplings of scalars as was mentioned in the
discussion of the WZ model). These ''miraculous cancellations''
protect the Higgs mass from becoming arbitrarily large.

\noindent
Another way to see this is that even though a mass term is still
allowed for the Higgs by the coupling in the superpotential $\mu H_1 H_2$, the
non-renormalization of the superpotential guarantees that, as long
as supersymmetry is not broken, the mass parameter $\mu$ will not be
corrected by loop effects.

\noindent
Therefore if supersymmetry was exact the fermions and bosons would be
degenerate, but if supersymmetry breaks at a scale close to the 
electroweak scale then it will protect the Higgs from becoming too large.
This is the main reason to expect supersymmetry to be broken at low
energies of order $10^2-10^3$ GeV to solve the naturalness part of the
hierarchy problem.

\noindent
Furthermore, the fact that we expect supersymmetry to be broken by
non-perturbative effects (of order $e^{-1/g^2})$ is very promising as a
way to explain the existence of the hierarchy (first part of the
hierarchy problem). That is that if we start at a scale $M \gg M_{\te{ew}}$ (e.g. $M\approx M_{\te{pl}}$ in
string theory or GUT's), the supersymmetry breaking scale can be
generated as $M_{\te{SUSY}} \approx M e^{-1/g^2}$, for a small gauge coupling, say
$g\approx 0.1$, this would naturally explain why $M_{\te{SUSY}} \ll M$.

\section{The cosmological constant problem}
\label{sec:CosmologicalConstantProblem}

This is probably a more difficult problem as explained in section \ref{sec:ProblemsOfTheStandardModel}. The recent evidence of an accelerating universe indicates a new scale in physics which is the cosmological constant scale $M_\La$, with 
$\frac{M_\La}{M_{\te{ew}}}  \sim \frac{M_{\te{ew}}}{M_{\te{pl}}} \approx 10^{-15}$. Explaining 
why $M_\La$ is so small is the cosmological constant problem. Again it
can expressed in two parts, why the ratio is so small and (more
difficult) why this ratio is stable under quantum corrections.

\noindent
Supersymmetry could in principle solve this problem, since it is 
easy to keep the vacuum energy $\La$ to be zero in a supersymmetric
theory. However keeping it so small would require a supersymmetry
breaking scale of order $\Delta m \approx M_\La \sim 10^{-3}$ eV. But that
would imply that the superpartner of the electron would be essentially
of the same mass as the electron and should have been seen
experimentally long ago. Therefore at best supersymmetry
keeps the cosmological constant $\La$ small until it breaks. If it
breaks at the electroweak scale $M_{\te{ew}}$ that would lead to $M_\La \approx
M_{\te{ew}}$ which is not good enough.

\noindent
Can we address both the hierarchy- and the cosmological constant
problem at the same time? Some attempts are recently put forward in terms of the
string theory ''landscape'' in which our universe is only one of a
set of a huge number of solutions (or vacua) of the theory. This
number being greater than $10^{500}$ would indicate that a few of
these universes will have the value of the cosmological constant we
have today, and we happen to live in one of those (in the same way
that there are many galaxies and planets
 in the universe and we just happen to live
in one).

\noindent
This is still very controversial, but has lead to
speculations that if this is a way of solving the cosmological
constant problem, it would indicate a similar solution of the
hierarchy problem and the role of supersymmetry would be diminished in
explaining the hierarchy problem. This would imply that the scale of
supersymmetry breaking could be much larger. It is fair to say that
there is not at present a satisfactory approach to both the hierarchy
and cosmological constant problems. It is important to keep in mind
that even though low energy supersymmetry solves the hierarchy problem
in a very elegant way, the fact that it does not address the
cosmological constant problem is worrisome in the sense that any
solution of the cosmological constant problem could affect our
understanding of low energy physics to change the nature of the
hierarchy problem and then the importance of low energy 
supersymmetry. This is a very active area of
research at the moment.

\chapter{Extra dimensions}
\label{sec:ExtraDimensions}

It is important to look for alternative ways to address the problems
that supersymmetry solves and also to address other trouble spots of the
Standard Model. We mentioned in the first lecture that supersymmetry
and extra dimensions are the natural extensions of spacetime
symmetries that may play an important role in our understanding of
nature. Here we will start the discussion of physics in extra dimensions.

\section{Basics of Kaluza Klein theories}
\label{sec:BasicsOfKaluzaKleinTheories}

\subsection{History}

\begin{itemize}
\item In 1914 \textsc{Nordstrom} and 1919 - 1921 \textsc{Kaluza} independently tried to unify gravity and
electromagnetism. \textsc{Nordstrom} was attempting an unsuccessful theory of
gravity in terms of scalar fields, prior to \textsc{Einstein}. \textsc{Kaluza} used
general relativity extended to five dimensions. His concepts were
based on \textsc{Weyl}'s ideas.
\item 1926 \textsc{Klein}: cylindric universe with 5th dimension of small radius $R$
\item after 1926, several people developed the KK ideas (\textsc{Einstein}, \textsc{Jordan}, \textsc{Pauli}, \textsc{Ehrenfest},...)
\begin{figure}[ht]
    \centering
       \includegraphics[width=0.70\textwidth]{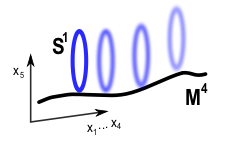}
       \caption{ \footnotesize{Example of a five dimensional spacetime $M^4\times S^1$ where $S^1$ is a circular extra dimension in addition to four dimensional $M^4$.}}
\end{figure}
\item 1960's: \textsc{de Witt} obtaining 4 dimensional Yang Mills theories in 4d from $D>5$. Also strings with $D = 26$.
\item In 1970's and 1980's: Superstrings required $D=10$. Developments in supergravity required extra dimensions and possible maximum numbers of dimensions for SUSY were discussed:
$D = 11$ turned out to be the maximum number of dimensions (\textsc{Nahm}).
\textsc{Witten} examined the coset
    \begin{align*}
    G/H \ \ &= \ \ \frac{SU(3) \times SU(2) \times U(1)}{SU(2) \times U(1) \times U(1)} \\ \dim(G/H) \ \ &= \ \ (8 \, + \, 3 \, + \, 1) \ - \ (3 \, + \, 1 \, + \, 1) \eq 7
\end{align*}
which also implied $D = 11$ to be the minimum. 11 dimensions, however, do not admit chirality since in odd dimensions, there is no analogue of the $\ga_5$ matrix in four dimensions.
\item 1990's: Superstrings revived $D = 11$ (M theory) and brane world scenario (large extra dimensions).
\end{itemize}

\paragraph{Exercise 7.1:}{\it Consider the Schr\"odinger equation for a particle moving in two dimensions $x$ and $y.$ The second dimension is a circle or radius $r.$ The potential corresponds to a square well $(V(x)=0$ for $x\in(0,a)$ and $V=\infty$ otherwise). Derive the energy levels for the two-dimensional Schr\"odinger equation and compare the result with the standard one-dimensional situation in the limit $r\ll a.$} 

\subsection{Scalar field in 5 dimensions}
\label{sec:ScalarFieldIn5Dimensions}
Before discussing higher dimensional gravity, we will start with the simpler cases of scalar fields in
extra dimensions, followed by vector fields and other bosonic fields
of helicity $\la\le 1$. This will illustrate the
effects of having extra dimensions in simple terms. We will be building up on the
level of complexity to reach gravitational theories in five and
higher dimensions. In the next chapter we extend the discussion to
include fermionic fields.

\noindent
Consider a massless 5D scalar field $\vph(x^{M})\ , \ M = 0,1,...,4$
with action
    \[{\cal S}_{5 \te{D}} \eq \int \dd^{5}x \ \pa^{M}\vph \, \pa_{M} \vph \ .
\]
Set the extra dimension $x^{4} = y$ defining a circle of radius
 ${r}$ with $y\equiv y+2\pi r$.

\noindent
Our spacetime is now $\mathbb M_{4} \times S^{1}$. Periodicity in $y$ direction implies discrete Fourier expansion
    \[\vph(x^{\mu},y) \eq \sum^{\infty}_{n = -\infty} \vph_{n}(x^{\mu}) \, \exp \left(\frac{iny}{r} \right) \ .
\]
Notice that the Fourier coefficients are functions of the standard 4D
coordinates and therefore are (an infinite number of) 4D scalar fields.
The equations of motion for the Fourier modes are (in general massive) wave equations
    \[\pa^{M}\pa_{M} \vph \eq 0 \so \sum^{\infty}_{n = -\infty} \left(\pa^{\mu}\pa_{\mu} \ - \ \frac{n^{2}}{r^{2}} \right) \, \vph_{n}(x^{\mu}) \, \exp \left(\frac{iny}{r} \right) \eq 0
\]
    \[\Longrightarrow \ \ \ \framebox{ $ \displaystyle \Bigl. \Bigr. \pa^{\mu} \pa_{\mu} \vph_{n}(x^{\mu}) \ - \ \frac{n^{2}}{r^{2}} \; \vph_{n}(x^{\mu}) \eq 0 \ . $ }
\]
These are then an infinite number of Klein Gordon equations for 
massive 4D fields. This means that each Fourier mode $\vph_{n}$ is a 4D particle
with mass $m_{n}^{2} = \frac{n^{2}}{r^{2}}$. Only the zero mode ($n=0$)
is massless. One can visualize the states as an infinite tower of massive states (with
increasing mass proportional to $n$). This is called {\em Kaluza Klein tower}
and the massive states ($n\neq 0$) are called {\em Kaluza Klein-} or
{\em momentum states}, since they come from the momentum in the extra dimension:
\begin{figure}[ht]
    \centering
            \includegraphics[width=0.40\textwidth]{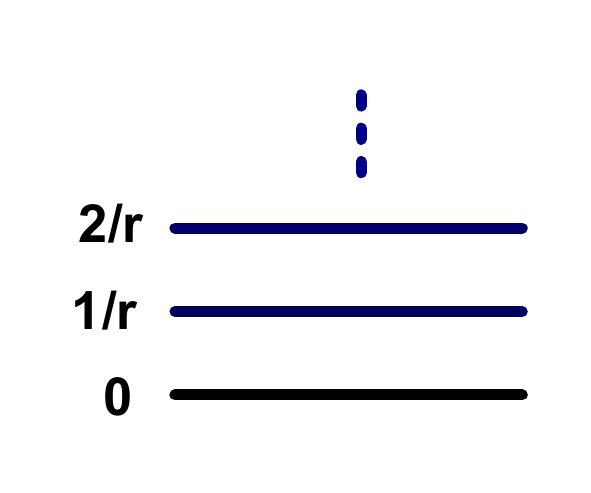}
             \caption{ \footnotesize{The Kaluza Klein tower of massive states due to an extra $S^1$ dimension. Masses $m_n = | n | /r$ grow linearly with the fifth dimension's wave number $n \in \mathbb Z$.}}
\end{figure}

\noindent
In order to obtain the effective action in 4D for all these particles,
let us plug the mode expansion of $\vph$ into the original 5D action,
\begin{align*}
{\cal S}_{5 \te{D}} \ \ &= \ \ \int \dd^{4}x \int \dd y \ \sum^{\infty}_{n = -\infty} \left( \pa^{\mu}\vph_{n}(x^{\mu}) \, \pa_{\mu} \vph_{n}(x^{\mu})^{*} \ - \ \frac{n^{2}}{r^{2}} |\vph_{n}|^{2} \right) \\
&= \ \ 2\, \pi \, r \int \dd^{4}x \ \bigl( \pa^{\mu}\vph_{0}(x^{\mu})  \, \pa_{\mu} \vph_{0}(x^{\mu})^{*} \ + \ ...  \bigr) \eq {\cal S}_{4 \te{D}} \ + \ ... \ .
\end{align*}
This means that the 5D action reduces to one 4D action for a massless
scalar field plus an infinite sum of massive scalar actions in 4D. If we are only interested in energies smaller than the $\frac{1}{r}$ scale, we may concentrate
only on the $0$ mode action. If we restrict our attention to the zero mode (like \textsc{Kaluza} did), then $\vph(x^{M}) =
\vph(x^{\mu})$. This would be equivalent to just truncating all the massive fields. In this case speak of {\em dimensional reduction}. More generally, if we
 keep all the massive modes we talk about {\em compactification}, meaning that the extra dimension is compact
 and its existence is taken into account as long as the Fourier modes are included.

\subsection{Vector fields in 5 dimensions}
\label{sec:VectorFieldIn5Dimensions}

Let us now move to the next simpler case of an abelian vector field in
5D, similar to an electromagnetic field in 4D.
We can split a massless vector field $A_{M}(x^{M})$ into
    \[A_{M} \eq \left\{ \begin{array}{lr} A_{\mu} &(\te{vector in 4 dimensions}) \\ A_{4} =: \rho &(\te{scalar in 4 dimensions}) \end{array} \right. \ .
\]
Each component has a discrete Fourier expansion
    \[A_{\mu} \eq \sum^{\infty}_{n = -\infty} A_{\mu}^{n} \, \exp \left( \frac{iny}{r} \right) \co \rho \eq \sum^{\infty}_{n = -\infty} \rho_{n} \, \exp \left( \frac{iny}{r} \right) \ .
\]
Consider the action
    \[{\cal S}_{5\te{D}} \eq \int \dd^{5}x \ \frac{1}{g_{5 \te{D}}^{2}} \; F_{MN} \, F^{MN}
\]
with field strength
    \[F_{MN} \ \ := \ \ \pa_{M}A_{N} \ - \ \pa_{N}A_{M}
\]
implying
    \[\pa^{M} \pa_{M} A_{N} \ - \ \pa^{M} \pa_{N} A_{M} \eq 0 \ .
\]
Choose a gauge, e.g. transverse
    \[\pa^{M} A_{M} \eq 0 \co A_{0} \eq 0 \so \pa^{M} \pa_{M} A_{N} \eq 0 \ ,
\]
then this obviously becomes equivalent to the scalar field case (for each
component $A_M$) indicating an infinite tower of massive states for
each massless state in 5D. In order to find the 4D effective action we
once again plug this into the 5D action:
    \[{\cal S}_{5 \te{D}} \ \ \mapsto \ \ {\cal S}_{4 \te{D}} \eq \int \dd^{4}x \ \left(\frac{2\pi r}{g_{5 \te{D}}^{2}} \; F_{(0)}\,^{\mu \nu} \; F_{(0)\mu \nu} \ + \ \frac{2\pi r}{g_{5 \te{D}}^{2}} \; \pa_{\mu} \rho_{0} \, \pa^{\mu} \rho_{0} \ + \ ... \right) \ ,
\]
Therefore we end up with a 4D theory of a gauge particle (massless), a
massless scalar and infinite towers of massive vector and scalar fields.
Notice that the gauge couplings of 4- and 5 dimensional
actions (coefficients of $F_{MN} F^{MN}$ and $F_{\mu\nu}F^{\mu\nu}$) are related by
    \[\frac{1}{g_{4 }^{2}} \eq \frac{2\pi r}{g_{5 }^{2}} \ .
\]
In $D$ spacetime dimensions, this generalizes to
    \[\frac{1}{g_{4}^{2}} \eq \frac{V_{D - 4}}{g_{D}^{2}}
\]
where $V_{n}$ is the volume of the $n$ dimensional compact space (e.g. an $n$ sphere of radius $r$). Higher dimensional electromagnetic fields have further interesting
issues that we pass to discuss.

\subsubsection{Electric (and gravitational) potential}
\label{sec:ElectricAndGravitationalPotential}

Gauss' law implies for the electric field $\ve{E}$ and its potential $\Phi$ of a point charge $Q$:
\begin{align*}
\oint^{}\limits_{S^{2}} \ve{E} \cdot d\ve{S} \ \ &= \ \ Q \so \|\ve{E}\| \ \ \propto \ \ \frac{1}{R^{2}} \co \Phi \ \ \propto \ \ \frac{1}{R} &4 \ \te{dimensions} \\
\oint^{}\limits_{S^{3}} \ve{E} \cdot d\ve{S} \ \ &= \ \ Q \so \|\ve{E}\| \ \ \propto \ \ \frac{1}{R^{3}} \co \Phi \ \ \propto \ \ \frac{1}{R^{2}} &5 \ \te{dimensions}
\end{align*}
So in $D$ spacetime dimensions
    \[\|\ve{E}\| \ \ \propto \ \ \frac{1}{R^{D - 2}} \co \Phi \ \ \propto \ \ \frac{1}{R^{D - 3}} \ .
\]
If one dimension is compactified (radius $r$) like in $\mathbb M_{4} \times S^{1}$, then
    \[\|\ve{E}\| \ \ \propto \ \ \left\{ \begin{array}{ll} \displaystyle \frac{1}{R^{3}} &: \ R < r \\ \displaystyle \frac{1}{R^{2}} &: \ R \gg r \end{array} \right. \ .
\]
Analogues arguments hold for gravitational fields and their potentials.

\subsubsection{Comments on spin and degree of freedom counting}
\label{sec:CommentsOnSpin}

We know that a gauge particle in 4 dimensions has spin one and carries two
degrees of freedom. We may ask about the generalization of these
results to a higher dimensional gauge field.

\noindent
Recall Lorentz algebra in 4 dimension
    \[\Bigl[M^{\mu \nu} \ , \ M^{\rho \si} \Bigr] \eq i \bigl(\eta^{\mu \si} \, M^{\nu \rho} \ + \ \eta^{\nu \rho} \, M^{\mu \si} \ - \ \eta^{\nu \si} \, M^{\mu \rho} \ - \ \eta^{\mu \rho} \, M^{\nu \si} \bigr)
\]
    \[J_{i} \eq \ep_{ijk} \, M_{jk} \co J \ \ \propto \ \ M_{23} \ .
\]
For massless representations in $D$ dimensions, $O(D - 2)$ is little group:
    \[P^{\mu} \eq (E , \ E \ , \underbrace{0 \ , \ ... \ , \ 0}_{O(D - 2)})
\]
The Lorentz algebra is just like in 4 dimensions, replace $\mu$, $\nu$, ... by $M$, $N$, ..., so $M_{23}$ commutes with $M_{45}$ and $M_{67}$ for example. Define the spin to be the maximum eigenvalue of any $M^{i(i+1)}$. The number of degrees of freedom in 4 dimensions is 2 ($A_{\mu} \mapsto A_{i}$ with $i = 2,3$) corresponding to the 2 photon polarizations and $(D - 2)$ in $D$ dimension, $A_{M} \mapsto A_{i}$ where $i = 1,2,...,D - 2$.

\subsection{Duality and antisymmetric tensor fields}
\label{sec:CommentsOnDuality}

So far we considered scalar- and vector fields:
    \[\begin{array}{c|ccc} &\te{scalar} &\te{vector} &\te{index - range} \\\hline D = 4 &\vph(x^{\mu}) &A_{\mu}(x^{\mu}) &\mu = 0,1,2,3 \\ D > 4 &\vph(x^{M}) &A_{M}(x^{M}) &M = 0,1,...,D - 1 \end{array}
\]
We will see now that in extra dimensions there are further fields
corresponding to bosonic particles of helicity $\lambda\leq 1$.
These are antisymmetric tensor fields, which in 4D are just equivalent
to scalars or vector fields by a symmetry known as {\em duality}.
 But in extra dimensions these will be new types of particles (that
 play an important role in string theory for instance).

\noindent
In 4 dimensions, define a dual field strength to the Faraday tensor $F^{\mu \nu}$ via
    \[\tilde{F}^{\mu \nu} \ \ := \ \ \frac{1}{2} \; \ep^{\mu \nu \rho \si} \, F_{\rho \si} \ ,
\]
then Maxwell's equations in vacuo read:
\begin{align*}
\pa^{\mu} F_{\mu \nu} \ \ &= \ \ 0 &(\te{field equations}) \\
\pa^{\mu} \tilde{F}_{\mu \nu} \ \ &= \ \ 0 &(\te{Bianchi identities})
\end{align*}
The exchange $F \leftrightarrow \tilde{F}$ (the {\em electromagnetic duality}) corresponding to $\ve{E} \leftrightarrow \ve{B}$ swaps field equations and Bianchi identities.

\noindent
In 5 dimensions, one could define in analogy
    \[\tilde{F}^{MNP} \eq \ep^{MNPQR} \, F_{QR} \ .
\]
One can generally start with an antisymmetric $(p + 1)$ - tensor $A_{M_{1}...M_{p + 1}}$ and derive a field strength
    \[F_{M_{1}...M_{p + 2}} \eq \pa_{[M_{1}} A_{M_{2}...M_{p + 2}]}
\]
and its dual (with $D - (p + 2)$ indices)
    \[\tilde{F}_{M_{1}...M_{D - p - 2}} \eq \ep_{M_{1}...M_{D}} \, F^{M_{D - p - 1}...M_{D}} \ .
\]
Consider for example
\begin{itemize}
\item $D = 4$
    \[F_{\mu \nu \rho} \eq \pa_{[\mu} B_{\nu \rho]} \so \tilde{F}_{\si} \eq \ep_{\si \mu \nu \rho} \, F^{\mu \nu \rho} \eq \pa_{\si}a
\]
The dual potentials that yield field strengths $F_{\mu \nu} \leftrightarrow \tilde{F}_{\mu \nu}$ have different number of indices, 2 tensor $B_{\nu \rho} \leftrightarrow a$ (scalar potential).
\item $D = 6$
    \[F_{MNP} \eq \pa_{[M} B_{NP]} \so \tilde{F}_{QRS} \eq \ep_{MNPQRS} \, F^{MNP} \eq \pa_{[Q} \tilde{B}_{RS]}
\]
Here the potentials $B_{NP} \leftrightarrow \tilde{B}_{RS}$ are of the same type.
\end{itemize}
Antisymmetric tensors carry spin 1 or less, in 6 dimensions:
    \[B_{MN} \eq \left\{ \begin{array}{lr} B_{\mu \nu} &: \ \te{rank two tensor in 4 dimensions} \\ B_{\mu5} \ , \ B_{\mu6} &: \ \te{2 vectors in 4 dimensions} \\ B_{56} &: \ \te{scalar in 4 dimensions} \end{array} \right.
\]
To see the number of degrees of freedom, consider little group
    \[B_{M_{1}...M_{p + 1}} \ \ \mapsto \ \ B_{i_{1}...i_{p + 1}} \co i_{k} \eq 1,...,(D - 2) \ .
\]
These are $\left( \begin{smallmatrix} D-2 \\ p+1 \end{smallmatrix} \right)$ independent components. Note that under duality, couplings $g$ are mapped to (multiples of) their inverses,
    \[{\cal L} \eq \frac{1}{g^{2}} \; (\pa_{[M_{1}} B_{M_{2}...M_{p + 2}]})^{2} \ \ \ \leftrightarrow \ \ \ g^{2} \, (\pa_{[M_{1}} \tilde{B}_{M_{2}...M_{D - (p + 2)}]})^{2} \ .
\]

\subsubsection{$p$ branes}
\label{sec:Couplings}

Electromagnetic fields couple to the worldline of particles via
    \[{\cal S} \ \ \sim \ \ \int A_{\mu} \ \dd x^{\mu} \ ,
\]
This can be seen as follows: the electromagnetic field couples to a
conserved current in 4 dimensions as $\int \dd^4 x A_\mu J^\mu$
(with Dirac current $J^\mu =\bar{\psi}\gamma^\mu \psi $ for an electron field for
instance). For a particle of charge $q$, the current can be written as an
integral over the world line of the particle $J^\mu=q\int
\dd \xi^\mu  \delta^4(x-\xi)$ such that $\int J^0 \dd^3x =q$ and so the
coupling becomes $\int \dd^4 x J^\mu A_\mu = q \int \dd\xi^\mu A_\mu$. 

\noindent
We can extend this idea for higher dimensional objects.  
For a potential $B_{[\mu \nu]}$ with two indices, the analogue is
    \[\int B_{\mu \nu} \ \dd x^{\mu} \wedge \dd x^{\nu} \ ,
\]
i.e. need a string with 2 dimensional worldsheet to couple. Further generalizations are
\begin{align*}
&\int B_{\mu \nu \rho} \ \dd x^{\mu} \wedge \dd x^{\nu} \wedge \dd x^{\rho} &(\te{membrane}) \\
&\int B_{M_{1}...M_{p + 1}} \ \dd x^{M_{1}} \wedge ... \wedge \dd x^{M_{p + 1}} &(p \ \te{brane})
\end{align*}
Therefore we can see that antisymmetric tensors of higher rank coupled
naturally to extended objects. This leads to the
concept of a $p$ brane as a generalization of a particle that couples
to antisymmetric tensors of rank $p+1$. A particle carries charge
under a vector field, such as electromagnetism. In the same sense, $p$
branes carry a \textbf{new} kind of charge with respect to a higher rank
antisymmetric tensor.

\paragraph{Exercise 7.2:}{\it Consider the following Lagrangian
$${\cal S}  \eq \int \dd^4 x\ \left(\frac{1}{g^2} \; H_{\mu\nu\rho} \, H^{\mu \nu \rho} \ + \ a \,  \epsilon^{\mu\nu\rho\sigma} \, \partial_{\mu}H_{\nu\rho\sigma}\right) \ .$$
Solve the equation of motion for the Lagrange multiplier $a$ to obtain an action for a propagating massless Kalb-Ramond field $B_{\mu\nu}.$ Alternatively, solve the equation of motion for the field $H_{\nu\rho\sigma},$ to obtain an action for the propagating axion field $a.$  What happens to the coupling $g$ under this transformation?}
\subsection{Gravitation in Kaluza Klein theory}
\label{sec:Graviton}

After discussing scalar-, vector- and antisymmetric tensor fields
    \[\begin{array}{c|cc} &\te{spin} &\te{deg. of freedom} \\\hline \te{scalar} \ \vph &0 &1 + 1 \\ \te{vector} \ A_{M} &0 \ , \ 1 &D - 2 \\ \te{antisymmetric tensor} \ A_{M_{1}...M_{p + 1}} &0 \ , \ 1 &\left( \begin{smallmatrix} D-2 \\ p+1 \end{smallmatrix} \right) \end{array}
\]
we are now ready to consider the graviton $G_{MN}$ of Kaluza Klein theory in $D$ dimensions
    \[G_{MN} \eq \left\{ \begin{array}{lc} G_{\mu \nu} &\te{graviton} \\ G_{\mu n} &\te{vectors} \\ G_{mn} &\te{scalars} \end{array} \right.
\]
where $\mu, \nu = 0,1,2,3$ and $m,n = 4,...,D - 1$.

\noindent
The background metric appears in the 5 dimensional {\em Einstein Hilbert action}
    \[{\cal S} \eq \int \dd^{5}x \ \sqrt{|G|} \, ^{(5)}R \co ^{(5)}R_{MN} \eq 0 \ .
\]
One possible solution is the 5 dimensional Minkowski metric $G_{MN} =
 \eta_{MN}$, another one is that of 4 dimensional Minkowski spacetime $M_4$ times a circle
 $S^1$, i.e. the metric is of the $\mathbb M_{4} \times S^{1}$ type
    \[\dd s^{2} \eq W(y) \, \eta_{\mu \nu} \, \dd x^{\mu} \, \dd x^{\nu} \ - \ \dd y^{2}
\]
where $\mathbb M_{3} \times S^{1} \times S^{1}$ is equally valid. 
In this setting, $W(y)$ is a \te{warped factor} that is allowed by the symmetries of
the background and $y$ is 
restricted to the interval $[0 , 2\pi r]$. For simplicity we will set
the warp factor to a constant but will consider it later where it
will play an important role. 

\noindent
Consider excitations in addition to the background metric
    \[G_{MN} \eq \phi^{-\frac{1}{3}} \, \ccb \bigl( g_{\mu \nu}  \ - \ \ka^{2} \, \phi \, A_{\mu} \, A_{\nu} \bigr) &-\ka \, \phi \, A_{\mu} \\ -\ka \, \phi \, A_{\nu} &\phi \cce
\]
in Fourier expansion
    \[G_{MN} \eq \underbrace{\phi^{(0)-\frac{1}{3}} \ccb \bigl( g^{(0)}_{\mu \nu} \ - \ \ka^{2} \, \phi^{(0)} \, A^{(0)}_{\mu} \, A^{(0)}_{\nu} \bigr) &-\ka \, \phi^{(0)} \, A^{(0)}_{\mu} \\ -\ka \, \phi^{(0)} \, A^{(0)}_{\nu} &\phi^{(0)} \cce}_{\te{Kaluza Klein ansatz}} \ + \ \infty \ \te{tower of massive modes}
\]
and plug the zero mode part into the Einstein Hilbert action:
    \[{\cal S}_{4D} \eq \int \dd^{4}x \ \sqrt{|g|} \, \left\{ M_{\te{pl}}^{2 \ (4)}R \ - \ \frac{1}{4}\phi^{(0)} \; F^{(0)}_{\mu \nu} \, F^{(0)\mu \nu} \ + \ \frac{1}{6} \; \frac{\pa^{\mu}\phi^{(0)} \, \pa_{\mu} \phi^{(0)}}{(\phi^{(0)})^{2}} \ + \ ... \right\}
\]
This is the unified theory of gravity, electromagnetism and scalar
fields!

\paragraph{Exercise 7.3:}{\it Show that the last equation follows from a pure gravitational theory in five-dimensions, using $^{(5)}R = \, ^{(4)}R -2e^{-\s}\nabla^2e^\s-\frac{1}{4}e^{2\s}F_{\mu\nu}F^{\mu\nu}$ where $G_{55}=e^{2\s}.$ Relate the gauge coupling to the $U(1)$ isometry of the compact space.}
\subsubsection{Symmetries}

\begin{itemize}
\item general 4 dimensional coordinate transformations
    \[x^{\mu} \ \ \mapsto \ \ x'^{\mu}(x^{\nu}) \co g_{\mu \nu}^{(0)} \ \te{(graviton)} \co A_{\mu}^{(0)} \ \te{(vector)}
\]
\item $y$ transformation
    \[y \ \ \mapsto \ \ y' \eq F(x^{\mu} , y)
\]
Notice that
    \[ \dd s^{2} \eq \phi^{(0)-\frac{1}{3}} \, \left\{ g_{\mu \nu}^{(0)} \, \dd x^{\mu} \, \dd x^{\nu} \ - \ \phi^{(0)}  \, \bigl( \dd y \ - \ \ka \, A_{\mu}^{(0)} \, \dd x^{\mu} \bigr)^{2} \right\} \ ,
\]
so, in order to leave $\dd s^{2}$ invariant, need
    \[F(x^{\mu},y) \eq y \ + \ f(x^{\mu}) \so \dd y' \eq \dd y \ + \ \frac{\pa f}{\pa x^{\mu}} \; \dd x^{\mu} \co A_{\mu}^{'(0)} \eq A_{\mu}^{(0)} \ + \ \frac{1}{\ka} \; \frac{\pa f}{\pa x^{\mu}}
\]
which are gauge transformation for a massless field
$A_{\mu}^{(0)}$! This is the way to understand that standard gauge
symmetries can be derived from general coordinate transformations in
extra dimensions, explaining the Kaluza Klein programme of unifying
all the interactions by means of extra dimensions.
\item overall scaling
    \[y \ \ \mapsto \ \ \la \, y \co A_{\mu}^{(0)} \ \ \mapsto \ \ \la \, A_{\mu}^{(0)} \co \phi^{(0)} \ \ \mapsto \ \ \frac{1}{\la^{2}} \; \phi^{(0)} \so \dd s^{2} \ \ \mapsto \ \ \la^{\frac{2}{3}} \, \dd s^{2}
\]
$\phi^{(0)}$ is a massless {\em modulus field}, a flat direction in
the potential, so $\langle \phi^{(0)} \rangle$ and therefore the size of
the 5th dimension is arbitrary. $\phi^{(0)}$ is called breathing mode,
radion or {\em dilaton}. This is a major problem for these theories: It
looks like \textbf{all} the values of the radius (or volume in general)
of the extra dimensions are equally good and the theory does not
provide a way to fix this size. It is a manifestation of the problem
that the theory cannot prefer a flat 5D Minkowski space (infinite
radius) over $\mathbb M_4\times S^1$ (or $\mathbb M_3\times S^1\times S^1$, etc.).
This is the {\em moduli problem} of extra dimensional theories. String
theories share this problem. Recent developments in string theory
allows to fix the value of the volume and shape of the extra
dimension, leading to a large but discrete set of solutions. This is
the so-called ''landscape'' of string solutions (each one describing a
different universe and ours is only one among a huge number of them).
\end{itemize}

\subsubsection{Comments}

\begin{itemize}
\item The Planck mass $M_{\te{pl}}^{2} = M_{*}^{3} \cdot 2\pi r$ is a
  derived quantity. We know experimentally that 
 $M_{\te{pl}} \approx 10^{19} \ \te{GeV}$, therefore we can adjust $M_{*}$
  and $r$ to give the right result. But there is no other constraint
  to fix $M_*$ and $r$.
\item Generalization to more dimensions
    \[G_{MN} \eq \ccb \bigl(g_{\mu \nu} \ - \ \ka^{2} \, A_{\mu}^{i} \, A_{\nu}^{j} \, h_{ij} \bigr) &-\ka \, \ga_{mn} \, K^{n}_{i} \, A^{i}_{\mu} \\ -\ka \, \ga_{mn} \, K^{m}_{i} \, A^{i}_{\nu} &\ga_{mn} \cce
\]
The $K^{m}_{i}$ are Killing vectors of an internal manifold ${\cal M}_{D - 4}$ with metric $\ga_{mn}$. The theory corresponds to Yang Mills in 4 dimensions. Note that the Planck mass now behaves like
    \[\framebox{ $ \Bigl. \Bigr. M_{\te{pl}}^{2} \eq M_{*}^{D - 2} \, V_{D - 4} \ \ \sim \ \ M_{*}^{D - 2} \, r^{D - 4} \eq M_{*}^{2} \, (M_{*} \, r)^{D - 4} .$} \
\]
In general we know that the highest energies explored so far require 
 $M_{*} > 1 \ \te{TeV}$ and $r < 10^{-16} \ \te{cm}$ since no
 signature of extra dimensions has been seen in any experiment.
In Kaluza Klein theories there is no reason to expect a large value of
 the volume and it has been usually assumed that  $M_{*} \approx M_{\te{pl}}$.
\end{itemize}

\section{The brane world scenario}
\label{sec:TheBraneWorldScenario}

So far we have been discussion the standard Kaluza Klein theory in
which our universe is higher dimensional. We have not seen the extra
dimensions because they are very small (smaller than the smallest
scale that can be probed experimentally at colliders which is $10^{-16}$ cm).

\noindent
We will introduce now a different and more general higher dimensional
scenario. The idea here is that our universe is a $p$ brane, or a
surface inside a higher dimensional {\em bulk spacetime}. A typical example
of this is as follows: all the Standard Model particles (quarks,
leptons but also gauge fields) are trapped on a 3 dimensional
spatial surface (the brane) inside a higher dimensional spacetime (the
bulk). Gravity on the other hand lives on the full bulk spacetime and
therefore only gravity probes the extra dimensions.

\noindent
Therefore we have to distinguish the $D$ dimensional bulk
space (background spacetime) from the $(p + 1)$ world volume
coordinates of a $p$ brane. Matter lives in the $d (= 4)$ dimensions
of the brane, whereas gravity takes place in the $D$ bulk
dimensions. This scenario seems very ad hoc at first sight but it is
naturally realized in string theory where matter tends to live on
D branes (a particular class of $p$ branes corresponding to surfaces
where ends of open strings are attached to). Whereas gravity, coming
from closed strings can leave in the full higher dimensional ($D=10$) spacetime.
Then the correspondence is as follows:
    \[\begin{array}{rcl} \te{gravity} &\longleftrightarrow &\te{closed strings} \\ \te{matter} &\longleftrightarrow &\te{open strings} \end{array}
\]
\begin{figure}[ht]
    \centering
 \includegraphics[width=0.70\textwidth]{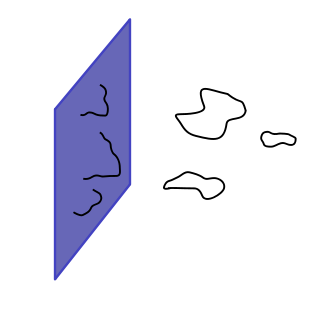}
       \caption{ \footnotesize{Brane world scenario with matter corresponding to open strings which start and end on the brane and gravity incorporated by closed strings probing the full bulk spacetime.}}
\end{figure}
\subsubsection{}

For phenomenological purposes we can distinguish two different classes of brane world scenarios.

\subsection{Large extra dimensions}
\label{sec:LargeExtraDimensions}

Let us first consider an unwarped compactification, that is a constant
warp factor $W(y)$. We have remarked that the fundamental higher dimensional scale $M_*$
is limited to be $M_{*} \geq 1 \ \te{TeV}$ in order to not contradict
experimental observations which can probe up to that energy.
By the same argument we have constrained the size of the extra
dimensions $r$ to be $r<10^{-16}$ cm because this is the length
associated to the TeV scale of that accelerators can probe. However,
in the brane world scenario, if only gravity feels the extra
dimensions, we have to use the constraints for gravity only. Since
gravity is so weak, it is difficult to test experimentally and so far
the best experiments can only test it to scales larger than $\approx 0.1$
mm. This is much larger than the $10^{-16}$ cm of the Standard
Model. Therefore, in the brane world scenario it is possible to have
extra dimensions as large as $0.1$ mm without contradicting any
experiment!

\noindent
This has an important implication also as to the value of $M_*$ (which
is usually taken to be of order $M_{\te{pl}}$) in Kaluza Klein theories.
From the Einstein Hilbert action, the Planck mass $M_{\te{pl}}$ is still given by
    \[M_{\te{pl}}^{2} \eq M_{*}^{D - 2} \, V_{D - 4}
\]
with $V_{D-4}\sim r^{D-4}$ denoting the volume of the extra dimensions. But now
we can have a much smaller fundamental scale $M_*$ if we allow the
volume to be large enough. We may even try to have the fundamental
scale to be of order $M_*\sim 1$ TeV. In five dimensions, this will
require a size of the extra dimension to be of order $r\approx 10^8$ km
in order to have a Planck mass of the observed value $M_{\te{pl}} \approx
10^{18}$ GeV (where we have used $r=M_{\te{pl}}^2/M_*^3$). 
This is clearly ruled out by experiments. However,
starting with a 6 dimensional spacetime we get $r^2=M_{\te{pl}}^2/M_*^4$,
which gives $r\approx 0.1$mm for $M_*=1$ TeV. This is then consistent
with all gravitational experiments as well as Standard Model
tests. Higher dimensions would give smaller values of $r$ and will
also be consistent. The interesting thing about the 6 dimensional
case is that it is possible to be tested by the next round of
experiments in both, the accelerator experiments probing scales of
order TeV and gravity experiments, studying deviations of the squared
law at scales smaller than $0.1$mm. 

\noindent
Notice that this set up changes the nature of the hierarchy problem
because now the small scale (i.e. $M_{\te{ew}}\approx M_{*} \approx 1$ TeV) is
fundamental whereas the large Planck scale is a derived quantity. The
hierarchy problem now is changed to explain why the size of the extra
dimensions is so large to generate the Planck scale of $10^{18}$ GeV
starting from a small scale $M_*\approx 1$ TeV. This changes the nature
of the hierarchy problem, because it turns it into a dynamical
question of how to fix the size of the extra dimensions. Notice that
this will require exponentially large extra dimensions (in units of
the inverse fundamental scale $M_*$). The hierarchy problem then
becomes the problem of finding a mechanism that gives
rise to exponentially large sizes of the extra dimensions.

\subsection{Warped compactifications}
\label{sec:WarpedCompactifications}

 This is the so-called
  {\em Randall Sundrum scenario}. The simplest case is again a
  5 dimensional theory but with the following properties. Instead
  of the extra dimension being a circle $S^1$, it is now an interval
  $I$ (which can be defined as an {\it orbifold} of $S^1$ by identifying the
  points $y \equiv -y$, if the original circle had length $2\pi r$, the
  interval $I$ will have half that size,  $\pi r$). The surfaces at each
  end of the interval play a
  role similar to a brane, being 3 dimensional surfaces inside a
  5 dimensional spacetime. The second important ingredient is that
  the warp factor $W(y)$ is not as
  determined by solving Einstein's equations in this background.  
We then have warped geometries with a $y$ dependent warp factor 
$\exp \bigl(W(y) \bigr)$, in 5 dimensions
    \[\dd s^{2} \eq \exp \bigl(W(y) \bigr) \, \eta_{\mu \nu} \, \dd x^{\mu} \, \dd x^{\nu} \ + \ \dd y^{2} \ .
\]
The volume $V_{D - 4}$ has a factor
    \[V_{D - 4} \ \ \sim \ \ \int^{+\pi}\limits_{-\pi} \dd y \ \exp \bigl( W(y) \bigr) \ .
\]
Consider then the two branes,one at $y = 0$ (''the Planck brane'')
and one at $y = \pi r$ (''the Standard Model brane''), the total 
action has contributions from the two branes and the bulk itself:
\begin{figure}[ht]
    \centering
\includegraphics[width=0.70\textwidth]{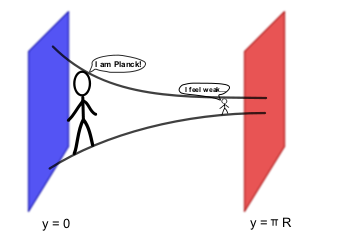}
    \caption{ \footnotesize{Brane configuration in the Randall-Sundrum scenario: The warped geometry in the $y$ direction gives rise to a mass hierarchy between the Planck brane at $y=0$ and the Standard Model brane at $y=\pi r$
       }}
\end{figure}
    \[{\cal S} \eq {\cal S}_{y = 0} \ + \ {\cal S}_{y = \pi r} \ + \ {\cal S}_{\te{bulk}}
\]
Einstein's equations imply $W(y) \propto e^{- |ky|}$ with $k$ a
constant (see \cite{RS} and example sheet 4), so the metric changes from $y = 0$ 
to $y = \pi r$ via $\eta_{\mu \nu} \longmapsto \exp(-k\pi r) \eta_{\mu \nu}$.
This means that all the length and energy scales change with $y$.
If the fundamental scale is $M_* \approx M_{\te{pl}}$, 
the $y = 0$ brane carries physics at $M_{\te{pl}}$, but as long as we
move away from this end of the interval, all the energy
scales will be ''red shifted'' by the factor $e^{-|ky|}$ until we reach
the other end of the interval in which $y = \pi r$ . This exponential
changes of scales is appropriate for the hierarchy problem. If the
fundamental scale is the Planck scale, at $y=0$ the physics will be
governed by this scale but at $y=r$ we will have an exponentially
smaller scale. In particular we can have the electroweak scale 
 $M_{\te{ew}}\approx M_{\te{pl}} \cdot e^{-\pi kr} \approx 1 \ \te{TeV}$ if $r$ is only
slightly bigger than the Planck length  $r \geq 50 \, \ell_{\te{pl}}$. This is a
more elegant way to ''solve' the hierarchy problem. We only need to find
a mechanism to fix the value of $r$ of order $50\,  \ell_{\te{pl}}$!
Notice that in this scenario 5 dimensions are compatible with
experiment (unlike the unwarped case that required a radius many
kilometers large). 

\paragraph{Exercise 7.4:}{\it Consider a five dimensional gravity theory with a negative cosmological constant $\Lambda<0,$ compactified on an interval $(0,\pi).$ Each end of the interval corresponds to a '3-brane' which we choose to have tension $\pm\Lambda/k$ respectively. Here $k$ is a common scale to be determined later in terms of the fundamental scale in 5D $M$ and $\Lambda.$ Verify that the warped metric
$$\dd s^2 \eq e^{-2W(\theta)} \, \eta_{\mu\nu} \, \dd x^\mu \, \dd x^\nu \ + \ r^2 \, \dd \theta^2$$
satisfies Einstein's equations. Here $e^{-2 W(\theta)}$ is the warp factor and $r$ is a constant measuring the size of the interval. You can use that Einstein's equations reduce to
$$\frac{6 \, W'^2}{r^2} \eq - \, \frac{\Lambda}{2 M^3} \co \frac{3 \, W''}{r^2} \eq \frac{\Lambda}{2  \, M^3 \,  kr} \; \bigl[\delta(\theta-\pi) \, - \, \delta(\theta) \bigr] \, .$$
Solve for $W(\theta)$ and use the warp factor to show that the effective $4D$ Planck scale is now}
$$M_{\textrm{pl}}^2 \eq M^3 \,   r \,  \int_{-\pi}^{\pi} \dd \theta \ e^{-2W} \eq \frac{M^3}{k}\; \bigl(1 \ - \ e^{-2kr} \bigr)\ .$$
{\it Find the value of the constant $k.$ Consider the Higgs Lagrangian on the brane at $\theta=\pi,$ bring it into canonical form and show that the mass is proportional to the factor $e^{-k\pi r}.$ How large can $r$ be in order to reproduce the electroweak scale from the Planck scale? Does this solve the hierarchy problem? How does the Planck scale differ from the 5D scale $M?$

}
\subsection{Brane world scenarios and the hierarchy problem}
\label{sec:BraneWorldScenariosAndTheHierarchyProblem}

Notice that in both scenarios, the problem of solving the hierarchy
problem has been turned into the problem of fixing the size of the
extra dimensions. It is worth remarking that both mechanisms have been
found to be realized in string theory (putting them on firmer
grounds). Studying mechanisms to fix the moduli that determines the
size and shape of extra dimensions is one of the most active areas of
research within string theory.

\chapter{Supersymmetry in higher dimensions}
\label{sec:SupersymmetryInHigherDimensions}

So far we have been discussed the possible bosonic fields in extra dimensions (scalars, vectors, antisymmetric tensors and metrics). What about fermionic fields in extra dimensions? Good references for the technical aspects are \cite{gamma1,gamma2,gamma3}.

\section{Spinors in higher dimensions}
\label{sec:SpinorsInHigherDimensions}

For a theory of fermions in more than four dimensions, need some analogue of the four dimensional Dirac $\ga$ matrices, i.e. representations of the Clifford algebra
    \[\Bigl\{ \Ga^{M} \ , \ \Ga^{N} \Bigr\} \eq 2 \, \eta^{MN} \co \Si^{MN} \eq \frac{i}{4} \; \Bigl[ \Ga^{M} \ , \ \Ga^{N} \Bigr] \ ,
\]
where the $\Si^{MN}$ are generators of $SO(1 , D-1)$ subject to the Lorentz algebra
	\[\Bigl[\Si^{MN} \ , \ \Si^{PQ } \Bigr] \ \ = \ \ i\, \bigl(\Si^{M Q} \, \eta^{N P} \ + \ \Si^{N P} \, \eta^{M Q} \ - \ \Si^{M P} \, \eta^{N Q} \ - \ \Si^{N Q} \, \eta^{M P} \bigr) \ .
\]

\subsection{Spinor representations in even dimensions $D = 2n$}

Define $n$ pairs of ladder operators
\begin{align*}
a^{0} \ \ &:= \ \ \frac{i}{2} \; \bigl( \Ga^{0} \ + \ \Ga^{1} \bigr) \ \ \ \ \ \ \so (a^{0})^\dag \eq \frac{i}{2} \; \bigl( -\Ga^{0} \ + \ \Ga^{1} \bigr) \\
a^{j} \ \ &:= \ \ \frac{i}{2} \; \bigl( \Ga^{2j} \ - \ i\Ga^{2j+1} \bigr) \so (a^{j})^\dag \eq \frac{i}{2} \; \bigl( \Ga^{2j} \ + \ i\Ga^{2j+1} \bigr) \co j=1,...,n-1 \ ,
\end{align*}
whose hermiticity properties are due to $(\Ga^{0})^\dag = + \Ga^{0}$ and $(\Ga^{M \neq 0})^\dag = -\Ga^{M \neq 0}$. From the Clifford algebra in $\eta^{MN} = \te{diag}(+1, \ -1, \ ... , \ -1)$ signature, it follows that the $a^{j}$ (where $j=0,1,...,n-1$ now) furnish a set of $n$ fermionic oscillators
	\[\Bigl\{ a^{i} \ , \ (a^{j})^\dag \Bigr\} \eq \de^{ij} \co \Bigl\{ a^{i} \ , \ a^{j} \Bigr\} \eq \Bigl\{ (a^{i})^\dag \ , \ (a^{j})^\dag \Bigr\}  \eq 0 \ .
\]
Let $|0 \rangle$ denote the vacuum such that $a^{i} |0 \rangle = 0$, then there are states
    \[\begin{array}{c|ccccc} \te{states} &|0 \rangle &(a^{i})^\dag \, |0 \rangle &(a^{i})^\dag \, (a^{j})^\dag \,|0 \rangle &\cdots & (a^{n})^\dag \, (a^{n - 1})^\dag \, ... \, (a^{1})^\dag \, |0 \rangle \\\hline \te{number} &1 &n &\left( \begin{smallmatrix} n \\ 2 \end{smallmatrix} \right) &\cdots &1 \end{array}
\]
of total number
    \[1 \ + \ n \ + \ \vecb n \\ 2 \vece \ + \ ... \ + \ 1 \eq \sum_{k = 0}^{n} \vecb n \\ k \vece \eq 2^{n} \eq 2^{\frac{D}{2}} \ .
\]
States in the spinor representations are defined by $n=D/2$ quantum numbers $s_{i} = \pm \frac{1}{2}$
    \[|s_{0}, \, ... \, , s_{n-1} \rangle \ \ := \ \ (a^{0})^{\dag \left(s_{0} + \frac{1}{2}\right)} \, ... \, (a^{n-1})^{\dag \left(s_{n-1} + \frac{1}{2}\right)} \, |0 \rangle \ .
\]
Note that the generators $\Si^{(2i)( 2i + 1)}$ mutually commute. So we diagonalize all of
\begin{align*}
(a^{0})^{\dag} \, a^{0} \ - \ \frac{1}{2} \ \ &= \ \ +\frac{1}{4} \; \Bigl[ \Ga^{0} \ , \ \Ga^{1} \Bigr] \ \; \ \eq -i \Si^{01} \\
(a^{j})^{\dag} \, a^{j} \ - \ \frac{1}{2} \ \ &= \ \ \frac{i}{4} \; \Bigl[ \Ga^{2j} \ , \ \Ga^{2j+1} \Bigr] \eq  \Si^{(2j)(2j+1)}
\end{align*}
and find the $|s_{0},...,s_{n-1} \rangle$ defined above to be the simultaneous eigenstates of
  \[S^{i} \ \ := \ \ \left\{ \begin{array}{ll} (a^{0})^\dag \, a^{0} \, \ - \ \frac{1}{2} \eq -i\Si^{01} &: \ i=0 \\ (a^{i})^\dag \, a^{i} \; \ - \ \frac{1}{2} \eq \Si^{(2i)( 2i + 1)} &: \ i=1,...,n-1 \end{array} \right.
\]
in the sense that
    \[S^{i} \, |s_{0}, \, ... \, , s_{n-1} \rangle \eq s_{i} \, |s_{0} , \, ... \, , s_{n-1} \rangle \ .
\]
Call those $|s_{0},...,s_{n-1} \rangle$ {\em Dirac spinors}. In $D = 4$ dimensions with $n = 2$, for instance, the states $|\pm \frac{1}{2}, \pm \frac{1}{2} \rangle$ form a 4 component spinor.

\noindent
Representations in even dimensions are reducible, since the generalization of $\ga^{5}$,
    \[\Ga^{2n + 1} \ \ := \ \ i^{n-1} \, \Ga^{0} \, \Ga^{1} \, ... \, \Ga^{2n-1} \ ,
\]
satisfies
    \[\Bigl\{ \Ga^{2n+1} \ , \ \Ga^{M} \Bigr\} \eq 0 \co \Bigl[ \Ga^{2n + 1} \ , \ \Si^{MN} \Bigr] \eq 0 \co (\Ga^{2n + 1})^{2} \eq \mathds{1} \ .
\]
It follows from
    \begin{align*}
    2^{n} \, S^{0} \, S^{1} \, ... \, S^{n-1} \ \ &= \ \ 2^{n} \; \frac{1}{4} \; \left( +\frac{i}{4} \right)^{n-1} \; \Bigl[ \Ga^{0} \ , \ \Ga^{1} \Bigr] \, ... \, \Bigl[ \Ga^{2n-2} \ , \ \Ga^{2n-1} \Bigr] \\
    &= \ \ i^{n-1} \Ga^{0} \, \Ga^{1} \, ... \, \Ga^{2n-1} \eq \Ga^{2n+1} \ .
\end{align*}
that all the $|s_{0} , ... ,s_{n-1} \rangle$ are eigenstates to $\Ga^{2n + 1}$
    \[\Ga^{2n + 1} \, |s_{0} , \, ... \, , s_{n-1} \rangle \eq \pm |s_{0}, \, ... \, ,s_{n-1} \rangle
\]
with eigenvalue $+1$ for even numbers of $s_{i} = -\frac{1}{2}$ and $-1$ for odd ones. This property is called {\em chirality}, and spinors of definite chirality are referred to as {\em Weyl spinors}.

\subsection{Spinor epresentations in odd dimensions $D = 2n + 1$}

Just add $\Ga^{2n + 1} = i^{n-1}  \Ga^{0} \Ga^{1} \, ... \, \Ga^{2n-1}$ to the $\Ga^{M}$ matrices of $D=2n$ dimensions. From its properties $\{ \Ga^{2n+1} , \Ga^M \} = 0$ and $(\Ga^{2n+1})^2 = 1$, it perfectly extends the Clifford algebra in $D=2n$ dimensions to $D=2n+1$ with extended metric $\eta^{\mu \nu}= (-1, +1,...,+1)$.

\noindent
Since there is no further $\Ga$ matrix with which $\Ga^{2n+1}$ could be paired to a further $a^{i}$ operator, the representation is the same as for $D = 2n$, but now irreducible. The $SO(1,2n)$ generators in addition to those of $SO(1,2n-1)$ are given by $\frac{i}{2} \Ga^M \Ga^{2n+1}$ with $M=0,1,...,2n-1$. Since odd dimensions do not have a ''$\ga^{5}$'', there is no chirality. The spinor representations' dimension is $2^{\frac{D - 1}{2}}$.

\noindent
In general, define $N_D$ to give the number of spinor components:
	\[N_D \ \ := \ \ \left\{ \begin{array}{ll}  2^{n} = 2^{\frac{D}{2}} &: \ D = 2n \ \te{even} \\ 2^{n} = 2^{\frac{D-1}{2}} &: \ D =2n+1 \ \te{odd} \end{array} \right.
\]

\subsection{Majorana spinors}
\label{sec:Maj}

Let us now introduce the notion of {\em reality} for spinors in Minkowski spacetime. Under infinitesimal Lorentz transformations, spinors $\psi$ transform into $\psi ' = \psi +  i \om_{MN} \Si^{MN} \psi$. Since the $\Si^{MN}$ are in general complex, it is not guaranteed that relations between $\psi$ and its complex conjugate $\psi^\ast$ are consistent with Lorentz transformations.

\noindent
A relation between $\psi \leftrightarrow \psi^\ast$ is referred to as the {\em Majorana condition}. It has to be of the form $\psi^\ast = C \Ga^0 \psi$ where $C$ is the {\em charge conjugation matrix}. Consistency requires $(C \Ga^0)\ast C \Ga^0=1$ which is possible in dimensions $D=0,1,2,3,4 \, \te{mod} \, 8$. In other words, among the physically sensible dimensions, $D=5,6,7$ do not admit a Majorana condition.

\noindent
A Majorana condition can be imposed on a Weyl spinor if $D=0,1,2,3,4 \, \te{mod} \, 8$ and the Weyl representation is conjugate to itself. Weyl spinors exist in even dimensions $D=2n$, and by analyzing the complex conjugate of the chirality matrix
\[ (\Ga^{2n+1})^\ast \eq (-1)^{n+1}  \, C^{-1} \, \Ga_0^{-1} \, \Ga^{2n+1} \, \Ga_0 \, C \ , \]
it turns out that charge conjugation only preserves the spinors' chirality if $(-1)^{n+1}=+1$. If $n$ is even, i.e. in $D=4,8,12,...$ dimensions, the two inequivalent Weyl representations are complex conjugate to each other, and one can either impose the Weyl or Majorana condition, but not both! In dimensions $D=2 \, \te{mod} \, 8$, the Weyl representations are self conjugate and compatible with the Majorana condition, so Majorana Weyl spinors are possible in dimensions $D=2$ and $D=10$.

\section{Supersymmetry algebra}

The SUSY algebra in $D$ dimensions consists of generators $M_{MN}$,
$P_{M}$, $Q_{\al}$ last of which are spinors in $D$ dimensions. The
algebra has the same structure as in 4 dimensions, with the bosonic
generators defining a standard Poincar\'e algebra in higher dimensions and 
    \[\Bigl\{ Q_{\al} \ , \ Q_{\be} \Bigr\} \eq a_{\al \be}^{M} \, P_{M} \ + \ Z_{\al \be}
\]
where $a_{\al \be}^{M}$ are constants and the central charges $Z_{\al \be}$
 now can also include brane charges. This is the  $D > 4$ Coleman
 Mandula- or HLS generalization of the $D=4$ algebra. The
 arguments for the proof are identical to those in 4 dimensions and we will skip
 them here.

\noindent
A new feature of the Poincar\'e algebra is that all the generators
$M^{(2j)(2j+1)}$ commute with each other and can thus be simultaneously
diagonalized as we have seen in the discussion of the higher
dimensional spinorial representation. Then we can have several ''spins'' defined as the eigenvalues of these
operators. Of particular relevance is the generator $M^{01}$. This is
used to define a {\it weight} $w$ of an operator $\cal{O}$ by 
\[
\Bigl[M^{01} \ , \ {\cal{O}} \Bigr] \eq -i w \, {\cal{O}}
\]
where $\cal{O}$ and ${\cal O}^*$ have the same weight.

\subsection{Representations of supersymmetry algebra in higher dimensions}
\label{sec:RepresentationsOfSupersymmetryAlgebraInHigherDimensions}

Consider massless states $P^{\mu} = (E \ , E \ , 0 \ , ... \ ,
 0)$ with little group $SO(D - 2)$. We define the spin to be the
 maximum eigenvalue of $M_{MN}$ in the representation. Notice that for
 the momentum of a massless particle $P^1-P^0=0$ and that 
\[
\Bigl[M^{01} \ , \ P^1 \, \pm \, P^0 \Bigr] \eq \mp i  \, \bigl(P^1 \ \pm \ P^0 \bigr) \ .
\]
Therefore the weight of $P^1\pm P^0$ is $w=\pm 1 $. As the ''$-$'' combination $P^{1} - P^{0}$ is zero in massless representations, the weight $w=-1$ can be excluded and we only need to consider combinations of $\{Q, Q\}$ in which both $Q$'s have weight $w=+\frac{1}{2}$.

\noindent
So if we start with arbitrary spinors $Q_\alpha$ of the form
\[
Q_{\al} \eq |\pm \tfrac{1}{2}, \ \pm \tfrac{1}{2}, \ \pm \tfrac{1}{2}, \ \cdots, \ \pm \tfrac{1}{2} \rangle \co \al = 1,...,N_D
\]
with $N_D$ components (recall that $N_D = 2^{\frac{D}{2}}$ for even and $N_D = 2^{ \frac{D-1}{2} }$ for odd dimensionality respectively), requiring weight $+\frac{1}{2}$ means that (, as a special case of $[M^{MN},Q_{\al}] = -\Si^{MN} Q_{\al}$,)
	\[\Bigl[M^{01} \ , \ Q_{\al} \Bigr] \eq -\Si^{01} \, Q_{\al} \eq -iS^{0} \, Q_{\al} \ \ \stackrel{!}{=} \ \ -\frac{i}{2} \; Q_{\al} \ ,
\]
so $Q_\al$ has to be of the form
\[
Q_{\al} \, \Bigl. \Bigr|_{w=+\frac{1}{2}} \eq |\stackrel{\downarrow}{+} \tfrac{1}{2}, \ \pm \tfrac{1}{2},  \ \pm \tfrac{1}{2}, \ \cdots, \ \pm \tfrac{1}{2} \rangle \co \al = 1,..., \frac{N_D}{2} \ .
 \]
This leads to half of the number of components of $Q_\al$ in the massless case, namely $\frac{N_D}{2}$.

\noindent
Furthermore, we can separate the $Q$'s into $Q^+$ and $Q^-$ according
 to eigenvalues of $M_{23}$ (standard spin in 4d). They furnish an algebra of
 the form $\{Q^+,Q^+\}=\{Q^-,Q^-\}=0$ and $\{Q^+,Q^-\}\neq 0$ corresponding to creation- and annihilation operators. To see this, consider the commutator
	\[\Bigl[ M^{(2j)(2j+1)} \ , \ Q_{( \al} \, Q_{\be )} \Bigr] \eq -Q_{(\al} \, S^{j} \, Q_{\be)} \ - \ S^{j} \, Q_{(\al} \,  Q_{\be)} \eq - (s^{(\al)}_{j} \ + \ s^{(\be)}_{j}) \, Q_{( \al} \, Q_{\be )} \ .
\]
Using the super Poincar\'e algebra, we can also show this expression to be a linear combination of the $P^{2}$...$P^{D-1}$ which are all zero in our case $P^{\mu} = (E \ , E \ , 0 \ , ... \ ,
 0)$. Consequently, all the combinations $s^{(\al)}_{j} + s^{(\be)}_{j}$ have to vanish leaving $\{ Q^+_{\al} , Q^-_{\be=\al} \}$ as the only nonzero anticommutators.

\noindent
This implies that a supersymmetric multiplet can be constructed
starting from a ''vacuum'' state $|\la \rangle$ of helicity $\lambda$ 
annihilated by the $Q^-$ operators, $Q^- |\lambda \rangle=0$, and the rest of the states in the multiplet are generated by acting on $Q^+$. Therefore they will be of the form
\[
Q^+_{\al} \Bigl. \Bigr|_{w=+\frac{1}{2}} \eq |+ \tfrac{1}{2}, \ \stackrel{\downarrow}{+} \tfrac{1}{2}, \ \pm \tfrac{1}{2}, \ \cdots, \ \pm \tfrac{1}{2} \rangle \co \al = 1,...,\frac{N_D}{4}
\]
and the total number will be $\frac{N_D}{4}$. 

\noindent
Given some state $| \la \rangle$ of helicity $\la$ (i.e. $M_{23} |\lambda \rangle = \la |\lambda \rangle$), the action of any $Q^+_{\al}$ will lower the $M^{23}$ eigenvalue:
	\begin{align*}
	M^{23} \, Q_{\al}^{+} \, |\la \rangle \ \ &= \ \ \Bigl[M^{23} \ , \ Q_{\al}^{+} \Bigr] \, | \la \rangle \ + \ Q_{\al}^{+} \, M^{23} \, |\la \rangle \eq -\Si^{23} \, Q_{\al}^{+} \, |\la \rangle \ + \ \la \, Q_{\al}^{+} \, |\la \rangle \\
	&= \ \ \left( \la \ - \  \frac{1}{2} \right) \, Q_{\al}^{+} \, |\la \rangle
\end{align*}
We therefore obtain the follwing helicities by application of the $Q^+_{\al} \bigl. \bigr|_{w=+\frac{1}{2}}$
\[
|\lambda \rangle \co |\lambda-\tfrac{1}{2} \rangle \ , \ \ \ \ \ \ ... \ , \ |\lambda -\tfrac{1}{2} \cdot \tfrac{N_D}{4} \rangle \ .
\]
It follows for the range of occurring $\la$'s that
\[
\lambda_{\te{max}} \ - \ \lambda_{\te{min}} \eq \lambda \ - \ \left(\lambda \,- \, \tfrac{N_D}{8} \right) \eq \frac{N_D}{8} \ ,
\]
imposing $|\lambda| \leq 2$ thus requires $N_D \leq 32$. But remembering
that $N_D = 2^{\frac{D}{2}}, 2^{ \frac{D-1}{2} }$ for even and odd dimensionality,
this implies a maximum number of spacetime dimensions $D=10, 11$.

\noindent
Notice the similarity of this argument with the previous proof that
the maximum number of supersymmetries in 4 dimensions was ${\cal N}=8$. We will see later that precisely ${\cal N}=8$ supergravity is obtained from the
supersymmetric theories in $D=10$ and $D=11$.

\noindent
Let us take a closer look at the spectrum of $D= 11$ and $D = 10$:
\begin{itemize}
\item D = 11

Only ${\cal N} = 1$ SUSY is possible. The only multiplet consists of
    \[\underbrace{g_{MN}}_{\te{graviton}} \co \underbrace{\psi_{M}^{\al}}_{\te{gravitino}} \co \underbrace{A_{MNP}}_{\te{antisymmetric tensor (non-chiral)}}
\]
In order to count the (on shell) degrees of freedom for each field we have to perform the analysis based on the little group $O(D-2)$. The
graviton in $D$ dimensions carries $\frac{(D-2)(D-1)}{2} -1$ components, corresponding to
a symmetric tensor in $D-2$ dimensions minus the trace, which is $45-1=44$ in the $D=11$ case. An antisymmetric tensor of rank $p+1$ in $D$ dimensions
has $\left( \begin{smallmatrix} D-2 \\ p+1 \end{smallmatrix} \right)$ degrees of freedom, in the case of $A_{MNP}$ with $p+1=3$, this is $\left( \begin{smallmatrix} 9 \\ 3 \end{smallmatrix} \right)=84$.

\noindent
For the gravitino spinor $\psi^{\mu}_{\al}$, we have
$2^{\frac{D-3}{2}} \cdot (D-2) - 2^{\frac{D-3}{2}}$ independent components: The first factor is the product of the spinor components times the vector components of the
gravitino (since it carries both indices), and the subtraction of the
$2^{\frac{D-3}{2}}$ degrees of freedom of a spin $\frac{1}{2}$ particle is similar to the
subtraction of the trace for the graviton. In terms of $su(2)$ representations $(1) \otimes \left(\frac{1}{2} \right) = \left( \frac{3}{2} \right) \oplus \left( \frac{1}{2} \right)$, one can say that the spin $\frac{1}{2}$ contribution on the right hand side is discarded. More generally, a vector spinor $\Psi_M^\al$ only furnishes an  irreducible Lorentz representation if contractions with any invariant tensor (such as the metric and the higher dimensional $\Ga$ matrices) vanish. If the ''gamma trace'' $ \Psi_M^\al \Ga^M_{\al \be}$ was nonzero, then it would be a lower irreducible representation on its own right. In $D=11$, we obtain $9\cdot 2^{4} - 2^{4} = 128$ components for the gravitino which matches the number of bosonic degrees of freedom $84+44$.

\item D = 10

This allows ${\cal N} = 2$:
    \[\begin{array}{c|cccccc} \te{IIA} &g_{MN} &2 \times \psi_{M}^{\al} &B_{MN} &\phi &A_{MNP} &\la \\\hline \te{IIB} &g_{MN} &2 \times \psi^{\al}_{M} &2 \times B_{MN} &2 \times \phi &A^\dag_{MNPQ} &\la \\\hline \te{I} &(g_{MN} &B_{MN} &\phi &\psi^{\al}_{M}) &(A_{M} &\la) \ (\te{chiral}) \end{array}
\]
\end{itemize}
About antisymmetric tensors $A_{M_{1}...M_{p+1}}$ of spin 0 or 1, we know:
\begin{itemize}
\item $A_{M}$ couples to a particle $\int A^{M} \ \dd x_{M}$, where $\dd x_{M}$ refers to the world line
\item $A_{MN}$ couples to a string $\int A^{MN} \ \dd x_{M} \wedge \dd x_{N}$ (world sheet)
\item $A_{MNP}$ to a membrane ...
\item $A_{M_{1}...M_{p+1}}$ to a $p$ brane
\end{itemize}
The coupling is dependent of the object's charges:
    \[\begin{array}{r|cc} \te{object} &\te{charge} &\te{couples to} \\\hline \te{particle} &q &A_{M} \\ \te{string} &q_{M} &A_{MN} \\ p \ \te{brane} &q_{M_{1}...M_{p}} &A_{M_{1}...M_{p + 1}} \end{array}
\]
Charges are new examples of central charges in the SUSY algebra:
    \[\Bigl\{Q \ , \ Q\Bigr\} \ \ \propto \ \ a \, P \ + \ b^{M_{1}...M_{p}} \, q_{M_{1}...M_{p}}
\]

\section{Dimensional Reduction}
\label{sec:DimensionalReduction}

Let us review the general procedure of reducing any number of dimensions bigger than 4 to $d=4$. Recall the example of a scalar in 5 dimensions $M_{5} \eq \mathbb M_{4} \times S^{1}$ (the last of which has radius $R$) where field in 5 dimensions could be replaced by $\infty$ many fields in $d=4$. If $\vph$ is massless,
    \[\pa_{M} \pa^{M} \vph \eq 0 \so \pa_{\mu} \pa^{\mu} \vph_{n} \ - \ \frac{n^{2}}{R^{2}} \; \vph_{n} \eq 0 \ ,
\]
then the Fourier mode $\vph_{n}$ with respect to the $S^{1}$ dimension has a mass of $\frac{n}{R}$.

\noindent
For dimensional reduction, only keep the $n = 0$ mode,
\begin{align*}
\vph(x^{M}) \ \ &\mapsto \ \ \vph(x^{\mu}) \\
A_{M}(x^{M}) \ \ &\mapsto \ \ A_{\mu}(x^{\mu}) \co \underbrace{A_{m}(x^{\mu})}_{\te{scalars}} \co m \eq 4,...,D-1 \\
B_{MN} \ \ &\mapsto \ \ B_{\mu \nu} \co \underbrace{B_{\mu n}}_{\te{vectors}} \co \underbrace{B_{mn}}_{\te{scalars}} \\
\underbrace{\psi}_{2^{n}} \ \ &\mapsto \ \ \underbrace{\psi}_{\frac{1}{4}2^{n} \ 4\te{D} - \te{spinors}} \ .
\end{align*}
Consider e.g. the reduction of $D=11$ to $d=4$: The fundamental fields are graviton $g_{MN}$ that carries $\frac{9 \cdot 10}{2}-1=44$ degrees of freedom and the gravitino $\psi_M^\alpha$ with $9 \cdot 2^{\frac{9-1}{2}} - 2^{\frac{9-1}{2}} = 8 \cdot 16= 128$ components. Again, the
subtraction is an extra spinor degree of freedom. The final field is
an antisymmetric tensor $A_{MNP}$ that carries $\left( \begin{smallmatrix} 9 \\ 3 \end{smallmatrix} \right)=84$ degrees of
freedom. Note that we have $128$ bosonic degrees of freedom and $128$ fermionic
degrees of freedom. Dimensional reduction to $d=4$ leads to:
\begin{align*}
g_{MN} \ \ &\mapsto \ \ \underbrace{g_{\mu \nu}}_{\te{graviton}} \co \underbrace{g_{\mu m}}_{7 \ \te{vectors}} \co \underbrace{g_{mn}}_{\frac{7 \cdot 8}{2} = 28 \ \te{scalars (symmetry!)}} \\
A_{MNP} \ \ &\mapsto \ \ A_{\mu \nu \rho} \co \underbrace{A_{\mu \nu m}}_{7 \ \te{tensors}} \co \underbrace{A_{\mu mn}}_{21 \ \te{vectors}} \co \underbrace{A_{mnp}}_{\frac{7 \cdot 6 \cdot 5}{1 \cdot 2 \cdot 3} = 35 \ \te{scalars (antisymmetry!)}} \\
\psi_{M}^{\al} \ \ &\mapsto \ \ \underbrace{\psi_{\mu}^{\al}}_{\frac{32}{4} = 8} \co \underbrace{\psi_{m}^{\al}}_{7 \cdot 8 = 56 \ \te{fermions}}
\end{align*}
Recall here that a three index antisymmetric tensor $A_{\mu \nu \rho}$ in 4 dimensions carries no degrees of freedom and that two index antisymmetric tensors $A_{\mu \nu m}$ are dual to scalars. The spectrum is the same as the ${\cal N}=8$ supergravity in 4 dimensions:
	\[ \begin{array}{c|c|c|c} \te{number} &\te{helicity} &\te{particle type} &\te{on shell degrees of freedom in} \ d=4 \\\hline
1 &2 &\te{graviton} &1 \cdot \left( \frac{(4-2)(4-1)}{2}-1 \right) = 1 \cdot 2 = 2 \\ 8 &\frac{3}{2} &\te{gravitino} &8 \cdot \left( 2^{ \frac{4-2}{2} } \cdot (4-2) - 2^{ \frac{4-2}{2} } \right) = 8 \cdot 2 = 16 \\
28 &1 &\te{vector} &28 \cdot (4-2) = 28 \cdot 2 = 56 \\ 56 &\frac{1}{2} &\te{fermion} &56 \cdot 2^{ \frac{4-2}{2} } = 56 \cdot 2 = 112 \\
70 &0 &\te{scalar} &70 \cdot 1 = 70 \end{array}
\]
There is a theory of ${\cal N} = 8$ supergravity based on the $g_{MN}$ and
$A_{MNP}$. Reducing the dimension from 11 to 4 has an effect of ${\cal N} = 1
\mapsto {\cal N} = 8$. This ${\cal N} = 8$ model is non-chiral, but other
compactifications and $p$ branes in a 10 dimensional string theory can provide chiral ${\cal N} = 1$ models close to the MSSM. Notice
that the statement of why the maximum dimensionality of
supersymmetric theories is $11$ is identical to the statement that the
maximum number of supersymmetries in 4 dimensions is ${\cal N}=8$ since
both theories are related by dimensional reduction. Actually, the
explicit construction of extended supergravity theories was originally
done by going to the simpler theory in extra dimensions and
dimensionally reduce it.

\section{Summary}
\label{sec:Summary}

This is the end of these lectures. We have seen that both
supersymmetry and extra dimensions provide the natural way to extend
the spacetime symmetries of standard field theories.

\noindent
They both have a set of beautiful formal properties, but they also
address important unsolved physical questions such as the hierarchy problem for instance.

\noindent
For supersymmetry we can say that it is a very elegant  and unique extension of spacetime symmetry:
\begin{itemize}
\item It may be realized at low energies, the energy of SUSY breaking of 1 TeV is within experimental reach (hierarchy, unification, dark matter)
\item It may be an essential ingredient of fundamental theory (M theory, strings).
\item It is a powerful tool to understand QFTs, especially non-perturbatively (S-duality, Seiberg-Witten, AdS/CFT).
\end{itemize}
Both supersymmetry and extra dimensions may be subject to be tested soon in
experiments. They are both basic ingredients of string theory but may be relevant only at higher energies, we need to remain patient. Independent of any experimental verification they have expanded our understanding of physical theories which is a good argument to continue their study.

\vskip1cm
\goodbreak


\appendix

\chapter{Useful spinor identities}
\label{sec:UsefulSpinorIdentities}

Identities involving spinors and vectors of the Lorentz group $SO(3,1)$ depend on the conventions chosen, in particular the signs and factors of $i$ and $\frac{1}{2}$ involved. Let us therefore list the conventions chosen throughout these notes:
\begin{itemize}
\item the Minkowski metric is ''mostly negative''
\[ \eta^{\mu \nu} \eq \te{diag}(+1, \, -1, \, -1, \, -1) \]
\item the invariant $SL(2,\mathbb C)$ tensors $\ep_{\al \be}, \ep_{\dot{\al} \dot{\be}}$ and their inverses $\ep^{\al \be}, \ep^{\dot{\al} \dot{\be}}$ have entries
\[ \ep_{\al \be} \eq \ccb 0 &-1 \\ 1 &0 \cce \eq \ep_{\dot{\al} \dot{\be}} \co \ep^{\al \be} \eq \ccb 0 &1 \\ -1 &0 \cce \eq \ep^{\dot{\al} \dot{\be}}  \]
\item the nonzero components of the totally antisymmetric $\ep$ tensor in four dimensions are determined by
\[ \ep_{0123} \eq - \, \ep^{0123} \eq +1 \]
\end{itemize}

\section{Bispinors}
\label{sec:PureBispinors}

Given the antisymmetric ''metric'' $\ep$ for bispinors, we have to define the way indices are contracted in spinor products:
\begin{align*}
  \psi \chi \ \ &:= \ \ \psi^{\al} \, \chi_{\al} \eq \chi \psi  \\
  \bar{\psi} \bar{\chi} \ \ &:= \ \ \bar{\psi}_{\dot{\al}} \,  \bar{\chi}^{\dot{\al}} \eq \bar{\chi} \bar{\psi}
\end{align*}
Left- and right handed bispinors follow opposite contraction rules to ensure that $(\psi \chi)^\dag = \bar{\psi} \bar{\chi}$, where the hermitian conjugation is assumed to reverse the order of the spinors without a minus sign due to anticommutation.

\noindent
The irreducible spinor representations of $SO(3,1)$ are two dimensional, so any antisymmetric expression $T_{[\al \be]}$ ($\bar{T}_{[\dot{\al} \dot{\be}]}$) is proportional to the unique antisymmetric rank two tensor $\ep_{\al \be}$ ($\ep_{\dot{\al} \dot{\be}}$). Hence, we find for anticommuting variables that
\begin{align*}
\theta_{\al} \, \theta_{\be} \ \ &= \ \ +\frac{1}{2} \; \ep_{\al \be} \, (\theta \theta) \co \theta^{\al} \, \theta^{\be} \ \ = \ \ -\frac{1}{2} \; \ep^{\al \be} \, (\theta \theta) \\
\bar{\theta}_{\dot{\al}} \, \bar{\theta}_{\dot{\be}} \ \ &= \ \ -\frac{1}{2} \; \ep_{\dot{\al} \dot{\be}} \, (\bar{\theta} \bar{\theta}) \co \bar{\theta}^{\dot{\al}} \, \bar{\theta}^{\dot{\be}} \ \ = \ \ +\frac{1}{2} \; \ep^{\dot{\al} \dot{\be}} \, (\bar{\theta} \bar{\theta}) \ ,
\end{align*}
the $\pm \frac{1}{2}$ factors can be determined by contraction with the inverse $\ep$ symbol. An easy corollory is
\begin{align*}
(\theta \chi) \, (\theta \xi) \ \ &= \ \ -\frac{1}{2} \; (\theta \theta) \, (\chi \xi) \\
(\bar{\theta} \bar{\chi}) \, (\bar{\theta} \bar{\xi}) \ \ &= \ \ -\frac{1}{2} \; (\bar{\theta} \bar{\theta}) \, (\bar{\chi} \bar{\xi}) \ .
\end{align*}

\section{Sigma matrices}
\label{sec:SigmaMatrices}

We work with the following four vectors of generalized sigma matrices
\[ \si^\mu \eq ( \mathds{1}, \, \ve{\si}) \co   \bar{\si}^\mu \eq ( \mathds{1}, \, -\vec{\si}) \]
where the spatial entries are simply given by the standard Pauli matrices
\[ \ve{\si} \eq \left\{ \ccb 0 &1 \\ 1 &0 \cce \ , \ \ccb 0 &-i \\ i &0 \cce \ , \ \ccb 1 &0 \\ 0 &-1 \cce \right\} \ . \]
They can be easily verified to satisfy the Dirac algebra
\[ \si^{\mu} \, \bar{\si}^{\nu} \ + \ \si^{\nu} \, \bar{\si}^{\mu} \ \ = \ \ 2 \, \eta^{\mu \nu} \, \mathds{1} \ . \]
One can regard the $\si$ matrices as the Clebsch Gordan coefficients converting the tensor product of left- and righthanded spinors into a Lorentz vector and vice versa. The following relations give the details of this dictionary:
\[\te{Tr} \Bigl\{ \si^{\mu} \, \bar{\si}^{\nu} \Bigr\} \ \ = \ \ 2 \ \eta^{\mu \nu} \co
(\si^{\mu})_{\al \dot{\al}} \, (\bar{\si}_{\mu})^{\dot{\be} \be} \ \ = \ \ 2 \, \de_{\al}\,^{\be} \, \de_{\dot{\al}}\,^{\dot{\be}} \]
Finally, the antisymmetric $\si$ products $\si^{\mu \nu} = \frac{i}{2} \si^{[\mu} \bar{\si}^{\nu]}$ and $\bar{\si}^{\mu \nu} = \frac{i}{2} \bar{\si}^{[\mu} \si^{\nu]}$ play an important role:
\begin{align*}
&\si^{\mu \nu} \ \ = \ \ +\frac{1}{2i} \; \ep^{\mu \nu \rho \si} \, \si_{\rho \si} \co \bar{\si}^{\mu \nu} \ \ = \ \ -\frac{1}{2i} \; \ep^{\mu \nu \rho \si} \, \bar{\si}_{\rho \si} \\
&\te{Tr} \Bigl\{ \si^{\mu \nu} \, \si^{\ka \tau} \Bigr\} \ \ = \ \ \frac{1}{2} \; \bigl( \eta^{\mu \ka} \, \eta^{\nu \tau} \ - \ \eta^{\mu \tau} \, \eta^{\nu \ka} \ + \ i \ep^{\mu \nu \ka \tau} \bigr) \\
&\te{Tr} \Bigl\{ \bar{\si}^{\mu \nu} \, \bar{\si}^{\ka \tau} \Bigr\} \ \ = \ \ \frac{1}{2} \; \bigl( \eta^{\mu \ka} \, \eta^{\nu \tau} \ - \ \eta^{\mu \tau} \, \eta^{\nu \ka} \ - \ i \ep^{\mu \nu \ka \tau} \bigr) \\
&(\si^{\mu \nu})_\al \, ^\be \, (\si_{\mu \nu})_\ga \,^\de \eq \ep_{\al \ga} \, \ep^{\be \de} \ + \ \de_\al^\de \, \de_\ga^\be
\end{align*}

\section{Bispinors involving sigma matrices}
\label{sec:BispinorsInvolvingSigmaMatrices}

To conclude this appendix, we give some identities to manipulate $\si$ matrices interacting with two spinors. The symmetry properties are given as follows:
\begin{align*}
\psi \, \si^{\mu} \, \bar{\chi} \ \ &= \ \ - \, \bar{\chi} \, \bar{\si}^{\mu} \, \psi \\
\psi \, \si^{\mu} \, \bar{\si}^{\nu} \, \chi \ \ &= \ \ \chi \, \si^{\nu} \, \bar{\si}^{\mu} \, \psi \\
\psi \, \si^{\mu \nu} \, \chi \ \ &= \ \ - \, \chi \, \si^{\mu \nu} \, \psi
\end{align*}

\paragraph{Exercise A.1:} Prove these symmetry properties.

\paragraph{Exercise A.2:} Show that
\begin{align*}
(\theta\psi) \, (\bc\bet) \ \ &= \ \ - \, \frac{1}{2} \; (\theta \, \s^\mu \, \bet) \, (\bc \, \bs_\mu \, \psi)\\
(\theta \, \s^\mu \, \bt) \, (\theta \, \si^\nu \, \bt) \ \ &= \ \ \frac{1}{2} \; \eta^{\mu\nu} \, (\theta\theta) \, (\bt\bt) \ .
\end{align*}

\chapter{Dirac spinors versus Weyl spinors}
\label{sec:Dirac}

In this appendix, we give the dictionary connecting the ideas of Weyl spinors with the more standard \textsc{Dirac} theory in $D=4$ dimensions. 

\section{Basics}
\label{sec:DiracSpinors}

A {\em Dirac spinor} $\Psi_D$ is defined to be the direct sum of two Weyl spinors $\psi, \bar \chi$ of opposite chiraliy, it therefore falls into a reducible representation of the Lorentz group,
    \[\Psi_{D} \ \ := \ \ \vecb \psi_{\al} \\ \bar{\chi}^{\dot{\al}} \vece \ .
\]
The Dirac analogue of the Weyl spinors' sigma matrices are the $4 \times 4$ gamma matrices $\ga^\mu$ subject to the {\em Clifford algebra}
 \[\ga^{\mu} \ \ := \ \ \ccb 0 &\si^{\mu} \\ \bar{\si}^{\mu} &0 \cce \co \Bigl\{ \ga^{\mu} \ , \ \ga^{\nu} \Bigr\} \eq 2 \, \eta^{\mu \nu} \, \mathds{1} \ .
\]
Due to the reducibility, the generators of the Lorentz group take block diagonal form
    \[\Si^{\mu \nu} \eq \frac{i}{4} \; \ga^{\mu \nu} \eq \ccb \si^{\mu \nu} &0 \\ 0 &\bar{\si}^{\mu \nu} \cce
\]
and naturally obey the same algebra like the irreducible blocks $\si^{\mu \nu}$, $\bar \si^{\mu \nu}$:
\[ \Bigl[ \Si^{\mu \nu} \ , \ \Si^{\la \rho} \Bigr] \eq \pm \, i \, \Si^{\nu \la} \, \eta^{\mu \rho} \ + \ ...
\]
To disentangle the two inequivalent Weyl representations, one defines the chiral matrix $\ga^{5}$ as
    \[\ga^{5} \ \ := \ \ i\ga^{0} \, \ga^{1} \, \ga^{2} \, \ga^{3} \eq \ccb -\mathds{1} &0 \\ 0 &\mathds{1} \cce \ ,
\]
such that the $\psi (\chi)$ components of a Dirac spinors have eigenvalues (chirality) $-1 \, (+1)$ under $\ga^5$,
    \[\ga^{5} \, \Psi_{D} \eq \ccb -\mathds{1} &0 \\ 0 &\mathds{1} \cce \, \vecb \psi_{\al} \\ \bar{\chi}^{\dot{\al}} \vece \eq \vecb -\psi_{\al} \\ \bar{\chi}^{\dot{\al}} \vece \ .
\]
Hence, one can define projection operators $P_{L}$, $P_{R}$,
    \[P_{L} \ \ := \ \ \frac{1}{2} \; \bigl(\mathds{1} \ - \ \ga^{5} \bigr) \co P_{R} \ \ := \ \ \frac{1}{2} \; \bigl(\mathds{1} \ + \ \ga^{5} \bigr) \ ,
\]
eliminating one part of definite chirality, i.e.
    \[P_{L} \, \Psi_{D} \eq \vecb \psi_{\al} \\ 0 \vece \co P_{R} \, \Psi_{D} \eq \vecb 0 \\ \bar{\chi}^{\dot{\al}} \vece \ .
\]
The fact that Lorentz generators preserve chirality can also be seen from $\{ \ga^5, \ga^\mu \} = 0$ implying $[ \ga^5 , \Si^{\mu \nu} ] = 0$.

\noindent 
Finally, define the {\em Dirac conjugate} $\Psib_{D}$ and {\em charge conjugate} spinor $\Psi_{D}\,^{C}$ by
\begin{align*}
\Psib_{D} \ \ &:= \ \ (\chi^{\al} , \ \bar{\psi}_{\dot{\al}}) \eq \Psi_{D}^\dag \, \ga^{0} \\
\Psi_{D}\,^{C} \ \ &:= \ \ C \, \Psib_{D}^{T} \eq \vecb \chi_{\al} \\ \bar{\psi}^{\dot{\al}} \vece \ ,
\end{align*}
where $C$ denotes the {\em charge conjugation matrix}
    \[C \ \ := \ \ \ccb \ep_{\al \be} &0 \\ 0 &\ep^{\dot{\al} \dot{\be}} \cce \ .
\]
{\em Majorana spinors} $\Psi_{M}$ have property $\psi_{\al} = \chi_{\al}$,
    \[\Psi_{M} \eq \vecb \psi_{\al} \\ \bar{\psi}^{\dot{\al}} \vece \eq \Psi_{M}\,^{C} \ ,
\]
so a general Dirac spinor (and its charge conjugate) can be decomposed as
    \[\Psi_{D} \eq \Psi_{M1} \ + \ i\Psi_{M2} \co \Psi_{D}\,^{C} \eq \Psi_{M1} \ - \ i\Psi_{M2} \ .
\]
Note that there can be no spinors in 4 dimensions which are both Majorana and Weyl, for more information see section \ref{sec:Maj}.

\section{Gamma matrix technology}
\label{sec:gamma}

Dirac notation allows to write many $\si$ matrix identities in a more compact form, in particular by means of the $\ga^5$. It first of all follows from cyclicity of the trace and the Dirac algebra that
    \[\te{Tr} \Bigl\{ \ga^\mu \, \ga^\nu \Bigr\} \eq 4 \, \eta^{\mu \nu} \co \te{Tr} \Bigl\{ \ga^5 \Bigr\} \eq \te{Tr} \Bigl\{ \ga^5 \,\ga^\mu \, \ga^\nu \Bigr\} \eq 0 \ .
\]
Duality properties of the Lorentz generators can be expressed in unified fashion as
 \[\Si^{\mu \nu} \eq \frac{i}{2} \; \ep^{\mu \nu \rho \si} \, \ga^5 \, \Si_{\rho \si} \ .
\]
Traces with four vectorial $\ga$ matrices split into parity odd- and even parts
    \begin{align*}
    \te{Tr} \Bigl\{ \ga^\mu \, \ga^\nu \, \ga^\la \, \ga^\rho \Bigr\} \ \ &= \ \ 4 \, \bigl( \eta^{\mu \nu} \, \eta^{\la \rho} \ - \ \eta^{\mu \la} \, \eta^{\nu \rho} \ + \ \eta^{\mu \rho} \, \eta^{\nu \la} \bigr) \\
    \te{Tr} \Bigl\{ \ga^5 \, \ga^\mu \, \ga^\nu \, \ga^\la \, \ga^\rho \Bigr\} \ \ &= \ \ - \, 4 i \, \ep^{\mu \nu \la \rho} \ .
\end{align*}
By doing the chiral projection and antisymmetrizing the Lorentz indices, one might extract the identities from section \ref{sec:SigmaMatrices}.

\section{The Supersymmetry algebra in Dirac notation}
\label{sec:Dirac}

Let us conclude this appendix by rewriting the (extended) supersymmetry algebra in 4 dimensions in Dirac language. First of all define generalized indices $_r \in \{ _\al , ^{\dot \al} \}, \,  ^s \in  \{ ^\be ,_{\dot \be} \}$ in the sense that
  \[(\ga^{\mu})_{r}\,^s (\Psi_D)_s \eq \ccb 0 &\si^\mu_{\al \dot \be} \\ \bar \si^{\mu \dot \alpha \beta} &0 \cce \, \vecb \psi_\be \\ \bar \chi^{\dot \alpha} \vece \eq \vecb (\si^\mu \, \bar \chi)_\al \\ (\bar \si^\mu \, \psi)^{\dot \alpha} \vece \eq (\ga^\mu \, \Psi_D)_r  \ ,
\]
then both the momentum term and the central part of the $Q$ anticommutator can be captured within one equation:
\[ \Bigl\{ Q_r^A \ , \ Q^{sB} \Bigr\} \eq 2 \, (\ga^{\mu})_{r}\,^s \, P_\mu \, \de^{AB} \ + \ \de_{r} ^{s} \, Z^{AB} \ ,
\]
The spinorial transformation properties of the $Q$'s are summarized as
\[ \Bigl[ Q_r^A \ , \ M^{\mu \nu} \Bigr] \eq (\Si^{\mu \nu})_r \,^s \, Q_s \ .
\]


\chapter{Solutions to the exercises}

\section{Chapter 2}

\subsection*{Exercise 2.1}

We want to show that the explicit map from $SL(2,C)$ to $SO(3,1)$ is given by
\[
\Lambda^\mu \, _\nu(N) \ \ = \ \ \frac{1}{2} \; {\rm Tr} \bigl\{ \bs^\mu \, N \, \s_\nu \, N^\dagger \bigr\}
\]
Given a vector $X^\mu,$ we define the associated $SL(2,C)$ matrix associated with it by
\[
\tilde{X} \eq X_\mu \,  \s^\mu \eq \left(\begin{array}{c c}
x_0+x_3 & x_1-i x_2\\
x_1+i x_2 & x_0-x_3
\end{array}\right)
\]
Let us look at the quantity $X^\mu\s_\mu$ for which we know how it transforms under both $SO(3,1)$ and $SL(2,C):$
\begin{eqnarray*}
\Lambda^\mu{}_{\nu} \, X^\nu \, \s_\mu&=& N \,  (X^\al \, \s_\al) \, N^\dagger\\
\Rightarrow \ \ \ \Lambda^\mu{}_{\nu} \, X^\nu \, \s_\mu \, \bs_\rho&=& X^\al \, N  \, \s_\al \, N^\dagger \, \bs_\rho \\
\Rightarrow \ \ \ \Lambda^\mu{}_{\nu} \, X^\nu \, \underbrace{{\rm Tr} \bigl\{ \s_\mu\, \bs_\rho \bigr\}}_{= \ 2\eta_{\mu\rho}}&=& X^\al \,  {\rm Tr} \bigl\{ N \, \s_\al \, N^\dagger \, \bs_\rho \bigr\} \\
\Rightarrow \ \ \ 2 \, \Lambda^\mu{}_{\nu} \, X^\nu \, \eta_{\mu\rho}&=& X^\al \, {\rm Tr} \bigl\{ N \,  \s_\al \,  N^\dagger \, \bs_\rho \bigr\} \\
\Lambda^\mu{}_{\nu} \, X^\nu&=& \frac{1}{2} \; X^\al \, {\rm Tr} \bigl\{ N \, \s_\al \, N^\dagger \, \bs^\mu \bigr\}
\end{eqnarray*}

\subsection*{Exercise 2.2}

We want to show that
$$ \s^{\mu\nu} \eq \frac{i}{4} \; (\s^\mu \, \bs^\nu \ - \ \s^\nu \, \bs^\mu)$$
satisfies the Lorentz algebra. Let's rewrite $\s^{\mu\nu}$ first:
\begin{eqnarray*}
\s^{\mu\nu}&=&\frac{i}{4} \; (\s^\mu \, \bs^\nu \ - \ \s^\nu \, \bs^\mu)\\
&=&\frac{i}{4} \; (\s^\mu \, \bs^\nu \ + \ \s^\nu \, \bs^\mu \ - \ 2 \, \s^\nu \, \bs^\mu)\\
&=&\frac{i}{4} \; (2 \,  \eta^{\mu\nu} \, {\bf 1} \ - \ 2 \, \s^\nu \, \bs^\mu)\\
&=&\frac{i}{2} \; (\eta^{\mu\nu} \, {\bf 1} \ - \ \s^\nu \, \bs^\mu)
\end{eqnarray*}
$\eta$ commutes with everything, which will be useful straight away:
\begin{eqnarray*}
\bigl [\s^{\mu\nu} \, , \, \s^{\al\be} \bigr]&=&-\frac{1}{4} \; \bigl[\s^\nu \, \bs^\mu \, , \, \s^\be \, \bs^\al \bigr] \\
&=&-\frac{1}{4} \; (\s^\nu \, \bs^\mu \, \s^\be \, \bs^\al \ - \ \s^\be \, \bs^\al \, \s^\nu\, \bs^\mu)\\
&=&-\frac{1}{4} \; (\s^\nu \, \bs^\mu \, \s^\be \, \bs^\al \ - \ \s^\nu \, \bs^\be \, \s^\mu\, \bs^\al \ + \ \s^\nu \, \bs^\be \, \s^\mu \, \bs^\al \\  && \ \ \ \ \ \ \ \ \ \ \ - \ \s^\be \, \bs^\nu \, \s^\al \, \bs^\mu\ + \ \s^\be \, \bs^\nu \, \s^\al \, \bs^\mu \ - \ \s^\be \, \bs^\al \, \s^\nu \, \bs^\mu)\\
&=& -\frac{1}{4} \; (2 \, \s^\nu \, \bs^\al \, \eta^{\mu\be} \ - \ 2 \, \s^\be \, \bs^\mu \, \eta^{\al\nu} \ + \ \s^\be \, \bs^\nu \, \s^\al \, \bs^\mu \ - \ \s^\nu \, \bs^\be \, \s^\mu \, \bs^\al)\\
&=& -\frac{1}{2} \; (\s^\nu \, \bs^\al \, \eta^{\mu\be} \ - \ \s^\be \, \bs^\mu \,\eta^{\al\nu} \ + \ \eta^{\be\nu} \, \s^\al \, \bs^\mu \ - \ \eta^{\al\mu} \, \s^\nu \, \bs^\beta)
\end{eqnarray*}
Using $\frac{1}{2}\s^\nu\bs^\al=-i\s^{\nu\al}+\frac{1}{2}\eta^{\nu\alpha}$ we get
\begin{eqnarray*}
\bigl[\s^{\mu\nu} \, , \, \s^{\al\be} \bigr]&=&i(\s^{\nu\al} \, \eta^{\mu\be} \ - \ \s^{\be\mu} \, \eta^{\al\nu} \ + \ \eta^{\be\nu} \, \s^{\al\mu} \ - \ \eta^{\al\mu} \, \s^{\nu\beta}) \\  && \ \ \ \ \ \ \ \ \ \ -\frac{1}{2} \; (\eta^{\nu \al} \, \eta^{\mu\be} \ - \ \eta^{\be\mu} \, \eta^{\al\nu} \ + \ \eta^{\be\nu} \, \eta^{\al\mu} \ - \ \eta^{\al\mu} \, \eta^{\nu\beta})\\
&=&i (\s^{\nu\al} \, \eta^{\mu\be} \ + \ \s^{\mu\be} \, \eta^{\nu\al} \ - \ \eta^{\nu\be} \, \s^{\mu\al} \ - \ \eta^{\mu\al} \, \s^{\nu\beta})
\end{eqnarray*}

\subsection*{Exercise 2.3}

Recall the Lorentz algebra:
\begin{eqnarray*}
\bigl[ P^\mu \, , \, P^\nu \bigr] & = & 0\\
\bigl[ M^{\mu\nu} \, , \, P^\al \bigr] & = & i (P^\mu \, \eta^{\nu\al} \ - \ P^\nu \,\eta^{\mu\al})\\
\bigl[ M^{\mu\nu} \, , \, M^{\rho\s}\bigr] & = & i (M^{\mu\s} \, \eta^{\nu\rho} \ + \ M^{\nu\rho} \, \eta^{\mu\s} \ - \ M^{\mu\rho} \, \eta^{\nu\s} \ - \ M^{\nu\s} \, \eta^{\mu\rho})
\end{eqnarray*}
The Pauli-Ljubanski vector is defined as
$$W_\mu \eq \frac{1}{2} \; \ep_{\mu\nu\rho\s} \, P^\nu \,  M^{\rho\s} \ .$$
Using this definition, we compute
\begin{eqnarray*}
0 \ \ \overset{!}{=} \ \ \bigl[W_\mu \, , \, P_\al \bigr]&=&\frac{1}{2} \; \ep_{\mu\nu\rho\s} \, [P^\nu \,  M^{\rho\s} \ , \ P_\al] \eq \frac{1}{2} \; \eta_{\al\be} \, \ep_{\mu\nu\rho\s} \, \bigl[P^\nu  \, M^{\rho\s} \, , \, P^\be \bigr]\\
&=&\frac{1}{2} \; \eta_{\al\be} \, \ep_{\mu\nu\rho\s} \, \Bigl(P^\nu \, \bigl[M^{\rho\s} \, , \, P^\be \bigr] \ + \ \underbrace{ \bigl[P^\nu \, , \, P^\be \bigr]}_{= \ 0} \,  M^{\rho\s} \Bigr)\\
&=&\frac{i}{2} \; \eta_{\al\be} \, \ep_{\mu\nu\rho\s} \, P^\nu \, (P^\rho \, \eta^{\s\be} \ - \ P^\s \, \eta^{\rho\be})\\
&=& i \ep_{\mu [\nu\rho ]\al} \,  P^{(\nu} \, P^{\rho)}
\end{eqnarray*}
$\ep$ is totally anti-symmetric in all indices but $P^\nu P^\rho$ is symmetric under commutation, so the expression vanishes.

\noindent
Next, turn to $W_\mu$'s commutator with Lorentz rotations: 
\begin{eqnarray*}
\bigl[W_\mu \, , \, M_{\rho\s} \bigr]&=&\frac{1}{2} \; \ep_{\mu\la\chi\theta} \, \bigl[M^{\la\chi} \, P^\theta \, , \, M_{\rho\s} \bigr]\\
&=&\frac{1}{2} \;\ep_{\mu\la\chi\theta} \, \Bigl(M^{\la\chi} \, \bigl[P^\theta \, , \, M_{\rho\s} \bigr] \ + \  \bigl[M^{\la\chi} \, , \, M_{\rho\s} \bigr] \, P^\theta \Bigr)\\
&=&\frac{1}{2} \; \ep_{\mu\la\chi\theta} \, \Bigl(M^{\la\chi} \, (i\de^\theta_\rho \, P_\s \ - \ i\de^\theta_\s \, P_\rho) \ + \ i(M^\la{}_\s \, \de^\chi_\rho \ + \ M^\chi{}_{\rho} \, \de^\la_\s \ - \ M^\la{}_\rho \, \de^{\chi}_{\s} \ - \ M^\chi{}_\s \, \de^\la_\rho) \, P^\theta \Bigr)\\
&=&\frac{i}{2} \; \ep_{\mu\la\chi\theta} \, \Bigl(M^{\la\chi} \, (\de^\theta_\rho  \, P_\s \ - \ \de^\theta_\s \, P_\rho) \ + \ (2M^\la{}_\s \, \de^\chi_\rho \ - \ 2M^\la{}_\rho \, \de^{\chi}_{\s}) \, P^\theta \Bigr)\\
&=&\frac{i}{2} \; \ep_{\mu\la\chi\theta} \, (\eta_{\s\tau} \, \de^\theta_\rho \ - \ \eta_{\rho\tau} \, \de^\theta_\s) \, (M^{\la\chi} \, P^\tau \ - \ 2M^{\la\tau} \, P^\chi)\\
&=&\frac{3i}{2} \; \ep_{\mu\la\chi\theta} \, (\eta_{\s\tau} \, \de^\theta_\rho \ - \ \eta_{\rho\tau} \, \de^\theta_\s) \, (M^{[\la\chi} \, P^{\tau]})\\
&\overset{(*)}{=}&\frac{i}{2} \; \ep_{\mu\la\chi\theta} \, (\eta_{\s\tau} \, \de^\theta_\rho \ - \ \eta_{\rho\tau}\, \de^\theta_\s) \, \ep^{\la\chi\tau\g} \, W_\g\\
&=&-i (\de_\mu^\tau \, \de_\theta^\g \ - \ \de_\mu^\g \, \de_\theta^\tau) \, (\eta_{\s\tau}\, \de^\theta_\rho \ - \ \eta_{\rho\tau} \, \de^\theta_\s) \, W_\g\\
&=&i \eta_{\mu\rho} \, W_\s \ - \ i \eta_{\mu\s}\, W_\rho
\end{eqnarray*}
In the process (*) we used the following identity $\ep^{\la\chi\tau\g}W_\g=3 M^{[\la\chi}P^{\tau]},$ which can be shown as follows:
\begin{eqnarray*}
\ep^{\la\chi\tau\g} \, W_\g&=&\frac{1}{2} \; \ep^{\la\chi\tau\g} \, \ep_{\g\al\be\de} \, M^{\al\be} \, P^\de\\
&=&3 \, \de_{[\al}^\la \, \de_\be^\chi \, \de_{\de]}^\tau \, M^{\al\be} \, P^\de\\
&=&3 M^{[\la\chi} \, P^{\tau]}\ .
\end{eqnarray*}
With this help we can easily show the commutation relation $[W_\mu,W_\nu]:$
\begin{eqnarray*}
\bigl[W_\mu \, , \, W_\nu \bigr]&=&\frac{1}{2} \; \ep_{\nu\rho\s\tau} \, \bigl[W_\mu \, , \, M^{\rho\s}\, P^\tau \bigr]\\
&=&\frac{1}{2} \; \ep_{\nu\rho\s\tau} \bigl[W_\mu \, , \, M^{\rho\s} \bigr] \, P^\tau\\
&=&\frac{i}{2} \; \ep_{\nu\rho\s\tau} \, (\de^\rho_\mu  \, W^\s \ - \ \de^\s_\mu \, W^\rho) \, P^\tau\\
&=&-i \ep_{\mu\nu\rho\s} \, W^\rho \, P^\s
\end{eqnarray*}
Finally, we are interested in the commutator of $W_\mu$ with the supersymmetry generators:
\begin{eqnarray*}
\nonumber -i(\s^{\mu\nu})_{\al} \, \! ^\be \,  Q_\be \,  P_\nu \ \ \overset{!}{=} \ \ \bigl[W^\mu \, , \, Q_\al \bigr]&=&\frac{1}{2} \; \eta^{\mu\nu} \, \ep_{\nu\rho\s\tau} \, P^\rho \, \bigl[M^{\s\tau} \, , \, Q_\al \bigr]\\
&=&- \, \frac{1}{2} \; \ep^{\mu}{}_{\rho\s\tau} \, P^\rho \, (\s^{\s\tau})_\al{}^{\be} \,Q_\be \\
&=&- i P^\rho \, (\sigma^\mu{} _\rho)_\al{}^\be \, Q_\be 
 \label{6ref}
\end{eqnarray*}
The last step follows from self duality of the $\si^{\mu \nu}$.

\subsection*{Exercise 2.4}

We want to show that $W^\mu W_\mu$ is a Casimir of the Poincare algebra but not of the Super-Poincare.
\begin{itemize}
\item Clearly $[W^\mu W_\mu,P^\nu]=0$ since $[W^\mu,P^\nu]=0.$
\item also Lorentz rotations commute with $W^\mu W_\mu$ since
\begin{eqnarray*}
0 \ \ \overset{!}{=} \ \ \bigl[W^\mu \,  W_\mu \, , \, M_{\rho\s} \bigr]&=& \bigl[W_\mu \, , \, M_{\rho\s} \bigr] \, W^\mu \ + \ W^\mu \, \bigl[W_\mu \, , \, M_{\rho\s} \bigr]\\
&=& i \eta_{\mu\rho} \,  W_\s \,  W^\mu \ - \ i\eta_{\mu\s} \, W_\rho \,  W^\mu \ + \ W^\mu \,  i \eta_{\mu\rho} \, W_\s \ - \ W^\mu \,  i \eta_{\s\mu} \, W_{\rho}\\
&=&i W_\s \, W_\rho \ - \ i W_\rho \,  W_\s \ + \  i W_\rho \,  W_\s \ - \ i W_\s \, W_\rho\\
&=&0
\end{eqnarray*}
\end{itemize}
$W^\mu W_\mu$ is Casimir of the Poincare algebra. Regarding the Super-Poincare algebra, for example consider:
\begin{eqnarray*}
\bigl[Q_\al \, , \, W^\mu \, W_\mu \bigr]&=& i(\s^{\mu\nu})_\al{}^\be \, Q_\be \, P_\nu \, W_\mu \ + \ W_\mu \, i (\s^{\mu\nu})_{\al} {}^{\be} \, Q_\be \, P_\nu\\
&=&2 i \, (\s^{\mu\nu})_\al{}^\be \, P_\nu \, Q_\be \, W_\mu 
\end{eqnarray*}
This clearly does not vanish and hence $W^\mu W_\mu$ is not a Casimir of the Super-Poincare algebra.

\section{Chapter 3}

\subsection*{Exercise 3.1}

The theta expansion of the most general superfield is 
\begin{align*}
S(x,\theta,\bt) \eq \vp(x) \ + \ \theta &\psi(x) \ + \ \bt \bc(x) \ + \ (\theta\theta) \, M(x) \ +\ (\bt\bt) \, N(x) \ + \ (\theta \s^\mu \bt) \, V_\mu(x) \\
&+ \ (\theta\theta) \, (\bt\bl)(x) \ + \ (\bt\bt) \, (\theta \rho)(x) \ + \ ( \theta\theta) \, (\bt\bt) \, D(x)
\end{align*}
We now act with a supersymmetry transformation
\begin{eqnarray*}\de S&=&i(\ep \, \mQ \ + \ \bep \,  \bmQ) \, S (x,\theta,\bt)\eq i(\ep^\al \, \mQ_\al \ + \ \bmQ_{\da} \, \bep^{\da}) \, S (x,\theta,\bt)\\
&=&\left[\ep^\al \; \left(\frac{\pa}{\pa\theta^{\al}} \ - \ i(\s^\mu)_{\al\db} \, \bt^{\db} \,\pa_\mu\right) \ + \ \left(-\frac{\pa}{\pa\bt^{\da}} \ + \ i\theta^\g \, (\s^\mu)_{\g\da} \, \pa_\mu\right) \, \bep^{\da}\right] \, S(x,\theta,\bt)
\end{eqnarray*}
Let's look at each component of the superfield individually:
\begin{eqnarray*}
&& \left[\ep^\al \; \left(\frac{\pa}{\pa\theta^{\al}} \ - \ i(\s^\mu)_{\al\db} \, \bt^{\db} \,\pa_\mu\right) \ + \ \left(-\frac{\pa}{\pa\bt^{\da}} \ + \ i\theta^\g \, (\s^\mu)_{\g\da} \, \pa_\mu\right) \, \bep^{\da}\right] \, S(x,\theta,\bt) \\ &=&-i(\ep\s^\mu\bt) \, \pa_\mu \vp \ + \ i (\theta \s^\mu\be) \, \pa_\mu\vp(x)\\
&&+ \ \ep^\la \, \psi_\la(x) \ - \ i (\ep\s^\mu\bt) \, \theta^\la \, \pa_\mu\psi_\la(x) \ + \ i (\theta \s^\mu\bep) \, \theta^\la \, \pa_\mu\psi_\la(x)\\
&&- \ i (\ep \s^\mu\bt) \, \bt_{\dl} \, \pa_\mu\bc^{\dl}(x) \ + \ \bep\bc(x) \ + \ i (\theta\s^\mu\bep) \, \bt_{\dl} \, \pa_\mu \bc^{\dl}(x)\\
&&+  \ 2 \, (\ep\theta) \, M(x) \ - \ i (\ep \s^\mu \bt) \, (\theta\theta) \, \pa_\mu M(x)\\
&&+  \ 2 \, (\bep\bt) \, N(x) \ + \ i (\theta \s^\mu \bep) \, (\bt\bt) \, \pa_\mu N(x)\\
&&+ \ (\ep\s^\mu\bt) \, V_\mu(x) \ - \ i (\ep\s^\mu\bt) \, (\theta\s^\la\bt) \, \pa_\mu V_\la(x) \ + \ (\theta\s^\mu\bep) \, V_\mu(x) \ + \ i (\theta\s^\mu\bep) \, (\theta\s^\la\bt) \,\pa_\mu V_\la(x)\\
&&+ \ 2 \, (\ep\theta) \, \bt\bl(x) \ - \ i(\ep\s^\mu\bt) \, (\theta\theta) \, \bt\pa_\mu \bl(x) \ + \ (\bep \bl) \, (\theta\theta)\\
&&+ \ (\bt\bt) \, \ep\rho(x) \ + \ 2 \, (\bt\bep) \, \theta\rho(x) \ - \ i(\theta\s^\mu\bep) \, (\bt\bt) \, \theta\pa_\mu\rho(x) \\
&&+ \ 2 \, (\ep\theta) \, (\bt\bt) \, D(x) \ + \ 2 \, (\bt\bep) \, (\theta\theta) \, D(x) \ .
\end{eqnarray*}
We now can collect all terms present at orders in $\theta$ and $\bt.$\\[0.2cm]
(i) Terms of $O(\theta^0,\bt^0):$
$$\ep \psi(x) \ + \ \bep \bc(x)$$
Hence $\de \vp=\ep \psi+\bep \bc.$\\[0.2cm]
(ii) Terms of $O(\theta^1,\bt^0)$ and $O(\theta^0,\bt^1):$
\begin{eqnarray*}&&2 \, (\ep\theta) \, M(x) \ + \ 2 \, (\bep\bt) \, N(x) \ + \ (\ep\s^\mu\bt) \, V_\mu(x) \ + \ (\theta\s^\mu\bep) \, V_\mu(x) \ - \ i(\ep\s^\mu\bt) \, \pa_\mu \vp\ + \ i (\theta \s^\mu\bep) \, \pa_\mu\vp(x)\\
&& \ \ \ \ \ \ \ \ \ \ \ \ \ =  \ \ \theta \, \bigl(2 \, M(x) \ + \ (\s^\mu\bep) \, (V_\mu \, + \, i\pa_\mu \vp) \, \bigr) \ + \ \bigl(2 \, \bep \, N(x) \ + \ (\ep\s^\mu) \, (V_\mu \, - \, i\pa_\mu\vp) \, \bigr) \, \bt
\end{eqnarray*}
Hence $\de\psi=2 M(x)+(\s^\mu\bep)(V_\mu+i\pa_\mu \vp)$ and $\de\bc=2\bep N(x)+\ep\s^\mu(V_\mu-i\pa_\mu\vp).$\\[0.2cm]
(iii) Terms of $O(\theta^2,\bt^0)$ and $O(\theta^0,\bt^2):$
\begin{eqnarray*}&&
(\bt\bt) \, \ep\rho(x) \ + \ (\bep\bl) \, (\theta\theta) \ + \ i (\theta \s^\mu\bep) \, \theta^\la \, \pa_\mu\psi_\la(x) \ - \ i (\ep \s^\mu\bt) \, \bt_{\dl} \, \pa_\mu\bc^{\dl}(x)\\
&=&(\theta\theta) \, (\bep\bl) \ - \ i \theta^\al \, \theta^\la \, (\s^\mu)_{\al\da} \, \bep^{\da} \, \pa_\mu\psi_\la(x) \ + \ (\bt\bt) \, \ep\rho(x) \ + \ i \ep^\al \, (\s^\mu)_{\al\da} \, \bt^{\da} \, \bt^{\dl} \, \pa_\mu\bc_{\dl}(x)\\
&=&(\theta\theta) \, (\bep\bl) \ + \ \frac{i}{2} \; \ep^{\al\la} \, (\theta\theta) \, (\s^\mu)_{\al\da} \, \bep^{\da} \, \pa_\mu\psi_\la(x) \ + \ (\bt\bt) \, \ep\rho(x) \ + \ \frac{i}{2} \; \ep^\al \,  (\s^\mu)_{\al\da} \, \ep^{\da\dl} \, (\bt\bt)  \, \pa_\mu\bc_{\dl}(x)\\
&=&(\theta\theta) \, (\bep\bl) \ + \ \frac{i}{2} \; (\theta\theta) \, (\s^\mu)_{\al\da} \, \bep^{\da} \, \pa_\mu\psi^\al(x) \ + \ (\bt\bt) \, \ep\rho(x) \ + \ \frac{i}{2} \; (\bt\bt) \, \ep^\al \,(\s^\mu)_{\al\da} \, \pa_\mu\bc^{\da}(x)
\end{eqnarray*}
Hence $\de M=(\bep \bl)+\frac{i}{2}\pa_\mu\psi(x) \s^\mu\bep$ and $\de N=\ep\rho(x)+\frac{i}{2}  \ep \s^\mu\pa_\mu\bc(x).$\\[0.2cm]
(iv) Terms of $O(\theta^1,\bt^1):$
\begin{eqnarray*}
&& 2 \, (\bt\bep) \, \theta\rho(x) \ + \ 2 \, (\ep\theta) \, \bt\bl(x) \ - \ i (\ep\s^\mu\bt)\, \theta^\la \, \pa_\mu\psi_\la(x) \ + \ i (\theta\s^\mu\bep) \, \bt_{\dl} \, \pa_\mu \bc^{\dl}(x)\\
&\overset{\text{Fierz id.}}{=}&- \, (\theta\s^\mu\bt) \, (\bl\bs_\mu\ep \ + \ \bep\bs_\mu\rho) \ - \ i (\ep\s^\mu)_{\da} \, \bt^{\da} \, \theta^\la \, \pa_\mu\psi_\la(x) \ + \ i \theta^\al \, (\s^\mu \bep)_\al \, \bt_{\dl} \, \pa_\mu \bc^{\dl}(x)\\
&\overset{\text{Fierz id.}}{=}&- \, (\theta\s^\mu\bt) \, (\bl\bs_\mu\ep \ + \ \bep\bs_\mu\rho) \ + \ \frac{i}{2} \; (\theta\s^\nu\bt) \, \bigl(\ep\s^\mu \bs_\nu \pa_\mu\psi (x)\bigr)\ - \ \frac{i}{2} \; (\theta\s^\nu\bt) \bigl(\pa_\mu \bc(x)\bs_\nu\s^\mu \bep \bigr)\\
&\overset{*}{=}&(\theta\s^\mu\bt) \, (\ep\s_\mu\bl \ + \ \rho\s_\mu\bep) \ + \ \frac{i}{2} \; (\theta\s^\nu\bt) \, \bigl(\pa^\mu\psi (x)\s_\nu \bs_\mu\ep \bigr) \ - \ \frac{i}{2} \;(\theta\s^\nu\bt) \, \bigl( \bep\bs_\mu\s_\nu\pa^\mu \bc(x) \bigr)\\
&=&(\theta\s^\mu\bt) \, \left(\ep\s_\mu\bl \ + \ \rho\s_\mu\bep \ + \ \frac{i}{2} \; \bigl(\pa^\nu\psi (x)\s_\mu \bs_\nu\ep \ - \ \bep\bs_\nu\s_\mu\pa^\nu \bc(x) \bigr)\right)
\, ,
\end{eqnarray*}
where in $(*)$ we rewrote the expressions according to the bispinor symmetry properties of appendix \ref{sec:BispinorsInvolvingSigmaMatrices}. Hence $\de V_\mu=\ep\s_\mu\bl+\rho\s_\mu\bep+\frac{i}{2}(\pa^\nu\psi (x)\s_\mu \bs_\nu\ep- \bep\bs_\nu\s_\mu\pa^\nu \bc(x)).$\\[0.2cm]
(v) Terms of $O(\theta^2,\bt^1)$ and $O(\theta^2,\bt^1):$
\begin{eqnarray*}
&&- \, i (\ep \s^\mu \bt) \, (\theta\theta) \, \pa_\mu M(x) \ + \ i (\theta \s^\mu \bep) \, (\bt\bt) \, \pa_\mu N(x)\\ &&
-\ i (\ep\s^\mu\bt) \, (\theta\s^\la\bt) \, \pa_\mu V_\la(x) \ + \ i (\theta\s^\mu\bep) \,(\theta\s^\la\bt) \, \pa_\mu V_\la(x)\\ &&
+ \ 2 \, (\ep\theta) \, (\bt\bt) \, D(x) \ + \ 2 \,  (\bt\bep) \, (\theta\theta) \, D(x)\\
&=&(\theta\theta) \, \bigl(2 \,  (\bt\bep) \, D(x) \ - \ i (\ep \s^\mu \bt) \,\pa_\mu M(x)\bigr) \ + \ i (\theta\s^\mu\bep) \, (\theta\s^\la\bt) \, \pa_\mu V_\la(x)\\
&&+ \ (\bt\bt) \, \bigl(2 \, (\ep\theta) \, D(x) \ + \ i (\theta \s^\mu \bep) \, \pa_\mu N(x)\bigr) \ - \ i (\ep\s^\mu\bt) \, (\theta\s^\la\bt) \, \pa_\mu V_\la(x)\\
&=&(\theta\theta) \, \bigl(2 \, (\bt\bep) \, D(x) \ + \ i (\bt \bs^\mu \ep) \, \pa_\mu M(x)\bigr) \ + \ i  \bigl(\theta^\al \, (\s^\mu)_{\al\da} \, \bep^{\da} \bigr) \, \bigl(\theta^\be\, (\s^\la)_{\be\db} \, \bt^{\db} \bigr) \, \pa_\mu V_\la(x)\\
&&+ \ (\bt\bt) \, \bigl(2 \, (\theta\ep) \, D(x) \ + \ i (\theta \s^\mu \bep) \, \pa_\mu N(x)\bigr) \ - \ i  \bigl(\ep^\al \, (\s^\mu)_{\al\da} \, \bt^{\da} \bigr) \, \bigl(\theta^\be \, (\s^\la)_{\be\db} \, \bt^{\db} \bigr) \,\pa_\mu V_\la(x)\\
&=&(\theta\theta) \, \bigl(2 \, (\bt\bep) \, D(x) \ + \ i (\bt \bs^\mu \ep) \, \pa_\mu M(x)\bigr) \ + \ \frac{i}{2} \; (\theta\theta) \, \bigl(\ep^{\al\be} \, (\s^\mu)_{\al\da} \, \bep^{\da}\bigr) \, \bigl((\s^\la)_{\be\db} \, \bt^{\db} \bigr) \, \pa_\mu V_\la(x)\\
&&+ \ (\bt\bt) \, \bigl(2 \, (\theta\ep) \, D(x) \ + \ i (\theta \s^\mu \bep) \, \pa_\mu N(x)\bigr) \ + \ \frac{i}{2} \; (\bt\bt) \, \ep^{\da\db} \, \bigl(\ep^\al \, (\s^\mu)_{\al\da} \bigr) \, \bigl(\theta^\be \, (\s^\la)_{\be\db} \bigr) \, \pa_\mu V_\la(x)\\
&=&(\theta\theta) \, \bigl(2 \, (\bt\bep) \, D(x) \ + \ i (\bt \bs^\mu \ep) \,\pa_\mu M(x)\bigr) \ + \ \frac{i}{2} \; (\theta\theta) \, \bigl(\ep^{\al\be} \, (\s^\mu)_{\al\da} \, \bep^{\da} \bigr) \, \bigl( (\s^\la)_{\be\db} \, \ep^{\db\dg} \, \bt_{\dg} \bigr) \, \pa_\mu V_\la(x)\\
&&+ \ (\bt\bt) \, \bigl(2 \, (\theta\ep) \, D(x) \ + \ i (\theta \s^\mu \bep) \, \pa_\mu N(x)\bigr) \ + \ \frac{i}{2} \; (\bt\bt) \, \ep^{\da\db} \, \bigl(\ep_\g \, \ep^{\al\g} \, (\s^\mu)_{\al\da} \bigr) \, \bigl(\theta^\be \, (\s^\la)_{\be\db} \bigr) \, \pa_\mu V_\la(x)\\
&=&(\theta\theta) \, \bt \, \bigl(2 \, \bep \, D(x) \ + \ i (\bs^\mu \ep) \, \pa_\mu M(x)\bigr) \ + \ \frac{i}{2} \; (\theta\theta) \, (\s^\mu)_{\al\da} \, \bep^{\da} \, (\bs^\la)^{\dg\al} \, \bt_{\dg} \, \pa_\mu V_\la(x)\\
&&+ \ (\bt\bt) \, \theta \, \bigl(2 \,  \ep\, D(x) \ + \ i (\s^\mu \bep) \, \pa_\mu N(x)\bigr) \ + \ \frac{i}{2} \; (\bt\bt) \,  \ep_\g \, (\bs^\mu)^{\db\g} \, \theta^\be \, (\s^\la)_{\be\db} \,\pa_\mu V_\la(x)\\
&=&(\theta\theta) \, \bt \, \left(2 \, \bep \, D(x) \ + \ i (\bs^\mu \ep) \, \pa_\mu M(x) \ +\ \frac{i}{2} \; (\bs^\la \s^\mu\bep ) \, \pa_\mu V_\la(x)\right)\\
&&+\  (\bt\bt) \, \theta \, \left(2 \, \ep \, D(x) \ + \ i (\s^\mu \bep) \, \pa_\mu N(x) \ - \ \frac{i}{2} \; (\s^\la\bs^\mu\ep) \, \pa_\mu V_\la(x)\right)
\end{eqnarray*}
Hence we find $\de\rho=2 (\bep)D(x)+i (\bs^\mu \ep)\pa_\mu M(x)+\frac{i}{2} (\bs^\la)(\s^\mu)\bep\pa_\mu V_\la(x)$\\ and $\de\bl=2 (\ep_\al)D(x)+i (\s^\mu \bep)\pa_\mu N(x)-\frac{i}{2} \s^\la\bs^\mu\ep\pa_\mu V_\la(x).$\\[0.2cm]
(vi) Terms of $O(\theta^2,\bt^2):$
\begin{eqnarray*}
&&- \, i(\ep\s^\mu\bt) \, (\theta\theta) \, \bigl(\bt\pa_\mu \bl(x) \bigr) \ - \ i(\theta\s^\mu\bep) \, (\bt\bt)\, \bigl(\theta\pa_\mu\rho(x) \bigr)\\
&=&i(\theta\theta) \, \ep^\al \, (\s^\mu)_{\al\da} \, \bt^{\da} \, \bt^{\db} \, \pa_\mu \bl_{\db}(x) \ - \ i(\bt\bt) \, \theta^\al \, (\s^\mu)_{\al\da} \, \bep^{\da} \, \theta^\be \, \pa_\mu\rho_\be(x) \\
&=&\frac{i}{2} \; (\theta\theta) \, (\bt\bt) \ \ep^\al \, (\s^\mu)_{\al\da} \, \ep^{\da\db} \,\pa_\mu \bl_{\db}(x) \ - \ \frac{i}{2} \; (\bt\bt) \, (\theta\theta) \, \ep^{\al\be} \, (\s^\mu)_{\al\da} \, \bep^{\da} \, \pa_\mu\rho_\be(x)\\
&=&\frac{i}{2} \; (\theta\theta) \, (\bt\bt) \, \bigl(\ep \s^\mu \pa_\mu \bl(x) \ + \ \pa_\mu\rho(x) \s^\mu \bep\bigr)\\
\end{eqnarray*}
Hence we find $\de D=\frac{i}{2}\left(\ep(\s^\mu)\pa_\mu \bl(x)+\pa_\mu\rho(x)(\s^\mu)\bep\right).$

\subsection*{Exercise 3.2}

Start from the component expansion of the generic chiral superfield
$$\Phi(x,\theta,\bt) \eq \vp \ + \ \sqrt{2} \, \theta\psi \ + \ (\theta\theta) \, F \ + \ i(\theta\s^\mu\bt) \, \pa_\mu\vp \ - \ \frac{1}{4} \; (\theta\theta) \, (\bt\bt) \, \pa_\mu\pa^\mu \vp \ - \ \frac{i}{\sqrt{2}} \; (\theta\theta) \, \pa_\mu \psi \s^\mu\bt$$
\begin{eqnarray*}
0&\overset{!}{=}&-\bmD_{\da}\Phi \eq \bpa_{\da}\Phi \ + \ i\theta^\be \, (\s^\nu)_{\be\da} \, \pa_\nu \Phi\\
&=&\bpa_{\da} \, \left(i(\theta\s^\mu\bt) \, \pa_\mu\vp \ - \ \frac{1}{4} \; (\theta\theta) \, (\bt\bt) \, \pa_\mu\pa^\mu \vp \ - \ \frac{i}{\sqrt{2}} \; (\theta\theta) \, (\pa_\mu \psi \s^\mu\bt ) \right)\\
&&+ \ i\theta^\be \, (\s^\nu)_{\be\da} \, \left(\pa_\nu\vp \ + \ \sqrt{2} \, (\theta\pa_\nu\psi) \ + \ i(\theta\s^\mu\bt) \, \pa_\nu\pa_\mu \vp\right)\\
&=&\left(-i\theta^\al \, (\s^\nu)_{\al\da} \, \pa_\mu\vp \ + \ \frac{1}{2} \; (\theta\theta) \,\bt_{\da} \, \pa_\mu\pa^\mu \vp \ + \ \frac{i}{\sqrt{2}} \; (\theta\theta) \, \pa_\mu \psi^\al \, (\s^\mu)_{\al\da}\right)\\
&&+ \ i\theta^\be \, (\s^\nu)_{\be\da} \, \left(\pa_\nu\vp \ + \ \sqrt{2} \, \theta \, \pa_\nu\psi \ + \ i(\theta\s^\mu\bt) \, \pa_\nu\pa_\mu\vp\right)\\
&=&\frac{1}{2} \; (\theta\theta) \, \bt_{\da} \, \pa_\mu\pa^\mu \vp \ + \ \frac{i}{\sqrt{2}} \; (\theta\theta) \, \pa_\mu \psi^\al \, (\s^\mu)_{\al\da} \ + \ \left(\sqrt{2}i\theta^\be \, (\s^\nu)_{\be\da} \, \theta^\al \, \pa_\nu\psi_\al \ - \ \theta^\be \, (\s^\nu)_{\be\da} \, (\theta\s^\mu\bt) \, \pa_\nu\pa_\mu\vp\right)\\
&=&\frac{1}{2} \; (\theta\theta) \, \bt_{\da} \, \pa_\mu\pa^\mu \vp \ + \ \frac{i}{\sqrt{2}} \; (\theta\theta) \, \pa_\mu \psi^\al \, (\s^\mu)_{\al\da} \ - \ \left( \frac{i\ep^{\be\al} }{\sqrt{2}} \;  (\theta\theta) \, (\s^\nu)_{\be\da} \, \pa_\nu\psi_\al \ + \ \theta^\be \, (\s^\nu)_{\be\da} \, (\theta\s^\mu\bt) \, \pa_\nu\pa_\mu\vp\right)\\
&=&\frac{1}{2} \; (\theta\theta) \, \bt_{\da} \, \pa_\mu\pa^\mu \vp \ - \ \bigl(\theta^\be\,(\s^\nu)_{\be\da} \, (\theta\s^\mu\bt) \, \pa_\nu\pa_\mu\vp\bigr)\\
&=&0\ ,
\end{eqnarray*}
where in the last step we used the following identity
\begin{eqnarray*}
\theta^\be \, (\s^\nu)_{\be\da} \, (\theta\s^\mu\bt) \, \pa_\mu \pa_\nu \varphi &=&\theta^\be \, (\s^\nu)_{\be\da} \, \theta^\g \, (\s^\mu)_{\g\dg} \,\bt^{\dg} \, \pa_{(\mu} \pa_{\nu)} \varphi \\
&=&- \, \frac{\theta\theta}{2} \; \ep^{\be \ga} \, (\si^\nu)_{\be \dot \al} \, (\si^\mu)_{\ga \dot \ga}  \, \bar \theta^{\dot \ga} \, \pa_{(\mu} \pa_{\nu)} \varphi\\
&=&\frac{\theta\theta}{2} \; \ep_{\dot \al \dot \tau} \, (\bar \si^{(\nu} \si^{\mu )})^{\dot \tau}{}_{\dot \ga} \, \bar \theta^{\dot \ga} \, \pa_{(\mu} \pa_{\nu)} \varphi \\
&=&\frac{\theta\theta}{2} \; \ep_{\dot \al \dot \tau} \, \eta^{\mu \nu} \, \de_{\dot \ga}^{\dot \tau} \, \bar \theta^{\dot \ga} \, \pa_{(\mu} \pa_{\nu)} \varphi \\
&=&\frac{\theta\theta}{2} \; \bar \theta_{\dot \al} \, \pa^2 \varphi \ .
\end{eqnarray*}

\subsection*{Exercise 3.3}

\begin{itemize}
\item chirality

\noindent
One can write the right handed supercovariant derivative $\bar{{\cal D}}_{\dot{\al}} W_{\al}$ as $\ep^{\dot{\be}\dot{\ga}} \bar{{\cal D}}_{\dot{\al}} \bar{{\cal D}}_{\dot{\be}} \bar{{\cal D}}_{\dot{\ga}}$ acting on some superfield. Since the $\bar{{\cal D}}$ anticommute, the expression $\bar{{\cal D}}_{\dot{\al}} \bar{{\cal D}}_{\dot{\be}} \bar{{\cal D}}_{\dot{\ga}}$ can be regarded as totally antisymmetrized. But the $\dot{\al},\dot{\be},\dot{\ga}$ indices can only take two distict values, so any totally antisymmetric rank three tensor vanishes, $T_{[\dot{\al} \dot{\be} \dot{\ga}]} = 0$. In short:
\[\bar{{\cal D}}_{\dot{\al}} W_{\al} \eq -\frac{1}{4} \; \ep^{\dot{\be}\dot{\ga}} \, \bar{{\cal D}}_{[\dot{\al}} \bar{{\cal D}}_{\dot{\be}} \bar{{\cal D}}_{\dot{\ga}]} \, \bigl(  {\cal D}_{\al}V \bigr) \eq 0 \]
\item gauge invariance

\noindent
With $\La^\dag$ being antichiral, it is quite obvious that only the $\La$ contribution of the $V$ tranformation law can contribute. But the anticommutator $\{ {\cal D}_{\al} , \bar{{\cal D}}_{\dot{\be}} \} = - 2i  (\si^{\mu})_{\al \dot{\be}} \pa_{\mu}$ implies that under $V \mapsto V - \frac{i}{2} (\La - \La^\dag)$
\[ \de W_{\al} \eq  \frac{i}{8} \; \ep^{ \dot{\be} \dot{\ga} } \, \bar{{\cal D}}_{\dot{\be}} \, \Bigl\{ \bar{{\cal D}}_{\dot{\ga}} \ , \ {\cal D}_{\al} \Bigr\} \, \La \eq - \, \frac{1}{4} \; (\si^{\mu})_{\al \dot{\ga}} \, \bar{ {\cal D} }^{\dot{\ga}} \, ( \pa_{\mu} \La ) \eq 0  \]
since also $\pa_\mu \La$ is a chiral superfield.
\end{itemize}

\subsection*{Exercise 3.4}

We want to find an expression for $W_\al$ in components. For this purpose, rewrite the vector field in terms of the shifted spacetime variable $y^\mu = x^\mu + i \theta \si^\mu \bar \theta$:
\[
V_{\te{WZ}}(x,\theta,\bar \theta ) \eq (\theta \, \si^\mu \, \bar \theta) \, V_\mu(y) \ + \ (\theta \theta) \, ( \bar \theta \bar \la)(y) \ + \ (\bar \theta \bar \theta) \, (  \theta \la) (y) \ + \ \frac{1}{2} \,  (\theta \theta) \, ( \bar \theta \bar \la)  \, \bigl[ \, D(y) \ - \ i \, \pa_\mu V^\mu(y) \, \bigr]
\]
Next act with $-\frac{1}{4} {\cal D}_\al = -\frac{1}{4} \pa_\al - \frac{i}{2} (\si^\rho \bar \theta)_\al \pa_\rho$, leaving the $y$ argument implicit:
\begin{align*}
-\frac{1}{4} \; {\cal D}_\al \, V_{\te{WZ}} \ \ &= \ \ \frac{1}{4} \, \Bigl\{ \,- \, \si^\mu_{\al \dot \be} \, \bar \theta^{\dot \be} \, V_\mu \ - \ 2 \, \theta_\al\, (\bar \theta \bar \la) \ - \ \la_\al \, (\bar \theta \bar \theta) \ - \, \theta_\al \, (\bar \theta \bar \theta)  \bigl[ \, D \ - \ i \, \pa_\mu V^\mu \, \bigr] \Bigr. \\
& \ \ \ \ \ \ \Bigl. - \ 2i \, \si^\rho_{\al \dot \be} \, \bar \theta^{\dot \be} \, \theta^\ga \, \si^{\mu}_{\ga \dot \de} \, \bar \theta^{\dot \de} \, \pa_\rho V_\mu \ + \ 2i \, (\theta \theta) \, \si^\rho_{\al \dot \be} \, \bar \theta^{\dot \be} \, \bar \theta^{\dot \de} \, \pa_\rho \bar \la_{\dot \de} \, \Bigr\}
\end{align*}
In the $y$ variable, the antichiral covariant derivative simply acts as $\bar \pa_{\dot \al}$ such that the first two terms drop out and the rest gives
\begin{align*}
-\frac{1}{4} \; (\bar {\cal D} \bar {\cal D}) \, {\cal D}_\al \, V_{\te{WZ}} \ \ &= \ \ - \, \underbrace{\frac{1}{4} \, \bar \pa_{\dot \al} \, \bar \pa^{\dot \al} \, (\bar \theta \bar \theta)}_{= \ 1} \, \Bigl\{ \, \la_\al \ + \ \theta_\al \, D \ - \ i \, \eta^{\mu \rho} \, \theta_\al \, \pa_\rho V_\mu \Bigr. \\
& \ \ \ \ \ \ \ \ \ \ \ \ \ \ \ \ \ \ \ \Bigl. - \ i \, (\theta \theta) \, \si^\rho_{\al \dot \be} \, \ep^{\dot \be \dot \de} \, \pa_\rho \bar \la_{\dot \de} \ - \ i \, (\theta \, \si^\mu \, \bar \si^\rho)^\ga \, \ep_{\ga \al} \, \pa_\rho V_\mu \, \Bigr\} \\
&= \ \ \la_\al \ + \ \theta_\al \, D \ + \ (\si^{\mu \nu})_\al \,^\be \, \ep_{\be \ga} \, \theta^\ga \, F_{\mu \nu} \ - \ i \, (\theta \theta) \, \si^\rho_{\al \dot \be} \, \ep^{\dot \be \dot \de} \, \pa_\rho \bar \la_{\dot \de} \\
&= \ \ W_\al
\end{align*}
Due to $\eta^{\mu \rho} = \frac{1}{2} ( \si^\mu \bar \si^\rho + \si^\rho \bar \si^\mu)$, the two $\pa V$ terms nicely combine to the antisymmetric expression $(\si^\mu \bar \si^\rho - \si^\rho \bar \si^\mu) \pa_\rho V_\mu = 2i \si^{\mu \rho} F_{\mu \rho}$

\subsection*{Exercise 3.5}

Plug the transformed vector field $e^{\pm 2qV'}$ into $W_\al'$
\begin{align*}
W'_{\al} \ \ &= \ \ -\frac{1}{8 \,q} \; (\bar{{\cal D}}\bar{{\cal D}}) \, \bigl(e^{iq\La} \, \exp(-2qV) \, e^{-iq\La^\dag} \,  {\cal D}_{\al} \, e^{iq\La^\dag} \, \exp(2qV) \, e^{-iq\La} \bigr) \\
&= \ \ - \; \frac{1}{8 \,q} \; e^{iq\La} \, (\bar{{\cal D}}\bar{{\cal D}}) \, \Bigl( \exp(-2qV) \, \underbrace{e^{-iq\La^\dag} \, e^{iq\La^\dag}}_{=\ 1} \, {\cal D}_{\al}  \, \bigl( \exp(2qV) \,  e^{-iq\La} \bigr) \Bigr) \\
&= \ \ e^{iq\La} \, \left( - \frac{1}{8 \,q} \;  (\bar{{\cal D}}\bar{{\cal D}}) \, \exp(-2qV) \, {\cal D}_{\al}  \,  \exp(2qV) \right) \, e^{-iq\La}
- \; \frac{1}{8 \,q} \; e^{iq\La} \, \underbrace{(\bar{{\cal D}}\bar{{\cal D}}) \,  {\cal D}_{\al}  \,  e^{-iq\La}}_{= \ 0 } \\
&= \ \ e^{iq\La} \, W_{\al} \, e^{-iq\La}
\end{align*}
Firstly, we have used  $\bar{{\cal D}}_{\dot{\al}} e^{iq\La} = e^{iq\La} \bar{{\cal D}}_{\dot{\al}}$ and secondly, ${\cal D}_{\al} e^{iq\La^\dag} = e^{iq\La^\dag} {\cal D}_{\al}$ for the chiral superfield, and $(\bar{{\cal D}}\bar{{\cal D}})  {\cal D}_{\al}  e^{iq\La}=0$ can be checked by means of $\bar{{\cal D}}_{\dot{\be}} e^{-iq\La} = 0$ and the anticommutator $\{ {\cal D}_{\al} , \bar{{\cal D}}_{\dot{\be}} \} = - 2i  (\si^{\mu})_{\al \dot{\be}} \pa_{\mu}$.

\section{Chapter 4}

\subsection*{Exercise 4.1}

Starting from
$$\Phi^\dagger(x,\theta,\bt) \eq \vp^* \ + \ \sqrt{2} \, \bt\bar{\psi} \ + \ (\bt\bt) \, F^* \ - \ i(\theta\s^\mu\bt) \, \pa_\mu\vp^* \ - \ \frac{(\theta\theta) \, (\bt\bt)}{4} \; \pa_\mu\pa^\mu \vp^* \ + \ \frac{i(\bt\bt)}{\sqrt{2}} \; \theta\s^\mu\pa_\mu \bar{\psi}$$
let us look at $\Phi^\dagger\Phi$ and identify the $(\theta\theta)(\bt\bt)$ component:
\begin{eqnarray*}
\Phi^\dagger \Phi&=&\left( \vp^* \ + \ \sqrt{2} \, \bt\bar{\psi} \ + \ (\bt\bt) \, F^* \ - \ i(\theta\s^\mu\bt) \, \pa_\mu\vp^* \ - \ \frac{(\theta\theta) \, (\bt\bt)}{4} \; \pa_\mu\pa^\mu \vp^* \ + \ \frac{i(\bt\bt)}{\sqrt{2}} \; \theta\s^\mu\pa_\mu \bar{\psi} \right)\\
&&\left(\vp \ + \ \sqrt{2} \, \theta\psi \ + \ (\theta\theta) \, F \ + \ i(\theta\s^\mu\bt) \, \pa_\mu\vp \ - \ \frac{(\theta\theta) \,(\bt\bt)}{4} \; \pa_\mu\pa^\mu \vp \ - \ \frac{i(\theta\theta)}{\sqrt{2}} \; \pa_\mu \psi \s^\mu\bt\right)\\
&\supset& (\theta\theta) \, (\bt\bt) \, \left[-\frac{1}{4} \; \vp^* \, \pa_\mu\pa^\mu \vp \ - \ \frac{1}{4} \; \vp \, \pa_\mu\pa^\mu \vp^* \ + \ |F|^2\right] \ + \ (\theta\s^\mu\bt) \, (\theta\s^\nu\bt) \, \pa_\nu\vp \, \pa_\mu\vp^*\\&&- \ i\bt\bar{\psi} \, (\theta\theta) \, \pa_\mu \psi \s^\mu\bt \ + \ i(\bt\bt) \, (\theta\s^\mu\pa_\mu \bar{\psi}) \, ( \theta\psi)\\
&=& (\theta\theta) \, (\bt\bt) \, \left[-\frac{1}{4} \; \vp^* \, \pa_\mu\pa^\mu \vp \ - \ \frac{1}{4} \; \vp \, \pa_\mu\pa^\mu \vp^* \ + \ |F|^2\right] \ + \ \frac{1}{2} \; (\theta\theta) \, (\bt\bt) \, \pa^\mu\vp \, \pa_\mu\vp^*\\
&&+ \ i\bt^{\da} \, \bar{\psi}_{\da} \, (\theta\theta) \, \pa_\mu \psi^\be \, (\s^\mu)_{\be\db} \, \bt^{\db} \ + \ i(\bt\bt) \, \theta^\al \, (\s^\mu)_{\al\da} \, \pa_\mu \bar{\psi}^{\da} \, \theta^\be \, \psi_\be \\
&=& (\theta\theta) \, (\bt\bt) \, \left[-\frac{1}{4} \; \vp^* \, \pa_\mu\pa^\mu \vp \ - \ \frac{1}{4} \; \vp \, \pa_\mu\pa^\mu \vp^* \ + \ |F|^2\right] \ + \ \frac{1}{2} \; (\theta\theta) \,(\bt\bt) \, \pa^\mu\vp \, \pa_\mu\vp^*\\
&&+ \ \frac{i}{2} \; \ep^{\da\db} \, (\bt\bt) \, \bar{\psi}_{\da} \, (\theta\theta) \, \pa_\mu \psi^\be \, (\s^\mu)_{\be\db} \ + \ \frac{i}{2} \; (\bt\bt) \, (\theta\theta) \, \ep^{\al\be} \, (\s^\mu)_{\al\da} \, \pa_\mu \bar{\psi}^{\da} \, \psi_\be \\
&=& (\theta\theta) \, (\bt\bt) \, \left[-\frac{1}{4} \; \vp^* \, \pa_\mu\pa^\mu \vp \ - \ \frac{1}{4} \; \vp \, \pa_\mu\pa^\mu \vp^* \ + \ |F|^2 \ + \ \frac{1}{2} \; \pa^\mu\vp \, \pa_\mu\vp^* \right. \\
&& \ \ \ \ \ \ \ \ \ \left. \ + \ \frac{i}{2} \; \pa_\mu \psi (\s^\mu)\bar{\psi}  \ - \ \frac{i}{2} \; \psi(\s^\mu) \, \pa_\mu \bar{\psi} \right] \\
&=& (\theta\theta) \, (\bt\bt) \, \Bigl[ |F|^2 \ + \ \pa^\mu\vp \, \pa_\mu\vp^* \ - \ i \psi(\s^\mu)\pa_\mu \bar{\psi} \Bigr] \ + \ \te{total derivatives}
\end{eqnarray*}

\subsection*{Exercise 4.2}

We want to look for the $\theta\theta$ component of the combination $\frac{1}{2}m\Phi^2+\frac{1}{3}g\Phi^3.$ Let's start with $\frac{1}{2}m\Phi^2:$
\begin{eqnarray*}
\frac{1}{2} \; m \, \Phi^2&=&\frac{1}{2} \; m \,  \left(\vp \ + \ \sqrt{2} \, \theta\psi \, +\, (\theta\theta) \, F \ + \ i(\theta\s^\mu\bt) \, \pa_\mu\vp \ - \ \frac{1}{4} \; (\theta\theta) \;(\bt\bt) \, \pa_\mu\pa^\mu \vp \ - \ \frac{i}{\sqrt{2}} \; (\theta\theta) \, \pa_\mu \psi \s^\mu\bt\right)\\ &&\left(\vp \ + \ \sqrt{2} \, \theta\psi \ + \ (\theta\theta)F \ + \ i(\theta\s^\mu\bt) \, \pa_\mu\vp \ - \ \frac{1}{4} \; (\theta\theta) \, (\bt\bt) \, \pa_\mu\pa^\mu \vp \ - \ \frac{i}{\sqrt{2}} \; (\theta\theta) \, \pa_\mu \psi \s^\mu\bt\right)\\
&\supset&\frac{m}{2} \; \bigl(\vp \ + \ \sqrt{2} \, \theta\psi \ + \ (\theta\theta) \, F \bigr) \, \bigl(\vp\ + \ \sqrt{2} \, \theta\psi \ + \ (\theta\theta) \, F\bigr)\\
&\supset&\frac{m}{2} \; \bigl((\theta\theta) \, (\vp F \, + \, F \vp) \ + \ 2 \, \theta^\al \, \psi_\al \, \theta^\be \, \psi_\be\bigr)\\
&=&\frac{m}{2} \; \bigl((\theta\theta) \, (\vp F \, + \, F \vp) \ - \ 2 \, \theta^\al \, \theta^\be \, \psi_\al \, \psi_\be \, \bigr)\\
&=&\frac{m}{2} \; \bigl((\theta\theta) \, (2\vp F) \ + \ (\theta\theta) \, \ep^{\al\be} \, \psi_\al \, \psi_\be\bigr)\\
&=&\frac{m}{2} \; \bigl((\theta\theta) \, (2\vp F) \ - \ (\theta\theta) \, \ep^{\be\al} \, \psi_\al \, \psi_\be\bigr)\\
&=&m \, (\theta\theta) \, \left(\vp \,  F \ - \ \frac{1}{2} \; (\psi\psi)\right)\\
\end{eqnarray*}
Next, consider $\frac{1}{3}g\Phi^3:$
\begin{eqnarray*}
\frac{1}{3} \; g \, \Phi^3&\supset&\frac{g}{3} \; \bigl(\vp \ + \ \sqrt{2} \, \theta\psi \ + \ (\theta\theta) \, F \bigr) \,  \bigl(\vp \ + \ \sqrt{2} \, \theta\psi \ + \ (\theta\theta) \, F \bigr) \,  \bigl(\vp \ + \ \sqrt{2} \, \theta\psi \ + \ (\theta\theta) \, F \bigr) \\
&\supset&\frac{g}{3} \; \bigl((\theta\theta) \, (\vp^2 F \, + \, \vp F \vp \, + \, F \vp^2) \ +\ 2 \, \vp \, (3 \,  \theta^\al \, \psi_\al \, \theta^\be \, \psi_\be)\bigr)\\
&=&g \, (\theta\theta) \, \bigl(\vp^2 \,F \ + \ \vp \, (\psi\psi)\bigr)
\end{eqnarray*}

\subsection*{Exercise 4.3}

We want to determine the F-component of $\frac{1}{4}W^\al W_\al:$
\begin{eqnarray*}
\frac{1}{4} \;W^\al \, W_\al|_{F}&=&\frac{1}{4} \; (\theta\theta) \, (-2i\la^\al \, \s^\mu_{\al\da} \, \pa_\mu\bl^{\da} \ + \ D^2) \ - \ \frac{1}{16} \; (\s^\mu\bs^\nu\theta)^\al \, (\s^\rho\bs^\la\theta)_\al \, F_{\mu\nu} \, F_{\rho\la}\\
&& + \ \frac{i}{4} \; D \,  \theta^\al \, (\s^\mu\bs^\nu\theta)_\al \, F_{\mu\nu}\\
&=&\frac{1}{4} \; (\theta\theta) \, (-2i\la^\al \, \s^\mu_{\al\da} \, \pa_\mu\bl^{\da} \ + \ D^2) \ - \ \frac{1}{32} \; (\theta\theta) \, \text{Tr} \bigl\{ \s^\mu\bs^\nu\s^\la\bs^\rho \bigr\} \, F_{\mu\nu} \, F_{\rho\la}
\end{eqnarray*}
In the last step we used:
\begin{eqnarray*}
\frac{i}{4} \; D \, \theta^\al \, (\s^\mu\bs^\nu\theta)_\al \, F_{\mu\nu}&=&\frac{i}{4} \;D \, F_{\mu\nu} \, \theta^\al \, \theta^\g \, (\s^\mu)_{\al\da} \, (\bs^\nu)^{\da\be} \, \ep_{\be\g} \, F_{\mu\nu}\\
&=&- \, \frac{i}{8} \; D \, F_{\mu\nu} \, (\theta\theta) \, \ep^{\al\g} \, (\s^\mu)_{\al\da} \, (\bs^\nu)^{\da\be} \, \ep_{\be\g}\\
&=&\frac{i}{8} \; D \,  F_{\mu\nu} \, (\theta\theta) \, (\s^\mu)_{\al\da} \, (\bs^\nu)^{\da\be} \, \de^\al_{\be}\\
&=&\frac{i}{8} \; D \,F_{\mu\nu} \, (\theta\theta) \, (\s^\mu)_{\al\da} \, (\bs^\nu)^{\da\al}\\
&=&\frac{i}{4} \; D \, F_{\mu\nu} \, (\theta\theta) \, \eta^{\mu\nu}\\
&=&0\ .
\end{eqnarray*}
\begin{eqnarray*}
-\ \frac{1}{16} \; (\s^\mu\bs^\nu\theta)^\al \, (\s^\rho\bs^\la\theta)_\al  \, F_{\mu\nu} \, F_{\rho\la}&=&- \, \frac{1}{16} \; \ep_{\al\be} \,(\s^\mu\bs^\nu\theta)^\al \,(\s^\rho\bs^\la\theta)^\be \, F_{\mu\nu} \, F_{\rho\la}\\
&=&- \, \frac{1}{16} \; \ep^{\al\be} \, (\s^\mu)_{\al\da} \, (\bs^\nu)^{\da\g} \, \theta_\g \, (\s^\rho)_{\be\db} \, (\bs^\la)^{\db\de} \, \theta_\de \,  F_{\mu\nu} \, F_{\rho\la}\\
&=&- \, \frac{1}{32} \; (\theta\theta) \, \text{Tr} \bigl\{ \s^\mu\bs^\nu\s^\la\bs^\rho \bigr\} \, F_{\mu\nu} \, F_{\rho\la}\\
&=&- \, \frac{i}{16} \; (\theta\theta) \, \ep^{\mu\la\tau\rho} \, F_{\mu\la} \, F_{\rho\tau} \ - \ \frac{1}{8} \; (\theta\theta) \, F_{\mu\nu} \, F^{\mu\nu}
\end{eqnarray*}
where in the last step we used
$$\text{Tr} \bigl\{ \s^\mu\bs^\nu\s^\la\bs^\rho\bigr\}  \eq 2i \, \ep^{\mu\nu\la\rho}  \ + \ 2 \, \eta^{\mu\nu} \, \eta^{\la\rho} \ - \ 2 \, \eta^{\mu\la} \, \eta^{\nu\rho} \, + \, 2 \, \eta^{\mu\rho} \, \eta^{\nu\la}$$
If we write $\tilde F_{\mu\nu}=\frac{1}{2} \ep_{\mu\nu\rho\la}F^{\rho\la},$ we have
$$\frac{1}{4} \; W_\al \, W^\al|_{F} \eq - \, \frac{i}{2} \; \la\s^\mu\pa_\mu\bl \ + \ \frac{1}{4} \; D^2 \ - \ \frac{1}{8} \; F_{\mu\nu} \, F^{\mu\nu} \ + \ \frac{i}{8} \, \tilde F_{\mu\nu} \, F^{\mu\nu}$$

\subsection*{Exercise 4.4}

We start with the chiral scalar part of the supergravity Lagrangian in the conformally flat limit
\[
{\cal S} \eq -3 \int \dd^4x \ \dd^4 \theta\ \varphi \, \bar{\varphi}\,e^{-\frac{\kappa^2}{3}K} \ + \ \left(\int \dd^4x \ \dd^2\theta\ \varphi^3 \, W \ + \ {\rm h.c.}\right)
\]
Ignoring fermionic components we can integrate over half the superspace
\begin{eqnarray}
{\cal L}&=&-3 \int \dd^4 \theta\ \varphi \, \bar{\varphi}\,e^{-K/3} \ + \ \left(\int \dd^2\theta\ \varphi^3 \, W \ + \ {\rm h.c.}\right) \notag \\
\nonumber &=&-3 \int \dd^2 \bar{\theta} \ \left(\bar{\varphi} \, e^{-K/3} \, F^\varphi \ - \ \frac{1}{3} \; \bar{\varphi}\varphi \, e^{-K/3} \, K_i \, F^i \right) \ + \ 3 \, \varphi^2 \,  F^\varphi \, W \ + \ \varphi^3 \, F^i W_i \ +\ \int \dd^2 \bar{\theta} \ \bar{\varphi}^3 \, \bar{W} \notag \\
\nonumber &=&- \, e^{-K/3} \, \left( 3 \, F^{\bar{\vp}} \,F^{\vp} \ - \   \bar{\vp} \, K_{\bar{i}} \, F^{\bar{i}} \, F^{\vp} \ - \   \vp  \, K_{i} \, F^{i} \, F^{\bar{\vp}} \ - \   \vp \, \bar{\varphi} \, K_{i\bar{j}} \, F^{i} \, F^{\bar{j}} \ + \ \frac{1}{3} \; \bar{\varphi} \, \varphi \, K_i \, F^i \,  K_{\bar j} \,F^{\bar{j}}\right)\\
&& + \ 3 \, \varphi^2 \, F^\vp \, W \ + \  \varphi^3 \, F^i \, W_i \ + \ 3 \, \bar{\varphi}^2 \, F^{\bar{\varphi}} \, \bar{W} \ + \ \bar{\varphi}^3 \, F^{\bar{i}} \, \bar{W}_{\bar{i} } \notag
\end{eqnarray}
This gives us equations of motion for the auxiliary $F$ fields:
\begin{eqnarray}
0&=&-3 \, e^{-K/3} \, \left(F^\vp \ - \ \frac{1}{3} \, \vp \,  K_i \, F^i \right) \ + \ 3 \, \bar{\vp}^2 \, \bar{W} \notag \\
\nonumber 0&=& \vp^3 \, W_i \ - \ 3 \, e^{-K/3} \, \left(-\frac{1}{3} \; \vp \, K_i \, F^{\bar{\vp}} \ -\ \frac{1}{3} \, \vp \, \bar{\varphi} \, K_{i\bar{j}} \, F^{\bar{j}} \ + \ \frac{1}{9} \; \bar{\varphi} \, \varphi \,  K_i \,  K_{\bar j} \, F^{\bar{j}} \right)\\
\nonumber &=& \vp^3 \, W_i \ + \ \vp^3 \, W \, K_i \ + \ e^{-K/3} \, \vp \, \bar{\vp} \, K_{i\bar{j}} \, F^{\bar{j}}\\
&=& \vp^3 \, D_i \, W \ + \ e^{-K/3} \, \vp \, \bar{\vp} \,K_{i\bar{j}} \, F^{\bar{j}} \notag
\end{eqnarray}
This can be solved for the F-terms and we can plug the solution back into the Lagrangian and we find:
\[
{\cal L} \eq ... \ + \ \vp^2 \, \bar{\vp}^2 \, e^{K/3} \, (K^{i\bar{j}} \, D_i W  \, D_{\bar{j}}\bar{W} \ - \ 3 \, |W|^2)
\]
To determine the value for the chiral compensator, we shall need that
\[
-\, \frac{3}{\kappa^2}\int \dd^4 x\  \dd^4\theta \ \bar{E} \, e^{-\frac{\kappa^2}{3}K}
\]
includes the Einstein-Hilbert term at leading order in $\kappa$ along with a single power $\vp\bar{\vp}$. To get the canonical form this requires the scalar component of the chiral compensator to be $\vp=\bar{\vp}=e^{K/6}.$ We then obtain the standard F-term scalar potential in supergravity
\[
V \eq e^K \, (K^{i\bar{j}} \, D_i W \, D_{\bar{j}}\bar{W} \ - \ 3 \, |W|^2) \ .
\]

\section{Chapter 5}

\subsection*{Exercise 5.1}
We have a global\footnote{The supertrace constraint does not apply for local supersymmetry (i.e. supergravity), there ${\rm STr} \{ M^2 \} \sim m_{3/2}^2.$} ${\cal N}=1$ supersymmetric Lagrangian. The F-terms are
$$F_i \eq - \, \frac{\pa W^*}{\pa \vp_i^*} \co F_i^* \eq - \, \frac{\pa W}{\pa \vp_i} \ .$$
The scalar potential is
$$V \eq \sum_i |F_i|^2\ .$$
We first want the scalar mass matrix. If we split the complex field into real and imaginary parts $\vp_i=\vp_{i,1}+i\vp_{i,2}$ the scalar mass matrix is
\begin{equation}
M_{\al\be}^2 \eq \frac{1}{2}\left(\begin{array}{c c c c}
\frac{\pa^2 V}{\pa \vp_{1,1}\pa\vp_{1,1}} & \frac{\pa^2 V}{\pa \vp_{1,1}\pa\vp_{1,2}} & \ldots &\frac{\pa^2 V}{\pa \vp_{1,1}\pa\vp_{n,2}} \\
\frac{\pa^2 V}{\pa \vp_{1,2}\pa\vp_{1,1}} & \frac{\pa^2 V}{\pa \vp_{1,2}\pa\vp_{1,2}} & \ldots & \frac{\pa^2 V}{\pa \vp_{1,2}\pa\vp_{n,2}} \\
\vdots & & & \\
\frac{\pa^2 V}{\pa \vp_{n,2}\pa\vp_{1,1}} & \frac{\pa^2 V}{\pa \vp_{n,2}\pa\vp_{1,2}} & \ldots &\frac{\pa^2 V}{\pa \vp_{n,2}\pa\vp_{n,2}} 
\end{array}\right) \notag
\end{equation}
For the trace we only need  the trace
\begin{equation}
{\rm Tr} \{ M_{\al\be}^2 \}  \eq \frac{1}{2}\sum_{j=1,2}\sum_i \frac{\pa^2 V}{\pa\vp_{i,j}^2} \notag
\end{equation}
Now $\vp_i=\vp_{i,1}+i\vp_{i,2}$ and $\vp_{i}^*=\vp_{i,1}-i\vp_{i,2}$ which simplifies the derivatives:
\begin{equation}
\frac{\pa V}{\pa \vp_{i,1}}\eq \frac{\pa V}{\pa \vp_i} \ + \ \frac{\pa V}{\pa\vp_i^*} \co \frac{\pa V}{\pa \vp_{i,2}}\eq i\frac{\pa V}{\pa \vp_i} \ - \ i\frac{\pa V}{\pa\vp_i^*} \notag
\end{equation}
\begin{equation}
\frac{\pa^2 V}{\pa\vp_{i,1}^2} \eq \frac{\pa^2 V}{\pa\vp_i^2} \ + \ \frac{\pa^2 V}{\pa(\vp_i^*)^2} \ + \ 2 \; \frac{\pa^2 V}{\pa\vp_i\pa \vp_i^*} \notag
\end{equation}
\begin{equation}
\frac{\pa^2 V}{\pa\vp_{i,2}^2} \eq - \, \frac{\pa^2 V}{\pa\vp_i^2} \ - \ \frac{\pa^2 V}{\pa(\vp_i^*)^2} \ + \ 2 \; \frac{\pa^2 V}{\pa\vp_i\pa \vp_i^*} \notag
\end{equation}
Hence,
\begin{eqnarray*}
{\rm Tr} \bigl\{ M_{\al\be}^2 \bigr\} &=&\frac{1}{2}\sum_{j=1,2}\sum_i \frac{\pa^2 V}{\pa\vp_{i,j}^2} \eq 2 \sum_i\frac{\pa^2 V}{\pa \vp_i\pa\vp_i^*}
 \eq 2 \sum_i\frac{\pa^2}{\pa \vp_i\pa\vp_i^*}(\sum_j |F_j|^2)\\
&=& 2 \sum_{i,j}\frac{\pa^2}{\pa \vp_i\pa\vp_i^*}\frac{\pa W^*}{\pa \vp_j^*}\frac{\pa W}{\pa \vp_j}\eq  2\sum_{i,j}\frac{\pa^2 W^*}{\pa \vp_i^* \pa \vp_j^*}\frac{\pa^2 W}{\pa \vp_i\pa \vp_j}
\end{eqnarray*}
Now consider the fermions. The fermion mass matrix is
\begin{equation}
M_{ij} \eq \frac{\pa^2 W}{\pa\Phi_i\pa\Phi_j} \notag
\end{equation}
Generally this is a symmetric complex matrix with complex eigenvalues. We can diagonalise $M_{ij}$ with a unitary matrix $U$
\begin{equation}
M' \eq U \, M \, U^\dagger \eq
\left(\begin{array}{c c c c}m_1 e^{i\vp_1}&\\
& m_2 e^{i\vp_2}&\\
& & \ddots &\\
& & & m_n e^{i\vp_n}
\end{array}\right) \notag
\end{equation}
We then have
\begin{equation}
M' \, (M')^\dagger \eq \left(\begin{array}{c c c c}m_1^2&\\
& m_2^2&\\
& & \ddots &\\
& & & m_n^2
\end{array}\right) \notag
\end{equation}
The trace of the last expression is what we want.
\begin{align*}
\bigl( M' \, (M')^\dagger \bigr)_{ir} \ \ &= \ \ (U_{ij} \,M_{jk} \, U^\dagger_{kl}) \, (U_{lp}\, M_{pq}^\dagger  \, U_{qr}^\dagger)\overset{M \text{ symmetric}}=(U_{ij} \, M_{jk} \,U^\dagger_{kl}) \, (U_{lp} \, M_{pq}^* \, U_{qr}^\dagger) \\
&= \ \ U_{ij} \, M_{jk} \, M^*_{kl} \, U^\dagger_{lr}
\end{align*}
So we get
\begin{eqnarray*}
{\rm Tr} \bigl\{ M' (M')^\dagger \bigr\} &=&U_{ij} \, M_{jk} \, M^*_{kl} \, U^\dagger_{li}\eq M_{jk} \, M^*_{kj}\\
&=& \sum_{j,k}\left(\frac{\pa^2 W}{\pa\vp_j\pa\vp_k}\right)\left(\frac{\pa^2 W^*}{\pa\vp_j^*\pa\vp_k^*}\right)
\end{eqnarray*}
Overall, we have
$$\sum_{\text{fermions}}(-1)^{2\cdot \frac{1}{2}+1} \, \left(2\cdot \tfrac{1}{2} \, + \, 1 \right) \, M_{\text{fermion}}^2\eq 2\sum_{j,k}\left(\frac{\pa^2 W}{\pa\vp_j\pa\vp_k}\right)\left(\frac{\pa^2 W^*}{\pa\vp_j^*\pa\vp_k^*}\right)
$$
So the supertrace vanishes!
In the O'Raifertaigh model we have the following mass spectrum:
\begin{center}
\begin{tabular}{c c c  c c c}
Bosons: &$\vp_1:$ & $0,$ $0$ & Fermions: & $\Psi_1:$ & $0$\\
& $\vp_2:$ &  $M,$ $M$ & & $\Psi_2:$ & M\\
& $\vp_3:$ & $\sqrt{M^2-2gm^2},$ $\sqrt{M^2+2gm^2}$ & & $\Psi_3:$ & M
\end{tabular}
\end{center}
Plugging into the formula for the supertrace we then obtain the desired result
\begin{eqnarray*}
{\rm STr} \bigl\{M^2 \bigr\}&=&M^2+M^2+(M^2-2gm^2)+(M^2+2gm^2)-2\cdot (M^2+M^2)\\
&=&0.
\end{eqnarray*}
\subsection*{Exercise 5.2} 
We have a chiral superfield  $\Phi$ of charge q, coupled to an abelian vector superfield V. The Lagrangian then is
\begin{equation}
 \mL \eq (\Phi^\dagger \, e^{qV} \, \Phi)_D \ + \ \frac{1}{4} \; (W^\al \,  W_\al \, |_F \ + \ {\rm h.c.}) \ + \ \xi \, V_D \notag
\end{equation}
Recall the theta expansion of a vector superfield in Wess-Zumino gauge
\begin{equation*}
 V_{\rm WZ} \eq (\theta \sigma^\mu \bt) \, V_\mu \ + \ i (\theta\theta)  \, (\bt\bl) \ - \ i (\bt\bt) \, (\theta \la) \ + \ \frac{1}{2} \; (\theta \theta) \, (\bt\bt) \,  D
\end{equation*}
The chiral superfield in components is given by
\begin{eqnarray*}
 \Phi&=&\vp \ + \ \sqrt{2} \, (\theta \psi) \ + \ ( \theta\theta) \,  F \ + \ i(\theta\sigma^\mu\bt) \, \pa_\mu \vp \ - \ \frac{1}{4} \; (\theta\theta) \, (\bt\bt) \,  \pa_\mu\pa^\mu\vp \ - \ \frac{i}{\sqrt{2}} \; (\theta\theta) \, (\pa_\mu \psi \sigma^\mu \bt) \\
\Phi^\dagger&=&\vp^* \ + \ \sqrt{2} \, (\bt \bp) \ + \  (\bt\bt) F^* \ - \ i(\theta\sigma^\mu\bt) \, \pa_\mu \vp^* \ - \ \frac{1}{4} \; (\theta\theta) \, (\bt\bt) \,  \pa_\mu\pa^\mu\vp^*\ + \ \frac{i}{\sqrt{2}} \; (\bt\bt) \, ( \theta \sigma^\mu \pa_\mu \bp)
\end{eqnarray*}
As evaluated in exercise 4.1, $\Phi^\dagger \Phi|_D=-i(\bp \bs^\mu\pa_{\mu} \psi)+\pa_\mu\vp \pa^\mu \vp^*+|F|^2.$ Now
\begin{eqnarray*}
 \Phi^\dagger \, V \,  \Phi|_{D}&=&\vp^* \, (\theta \sigma^\mu \bt) \, V_\mu \, i(\theta\sigma^\nu\bt)\, \pa_\nu \vp \ - \ i\vp^* \, (\bt\bt) \,( \theta \la) \, (\sqrt{2}\theta \psi) \ +\ \frac{|\vp|^2}{2}  \; (\theta\theta) \, (\bt\bt) \, D\\
  && + \ \sqrt{2} \, (\bt \bp) \, (\theta \sigma^\mu \bt) \, V_\mu  \, \sqrt{2} \, (\theta \psi) \ + \ i \sqrt{2} \, (\bt \bp) \, ( \theta\theta ) \, (\bt\bl) \, \vp \ - \ i(\theta\sigma^\mu\bt) \,\pa_\mu \vp^* \,(\theta \sigma^\nu \bt) \,V_\nu \, \vp \ \bigl. \bigr|_D \\
&=&  \frac{1}{2} \; |\vp|^2 \, D \ + \ \frac{i}{2} \, (\vp^* \, V^\mu  \, \pa_\mu \vp \ - \ \vp \, V^\mu \, \pa_\mu \vp^*) \  + \ \frac{i}{\sqrt{2}} \; \bigl( (\bl \bp) \, \vp \ - \ (\la\psi) \, \vp^* \bigr) \ - \ \frac{1}{2} \; (\bp \bs^\mu \psi) \, V_\mu  
\end{eqnarray*}
To get the last expression we applied appropriate Fierz identities as used in previous exercises.
\begin{eqnarray*}
  \Phi^\dagger  \; \frac{V^2}{2}  \; \Phi \, |_D &=&\frac{1}{2} \; \vp^* \, (\theta \sigma^\mu \bt) \, V_\mu \ (\theta \sigma^\nu \bt) \, V_\nu \ \bigl. \bigr|_D \\
&=&\frac{1}{4} \; |\vp|^2  \, V_\mu \, V^\mu 
\end{eqnarray*}
In total we have
\begin{eqnarray*}
 \Phi^\dagger \, e^{2q V} \, \Phi|_D&=& \pa_\mu \vp  \, \pa^\mu \vp^* \ + \ |F|^2 \ - \ i(\bp \bs^\mu\pa_\mu \psi) \ + \ q \, V^\mu \bigl( - \, (\bp \bs_\mu \psi) \ + \ i (\vp^*  \, \pa_\mu \vp \ - \ \vp \,   \pa_\mu \vp^*) \bigr) \\ &&+ \ \sqrt{2} iq  \, \bigl( \vp \, (\bl\bp) \ - \ \vp^* \, (\la\psi ) \bigr) \ + \ q \, \left(D \ + \ q \,  V_\mu \, V^\mu \right) \, |\vp|^2
\end{eqnarray*}
Hence the D-term part of the Lagrangian is
\begin{equation}
{\cal L}_D \eq  q\, D  \, |\vp|^2 \ + \ \frac{1}{2} \; D^2 \ + \ \frac{1}{2} \; \xi \, D \notag
\end{equation}
where the term $\frac{1}{2}D^2$ comes from $\frac{1}{4}W_\al W^\al+{\rm h.c.}$
Solving the equation of motion for D gives the following condition
\begin{equation}
 D \eq - \, q \, |\vp|^2 \ - \ \frac{ \xi }{2} \notag
\end{equation}
Plugging this back into the Lagrangian yields the following D-term potential
\begin{eqnarray*}
 V_D&=& \frac{1}{8} \; \left(\xi \ + \ 2 \, q \, |\vp|^2 \right)^2
\end{eqnarray*}
As $( q |\vp|^2+\xi / 2)=-D\neq 0,$ supersymmetry is broken. However, if $\vp$ can relax to a supersymmetric minimum, it will. In order that this is not possible, we require  $\xi$ and $q$ to have the same sign. Then $V_D$ is minimised at $\vp=0$, but the potential is positive and supersymmetry is broken. In this case
$$V_D \eq \frac{1}{8} \; \xi^2 \  + \ \frac{q \, \xi}{2} \; |\vp|^2 \ + \ \frac{q^2}{2} \; |\vp|^4$$
If $\langle\vp\rangle=0,$ the mass of $\vp$ is $m^2_\vp=q \xi$ (as the kinetic terms are $\pa_\mu\vp \pa^\mu \vp^*).$ Since no other fields obtain vevs no mass is generated for the fermions. Therefore the mass splitting in the multiplet is $m_\vp= \sqrt{q\xi}$ and $m_\psi=0.$
\subsection*{Exercise 5.3}
We have a supersymmetric field theory with chiral superfields $\Phi_i=(\vp_i,\psi_i,F)$ and vector superfields $V_a=(\la_a,A_a^\mu,D),$ with both D- and F-term supersymmetry breaking (i.e. $D_a\neq 0,$ $F_i\neq 0).$ In the vacuum, $\frac{\pa V}{\pa \vp_i}=0$ by definition.
\begin{eqnarray*}
V&=&\sum |F_i|^2+\frac{1}{2}\sum_a D^a D^a\\
&=&\sum_i \left(\frac{\pa W}{\pa \vp_i}\right)\left(\frac{\pa W^*}{\pa \vp_j^*}\right)+\frac{1}{2}\sum_a (\sum_j \vp^\dagger_j T^a\vp_j)(\sum_k \vp^\dagger_k T^a\vp_k)\\
\end{eqnarray*}
Now
\begin{eqnarray*}
\frac{\pa V}{\pa \vp_j}&=&\sum_i \left(\frac{\pa^2 W}{\pa \vp_i\pa\vp_j}\right)F_i+\sum_a D^a (\sum_k \vp^\dagger_k (T^a)_{kj})\\
&=&\left(\begin{array}{c c} \frac{\pa^2 W}{\pa \vp_i\pa\vp_j} & \sqrt{2}\sum_k \vp^\dagger_k (T^a)_{kj}\end{array}\right)\left(\begin{array}{c} F_i\\ \frac{D_a}{\sqrt{2}} \end{array}\right)=0.
\end{eqnarray*}
Regarding the D-term potential we can absorb any prefactor in the potential in the definition of the generator.
The gauge invariance of the superpotential implies
\begin{eqnarray*}
0&=&\de^{(a)}_{\rm gauge}W\\
&=&\frac{\pa W}{\pa \vp^i}\de^{(a)}_{\rm gauge}\vp^i\\
&=&-F_i^\dagger (T^a)_{ij}\vp_j
\end{eqnarray*}
We are free to dagger this equation and multiply it with a non-vanishing complex number $c,$ it still has to hold. So we can write
\begin{eqnarray*}
0&=& \left(\begin{array}{c c} c \sum_k \vp^\dagger_k (T^a)_{kj} & 0\end{array}\right)\left(\begin{array}{c} F_i\\ \frac{D_a}{\sqrt{2}} \end{array}\right)
\end{eqnarray*}
Combining both equations, we obtain
\begin{eqnarray*}
0&=&\left(\begin{array}{c c} \frac{\pa^2 W}{\pa \vp_i\pa\vp_j} & \sqrt{2}\sum_k \vp^\dagger_k (T^a)_{kj}\\
\sqrt{2}\sum_k \vp^\dagger_k (T^a)_{kj} & 0
\end{array}\right)\left(\begin{array}{c} F_i\\ \frac{D_a}{\sqrt{2}} \end{array}\right)
\end{eqnarray*}
We want to show that this matrix is the same as that of the fermion mass matrix:
$$\left(\begin{array}{c} \psi_i\\ \la_a \end{array}\right)^T\left(M_{ia}\right)\left(\begin{array}{c} \psi_i\\ \la_a \end{array}\right)$$
To find the entries of the mass matrix, we know the standard contribution for the fermion mass matrix is given by $\frac{\pa^2 W}{\pa \vp_i\pa \vp_j}.$ The off-diagonal terms can be obtained from the structure of the kinetic terms $\int d^2 \theta d^2 \bt\, \Phi^\dagger e^{V^a T^a}\Phi.$
\begin{eqnarray*}
\Phi&\sim&\vp+\sqrt{2}\theta\psi+\theta\theta Ff\\
\Phi^\dagger &\sim&\vp^\dagger+\sqrt{2}\bt\bp+\bt\bt F^*\\
V^a&\sim&(\theta\s^\mu\bt)V^a_\mu+i\theta\theta\bt\bl^a-i\bt\bt\theta\la^a+\frac{1}{2}\theta\theta\bt\bt D^a
\end{eqnarray*}
As deduced for the abelian case in Exercise 5.2, the $\psi\la$ term arises from $\Phi_i^\dagger e^{V^aT^a}\Phi_i.$ We identify the following cross-term by looking at the superfield expansion:
$$\sqrt{2} \vp^\dagger_i T^a_{ij}\psi_j \la_a.$$
So we find the anticipated cross-term in the fermion mass matrix. From the superfield expansion no gaugino mass term $(\la\la)$ is generated.

Now we can write
$$M_{ia}\left(\begin{array}{c}F_i\\ \frac{D_a}{\sqrt{2}} \end{array}\right)=0,$$
where $M_{ia}$ is the fermion mass matrix. This implies that there is at least one zero eigenvalue with eigenvector
$$\left(\begin{array}{c}F_i\\ \frac{D_a}{\sqrt{2}} \end{array}\right).$$ This means there exists a massless Goldstone fermion, oriented along the direction of supersymmetry breaking.

\subsection*{Exercise 5.4}
The setup is
\begin{eqnarray*}
K&=&-\log{(S+S^*)}-3\log{(T+T^*+CC^*)}\\
W&=&C^3+a e^{-\al S}+b
\end{eqnarray*}
We compute the supergravity scalar potential
\begin{equation}
V \eq e^K \; (D_i W \, K^{i\bar{j}} \, D_{\bar{j}} \bar{W} \ - \ 3 \, |W|^2)\ , \notag
\end{equation}
where $K^{i\bar{j}}$ is the inverse of the K\"ahler metric 
\begin{equation}
K_{i\bar{j}} \eq \frac{\pa^2 K}{\pa \Phi_i \, \pa \Phi_{\bar{j}}} \eq 
\left(
\begin{array}{ccc}
 \frac{1}{(S+S^*)^2} & 0 & 0 \\
 0 & \frac{3}{(-C C^*+T+T^*)^2} & -\frac{3 C}{(-C C^*+T+T^*)^2} \\
 0 & -\frac{3 C^*}{(-C C^*+T+T^*)^2} & \frac{3 (T+T^*)}{(-C C^*+T+T^*)^2}
\end{array}
\right) \notag
\end{equation}
The inverse K\"ahler metric is given by
\begin{equation}
K^{i\bar{j}} \eq \left(
\begin{array}{ccc}
 (S+S^*)^2 & 0 & 0 \\
 0 & -\frac{1}{3} (C C^*-T-T^*) (T+T^*) & \frac{1}{3} C (-C C^*+T+T^*) \\
 0 & \frac{1}{3} C^* (-C C^*+T+T^*) & \frac{1}{3} (-C C^*+T+T^*)
\end{array}
\right) \notag
\end{equation}
The scalar potential now can be written as
\begin{eqnarray*}
V&=& e^K \, \left(D_S W \, K^{SS^*} \, D_{S^*}\bar{W} \ + \ D_T W \, K^{TT^*} \,D_{T^*}\bar{W} \ + \ D_T W \,  K^{TC^*} \, D_{C^*}\bar{W}\right. \\ &&\left.\; \ \ \ \ + \ D_C W \, K^{CT^*} \, D_{T^*}\bar{W} \ + \ D_C W \, K^{CC^*} \, D_{C^*}\bar{W} \ - \ 3 \, |W|^2\right)\\
&\overset{\pa_T W=0}{=}&e^K \, \left(D_S W \, K^{SS^*} \, D_{S^*}\bar{W} \ + \ |W|^2 \,  \pa_T K \,  K^{TT^*} \, \pa_{T^*}K \ + \ W \,  \pa_T K \, K^{TC^*} \, D_{C^*}\bar{W}\right. \\ &&\left.\;  \ \ \ \ + \ \bar{W} \, D_C W \, K^{CT^*} \, \pa_{T^*}K \ + \ D_C W \, K^{CC^*} \, D_{C^*}\bar{W} \ - \ 3 \, |W|^2\right)\\
&=& e^K\left(D_S W K^{SS^*}D_{S^*}\bar{W}+D_T W K^{TT^*}D_{T^*}\bar{W}+D_T W K^{TC^*}D_{C^*}\bar{W}\right. \\ &&\left.\;  \ \ \ \ +  \ D_C W K^{CT^*}D_{T^*}\bar{W}+D_C W K^{CC^*}D_{C^*}\bar{W} \ - \ 3 \, |W|^2\right)\\
&\overset{\pa_T W=0}{=}&e^K \, \left(D_S W \, K^{SS^*} \, D_{S^*}\bar{W} \ + \ |W|^2 \, \pa_T K \, K^{TT^*} \, \pa_{T^*}K \ + \ W \, \pa_T K \, K^{TC^*} \, D_{C^*}\bar{W}\right. \\ &&\left.\;  \ \ \ \ + \ \bar{W} \, D_C W \,  K^{CT^*} \, \pa_{T^*}K \ + \ D_C W  \, K^{CC^*} \, D_{C^*}\bar{W} \ - \ 3 \, |W|^2\right)\\
\end{eqnarray*}
We see that there is no mixing of covariant derivatives between $S$ and $T,C.$ Consider the part of the scalar potential that involves only derivatives with respect to the K\"ahler potential (depending on $C$ and $T):$
\begin{align*}
K^{TT^*} \, \pa_T K \,  \pa_{T^*} K \ + \ &K^{TC^*} \, \pa_T K \, \pa_{C^*} K\ + \ K^{CT^*} \, \pa_C K  \, \pa_{T^*} K \ + \ K^{CC^*} \, \pa_C K \, \pa_{C^*} K \\
& =\ \ \ldots \eq 3 \ .
\end{align*}
This result is true for all no-scale models and results in a large cancellation in the scalar potential, leading to
\begin{eqnarray*}
V&=&e^K \, \left(D_S W \,  K^{SS^*} \, D_{S^*}\bar{W} \ + \ W \, \pa_T K \, K^{TC^*}\, \pa_{C^*}\bar{W}\right. \\ &&\left.\; + \ \bar{W} \, \pa_C W \, K^{CT^*} \, \pa_{T^*}K\ + \ \pa_C W \, K^{CC^*} \, \pa_{C^*}\bar{W} \ + \ W  \, \pa_C K \,  K^{CC^*} \, \pa_{C^*}\bar{W} \ + \ \pa_C W \, K^{CC^*} \, \pa_{C^*}K \, \bar{W}\right)\\
&=&e^K \, \left(D_S W \, K^{SS^*} \, D_{S^*}\bar{W} \ + \ 3 \,  (T \, + \, T^* \, - \, C C^*) \, C^2 \, C^*{}^2\right)
\end{eqnarray*}
Now, $T+T^*>CC^*$ since $T$ is a modulus $(\langle T\rangle\neq 0)$ and $C$ is a matter field $(\langle C\rangle= 0)$ and so the minimum of the potential is $V=0$ with $D_S W=0$ and $C=0.$ The vacuum energy of the potential vanishes at its minimum. $T$ is called a modulus field since we can vary it freely and still remain at the minimum of the potential. Moduli fields denote flat directions of the potential.
To check whether SUSY is broken we need to determine the F-terms:
$$F_S \eq D_S W \eq \al \,  a \,  e^{-\al S} \ - \ \frac{1}{S+S^*} \; \left(C^3 \ + \ a \, e^{-\al S} \ + \ b\right).$$
$S$ adjusts such that $D_S W=0$ at the minimum, where $C=0.$
$$F_C \eq D_C W \eq 3 \, C^2 \ + \ \frac{3 \, \bar{C}}{T+T^*-CC^*} \; (C^3 \ + \ a \, e^{-\al S} \ + \ b) \ \overset{C=0}{=} \ 0 \ .$$
$$F_T \eq D_T W  \eq - \, \frac{3}{T+T^*-CC^*} \; (C^3 \ + \ a \, e^{-\al S} \ + \ b) \eq - \, \frac{3}{T+T^*} \; (a \,  e^{-\al S} \ + \ b) \ \ \neq \ \ 0 \ .$$
At the minimum $F_T\neq0$ and hence supersymmetry is broken. This is another feature of no-scale models.\footnote{Unfortunately, this does not solve the cosmological constant problem: higher order corrections (e.g. loops) always break the no-scale structure and regenerate a cosmological constant.}

\section{Chapter 7}
\subsection*{Exercise 7.1}
We start with the potential
\begin{equation}
V(x)=\begin{cases}
0 & x\in (0,a)\\
\infty & \text{otherwise}
\end{cases} \notag
\end{equation}
The $y$ direction is identified under $y\to y+2\pi r.$ The Schr\"odinger equation is
$$\left[V(x) \ - \ \frac{\hbar^2}{2m} \; \left(\frac{\pa^2}{\pa x^2} \, + \, \frac{\pa^2}{\pa y^2}\right)\right] \, \Psi_{n,m}(x,y) \eq E_{n,m} \, \Psi_n(x,y)$$
We require $\Psi_n(x,0)=\Psi_n(x,2\pi r),$ which allows us to write
$$\Psi_n(x,y) \eq \sum_{m=-\infty}^{\infty}\Psi_{n,m}(x) \, e^{\frac{im y}{r}} \ .$$
The square well potential requires $\Psi_{n,m}(a)=\Psi_{n,m}(0),$ which leads to the following ansatz
$$\Psi_{n,m}(x) \eq A_{n,m} \, \sin{\left(\frac{n\pi x}{a}\right)}.$$
In total the wavefunctions are
$$\Psi_{n,m}(x) \eq A_{n,m} \, \sin{\left(\frac{n\pi x}{a}\right)} \, e^{\frac{imy}{r}},$$
where $A_{n,m}$ is suitably normalised. We determine the energy levels by applying the Hamiltonian to this solution:
$$\left[-\frac{\hbar^2}{2m} \; \left(\frac{\pa^2}{\pa x^2} \, + \, \frac{\pa^2}{\pa y^2}\right)\right] \, \Psi_{n,m}(x,y) \eq - \, \frac{\hbar^2}{2m} \, \left(-\frac{n^2 \pi^2}{a^2} \  - \ \frac{m^2}{r^2}\right) \, \Psi_{n,m}(x,y)$$
So we obtain
\begin{equation}
E_{n,m} \eq \frac{\hbar^2}{2m} \; \left(\frac{n^2 \pi^2}{a^2} \ + \ \frac{m^2}{r^2}\right)
\notag
\end{equation}
In the limit $r\ll a $ the energy of any excited Kaluza-Klein state $m\neq 0$ is very much larger than the ordinary square-well states, which decouple from the physics at low-energy.

\subsection*{Exercise 7.2}
\begin{equation}
{\cal S}  \eq \int \dd^4 x  \ \left(\frac{1}{g^2} \; H_{\mu\nu\rho} \, H^{\mu \nu \rho} \ + \ a \, \ep^{\mu\nu\rho\s} \, \pa_\mu H_{\nu\rho\s}\right) \notag
\end{equation}
{\bf Variant A:}
Assuming all fields vanish at infinity, we can rewrite the action as 
\begin{equation}
{\cal S} \eq \int \dd^4 x \  \left(\frac{1}{g^2} \; H_{\mu\nu\rho} \, H^{\mu \nu \rho} \ - \ \pa_\mu a \, \ep^{\mu\nu\rho\s} \, H_{\nu\rho\s}\right) \notag
\end{equation}
Varying this action w.r.t. $H_{\nu\rho\s}$ gives
$$\frac{2}{g^2} \; H^{\nu\rho\s} \ - \ \pa_\mu a \,  \ep^{\mu\nu\rho\s} \eq 0$$
So we obtain
\begin{eqnarray*}
H^{\nu\rho\s} \eq \frac{g^2}{2} \; \pa_\mu a \, \ep^{\mu\nu\rho\s}\co
H_{\nu\rho\s} \eq \frac{g^2}{2} \; \pa^\la a \, \ep_{\la\nu\rho\s}
\end{eqnarray*} 
Substituting this in the action, we can write
\begin{eqnarray}
\nonumber {\cal S}[a]&=&\int \dd^4 x \ \left[ \frac{1}{g^2} \; \left(\frac{g^2}{2}\right) \,(\pa_\mu a \ep^{\mu\nu\rho\s}) \, (\pa^\la a \ep_{\la\nu\rho\s}) \ - \ (\pa_\mu a \, \ep^{\mu\nu\rho\s}) \; \frac{g^2}{2} \; (\pa^\la a \,  \ep_{\la\nu\rho\s})\right] \\
\nonumber &=&\int \dd^4 x\ \left(-\frac{g^2}{4}\right) \, (\pa_\mu a \,  \ep^{\mu\nu\rho\s}) \, (\pa^\la a \, \ep_{\la\nu\rho\s})\\
\nonumber &=&\int \dd^4 x\ \left(-\frac{g^2}{4}\right) \, (-6 \, \pa_\mu a \, \pa^\mu a )\\
&=&\int \dd^4 x\ \frac{3g^2}{2} \; \pa_\mu a \, \pa^\mu a \notag
\end{eqnarray}
{\bf Variant B:}
The e.o.m. for $a$ (in the original action) gives
$$\ep^{\mu\nu\rho\s} \, \pa_{\mu}H_{\nu\rho\s}\eq 0 \ .$$
In other words the exterior derivative of $H$ is vanishing $\dd H=0.$ Since $\mathbb{R}^4$ is topologically trivial, we can write $H= \dd B,$ with all propagating degrees of freedom embedded in $B.$ This means that the original action is equivalent to a Lagrangian
\begin{equation}
{\cal S}[B] \eq \int \dd^4 x\ \frac{1}{g^2} \; (\dd B)_{\mu\nu\rho} \, (\dd B)^{\mu\nu\rho} \notag
\end{equation}
Both actions ${\cal S}[a]$ and ${\cal S}[B]$ describe the same physics. Therefore in 4D a 2-form potential is dual to a scalar, and can be rewritten in terms of one. In this duality, the coupling constant transforms as $g\to \frac{1}{g}$ (up to numerical factors).

\subsection*{Exercise 7.3}
We start with a purely gravitational theory in 5D
\begin{equation}
{\cal S}  \eq \int \dd^5 x\  \sqrt{ |G|}\  ^{(5)}R\ . \notag
\end{equation}
The 5d metric is $G_{MN}$
\begin{equation}
\dd s^2 \eq G_{MN} \, \dd x^M \, \dd x^N \eq g_{\mu\nu} \, \dd x^\mu \, \dd \, x^\nu\ + \ 2 \, G_{\mu 5} \, \dd x^\mu  \, \dd x^5 \ + \ G_{55} \ \dd x^5 \,  \dd x^5\ . \notag
\end{equation}
We use the following decomposition of the 5-dimensional Ricci scalar $R:$
\begin{equation}
^{(5)}R \eq ^{(4)}R \ - \ 2 \, e^{-\s} \, \nabla^2 e^\s \ - \ \frac{1}{4} \; e^{2\s} \, F_{\mu\nu} \, F^{\mu\nu}\ . \notag
\end{equation}
Now in the vacuum $\langle A_\mu\rangle =0$, this implies $\sqrt{G}=\sqrt{g_4}e^\s,$ so we can rewrite the 5D action as follows
\begin{equation}
{\cal S}_{5D} \eq \int  \dd^5 x \ \sqrt{ | g |} \, \left(e^\s \, ^{(4)}R \ - \ 2 \, \nabla^2 \, e^\s \ - \ \frac{1}{4} \; e^{3\s} \, F_{\mu\nu} \, F^{\mu\nu}\right) \notag
\end{equation}
All quantities are calculated using $g_{\mu\nu}.$ We now want to rescale the metric so that the Einstein-Hilbert term is canonical. Let us make the following ansatz for rescaling
\begin{equation}
g_{\mu\nu} \ \ =: \ \  e^{-\s} \, \tilde{g}_{\mu\nu}\ . \notag
\end{equation}
This changes $R$ as follows
\begin{equation}
^{(4)}R \eq e^{-\s} \, \left( \, ^{(4)}\tilde{R} \ + \ 3 \, \tilde{\nabla}^2 \, \s \ - \ \frac{3}{2} \; \tilde{\pa}_{\mu}\s \, \tilde{\pa}^\mu\s \right) \notag
\end{equation}
One term in the action above can be discarded as a total derivative
\begin{equation}
-2 \int \dd^5 x \  \sqrt{ |g|} \, (\nabla_\mu\nabla^\mu) \, e^\s \eq - \,2 \int \dd^5 x \  \pa_\mu  \left(\sqrt{ |g |} \, \nabla^\mu\right) \, e^\s \eq 0  \notag
\end{equation}
Now we can collect all our results
\begin{eqnarray}
{\cal S}&=&\int \dd^5 x \ \sqrt{ | \tilde{g} | } \, \left( \, ^{(4)}\tilde{R} \ + \ \underbrace{3 \, \tilde{\nabla}^2\s}_{=0 \text{ as a total der.}} \ - \ \frac{3}{2} \; \tilde{\pa}_\mu\s \, \tilde{\pa}^\mu \s \ - \ \frac{1}{4} \; e^{3\s} \, \tilde{F}_{\mu\nu} \, \tilde{F}^{\mu\nu}\right) \notag \\
&=&\int \dd^5 x \ \sqrt{ | \tilde{g} |} \left(^{(4)} \tilde{R} \ - \ \frac{3}{2} \; \tilde{\pa}_\mu\s \, \tilde{\pa}^\mu \s \ - \ \frac{1}{4} \; e^{3\s} \, \tilde{F}_{\mu\nu} \, \tilde{F}^{\mu\nu}\right) \ . \notag
\end{eqnarray}
Recall that $\sqrt{|g|}g^{\mu\al}g^{\nu\be}F_{\mu\nu}F_{\al\be}=\sqrt{ | \tilde{g} |}\tilde{g}^{\mu\al}\tilde{g}^{\nu\be}F_{\mu\nu}F_{\al\be}.$ We ended up with the Einstein-Maxwell Lagrangian that we were looking for.

\noindent
Note that the gauge coupling depends on the size of the extra dimensions: Large gauge coupling implies small extra dimensions.

\noindent
In higher dimensions, every $U(1)$ isometry of the extra dimensions gives rise to a $U(1)$ gauge field. If for instance an $SU(2)$ isometry exists, then the lower-dimensional theory has an $SU(2)$ gauge theory.

\subsection*{Exercise 7.4}

We have a 5d gravity theory with a negative cosmological constant $\Lambda,$ compactified on an interval $(0,\pi).$ Each end of the interval corresponds to a 3-brane with tension $\pm\Lambda/k.$ The bulk metric is $G_{MN}.$ The induced metric on the visible and hidden branes is $g_{\mu\nu,{\rm} vis}$ and $g_{\mu\nu,{\rm} hidden}.$

\noindent
We would like to show that the warped metric
$$\dd s^2 \eq e^{-2W(y)} \, \eta_{\mu\nu} \, \dd x^\mu \,  \dd x^\nu \ + \ r^2 \, \dd y^2$$
satisfies Einstein's equations. We use the result that, for this metric ansatz, Einstein's equations reduce to
\begin{eqnarray}
\frac{6 \, W'^2}{r^2}&=&- \,\frac{\Lambda}{2 \, M^3}\notag \\
\frac{3 \, W''}{r^2}&=&\frac{\Lambda}{2 k \, M^3 \,  r} \; \bigl[ \de(y-\pi) \, - \, \de(y) \bigr] \notag
\end{eqnarray}
Let us try
$$ W(y) \eq r \,  |y| \, \sqrt{\frac{-\Lambda}{12  \, M^3}} \ .$$
Then $W'^2=r^2 \frac{-\Lambda}{12 M^3}$ is satisfied and well-defined at $r=0$ since $W'(y=0)$ is well-defined. We regard $W(y)$ as a periodic function in $y,$ only defined for $-\pi<y<\pi$ and then defined through periodicity. Near $x=0$
\begin{eqnarray*}
\frac{\dd}{\dd x} \; |x|&=&-1 \ + \ 2 H(x) \eq\begin{cases} 1 & x>0\\ -1 & x<0\, .\end{cases} \ \ \
\Rightarrow \ \ \ \frac{\dd^2}{\dd x^2} \; |x| \eq 2 \, \de(x)\ .
\end{eqnarray*}
For periodic $W(y)$ and $W(y)=r \sqrt{\frac{-\Lambda}{12 M^3}}|y|$ for $0<y<\pi$
\begin{eqnarray*}
W''(y)&=&2r \, \sqrt{\frac{-\Lambda}{12 M^3}}  \; \bigl[ \delta(y)\, - \, \delta(y-\pi)\bigr]
\end{eqnarray*}
This implies
\begin{eqnarray*}
-\sqrt{\frac{-\Lambda}{12 M^3}}&=&\frac{\Lambda}{12 k M^3}\ \ \ \Rightarrow \ \ \ k^2 \eq \frac{-\Lambda}{12 M^3}
\end{eqnarray*}
We can write $W=r k |y|$ and so the metric becomes
$$\dd s^2 \eq e^{-2rk|y|} \, \eta_{\mu\nu} \, \dd x^\mu \, \dd x^\nu \ + \ r^2 \,  \dd y^2$$
The 5-D action is
\begin{eqnarray*}
2M^3 \, \int \dd^5 x \ \sqrt{|g|}\, R&=&2M^3 \, \int \dd^4 x \, \int_{-\pi}^{\pi} \dd y \ \sqrt{|g|} \, R\ .
\end{eqnarray*}
The 4d curvature term is determined by the 'volume' of the extra dimension in dimensional reduction. Using that $g=r^2 e^{-2kr |y|}g_4$ and $\sqrt{-g} R=r e^{-2rk|y|} \sqrt{-g_4}R_4,$ we see that the following relation for the 4-dimension Planck mass has to hold:
\begin{equation}
2M_{\rm pl}^2 \eq 2M^3 \, r \, \int_{-\pi}^{\pi} \dd y \ e^{-2kr|y|}
\end{equation}
After integration we obtain:
$$M_{\rm pl}^2 \eq \frac{M^3}{k} \; ( 1 \, - \, e^{-2kr\pi} )$$
The Higgs Lagrangian is
\begin{eqnarray*}
{\cal S}_{\rm vis}&\sim & \int \dd^4 x \ \sqrt{ |g_{\rm vis} |} \ \Bigl[g_{\rm vis}^{\mu\nu} \, D_\mu H^\dagger \, D_\nu H \ - \ \lambda \, (|H|^2 \, - \, v_0^2)^2\Bigr]\\
&\overset{g_{\rm vis}=e^{-2kr\pi}g_4}{\sim}&\int \dd^4 x \  \sqrt{|g_4|} \, e^{-4kr \pi}\,\Bigl[ g_4^{\mu\nu} \, e^{2kr\pi} \, D_\mu H^\dagger \, D_\nu H \ - \ \lambda \, (|H|^2 \, - \, v_0^2)^2\Bigr]\\
&\overset{H e^{kr\pi}\to H}{=}&\int \dd^4 x \ \sqrt{ |g_4 |} \, \Bigl[g_4^{\mu\nu} \, D_\mu H^\dagger \, D_\nu H \ - \ \lambda \, (|H|^2 \, - \, e^{-2kr\pi}v_0^2)^2 \Bigr]
\end{eqnarray*}
In the last step we canonically normalised the Higgs field such that the kinetic terms are canonical. The Higgs mass then is given by $m_H=e^{-kr\pi}v_0$ and depends on the warp factor. The natural scale for $v_0$ is the Planck scale. To obtain a Higgs mass at the weak scale we need $\pi kr\sim 50:$
$$m_{\te H} \ \ \sim \ \ e^{-kr \pi} \, M_{\te{pl}} $$ 
This solves the hierarchy problem through warping. The 4-dim Planck scale and the 5-dim scale $M$ are here comparable as $e^{-2kr}$ is tiny.

\section{Appendix A}

\subsection*{Exercise A.1:}

Antisymmetry of the $\psi \si^\mu \bar \chi$ bilinear:
\begin{eqnarray*}
\bar \chi \bs^\nu \psi&=&\bar \chi_{\da} \, (\bs^\nu)^{\da\al} \, \psi_{\al}\\
&=&\bar \chi_{\da} \, \ep^{\al\be} \, \ep^{\da\db} \, (\s^\nu)_{\be\db} \, \psi_{\al}\\
&=&- \, \psi_{\al} \, \ep^{\al\be} \, \ep^{\da\db} \, (\s^\nu)_{\be\db} \, \bar \chi_{\da}\\
&=&- \, \psi^\be \, (\s^\nu)_{\be\db} \, \bar \chi^{\db}\\
&=&- \, \psi(\s^\nu)\bar \chi \\
\end{eqnarray*}
Inverting the $\psi \si^\mu \si^\nu \chi$ product:
\begin{eqnarray*}
\psi \s^\mu\bs_\nu \chi&=&\psi^\al \, (\s^\mu)_{\al\da} \, (\bs_\nu)^{\da\be} \, \chi_\be\\
&=&\ep^{\al\de} \, \psi_{\de} \, (\s^\mu)_{\al\da} \, \ep^{\da\db} \, \ep^{\be\g} \, (\s_\nu)_{\g\db} \, \chi_\be\\
&=&\psi_{\de} \, (\bs^\mu)^{\db\de} \, \ep^{\be\g} \, (\s_\nu)_{\g\db} \, \chi_\be\\
&=&\ep^{\g\be} \, \chi_\be \, (\bs^\mu)^{\db\de} \, (\s_\nu)_{\g\db} \, \psi_{\de}\\
&=&\chi^\g \, (\s_\nu)_{\g\db} \, (\bs^\mu)^{\db\de} \, \psi_{\de}\\
&=&\chi \s_\nu\bs^\mu\psi
\end{eqnarray*}
From $\si^{\mu \nu} = \frac{i}{2} \si^{[ \mu } \, \bar \si^{\nu ]}$, it easily follows that
\[
\psi \, \si^{\mu \nu} \, \chi \eq \frac{i}{2} \, \psi \s^{[ \mu }\bs^{\nu]} \chi \eq \frac{i}{2} \, \chi \s^{[ \nu} \bs^{\mu ] }\psi \eq - \, \chi \, \si^{\mu \nu} \, \psi \ .
\]

\subsection*{Exercise A.2:}

Firstly
\begin{eqnarray*}
(\theta\psi) \, (\bc\bet)&\overset{!}{=}&- \, \frac{1}{2} \;(\theta\s^\mu\bet) \, (\bc\bs_\mu\psi)\\
&=&- \, \frac{1}{2} \; \theta^\al \, (\s^\mu)_{\al\da} \, \bet^{\da} \, \bc_{\db} \, (\bs_\mu)^{\db\be} \, \psi_\be\\
&=&- \, \frac{1}{2} \; \theta^\al \, \bet^{\da} \, \bc_{\db} \, \psi_\be \, \underbrace{(\s^\mu)_{\al\da} \, (\bs_\mu)^{\db\be}}_{= \ 2\de_\be^\al\de_{\db}^{\da}}\\
&=&\theta^\al \, \bet^{\da} \, \bc_{\da} \, \psi_\al\\
&=&(\theta^\al\psi_\al) \, (\bet^{\da}\bc_{\da})\\
&=&(\theta\psi) \, (\bc\bet) \ ,
\end{eqnarray*}
and secondly
\begin{eqnarray*}
\frac{1}{2} \; \eta^{\mu\nu} \, (\theta\theta) \, (\bt\bt)&\overset{!}{=}&(\theta\s^\mu\bt)\, (\theta\s^\nu\bt)\\
&=&\theta^\al \, (\s^\mu)_{\al\da} \, \bt^{\da} \, \theta^\be \, (\s^\nu)_{\be\db} \, \bt^{\db}\\
&=&(\s^\mu)_{\al\da} \, (\s^\nu)_{\be\db} \, \theta^\al \, \bt^{\da} \, \theta^\be \, \bt^{\db}\\
&=&- \, (\s^\mu)_{\al\da} \, (\s^\nu)_{\be\db} \, \theta^\al \, \theta^\be \, \bt^{\da} \, \bt^{\db}\\
&=&\frac{1}{4} \; (\s^\mu)_{\al\da} \, (\s^\nu)_{\be\db} \, \ep^{\al\be} \, \ep^{\da\db} \, (\theta\theta) \, (\bt\bt)\\
&=&\frac{1}{4} \; (\s^\mu)_{\al\da} \, (\bs^\nu)^{\da\al} \, (\theta\theta) \, (\bt\bt)\\
&=&\frac{1}{2} \; \eta^{\mu\nu} \, (\theta\theta) \, (\bt\bt) \ .
\end{eqnarray*}




\begin{thebibliography}{MUR:66}
\bibitem{all}A list of books and online available lecture notes in alphabetical order:\\ I. Aitchison, {\em Supersymmetry in Particle Physics: An Elementary Introduction,} CUP (2007).\\
H. Baer, X. Tata: {\em Weak Scale Supersymmetry,} CUP (2006).\\
D. Bailin, A. Love: {\em Supersymmetric Gauge Field Theory and String Theory,} IOP (1994).\\
P.~Binetruy, {\em Supersymmetry: Theory, experiment and cosmology,}  OUP (2006).\\
I. Buchbinder, S. Kuzenko, {\em Ideas and Methods of Supersymmetry and Supergravity,} IOP (1998).\\
  C.~Csaki, {\em TASI lectures on extra dimensions and branes,} published in *Boulder 2002, Particle physics and cosmology* 605-698,  hep-ph/0404096.\\
C.~Csaki, J.~Hubisz and P.~Meade, {\em TASI lectures on electroweak symmetry breaking from extra dimensions,} published in *Boulder 2004, Physics in D $\geq$ 4* 703-776, hep-ph/0510275.\\
S.~P.~de Alwis, {\em Potentials for light moduli in supergravity and string theory,} hep-th/0602182.\\
  M. Dine, {\em Supersymmetry Phenomenology with a Broad Brush,} published in *Boulder 1996, Fields, strings and duality* 813-881 hep-ph/9612389.\\
    G.~Gabadadze, {\em ICTP lectures on large extra dimensions,} published in *Trieste 2002, Astroparticle physics and cosmology* 77-120, hep-ph/0308112.\\
   S.~J.~Gates, M.~T.~Grisaru, M.~Rocek and W.~Siegel, {\em Superspace, or one thousand and one lessons in supersymmetry,}  Front.\ Phys.\  {\bf 58} (1983) 1, hep-th/0108200.\\
  G.~D.~Kribs, {\em Phenomenology of extra dimensions,}\\  published in *Boulder 2004, Physics in D $\geq$ 4* 633-699, hep-ph/0605325.\\
  P LaBelle, {\em Supersymmetry DeMYSTiFied,} McGraw-Hill Professional (2009).\\
  M.~A.~Luty, {\em 2004 TASI lectures on supersymmetry breaking,} published in *Boulder 2004, Physics in D $\geq$ 4* 495-582, hep-th/0509029.\\
  S.~P.~Martin, {\em A Supersymmetry Primer,} in *Kane, G.L. (ed.): Perspectives on supersymmetry II* 1-153, hep-ph/9709356.\\
  H.~P.~Nilles, {\em Supersymmetry, Supergravity And Particle Physics,}  Phys.\ Rept.\  {\bf 110}, 1 (1984).\\
    R.~Rattazzi, {\em Cargese lectures on extra dimensions,} published in *Cargese 2003, Particle physics and cosmology* 461-517, hep-ph/0607055.\\
  R.~Sundrum, {\em To the fifth dimension and back. (TASI 2004),}  published in *Boulder 2004, Physics in D $\geq$ 4* 585-630, hep-th/0508134.\\
J. Terning, {\em Modern Supersymmetry: Dynamics and Duality,} OUP (2006).\\
S. Weinberg, {\em The quantum theory of fields, Volume III Supersymmetry,} CUP (2000).\\
J. Wess, J. Bagger, {\em Supersymmetry and Supergravity,} PUP (1992).\\
  P.~C.~West, {\em Introduction to supersymmetry and supergravity,} World Scientific (1990).
\bibitem{MKW} H.J.W. M\"uller-Kirsten, A. Wiedemann, {\em Supersymmetry, an introduction with conceptual and calculational details,} World Scientific (2010).
\bibitem{WB1} S. Weinberg, {\em The quantum theory of fields, Volume I Foundations,} CUP (1995).
\bibitem{pend} M.~T.~Grisaru and H.~N.~Pendleton,
 {\em Some Properties Of Scattering Amplitudes In Supersymmetric Theories,}
  Nucl.\ Phys.\  B {\bf 124} (1977) 81.
\bibitem{stup}  V.~Gates, E.~Kangaroo, M.~Roachcock and W.~C.~Gall, {\em Stuperspace,}
 Physica {\bf 15D} (1985) 289.
\bibitem{Salam:1974yz}
 A.~Salam and J.~A.~Strathdee,
{\em Supergauge Transformations,}
 Nucl.\ Phys.\  B {\bf 76} (1974) 477.
\bibitem{Salam:1974jj} 
 A.~Salam and J.~A.~Strathdee,
 {\em On Superfields And Fermi-Bose Symmetry,}
 Phys.\ Rev.\  D {\bf 11} (1975) 1521.
\bibitem{BZ} F.A. Berezin, A.A. Kirillov, D. Leites {\em Introduction to superanalysis,} Reidel (1987).
\bibitem{BDW} B. de Witt, {\em Supermanifolds,} CUP (1992).
\bibitem{sei}
 N.~Seiberg,
 {\em Naturalness Versus Supersymmetric Non-renormalization Theorems,}
 Phys.\ Lett.\  B {\bf 318} (1993) 469
 [arXiv:hep-ph/9309335].
\bibitem{WB3} S. Weinberg, {\em The quantum theory of fields, Volume III Supersymmetry}, CUP (2000).
\bibitem{Seiberg:1994rs}
 N.~Seiberg and E.~Witten,
 {\em Monopole Condensation, And Confinement In ${\cal N}=2$ Supersymmetric Yang-Mills,}
 Nucl.\ Phys.\  B {\bf 426} (1994) 19
 [Erratum-ibid.\  B {\bf 430} (1994) 485]
 [arXiv:hep-th/9407087].
\bibitem{Kaplunovsky:1994fg} 
 V.~Kaplunovsky and J.~Louis,
 {\em Field dependent gauge couplings in locally supersymmetric effective quantum field theories,}
 Nucl.\ Phys.\  B {\bf 422} (1994) 57
 [arXiv:hep-th/9402005].
\bibitem{seiberg} Z.~Komargodski and N.~Seiberg,
  {\em Comments on the Fayet-Iliopoulos Term in Field Theory and Supergravity,}
  JHEP {\bf 0906} (2009) 007
  [arXiv:0904.1159 [hep-th]].
\bibitem{RS} L.~Randall and R.~Sundrum,
  {\em A large mass hierarchy from a small extra dimension,}
  Phys.\ Rev.\ Lett.\  {\bf 83} (1999) 3370
  [arXiv:hep-ph/9905221].
\bibitem{gamma1}
  P.~C.~West,
  {\em Supergravity, brane dynamics and string duality,}
  arXiv:hep-th/9811101.
\bibitem{gamma2}
J.~Polchinski, {\em Superstring Theory and beyond, Volume II,} CUP 2005.
\bibitem{gamma3}
J.~A.~Strathdee,
  {\em Extended Poincar\'e Supersymmetry,}
  Int.\ J.\ Mod.\ Phys.\  A {\bf 2} (1987) 273.


\end{thebibliography}
\end{document}